\documentclass[12pt,a4paper]{article}
\usepackage[margin=1in]{geometry}
\usepackage{graphicx}
\usepackage[x11names]{xcolor}
\usepackage[bottom]{footmisc}
\definecolor{myblue}{rgb}{0,0,.5} 
\usepackage{titlesec}
\titleformat{\section}
    {\color{myblue}\normalfont\Large\bfseries}
    {\thesection}{1em}{}
\titleformat{\subsection}
    {\color{myblue}\normalfont\large\bfseries}
    {\thesubsection}{1em}{}
\titleformat{\subsubsection}
    {\color{myblue}\normalfont\normalsize\bfseries}
    {\thesubsubsection}{1em}{}
\usepackage{amsthm} \theoremstyle{definition}

\newtheorem{proposition}{Proposition}	\newtheorem{lemma}{Lemma}										\newtheorem{assumption}{Assumption}

\newtheorem{observation}{Observation}	\newtheorem{definition}{Definition}  	        			\newtheorem{remark}{Remark}      		\newtheorem{example}{Example}    	 	\usepackage[T1]{fontenc}
\usepackage{amsmath,amssymb,bm}
\usepackage{mathpazo,newpxtext}
\usepackage{mathtools}
\usepackage{multirow}
\usepackage{array}
\usepackage{arydshln}
\usepackage{setspace}
\usepackage{natbib}
\usepackage{bibunits}
\defaultbibliographystyle{aer}
\defaultbibliography{main}
\usepackage{etoolbox}
\newcommand{\addQEDstyle}[2]{\AtBeginEnvironment{#1}{\pushQED{\qed}\renewcommand{\qedsymbol}{#2}}\definecolor{QEDColor}{rgb}{.3,.3,.3}
    \AtEndEnvironment{#1}{\popQED}
}
\addQEDstyle{lemma}{{\color{QEDColor}$\blacksquare$}} 
\addQEDstyle{remark}{{\color{QEDColor}$\blacksquare$}}
\addQEDstyle{assumption}{{\color{QEDColor}$\blacksquare$}} 
\addQEDstyle{observation}{{\color{QEDColor}$\blacksquare$}}
\addQEDstyle{example}{{\color{QEDColor}$\blacksquare$}} 
\addQEDstyle{fact}{$\lozenge$}
\usepackage{array}
\usepackage{makecell}

\usepackage{enumitem}
\setlist[itemize]{itemsep=3pt}
\setlist[enumerate]{itemsep=3pt,label=(\alph*)}
\makeatletter
\renewcommand{\@pnumwidth}{2.5em}
\makeatother
\definecolor{linkcolor}{rgb}{0,0,.7}
\usepackage[setpagesize=false,bookmarks=true,colorlinks=true,linkcolor=linkcolor,urlcolor=linkcolor,citecolor=linkcolor,breaklinks=true,linktocpage]{hyperref}
\makeatletter
\providecommand{\diag}[1]{\operatornamewithlimits{diag}[#1]}
\providecommand\Gs{\Omega^\sharp}
\providecommand\Gf{\Omega^\flat}
\providecommand\Gn{\Omega^\natural}
\newcommand{\relmiddle}[1]{\mathrel{}\middle#1\mathrel{}}
\newcommand{\Mid}{\relmiddle{|}}

\newcommand\sgn{\operatorname{sgn}}
\definecolor{redlike}{rgb}{.8,.2,0}
\providecommand{\Red}[1]{{\color{redlike}#1}}
\makeatother

\allowdisplaybreaks

\providecommand\FigureNote[1]{\vskip .5em \parbox[c]{\hsize}{\footnotesize \emph{Note:} #1}}

\providecommand{\Heading}[1]{\medskip \noindent\textbf{#1.}}

\usepackage{overpic}
\usepackage{subcaption}
\usepackage[capitalise]{cleveref}
\crefname{fig}{Figure}{Figures}
\Crefname{fig}{Figure}{Figures}
\crefname{tab}{Table}{Tables}
\crefname{assumption}{Assumption}{Assumptions}
\crefname{observation}{Observation}{Observations} 
\crefname{footnote}{Footnote}{Footnotes} 
\crefname{appendix}{Appendix}{Appendices}
\Crefname{appendix}{Appendix}{Appendices}
\crefformat{footnote}{#2#1#3}

\makeatletter
\AtBeginDocument
 {
   \def\ltx@label#1{\cref@label{#1}}\def\label@in@display@noarg#1{\cref@old@label@in@display{#1}}\def\label@in@mmeasure@noarg#1{\begingroup \measuring@false \cref@old@label@in@display{#1}\endgroup}} \makeatother

\usepackage{quoting}

\providecommand{\PDF}[2]{\frac{\partial{#1}}{\partial{#2}}}

\providecommand{\ODF}[2]{\frac{\mathrm{d}{#1}}{\mathrm{d}#2}}

\providecommand{\Vt}[1]{\bm{#1}}
\providecommand{\Vta}{\bm{a}}

\providecommand{\Vte}{\bm{e}}
\providecommand{\Vtf}{\bm{f}}

\providecommand{\Vtl}{\bm{l}}

\providecommand{\Vts}{\bm{s}}

\providecommand{\Vtv}{\bm{v}}
\providecommand{\Vtw}{\bm{w}}
\providecommand{\Vtx}{\bm{x}}
\providecommand{\Vty}{\bm{y}}
\providecommand{\Vtz}{\bm{z}}
\providecommand{\VtA}{\mathbf{A}}

\providecommand{\VtC}{\mathbf{C}}
\providecommand{\VtD}{\mathbf{D}}
\providecommand{\VtE}{\mathbf{E}}
\providecommand{\VtF}{\mathbf{F}}
\providecommand{\VtG}{\mathbf{G}}

\providecommand{\VtI}{\mathbf{I}}

\providecommand{\VtM}{\mathbf{M}}

\providecommand{\VtP}{\mathbf{P}}

\providecommand{\VtS}{\mathbf{S}}

\providecommand{\VtV}{\mathbf{V}}
\providecommand{\VtW}{\mathbf{W}}
\providecommand{\VtX}{\mathbf{X}}

\providecommand{\VtTheta}{\bm{\Theta}}

\providecommand{\Rmd}{\mathrm{d}}

\providecommand{\RmA}{\mathrm{A}}

\providecommand{\RmH}{\mathrm{H}}

\providecommand{\RmM}{\mathrm{M}}

\providecommand{\ClI}{\mathcal{I}}

\providecommand{\ClK}{\mathcal{K}}

\providecommand{\ClO}{\mathcal{O}}

\providecommand{\ClX}{\mathcal{X}}

\providecommand{\BbR}{\mathbb{R}}

\providecommand{\Bra}{\bar{a}}

\providecommand{\Bre}{\bar{e}}

\providecommand{\Brv}{\bar{v}}
\providecommand{\Brw}{\bar{w}}
\providecommand{\Brx}{\bar{x}}
\providecommand{\Bry}{\bar{y}}

\providecommand{\Bralpha}{\bar{\alpha}}

\providecommand{\Tla}{\tilde{a}}

\providecommand{\Tlv}{\tilde{v}}

\providecommand{\Dtx}{\dot{x}}

\providecommand{\Is}{\equiv}

\providecommand{\HtVtx}{\hat{\Vtx}}

\providecommand{\HtVtV}{\hat{\VtV}}

\providecommand{\HtVtX}{\hat{\VtX}}

\providecommand{\BrVta}{\bar{\Vta}}

\providecommand{\BrVtx}{\bar{\Vtx}}

\providecommand{\TlVta}{\tilde{\Vta}}

\providecommand{\TlVtf}{\tilde{\Vtf}}

\providecommand{\TlVtv}{\tilde{\Vtv}}

\providecommand{\TlVtV}{\tilde{\VtV}}

\providecommand{\DtVtx}{\dot{\Vtx}}

\providecommand{\inI}{\in\ClI}

\providecommand{\inK}{\in\ClK}

\renewcommand{\Red}[1]{{#1}}

\makeatletter
\newcommand{\nomarkthanks}[1]{\begingroup
    \renewcommand\@makefnmark{}\thanks{#1}\endgroup
}
\makeatother

\providecommand{\TITLE}{\color{myblue}\Large Spatial Scale of Agglomeration and Dispersion:\\[.3em]Number, Spacing, and the Spatial Extent of Cities}
\title{\TITLE}
\author{
    \begin{tabular}{cp{2em}c}
        \color{myblue} Takashi Akamatsu & & \color{myblue} Tomoya Mori \\
      {\normalsize Tohoku Univ.} & & {\normalsize Kyoto Univ., RIETI}\\
      \\
      \color{myblue} Minoru Osawa & & \color{myblue} Yuki Takayama \\
      {\normalsize Kyoto Univ.} & & {\normalsize Science Tokyo} \\
    \end{tabular}
\protect\nomarkthanks{TA: 
    akamatsu@plan.civil.tohoku.ac.jp, 
    TM: 
    mori@kier.kyoto-u.ac.jp, 
    MO: 
    osawa.minoru.4z@kyoto-u.ac.jp, 
    YT: 
    takayama.y.cc65@m.isct.ac.jp. 
We are deeply grateful to Co-Editor Gilles Duranton for his guidance and support throughout the review process, which enormously improved the paper.  
We also thank two thoughtful referees for very helpful comments. 
We thank Treb Allen, Kristian Behrens, Donald Davis, Yasusada Murata, Jonathan Newton, Michael Pfl\"{u}ger, Diego Puga, Esteban Rossi-Hansberg, Shunsuke Segi, Akihisa Shibata, and Jacques-Fran\c{c}ois Thisse for their comments, including those for the earlier versions of this paper. 
We also thank all seminar and conference participants. 
This research has been supported by JSPS KAKENHI 17H00987, 18H01556, 19K15108, 21K04299, 22K04353, 23K22880, 23K22887, 24K00999, and 25H00543.
This research was conducted as part of the project, ``Agglomeration-based Framework for Empirical and Policy Analyses of Regional Economies,'' undertaken at the Research Institute of Economy, Trade and Industry.
We acknowledge financial support from the Kajima Foundations and the Murata Science Foundation.}}
\date{\normalsize\today}

\begin{document}
\maketitle

\renewcommand{\abstractname}{}

\begin{abstract}\onehalfspacing \noindent
How does transport cost affect the spatial organization of economic activities?  
This study develops a theoretical framework that distinguishes between two types of dispersion forces in spatial models: ``local'' dispersion forces acting \emph{within} cities, and ``global'' dispersion forces acting \emph{across} them.
The distinction leads to a systematic classification of spatial models into a few fundamental types, each with distinct endogenous spatial patterns and comparative statics in response to changes in transport costs. 
The framework reconciles empirical findings and clarifies how transport-induced reorganization of economic activities can depend on the spatial scale of dominant dispersion forces. 
\end{abstract}

\bigskip 



\begin{bibunit}

\clearpage 
\setstretch{1.3}

\section{Introduction}
\label{sec:introduction}

\begin{figure}[t!]
    \centering
	\begin{subfigure}[b]{.49\hsize}
		\centering
		\includegraphics[width=.85\hsize]{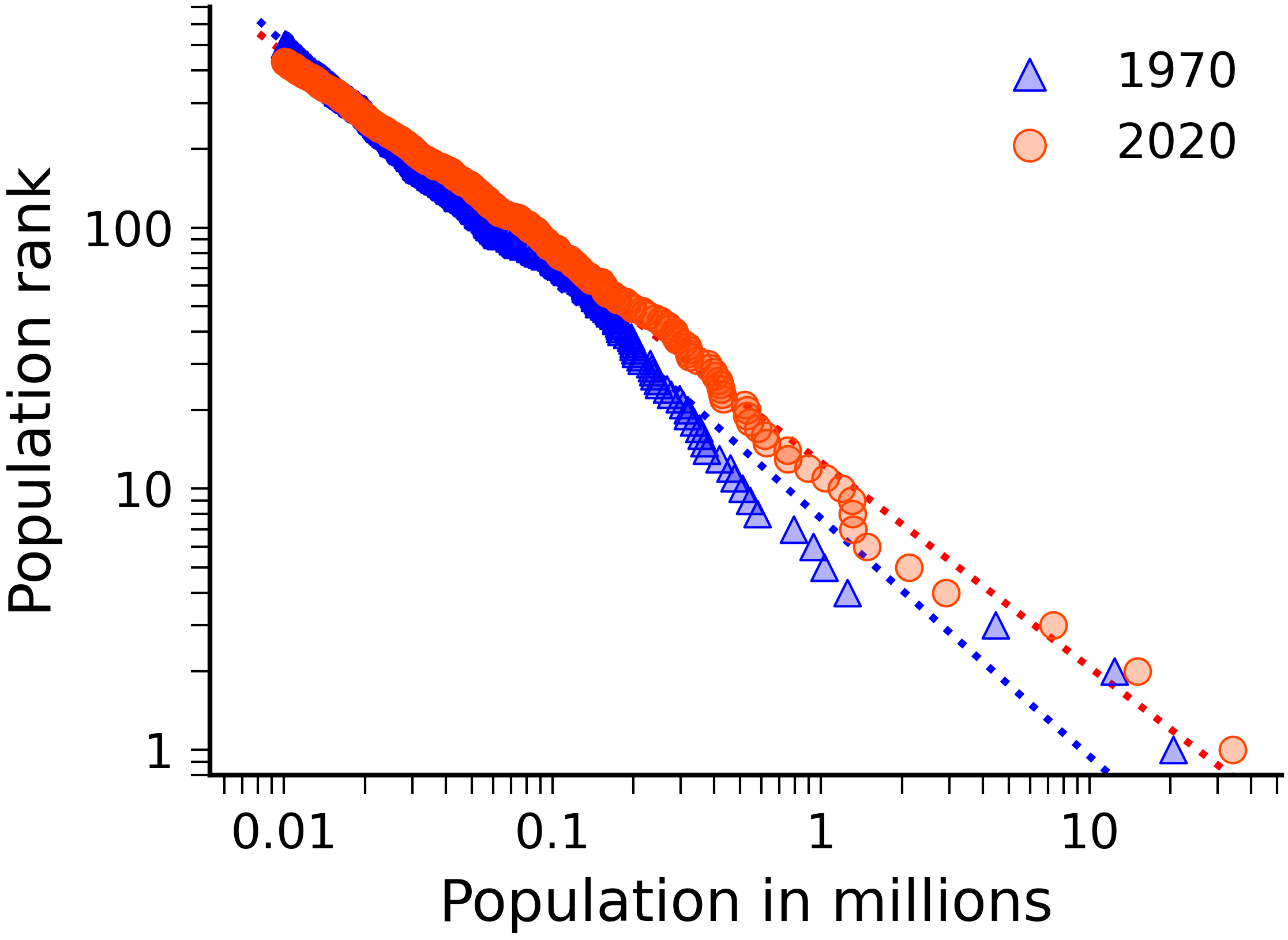}
		\caption{Population rank--size plot\label{fig:JP-GC}}
		
	\end{subfigure}
    \hfill 
	\begin{subfigure}[b]{.49\hsize}
		\centering
		\includegraphics[width=.85\hsize]{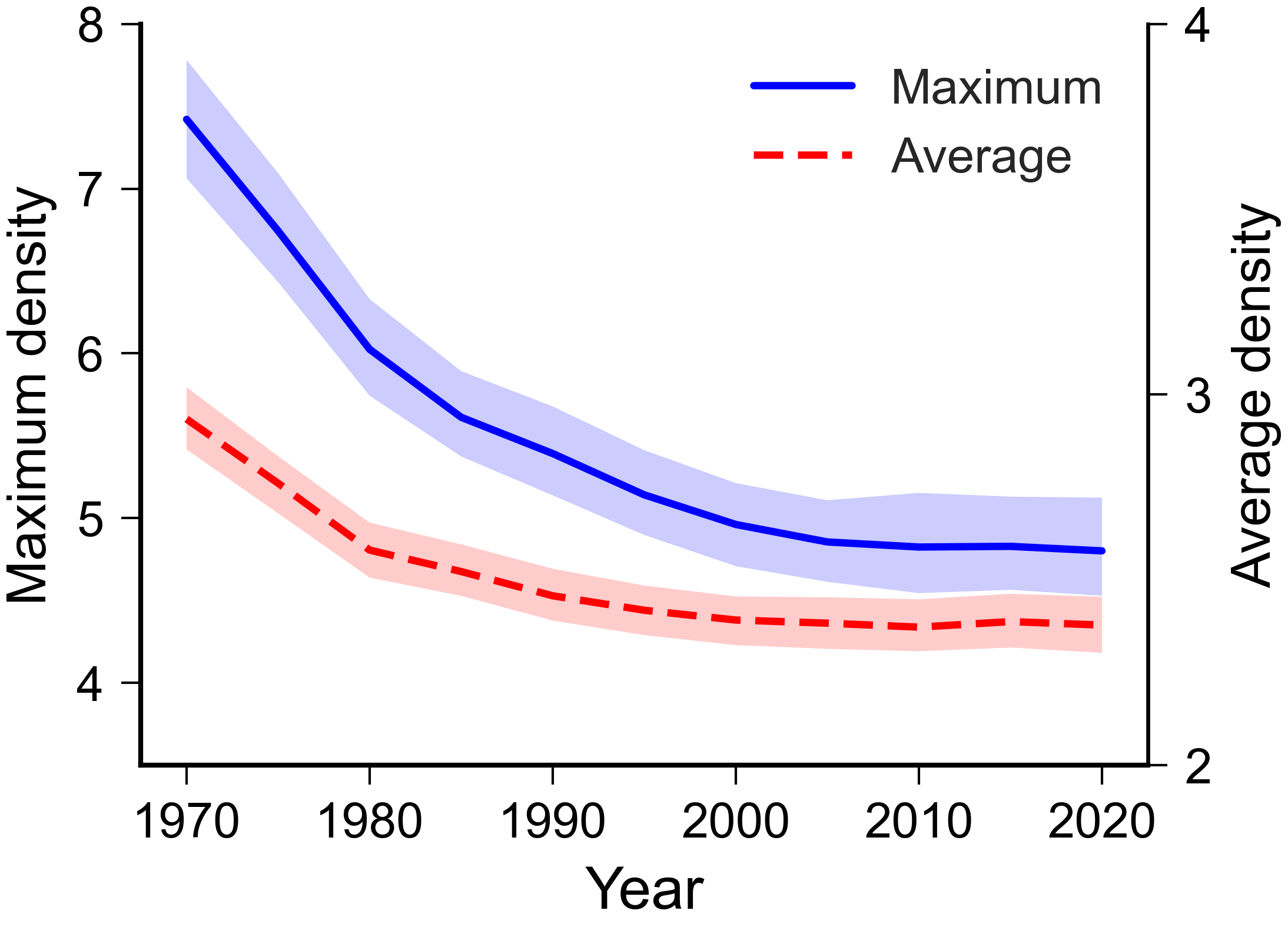}
		\caption{Population density  (1,000/km$^2$)\label{fig:JP-LD-1}}
	\end{subfigure}

    \bigskip 

	\begin{subfigure}[b]{\hsize}
		\centering
		\includegraphics[width=.8\hsize]{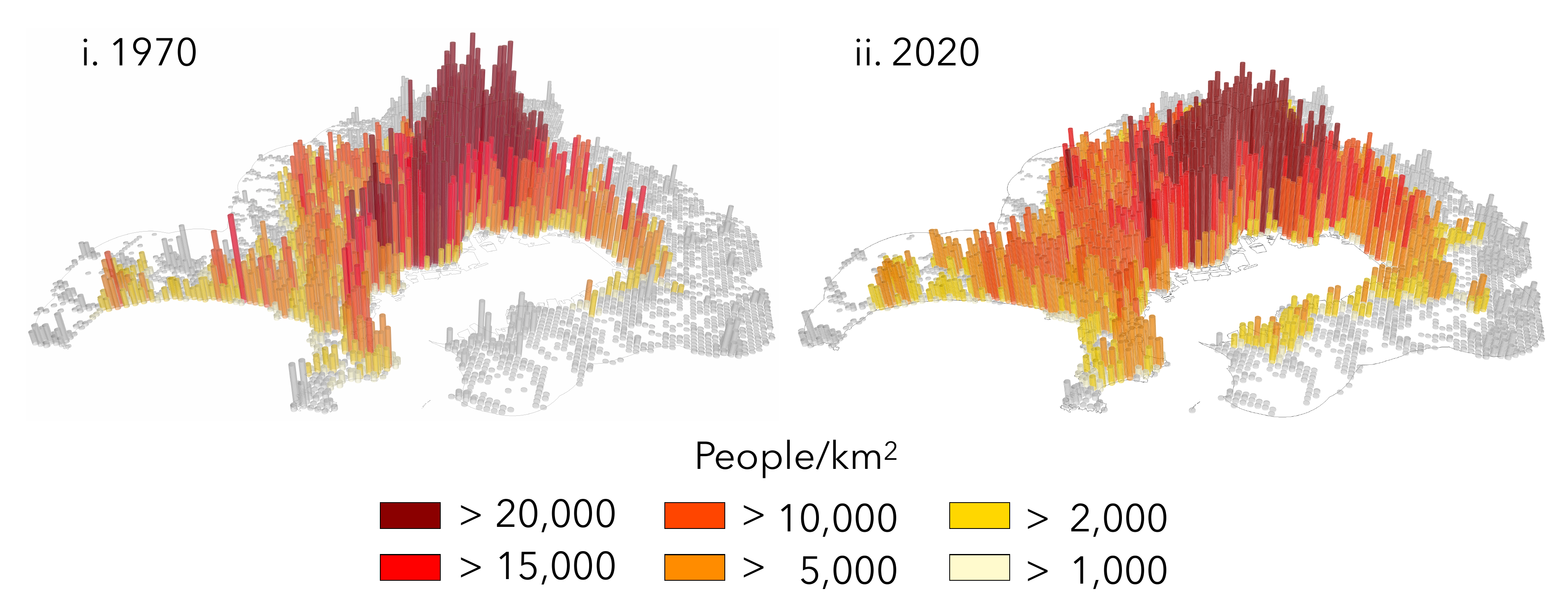}
		\caption{The population distribution within Tokyo in 1970 (left) and 2020 (right)\label{fig:JP-LD-Tokyo}}
	\end{subfigure}

    \caption{Global concentration and local dispersion in Japan from 1970 to 2020.}
    \label{fig:JP}
    \FigureNote{A \emph{city} is defined as a cluster of contiguous 1km-by-1km grid cells, each with a population density of at least 1,000/km\textsuperscript{2} and collectively comprising at least 10,000 residents. The number of cities decreased from 504 in 1970 to 431 in 2020. Panel~(B) shows the annual cross-city arithmetic means of maximum and average population densities along with the 95\% bootstrap confidence interval. Panel~(C) shows the within-city population distribution in Tokyo in 1970 and 2020. For further discussion, see Appendix \ref{app:ua}.} 

    \begin{subfigure}[b]{.32\hsize}
        \centering
        \includegraphics[width=\hsize]{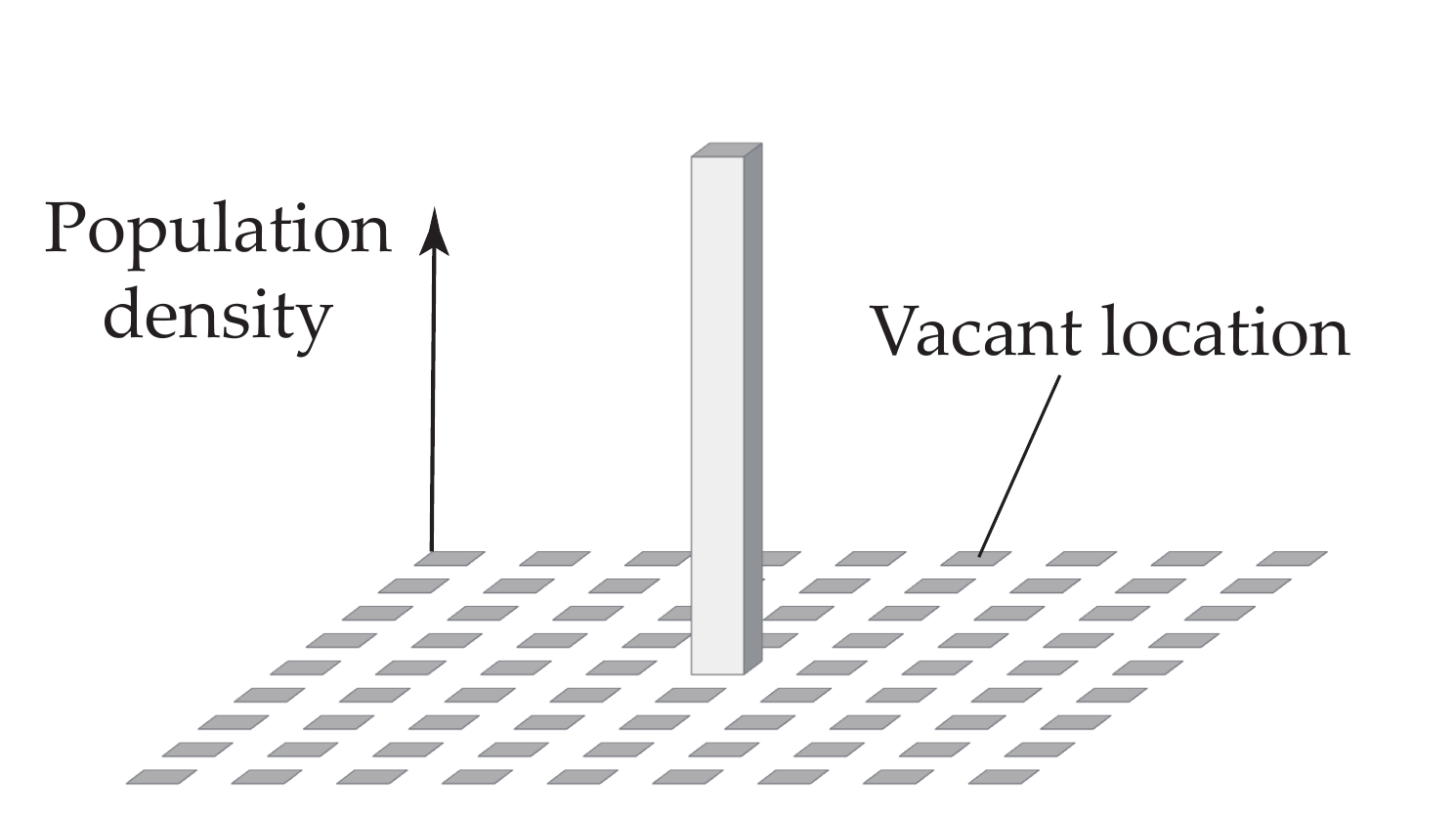} 
        \caption{Full agglomeration \label{fig:full-concentration}}
    \end{subfigure}
    \begin{subfigure}[b]{.32\hsize}
        \centering
        \includegraphics[width=\hsize]{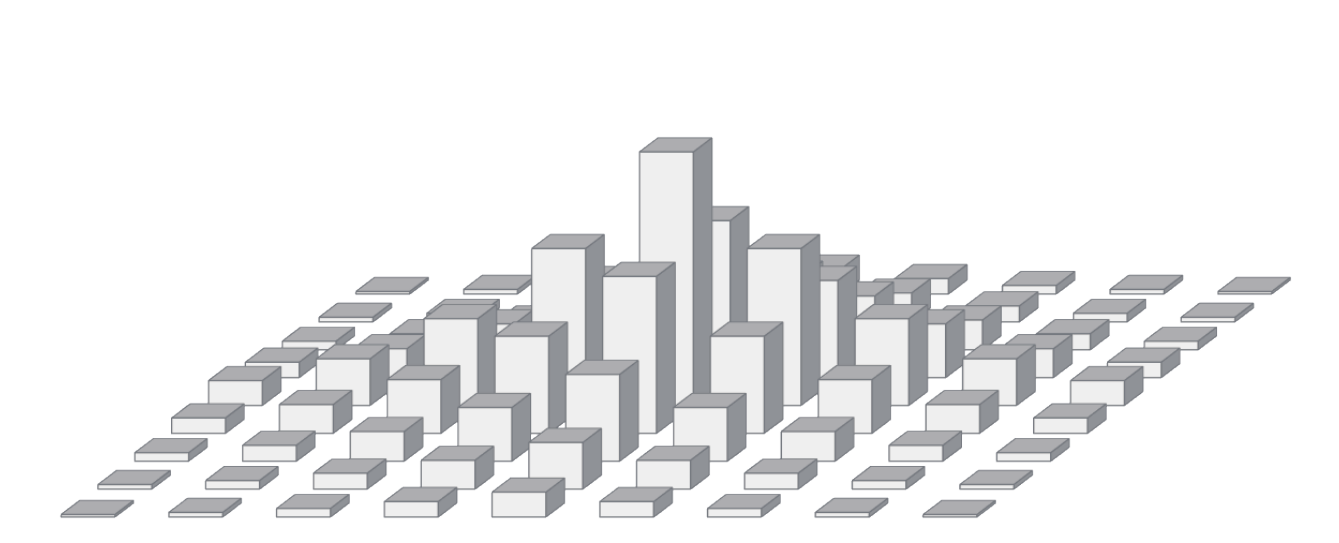}
        \caption{Monocentric distribution\label{fig:mono}}
    \end{subfigure}
    \begin{subfigure}[b]{.32\hsize}
        \centering
        \includegraphics[width=\hsize]{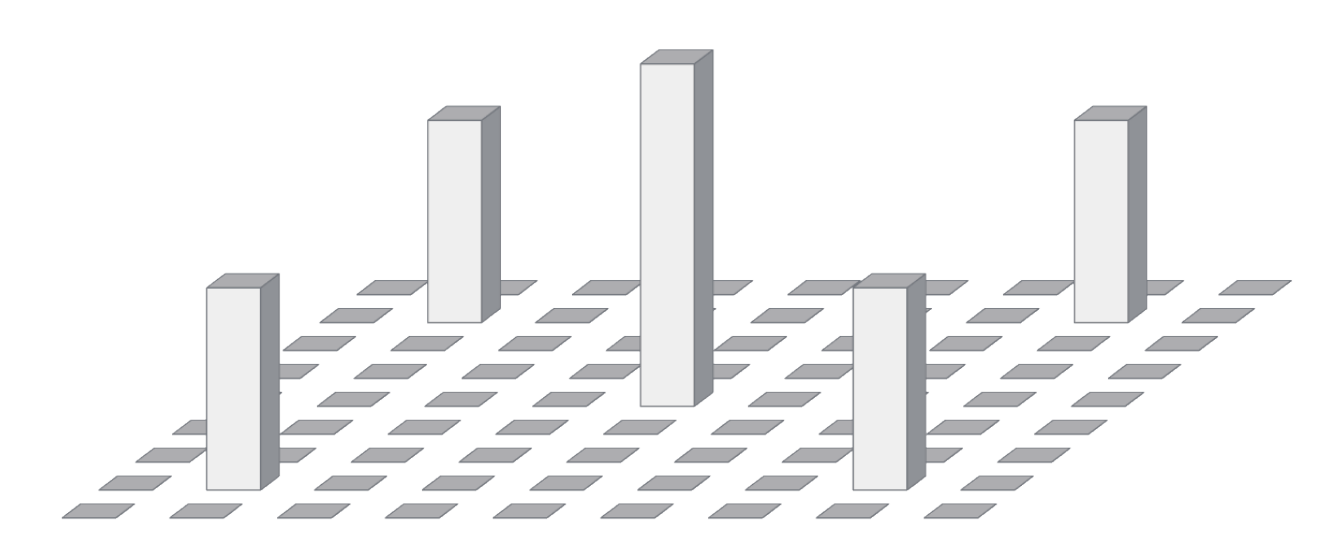}
        \caption{Polycentric distribution\label{fig:poly}}
    \end{subfigure}
    \caption{Spatial distributions in a square economy with uniform local fundamentals.}
    \label{fig:many-towns-single-city}
\end{figure}

Urban development often exhibits patterns that appear contradictory at first glance. 
In Japan, over the five decades from 1970 to 2020, urban populations became increasingly concentrated in larger cities.  
The population share of the top 100 cities rose by 19\,\%, while that of the remaining cities fell by 17\,\%, as reflected in the flatter slope of the rank--size plot (\cref{fig:JP-GC}).
Concurrently, however, populations became more spatially dispersed within individual cities. 
The average city experienced a 35\,\% decline in its maximum population density and a 24\,\% decline in average density (\cref{fig:JP-LD-1}), a trend evident in within-city distributions (e.g., \cref{fig:JP-LD-Tokyo}). Such a dual trend of economy-wide concentration and intra-urban decentralization is not unique to Japan but is observed across diverse contexts, including China, France, and the United States.\footnote{See Appendix \ref{app:ua} for more detailed discussion. 
\cite{Combes-et-al-DP2023} report similar evidence for France using newly constructed panel data spanning 1760--2020.} 

How can population simultaneously concentrate across cities and spread out within them? 
Transport costs have been central to theories of economic agglomeration and dispersion. Indeed, Japan's spatial reorganization coincided with the rapid expansion of nationwide high-speed railway and highway networks, which were essentially developed from scratch between 1970 and 2020.\footnote{Between 1970 and 2020, the total length increased from 1,119 \,km to 9,050 \,km for highways, and from 515 \,km to 3,106 \,km for high-speed railways. This is an increase of more than eight times and six times, respectively. See \cref{fig:network} in Appendix \ref{app:ua}.} 
However, interpreting this dual pattern through existing theory is not straightforward. 
This is because most analyses rely on highly stylized few-location settings, , which makes it difficult to analyze how agglomeration and dispersion occur in complex spatial economies with many locations.

To address this, the present study develops a unified theoretical framework that distinguishes between ``local'' dispersion forces acting \emph{within} cities and ``global'' dispersion forces acting \emph{across} them. 
The core intuition is that dispersion, or the tendency for economic activity to spread out, operates at two distinct spatial scales: one that pushes agents toward the fringes of their own city, and another that repels cities away from one another.
This distinction provides a coherent basis for classifying spatial models, clarifying how their comparative statics differ and how transport-induced reorganization depends on the dominant type of dispersion force in many-location settings.

To understand the core intuition, consider a hypothetical many-location economy with mobile agents, such as households or firms, making location choices. 
For exposition purposes, ``locations'' refer to generic discrete units such as regions, counties, cities, or grid cells. 
Suppose that  agents benefit from proximity to others through some agglomeration forces. 
In the absence of dispersion forces, everyone concentrates in a single location, as illustrated by \cref{fig:full-concentration}. 

First, consider the negative externalities that arise \emph{within} a location, depend on its own population, and affect only its residents. We term such externalities ``local'' dispersion forces. 
A representative example is crowding in the market for non-tradable goods. 
For instance, if the housing supply is inelastic, population growth within a location drives up housing prices, creating incentives for residents to relocate to nearby, less expensive areas. 
Agents move incrementally toward the periphery to mitigate congestion while still benefiting from proximity to the core. 
In equilibrium, this tension gives rise to a single-peaked monocentric spatial pattern, characterized by a dominant center surrounded by lower-density fringes, as \cref{fig:mono} illustrates. 

By contrast, some negative externalities spill \emph{across} locations. 
We refer to these as ``global'' dispersion forces.
A key example is market crowding through interregional trade in the presence of immobile factors such as land or other natural resources.
Through trade, a large central city can use its agglomeration advantages to dominate nearby markets, making surrounding locations unattractive for producers. 
When transport costs are sufficiently high, producers may instead locate farther from the center, where competition  is weaker and local demand can support entry. 
The result is a spatial pattern with multiple agglomerations that compete for market access and scarce resources and therefore repel each other across space, as \cref{fig:poly} illustrates.

As transport costs decline, local and global dispersion forces exert countervailing influences.
On a global scale, winners and losers can emerge. 
Improved transport access extends the spatial reach of firms and consumers, intensifying market crowding between economic centers. 
This undermines smaller agglomerations, driving economic activities to concentrate further in a limited number of major hubs. 
However, on  the local scale, lower transport costs reduce the relative advantages of proximity, fostering spatial spread within cities due to congestion forces.
These opposing mechanisms can jointly give rise to a dual pattern: economy-wide concentration alongside decentralization within each agglomeration. 

Distinguishing between local and global dispersion forces helps clarify the equilibrium spatial patterns implied by models of economic agglomeration and the comparative statics they generate. 
A wide range of spatial models can be classified into three broad types: (i) models with only local dispersion forces, (ii) models with only global dispersion forces, and (iii) models that incorporate both. These model classes yield fundamentally different predictions about how declining transport costs shape the spatial economy. 
In turn, quantitative spatial models can produce sharply divergent counterfactual outcomes depending on the dispersion forces they embed. 
For example, lower transport costs tend to promote spatial spread across locations when local dispersion forces dominate \citep[e.g.,][]{Helpman-Book1998,Allen-Arkolakis-QJE2014}, but promote further agglomeration toward central locations when global dispersion forces are primary \citep[e.g.,][]{Krugman-JPE1991}. 
In short, whether transport improvements lead to spatial spreading or further concentration depends on the spatial scale at which dispersion forces operate. 
Consequently, transport policies intended to support peripheral locations may succeed or backfire depending on whether the models guiding these policies adequately capture the relevant dispersion forces.

\section{The two-region economy}
\label{sec:model}

To introduce key concepts, we consider two-location models throughout this section, and proceed to many-location settings in later sections. 

\Heading{Preliminaries} 
There are perfectly mobile \emph{agents} (e.g., households) who choose their location to maximize utility. 
Throughout, locations are called \emph{regions} for convenience. 
Let $x_i \ge 0$ denote the continuous mass of agents in the region $i \in \{1,2\}$, where $x_1 + x_2 = 1$. 
The indirect utility of agents in the region~$i$ is a differentiable function of the spatial distribution $\Vtx = (x_1,x_2)$ and is indicated by $v_i(\Vtx)$. 
A \emph{spatial equilibrium} is a spatial distribution $\Vtx^*$ in which no agent is motivated to relocate. 
For example, if $x_1^*,x_2^* > 0$, then $\Vtx^*$ is a spatial equilibrium if and only if $v_1(\Vtx^*) = v_2(\Vtx^*)$. 

Transport between regions is costly, and $\phi \in (0,1)$ measures the \emph{ease of transport} between regions. 
Higher values of $\phi$ indicate better access. 
For later use, we also introduce the \emph{proximity matrix} $[\phi_{ij}]$ following \cite{Matsuyama-RIE2017}, where $\phi_{ij} \in (0,1]$ is the ease of transport from region~$i$ to $j$. 
For the two-region case in this section, 
\begin{align}
    \renewcommand{\arraystretch}{0.8}
        & \begin{bmatrix}
            \phi_{11} & \phi_{12} \\
            \phi_{21} & \phi_{22}
        \end{bmatrix}
        =
        \begin{bmatrix}
            1 & \phi \\
            \phi & 1
        \end{bmatrix}. 
        \label{eq:D_2}
\end{align}

To focus on forces driven by transport costs and the endogenous spatial distribution of agents, region-specific characteristics (e.g., innate amenity or productivity) are assumed to be homogeneous.  
With these common settings in place, we consider a series of specifications for $\Vtv(\Vtx) = (v_1(\Vtx),v_2(\Vtx))$, including general equilibrium models such as \cite{Krugman-JPE1991,Helpman-Book1998,Tabuchi-JUE1998,Redding-Sturm-AER2008,Allen-Arkolakis-QJE2014}, to illustrate the core ideas. 

\subsection{The Beckmann model: A ``local'' dispersion force} 
\label{sec:Bm}

We start with the following parsimonious specification: 
\begin{align}
    & v_i(\Vtx) 
    = 
    x_i^{- \beta} 
    \Big(\sum_{j} \phi_{ij} x_j \Big)^\alpha 
    & (\alpha,\beta > 0). 
    \label{eq:v-Bm}
\end{align}
The first term, $x_i^{-\beta}$, captures the localized congestion force within each region, while the second term is positive externalities between regions or the agents' desire to be close to each other. 
We call this the Beckmann model, following \cite{Beckmann-Book1976}.\footnote{Beckmann studied spatial agglomeration in a continuous one-dimensional space. To streamline the exposition, we consider its discrete-space and multiplicative analog.}  
 
The symmetric distribution of agents, $\BrVtx = (\tfrac{1}{2}, \tfrac{1}{2})$, is always a spatial equilibrium.  
In fact, if we define the utility difference between the two regions by 
\begin{align}
    \Delta (\Vtx) \Is v_1(\Vtx) - v_2(\Vtx), 
\end{align}
we confirm $\Delta (\BrVtx) = 0$. 

While $\BrVtx$ is an equilibrium, with both agglomeration and dispersion forces, $\BrVtx$ is not \emph{stable} if the former dominates the latter. 
To assess the stability of $\BrVtx$, we examine the incentive of movers.
Consider a small mass of agents moving from region 2 to 1, so that $x_1$ rises and $x_2$ falls symmetrically.
If $\Delta$ decreases under this perturbation, then $\Delta < 0$ after the move: utility in region 2 exceeds that in region 1, and the movers will prefer to return. 
This induces a restoring force and hence $\BrVtx$ is locally stable. 
If $\Delta$ increases, then $\Delta > 0$ after the move: the movers prefer to stay in region 1, and the deviation attracts additional movers now that the utility is higher in region $1$. Hence, $\BrVtx$ is unstable. 

\begin{figure}[t]
    \centering
    \includegraphics[width=7.5cm]{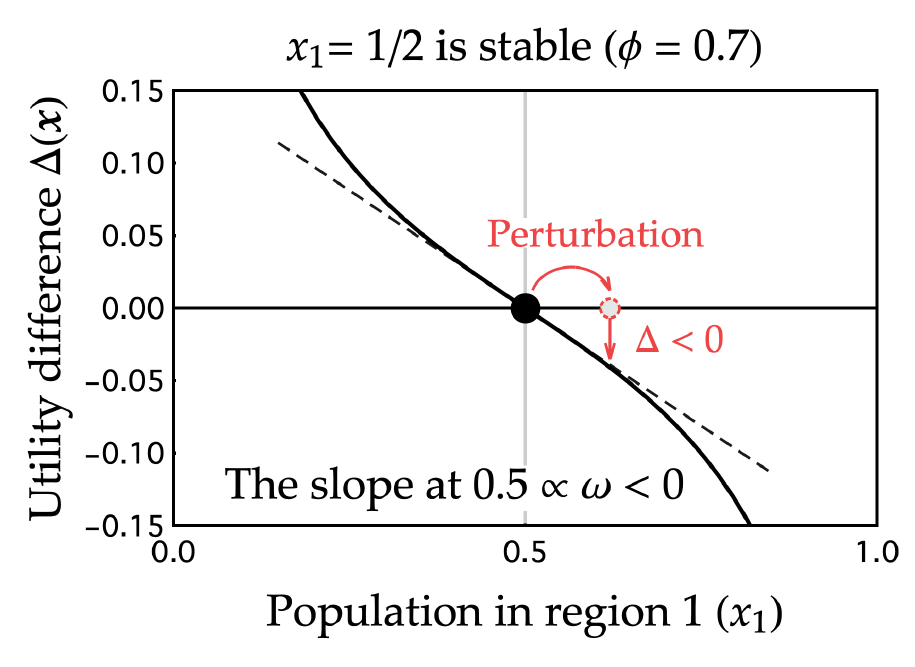} 
    \hfill
    \includegraphics[width=7.5cm]{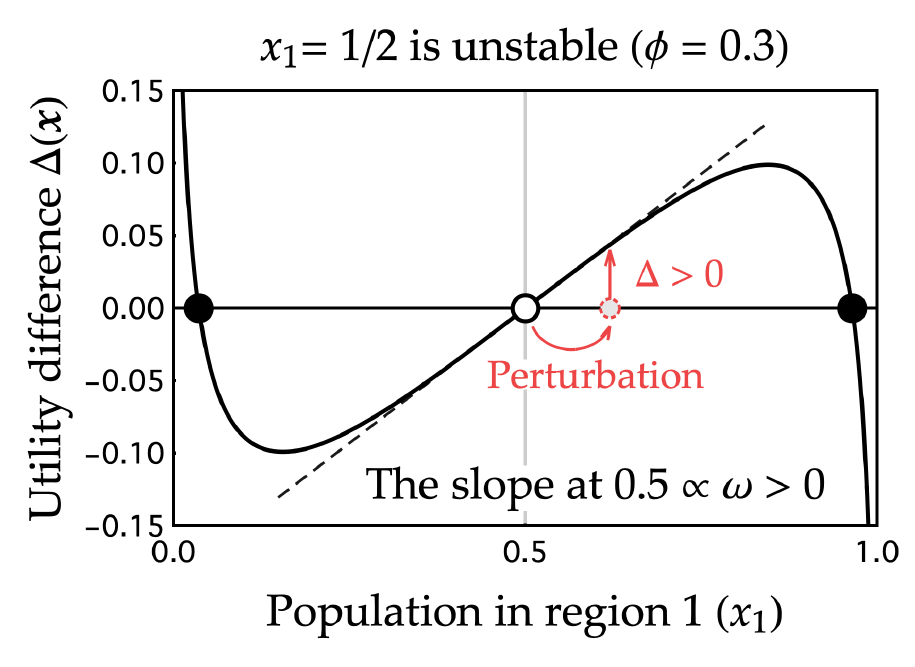} 
    \caption{Stability of the symmetry in the Beckmann model ($\alpha = 1/2, \beta = 1/6$)\label{fig:Bm-stab}}
\end{figure}
\Cref{fig:Bm-stab} illustrates the above discussion. 
The symmetric equilibrium $\BrVtx$ is stable in the left panel. 
In the right panel, $\BrVtx$ is unstable and two additional stable spatial equilibria arise, each exhibiting \emph{endogenous agglomeration}, in which one region becomes larger despite the perfect symmetry of the regional fundamentals. 

We can formalize the above discussion by a simple stability criterion.\footnote{All omitted proofs are in Appendix \ref{app:proofs}.}  
\begin{lemma}
    \label{lem:omega-stab}
Given a differentiable $\Vtv$, define the \emph{utility gain} for marginal movers by 
\begin{align}
	\label{eq:omega-two-reg}
	\omega
	&
	\Is
    \frac{\Brx}{\Brv}
    \cdot  
    \left. 
    \PDF{\Delta(\Vtx)}{x_1}
    \right|_{\Vtx = \BrVtx}
    =
    \frac{\Brx}{\Brv}
    \cdot  
    \left.
    \left( 
    \PDF{v_1(\Vtx)}{x_1}
    - 
    \PDF{v_2(\Vtx)}{x_1}
    \right)
    \right|_{\Vtx = \BrVtx}, 
\end{align}
where $\Brx \Is \tfrac{1}{2}$ and $\Brv = v_1(\BrVtx) = v_2(\BrVtx) > 0$. 
Then, the symmetric equilibrium $\BrVtx$ is stable if $\omega < 0$, and unstable if $\omega > 0$.  
\end{lemma}
\noindent 
In \cref{eq:omega-two-reg}, normalization by $\Brx / \Brv$ only simplifies the final expression of $\omega$. 

In the Beckmann model, we compute 
\begin{align}
    \omega 
    & 
    = 
    - \beta + \alpha \Theta, 
    \qquad \text{where} \qquad 
    \Theta \Is \frac{1 - \phi}{1 + \phi} \in (0,1). 
    \label{eq:omega-Bm}
\end{align} 
The negative term $-\beta$ represents the dispersion force, while the positive term $\alpha \Theta$ represents the agglomeration force. 
The sign of $\omega$, and hence the stability of $\BrVtx$, depends on which force dominates: $\BrVtx$ is stable if $\beta > \alpha \Theta$, and unstable if $\beta < \alpha \Theta$. 

In \cref{eq:omega-Bm}, the dispersion force does not depend on $\phi$ because it is not affected by interregional transport conditions. 
By contrast, the agglomeration force depends on $\phi$ as it incorporates interregional interactions. 
In particular, $\alpha \Theta$ decreases as $\phi$ increases. 
That is, the benefit of becoming close to others is small if transport is less costly.\footnote{If instead the regions are in autarky ($\phi = 0$), the agglomeration force reduces to local spillovers \`a la \cite{Henderson-AER1974}. 
Since $\Theta = 1$ and $\alpha \Theta = \alpha$, the stability condition is whether $\alpha < \beta$ or not.}  

Importantly, $\Theta$ itself is the \emph{proximity gain} for marginal movers at each $\phi$. 
To see this, set $\alpha = 1$ and $\beta = 0$. 
Then, $v_i(\Vtx) = \sum_{j} \phi_{ij} x_j$ corresponds to a parsimonious measure of proximity.  
Since $\omega = \Theta$ in this case, $\Theta$ indeed represents the proximity gain.  

With \cref{eq:omega-Bm}, we can determine the stability of $\BrVtx$ in the Beckmann model. 
If $\beta \geq \alpha$, congestion force is so strong that $\omega < 0$ for all $\phi\in(0,1)$. 
Agglomeration cannot occur, as $\BrVtx$ is always stable. 
If $0 < \beta < \alpha$, the level of $\phi$ matters. 
As \cref{fig:Bm-stab} illustrates, $\BrVtx$ is stable if $\phi$ is large (i.e., if the agglomeration force is small), and unstable otherwise.

\Heading{Gain functions}
Beyond the Beckmann model, the utility gain $\omega$ in \cref{lem:omega-stab} is well defined for any differentiable $\Vtv$. 
In many spatial models, $\omega$ is a simple function of $\Theta$,   
\begin{align}
\omega = \Omega(\Theta),  
\label{eq:gain-func}
\end{align}
as in the Beckmann model. 
For each model, we call this $\Omega$ the \emph{gain function} of the model.
The positive terms of $\Omega$ represent the model's agglomeration forces, whereas the negative terms represent the dispersion forces. 
How each term of $\Omega$ responds to changes in $\Theta$ then describes how transport conditions alter the strengths of these forces, and therefore their relative importance at each transport cost level.\footnote{Generalization is possible for the cases where the proximity gain is multi-dimensional such as $\VtTheta \in [0,1]^M$ where $M$ is the number of different interregional interactions. For clarity, we restrict our attention to the case $M=1$ throughout this study.} 

\subsection{The Braid model: A ``global'' dispersion force}
\label{sec:Br}

To illustrate what gain functions $\Omega$ look like under different specifications of the indirect utility function $\Vtv$, we employ another reduced-form model. 
Consider replacing the local congestion term in the Beckmann model as follows:
\begin{align}
    v_i(\Vtx) = y_i(\Vtx) \Big(\sum_{j} \phi_{ij} x_j\Big)^\alpha.  
    \label{eq:v-trade}
\end{align}
Here, $y_i(\Vtx)$ is the income of an agent in the region~$i$.   
Each region is endowed with one unit of fixed expenditure, which is allocated to all agents according to accessibility. 
Suppose that the share received by an agent in region~$i$ from region~$m$ is given by\footnote{Note that the total expenditure of the region~$m$ is equal to one: $\sum_{i\inI} x_i s_{i \mid m} = 1$. Possible microfoundations for this toy model are abstracted away for brevity.}
\begin{align}
    & s_{i \mid m}(\Vtx) \Is \frac{\phi_{im}}{\sum_{l} x_l \phi_{lm}}. 
\label{eq:share}
\end{align}
Then, $y_i(\Vtx) = \sum_{m} s_{i \mid m}(\Vtx)$. 
We refer to the model \eqref{eq:v-trade} as the Braid model after \cite{Braid-RSUE1988} who studied the case $\alpha = 0$. 

In this model, proximity to others has a negative impact on the utility of agents, as \cref{eq:share} embeds competition between agents in different regions over spatially dispersed expenditure. 
Concretely, for any combination of $i,j,m\in\{1,2\}$, 
\begin{align}
    & \frac{\partial s_{i \mid m}}{\partial x_j} 
    = 
    - 
    \frac{\phi_{im}}{\sum_{l} x_l \phi_{lm}}
    \cdot 
    \frac{\phi_{jm}}{\sum_{l} x_l \phi_{lm}}
    = 
    - s_{i\mid m} \cdot s_{j \mid m} < 0, 
    \label{eq:ds}
\end{align}
demonstrating that a marginal increase in agents in any region $j \in \{1,2\}$ has negative impacts on $s_{i\mid m}$. 
This is simply because it increases the denominator of $s_{i \mid m}$. 
Whether income $y_i = \sum_{j} s_{i \mid j}$ as a whole increases or not after a migration shock depends on the relative magnitudes of these impacts. 
Specifically, analogous to the utility gain $\omega$, we can compute the income gain of a marginal mover from region $2$ to $1$:\footnote{From \cref{eq:ds}, $\PDF{y_1}{x_1} = - (s_{1\mid 1})^2 - (s_{1\mid 2})^2 = - \frac{1 + \phi^2}{\Brx^2(1 + \phi)^2}$ and $\PDF{y_2}{x_1} = - s_{2\mid 1}\cdot s_{1\mid 1} - s_{2\mid 2} \cdot s_{1\mid 2} = - \frac{2\phi}{\Brx^2(1 + \phi)^2}$.} 
\begin{align}
    \frac{\Brx}{\Bry} \left.\left(\PDF{y_1(\Vtx)}{x_1} - \PDF{y_2(\Vtx)}{x_1}\right)\right|_{\Vtx = \BrVtx} 
    = - \frac{1 + \phi^2}{(1 + \phi)^2} + \frac{2\phi}{(1 + \phi)^2} 
    = - \frac{(1 - \phi)^2}{(1 + \phi)^2} 
    = - \Theta^2 < 0, 
    \label{eq:dydx-Br}
\end{align} 
where $\Bry = y_1(\BrVtx) = y_2(\BrVtx) = 1/\Brx$. 
This shows that a marginal migration shock always induces an income \emph{loss} for the movers. 
This force is on the second order of $\Theta$, reflecting that the proximity matrix $[\phi_{ij}]$ appears twice in the numerator of \cref{eq:ds}. 
  
From \cref{eq:omega-two-reg,eq:dydx-Br}, the utility gain for the Braid model is given as follows: 
\begin{align}
    \omega = \alpha \Theta  - \Theta ^2. 
    \label{eq:Br.omega}
\end{align}
The gain function for the model is therefore $\Omega(\Theta) = \alpha \Theta - \Theta^2$. 
The first term is the same agglomeration force as in the Beckmann model. 
The second term, $-\Theta^2$, represents the ``global'' dispersion force due to the income loss in \cref{eq:dydx-Br}.  

This dispersion force weakens as $\Theta$ falls (i.e., as $\phi$ increases).
If $\phi$ is very small, regions are effectively in autarky and  $y_i \approx s_{i\mid i} \approx 1/x_i$.
In this case, a migration shock that raises $x_i$ directly intensifies local competition and lowers income, creating a strong incentive to disperse ($-\Theta^2 \approx -1$).
By contrast, if $\phi$ is close to one, regions face nearly identical crowding conditions ($y_i \approx 1$), and migration shocks have little effect on income ($-\Theta^2 \approx 0$).

In the model, $\BrVtx$ can become unstable for some $\phi$  whenever $\alpha > 0$. 
If $0 < \alpha < 1$, $\BrVtx$ is stable for high transport costs (large $\Theta$ $\Leftrightarrow$ small $\phi$) and unstable for low transport costs (small $\Theta$ $\Leftrightarrow$ large $\phi$). 
If $\alpha \ge 1$, $\BrVtx$ is unstable for all $\phi$.

\begin{figure}[t]
    \centering
	\begin{subfigure}[b]{.48\hsize}
		\centering
        \begin{picture}(160,127)
            \put(0,0){\includegraphics[width=162pt]{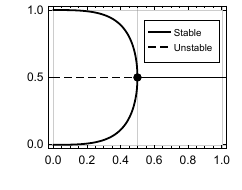}}
            \put(88,3){$\phi$}
            \put(5,28){\rotatebox{90}{\footnotesize Population of region 1}}
        \end{picture}
		\caption{The Beckmann model (\cref{eq:v-Bm})\label{fig:Toy-2}}
	\end{subfigure}
    \hfill 
    \begin{subfigure}[b]{.48\hsize}
        \centering
        \begin{picture}(160,127)
        \put(0,0){\includegraphics[width=162pt]{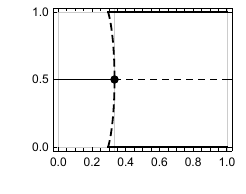}}
        \put(88,5){$\phi$}
        \put(5,28){\rotatebox{90}{\footnotesize Population of region 1}}
        \end{picture}
        \caption{The Braid model (\cref{eq:v-trade})\label{fig:Toy-1}}
        
    \end{subfigure}

    \bigskip 

	\begin{subfigure}[b]{.48\hsize}
		\centering
        \begin{picture}(160,127)
            \put(0,0){\includegraphics[width=162pt]{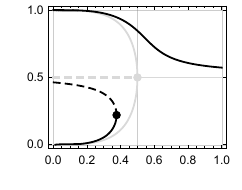}}
            \put(88,3){$\phi$}
            \put(5,28){\rotatebox{90}{\footnotesize Population of region 1}}
        \end{picture}
		\caption{The Beckmann model with asymmetry\label{fig:Toy-2-asym}}
		
	\end{subfigure}
    \hfill 
    \begin{subfigure}[b]{.48\hsize}
        \centering
        \begin{picture}(160,127)
        \put(0,0){\includegraphics[width=162pt]{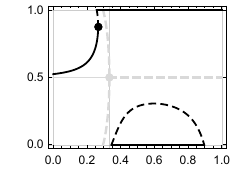}}
        \put(88,5){$\phi$}
        \put(5,28){\rotatebox{90}{\footnotesize Population of region 1}}
        \end{picture}
        \caption{The Braid model with asymmetry\label{fig:Toy-1-asym}}
        
    \end{subfigure}

    \caption{Equilibrium values of $x_1$ in the Beckmann model and the Braid model \label{fig:Toy}}

    \FigureNote{We set $\alpha = 1/2$ for both models, and $\beta = 1/6$ for the Beckmann model. 
    In Panels (A) and (B), the black markers indicate the points where the symmetry becomes unstable. In Panels (C) and (D), we multiply $v_1(\Vtx)$ by $1.05$. The black markers indicate the points at which stable and unstable equilibrium curves converge. For reference, the equilibria for the symmetric cases are shown in light gray.} 
\end{figure}

\Heading{Contrasting implications of transportation costs} 
The two reduced-form models yield fundamentally opposing implications. 
In the Beckmann model, $\BrVtx$ is stable for low transport costs (large $\phi$), and agglomeration occurs for high transport costs (small $\phi$). 
The Braid model exhibits the opposite behavior. 
\Cref{fig:Toy-1,fig:Toy-2} confirm this by showing the full equilibrium paths on the $\phi$-axis. 
The contrast persists in the presence of asymmetries. 
For instance, if region 1 possesses an exogenous advantage as in \cref{fig:Toy-1-asym,fig:Toy-2-asym}, a symmetric configuration is no longer an equilibrium.
Nevertheless, the qualitative findings remain consistent with the symmetric case: an increase in $\phi$ (declining transport costs) fosters dispersion in the Beckmann model but drives agglomeration in the Braid model. 

This divergence is rooted in the different spatial scales of dispersion forces. In the Beckmann model, dispersion under high $\phi$ is driven by the relative weakening of the agglomeration force, whereas the ``local'' dispersion force remains invariant to $\phi$. In the Braid model, endogenous agglomeration occurs due to the relative decline of the ``global'' dispersion force as $\phi$ increases. 
Thus, the spatial scale of the dominant dispersion force can alter the implications of declining transport costs.\footnote{In fact, the literature on two-region models has recognized that various dispersion mechanisms can have opposing effects. For example, \citet[Ch.8]{Fujita-Thisse-Book2013} compares the seminal models by \cite{Krugman-JPE1991} and \cite{Helpman-Book1998} and noted that ``Krugman's scenario is reversed'' (p.289) in Helpman-type frameworks with urban costs.}   
As noted in the introduction and further discussed in \cref{sec:N-region}, in many-region settings, these two types of dispersion forces lead to contrasting spatial patterns.

\subsection{Benefit matrix and the spatial scale of economic forces}
\label{sec:local-and-global-dispersion-forces}

While the insights from the stylized models considered so far are intuitive and nearly immediate, they rely on a reduced-form structure. In more comprehensive frameworks, particularly general equilibrium models where wages, prices, and land rents are endogenously determined, the impact of transport costs becomes considerably more complex. In such settings, multiple forces interact simultaneously, making it difficult to isolate whether ``local'' or ``global'' dispersion dominates.

To bridge the gap between our stylized insights and richer general equilibrium specifications, it is necessary to develop a formal method for evaluating how these endogenous forces respond to changes in transport costs. 
For this purpose, we introduce an analytical tool termed the \emph{benefit matrix}. 
This matrix allows us to systematically decompose the spatial externalities inherent in spatial models and classify them into the local and global forces in our earlier discussion. 

\begin{definition}[Benefit matrix]
For a differentiable indirect utility function $\Vtv$, its \emph{benefit matrix} $\VtV$ is its elasticity matrix at symmetric equilibrium $\VtV \Is \frac{\Brx}{\Brv}[\frac{\partial v_i}{\partial x_j}(\BrVtx)]_{i = 1,2; j = 1,2}$. 
\end{definition}

Then, the utility gain $\omega$ in the symmetric equilibrium $\BrVtx$ is the eigenvalue of the benefit matrix $\VtV$, associated with the eigenvector $\Vtz = (1 , -1)$. 
Intuitively, $\Vtz$ represents the direction of possible migration shocks because $(\Brx + \epsilon, \Brx - \epsilon) = \BrVtx + \epsilon \Vtz$. 
\begin{example}
    \label{ex:Bm-gain-func}
Consider the Beckmann model. Its benefit matrix is 
\begin{align}
    \VtV = - \beta \VtI + \alpha \VtD
    \label{eq:Bm.V}, 
\end{align}
where $\VtI$ is the identity matrix and $\VtD$ is the \emph{row-normalized proximity matrix}:
\begin{align}
    \label{eq:barD}
    \VtD = \frac{1}{1 + \phi}
    \begin{bmatrix}
        1 & \phi \\
        \phi & 1
    \end{bmatrix}. 
\end{align}
We confirm $\VtD\Vtz = \frac{1 - \phi}{1 + \phi} \Vtz = \Theta\Vtz$. That is, $\Vtz = (1,-1)$ is an eigenvector of $\VtD$, and the proximity gain $\Theta$ is the associated eigenvalue. 
Then, $\VtV \Vtz = (-\beta + \alpha \Theta) \Vtz = \omega \Vtz$ from \cref{eq:Bm.V}, showing that $\omega$ is indeed an eigenvalue of $\VtV$ corresponding to $\Vtz$. 
\end{example}
\begin{example}
    \label{ex:Br-gain-func}
The benefit matrix for the Braid model is
\begin{align}
    \VtV = \alpha \VtD - \VtD^2. 
    \label{eq:Br.V}, 
\end{align}
Since $\VtV\Vtz = (\alpha \Theta - \Theta^2)\Vtz$, $\omega = \alpha \Theta - \Theta^2$ is the eigenvalue of $\VtV$ associated with $\Vtz$. 
\end{example}

\Cref{ex:Bm-gain-func,ex:Br-gain-func} illustrate that the gain function $\Omega$ introduced earlier (\cref{eq:gain-func}) arises from the structure of $\VtV$. 
The utility gain takes the form $\omega = \Omega(\Theta)$ because the benefit matrix itself can be written in parallel form $\VtV = \Omega(\VtD)$. Here, the matrix function is interpreted in a straightforward way, e.g., matrix polynomials. 
We can thus think of $\VtV = \Omega(\VtD)$ and $\omega =\Omega(\Theta)$ interchangeably. 

In the two reduced-form examples, the gain function $\Omega(\Theta)$ allows us to formally distinguish the spatial scale of the economic forces. 
A negative constant term in the gain function $\Omega(\Theta)$ corresponds to a \emph{local} dispersion force.
In contrast, negative non-constant terms correspond to \emph{global} dispersion forces. 
The spatial scale for agglomeration forces can be similarly defined: positive constants are local agglomeration forces, and negative non-constant terms are global agglomeration forces. 

\subsection{General equilibrium models} 
\label{sec:ge-models}

We now examine specific models in the literature to illustrate that their benefit matrices $\VtV$ are simple functions of $\VtD$. 
We can then immediately obtain the associated gain functions and systematically determine the spatial scale of the dispersion forces in each model. 
Detailed derivations are provided in Online Appendix \ref{app:example_models}.

\providecommand{\CMA}{\textsc{CMA}} 

\medskip 

\Heading{The \cite{Helpman-Book1998} / \cite{Redding-Sturm-AER2008} model} 
Helpman considered an imperfectly competitive framework in which agents consume both differentiated tradable goods and local non-tradable goods (i.e., housing). 
We consider its variant by Redding and Sturm. 
The indirect utility of mobile workers in this model is 
\begin{align}
	&
	\label{eq:RS.payoff.main}
	v_i(\Vtx)
	=
	x_i^{-(1 - \mu)}
	w_i^{\mu}
    \cdot \CMA_i^{\frac{\mu}{\sigma - 1}} 
	,
\end{align}
where $\mu \in (0,1)$ is consumers' expenditure share on tradables and $1 - \mu \in (0,1)$ is that on non-tradables, $\sigma > 1$ is the elasticity of substitution of horizontally differentiated tradable varieties, $w_i$ is the nominal wage in region~$i$, and $\CMA_i \Is \sum_{j\inI} x_j w_j^{1 - \sigma}\phi_{ji}$ is the so-called ``consumer market access.'' 
The proximity matrix is $\phi_{ij} = \tau_{ij}^{1 - \sigma}$, where $\tau_{ij} \ge 1$ is the iceberg trade cost from region~$i$ to $j$. 
Given $\Vtx$, the wage $\Vtw = (w_i)$ is endogenously determined in general equilibrium with interregional trade. 

The benefit matrix for this model can be computed as follows: 
\begin{align}
    & \VtV = 
    C(\VtD) \cdot 
    \left( 
        - (1 - \mu) \VtI + c_1 \VtD
    \right), 
    \label{eq:RS.V.2}
\end{align} 
where $C(\VtD) \Is \big(\VtI + \frac{\sigma - 1}{\sigma} \VtD \big)^{-1}$ and $c_1 \Is \frac{\mu}{\sigma - 1} + \frac{\mu}{\sigma} - (1 - \mu)\frac{\sigma - 1}{\sigma}$.  
This then implies 
\begin{align}
    & \omega = C(\Theta) \cdot \left(- (1 - \mu) + c_1 \Theta\right) 
    \label{eq:RS.omega.2}
\end{align}
where $C(\Theta) \Is (1 + \frac{\sigma - 1}{\sigma} \Theta)^{-1} > 0$. 
Since $C(\Theta) > 0$, the sign of $\omega$ hinges on the sign of 
\begin{equation}
    \omega^\sharp \equiv - (1 - \mu) + c_1 \Theta, 
    \label{eq:RS.omega-s}
\end{equation}
which is similar to the Beckmann model. 
In particular, $-(1-\mu)$ corresponds to the \emph{local dispersion force} due to crowding in the non-tradables market in each region. 
If $\mu$ is sufficiently large, then agents' love for variety produces a strong \emph{global agglomeration force}: we have $c_1 > 0$ and hence $c_1 \Theta > 0$ if $\mu > (\frac{\sigma - 1}{\sigma})^2$, where we note $\frac{\sigma - 1}{\sigma} \in (0,1)$. 
If further 
$\mu > \frac{\sigma - 1}{\sigma}$, 
$\BrVtx$ is unstable for high transport costs (large $\Theta$) and stable for low transport costs (small $\Theta$), just as in the Beckmann model. 

We provide a slightly detailed derivation behind the final expressions \eqref{eq:RS.V.2} and \eqref{eq:RS.omega.2} for illustration. 
The definition of $\Vtv$ in \cref{eq:RS.payoff.main} yields
\begin{align}
	\VtV
    & = 
        - 
        (1 - \mu)\VtI
        +
        \mu 
        \VtW
        + 
        \frac{\mu}{\sigma - 1}
        \left( \VtD - (\sigma - 1) \VtD\VtW\right). 
        \label{eq:RS.V.1}
\end{align}The first term in \cref{eq:RS.V.1} represents the crowding in the housing markets. 
The second is the direct impact of nominal income on indirect utility, where 
\begin{align}
    \VtW 
    = \frac{1}{\sigma} C(\VtD) \cdot \VtD  
    \label{eq:RS.W.1}
\end{align}
is the elasticity matrix of nominal wages.\footnote{Concretely, $\VtW \Is \frac{\Brx}{\Brw}[\frac{\partial w_i}{\partial x_j}(\BrVtx)]_{i = 1,2; j=1,2}$ with $\Brw = w_1(\BrVtx) = w_2(\BrVtx)$. 
As $C(\VtD) = \big(\VtI + \frac{\sigma - 1}{\sigma} \VtD \big)^{-1}$ plays a role analogous to the Leontief inverse in input--output analysis, $\VtW$ captures the general equilibrium response of wages to marginal migration shocks taking into account interregional trade.}  
The third term in \cref{eq:RS.V.1} is the ``cost-of-living'' effects: 
the price index in a region falls when nearby regions offer a greater variety of goods but increases when they pay higher wages and thus charge higher prices \citep[see, e.g.,][]{Fujita-Thisse-Book2013,Baldwin-et-al-Book2003,Brakman-etal-Book2019}. 

Substituting \cref{eq:RS.W.1} into \cref{eq:RS.V.1} and rearranging, we obtain 
\begin{align}
    \VtV
    & = 
        - 
        (1 - \mu)\VtI
        + 
        \frac{\mu}{\sigma - 1}
        \VtD
        +
        \frac{\mu}{\sigma}
        (\VtI - \VtD)
        \cdot C(\VtD) \cdot \VtD
        \quad \text{and hence}
    \label{eq:RS.V.int}
    \\
    \omega
    & 
    = 
    \Omega(\Theta)
    = 
    \underbrace{- (1 - \mu)}_{< 0}
    \,
    + 
    \,
    \underbrace{\frac{\mu}{\sigma - 1}\Theta}_{>0}
    \,  
    +
    \,
    \underbrace{\frac{\mu}{\sigma} C(\Theta) (1 - \Theta)\Theta}_{>0}. 
    \label{eq:RS.omega.int}
\end{align}
The first two terms represent the partial equilibrium utility gains where wage adjustments are ignored. 
The third term in \cref{eq:RS.omega.int} summarizes the \emph{net} impact of wages on indirect utility in general equilibrium, both through the (positive) individual-level income gain and the (negative) cost-of-living effect. 
It is a net global agglomeration force because it is strictly positive for all $\Theta \in (0,1)$.  
Thus, the only dispersion force that can stabilize $\BrVtx$ is the local dispersion force captured by the first term of \cref{eq:RS.omega.int}. 

Further rearrangement of \cref{eq:RS.V.int,eq:RS.omega.int} yields the final expressions \eqref{eq:RS.V.2} and \eqref{eq:RS.omega.2}. 
By construction, $\omega^\sharp$ in \cref{eq:RS.omega-s} captures the \emph{net} utility gain considering all endogenous forces and their general equilibrium trade-offs. 
For example, despite there is no \emph{net} global dispersion force, $\omega^\sharp = - (1 - \mu) + c_1 \Theta$ has a negative term involving $\Theta$:  
\begin{align}
    c_1 \Theta = \frac{\mu}{\sigma} \Theta + \frac{\mu}{\sigma - 1} \Theta 
    \underbrace{- (1 - \mu) \frac{\sigma - 1}{\sigma} \Theta}_{ < 0}
    \label{eq:RS.c1}
\end{align}
This last term represents how the local dispersion force $-(1 - \mu)$ counteract, in each agent's migration incentives, the core agglomeration forces in the model, such as the love for variety and demand linkage (the first two terms of \cref{eq:RS.c1}).

\Heading{The \cite{Allen-Arkolakis-QJE2014} model} 
The model is a perfectly competitive framework with both positive and negative externalities. 
Appendix \ref{app:allen-arkolakis} shows that the benefit matrix for the Allen--Arkolakis model is 
\begin{align}
    \VtV = C(\VtD) \cdot \left(
            c_0 
            \VtI 
            + 
            c_1 
            \VtD
        \right),  
\end{align}
where $C(\VtD) \Is ((\sigma\VtI + (\sigma - 1)\VtD)(\VtI - \VtD))^{-1}$, 
$c_0 \Is \alpha - \beta - \frac{1 + \alpha}{\sigma}$, and $c_1 \Is \alpha - \beta + \frac{1 + \beta}{\sigma}$. 
The parameters $\alpha > 0$ and $\beta > 0$ are the magnitudes of local agglomeration and congestion effects with respect to local population, respectively, and $\sigma > 1$ is the elasticity of substitution across regionally differentiated goods. 
The gain function $\Omega$ is obtained accordingly. 
Analogous to the Helpman model, the sign of $\omega$ depends on a linear expression $\omega^\sharp \equiv c_0 + c_1 \Theta$ that captures the net migration incentive for agents. 

If the dispersion force is relatively strong (relatively large $\beta$), we have $c_0 < 0$, indicating a \emph{net} local dispersion force. 
Likewise, if the agglomeration force is relatively strong (relatively large $\alpha$), we  have $c_1 > 0$, indicating a \emph{net} global agglomeration force. 
In particular, if $\beta < \alpha < \Bralpha \Is \frac{\beta \sigma + 1}{\sigma - 1}$, agglomeration occurs if transport costs are high (small $\phi$) and dispersion occurs if transport costs are low (large $\phi$). Under the presence of externalities, the only stabilizing force in the Allen--Arkolakis model is the local crowding effect.\footnote{In the perfectly competitive case ($\alpha = 0$ and $\beta = 0$), the model reduces to the \cite{Armington-IMF1969} framework.
For this case, $\omega = \frac{1}{\sigma} C(\Theta)(-1 + \Theta) < 0$, and the net agglomeration term $(-1 + \Theta)$ does not contain any negative components in $\Theta$.
Nonetheless, $\omega$ is a negative function and can be interpreted as the underlying ``global dispersion force'' inherent to general equilibrium in the Armington framework.}  
That is, as discussed in \cite{Allen-Arkolakis-QJE2014}, this model bears a structural similarity to the Helpman model.

\Heading{The \cite{Krugman-JPE1991} model} 
Krugman's seminal model emphasizes the role of market crowding. 
In this model, there is always a nonzero demand for the manufacturing goods in each region due to the presence of immobile consumers. 
This discourages concentration of production in one region if transport costs are high, and this produces a global dispersion force as in the Braid model. 
For this model, the benefit matrix is
\begin{align}
    \VtV = \VtW + \frac{\mu}{\sigma - 1} \left(\VtD - (\sigma - 1) \VtD \VtW\right)
\end{align}
where the elasticity matrix for the nominal wage is 
\begin{align}
    \VtW = 
    \frac{1}{\sigma}
	C(\VtD)
    \cdot 
    \left(
        \mu \VtD - \VtD^2
    \right), 
    \quad\text{where}\quad 
    C(\VtD) \Is 
    \left( 
        \VtI
        - \frac{\mu}{\sigma}\VtD 
        - \frac{\sigma - 1}{\sigma}\VtD^2
    \right)^{-1}.
    \label{eq:Km.V.W.1}
\end{align} 
Compared to \cref{eq:RS.V.1}, the local dispersion force is absent. 
Compared with \cref{eq:RS.W.1,eq:Km.V.W.1}, the Krugman model has a negative term $-\VtD^2$, while the Helpman model does not. 
We can rearrange $\VtV$ to see 
\begin{align}
    \VtV = C(\VtD) \cdot 
    \left( 
        c_1 \VtD - c_2 \VtD^2 
    \right). 
\end{align}
where $c_1 \Is \frac{\mu}{\sigma - 1} + \frac{\mu}{\sigma} > 0$ and $c_2 \Is \frac{\mu^2}{\sigma - 1} + \frac{1}{\sigma} > 0$. 
The gain function is then 
$\omega = C(\Theta) \cdot ( c_1 \Theta - c_2 \Theta^2 )$ with $C(\Theta) \Is (1 - \frac{\mu}{\sigma}\Theta - \frac{\sigma - 1}{\sigma}\Theta^2)^{-1} > 0$, where the core trade-off is captured by $\omega^\sharp \Is c_1\Theta - c_2\Theta^2$. 
The first term $c_1\Theta > 0$ represents global agglomeration forces, and the second term $-c_2\Theta^2 < 0$ captures global dispersion forces augmented by immobile demands. 
As in the Braid model, dispersion is preferred if transport costs are high (small $\phi$), and agglomeration occurs otherwise (large $\phi$). 

\Heading{The \cite{Tabuchi-JUE1998} model} 
Tabuchi integrated urban costs into Krugman's framework with immobile consumers. 
In addition to the regional-scale component of Krugman, each region has an Alonso--Muth--Mills monocentric city structure. 
Within each region, agents commute to a central business district and face the trade-off between commuting costs and land rent. 
The benefit matrix for the Tabuchi model reduces to 
\begin{align}
    \VtV = C(\VtD) \cdot \left( -c_0 \VtI + c_1 \VtD - c_2 \VtD^2 \right),
    \label{eq:V-Tabuchi}
\end{align}
where $ C(\VtD) $ is a matrix factor that captures general equilibrium effects, and 
$c_0 = \gamma \epsilon_1 > 0$, $
c_1 = \frac{\mu}{\sigma - 1} + \frac{\mu}{\sigma} > 0$, and $
c_2 = \frac{\mu^2}{\sigma - 1} \epsilon_2 + \frac{1}{\sigma} \epsilon_3$. 
Here, $\gamma$ is the share of housing expenditures, and $\epsilon_1, \epsilon_2, \epsilon_3$ capture the effects of urban costs. 
For example, the larger the commuting cost parameter and/or the opportunity cost of the land, the larger $\epsilon_1 > 0$ becomes, so that the local dispersion force $-c_0$ becomes more pronounced. 
The stability of $\BrVtx$ depends on the quadratic expression $\omega^\sharp \Is -c_0 + c_1 \Theta - c_2 \Theta^2$. 

The agglomeration force ($c_1 \Theta > 0$) is the same as the Krugman model, while $ -c_0 < 0$ captures the local dispersion force due to urban costs. 
The term $c_2 \Theta^2$ captures net global forces that include the impacts of urban costs through general equilibrium. 
The model features both local and global dispersion forces, and the symmetric equilibrium can be stable for both high and low levels of transport costs. 

\Heading{Idiosyncratic taste shocks}
Idiosyncratic preference shocks \citep{McFadden-Book1974,McFadden-Book1978,McFadden-TRB1978} are important both in the quantitative and theoretical literature \citep{Hunt-Simmonds-EPB1993,Waddell-JAP2002,Anas-Liu-JRS2007,Redding-Rossi-Hansberg-ARE2017,anderson1992discrete,Tabuchi-Thisse-JDE2002,Murata-JUE2003}. 
Regarding taste shocks, \cite{Behrens-Murata-DP2018} demonstrated that the spatial equilibrium condition in models with idiosyncratic shocks can be equivalently represented by that in homogeneous preference models with local non-tradables markets as in \cite{Helpman-Book1998}. 
Thus, from the perspective of stabilizing forces at $\BrVtx$, introducing idiosyncratic shocks into a model with homogeneous preference is equivalent to adding a negative constant to the utility gain~$\omega$, i.e., to embedding an additional local dispersion force. 
Welfare implications are, however, different and care should be taken \citep{Behrens-Murata-DP2018}. 

\subsection{The three model classes}
\label{sec:model-class}

As the above examples illustrate, for a wide class of spatial models, 
\begin{align}
    & \VtV = C(\VtD) \cdot \bigl(c_0 \VtI + c_1 \VtD + c_2 \VtD^2\bigr), \quad \text{and hence} \quad
    \\ 
    & \omega = C(\Theta) \cdot \bigl(c_0 + c_1 \Theta + c_2 \Theta^2\bigr), 
\end{align}
where $C$ and the coefficients $\{c_0, c_1, c_2\}$ are model dependent. 
In this representation, the stability of $\BrVtx$ is governed by the quadratic term, since the sign of $\omega$ is entirely determined by the sign of $c_0 + c_1 \Theta + c_2 \Theta^2$. 

\begin{figure}[t]
    \centering
    \begin{subfigure}[c]{.32\linewidth}
		\centering
        \begin{picture}(115,107)
            \put(0,0){\includegraphics{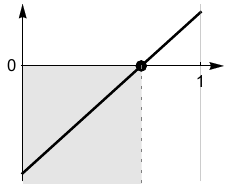}}
            \put(110,55){\footnotesize $\Theta$} 
            \put(25,98){\footnotesize \fbox{\parbox[c][3mm][c]{1.8cm}{\centering $c_0 + c_1 \Theta$}}}
            \put(-8,4){\footnotesize $-c_0$}
        \end{picture}
        \caption{Allen--Arkolakis (Type~L)}
    \end{subfigure}
    \hfill
    \begin{subfigure}[c]{.32\linewidth}
        \centering
        \begin{picture}(115,107)
            \put(0,0){\includegraphics{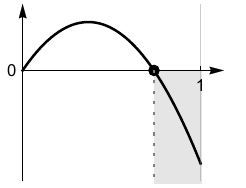}}
            \put(110,53){\footnotesize $\Theta$} 
            \put(25,98){\footnotesize \fbox{\parbox[c][3mm][c]{1.8cm}{\centering $c_1 \Theta + c_2 \Theta^2$}}}
        \end{picture}
        \caption{Krugman (Type~G)}
    \end{subfigure}
    \hfill
    \begin{subfigure}[c]{.32\linewidth}
		\centering
        \begin{picture}(115,107)
            \put(0,0){\includegraphics{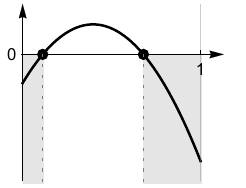}}
            \put(110,61){\footnotesize $\Theta$} 
            \put(10,98){\footnotesize \fbox{\parbox[c][3mm][c]{2.8cm}{\centering $c_0 + c_1 \Theta + c_2\Theta^2$}}}
            \put(-8,48){\footnotesize $-c_0$}
        \end{picture}
        \caption{Tabuchi (Type~LG)}
    \end{subfigure}
    \caption{Representative forms of the quadratic component of the utility gain} 
    \label{fig:G.three-classes}
    \FigureNote{For selected examples from the three model types, the ``net'' utility gain is plotted as a function of the proximity gain~$\Theta$. Larger $\Theta$ corresponds to smaller $\phi$. 
    In each panel, $\BrVtx$ is stable for $\Theta$ such that the curve lies below the horizontal axis. 
    The parameter values are chosen such that the stability of $\BrVtx$ depends on $\Theta$. 
    Local dispersion forces stabilize $\BrVtx$ for small $\Theta$, and global dispersion forces for large $\Theta$.} 
\end{figure}

\begin{table}[tb]
    \centering
    \makebox[\textwidth][c]{\footnotesize
    \begin{tabular}{ 
        >{\centering\arraybackslash}m{1.3cm} | >{\centering\arraybackslash}m{2.8cm} | >{\centering\arraybackslash}m{2.8cm} | m{7.2cm} }
        
        \hline
    
        \hline
        
        \thead{Model\\Class}
        & \thead{Stability of $\BrVtx$}
        & \thead{Dominant\\dispersion force}
        & \thead{Examples} 
        
        \\
        
        \hline
    
        \hline
    
        Type~L 
        & \shortstack[c]{Low\\transport costs}
        & Local 
        & \makecell[l]{
            \cite{Beckmann-Book1976}\\
            \cite{Helpman-Book1998}\\
            \cite{Murata-Thisse-JUE2005}\\
            \cite{Redding-Sturm-AER2008}\\
            \cite{Allen-Arkolakis-QJE2014}\\
            \cite{Redding-Rossi-Hansberg-ARE2017}, Section 3
        } \\

        \hline
    
        Type~G  
        & \shortstack[c]{High\\transport costs}
        & Global 
        &      
         \makecell[l]{
            \cite{Harris-Wilson-EPA1978}\\
            \cite{Krugman-JPE1991}\\
            \cite{Krugman-Venables-QJE1995}\\
            \cite{Puga-EER1999}, Section 3\\
            \cite{Forslid-Ottaviano-JoEG2003}\\
            \cite{Pfluger-RSUE2003}
        } 
        \\

        \hline
        Type~LG
        & \shortstack[c]{High and low\\transport costs}
        & \shortstack[c]{Both\\local and global}
        & \makecell[l]{
            \cite{Tabuchi-JUE1998}\\
            \cite{Fujita-Krugman-Venables-Book1999}, Section 14.4\\
                \cite{Puga-EER1999}, Section 4\\
                \cite{Pfluger-Suedekum-JUE2008}\\
                \cite{Pfluger-Tabuchi-RSUE2010}\\
                \cite{Kucheryavyy-etal-JIE2024}
            }
        \\
        \hline
    
        \hline
        \end{tabular}
        }\caption{Notable examples}
        \label{tab:model-class-examples}

    \FigureNote{The section numbers in the table correspond to those in the referenced papers. 
    For \cite{Krugman-Venables-QJE1995} and \cite{Puga-EER1999}, the spatial distribution of interest is the share of manufacturing sector. 
    In all the models in the table, the stability of $\BrVtx$ hinges on the sign of a quadratic function of the form $c_0 + c_1\Theta + c_2 \Theta^2$ with model-dependent coefficients $\{c_0,c_1,c_2\}$. 
    }
\end{table} 

For models of this form, we can define three prototypical classes according to the transport cost conditions under which the symmetric equilibrium is stable: low transport costs (small $\Theta$ or large $\phi$), high transport costs (large $\Theta$ or small $\phi$), or both. 
For convenience, we refer to the three model classes as \emph{Type~L}, \emph{Type~G}, and \emph{Type~LG}, where L and G stand for ``local''  and ``global,'' respectively.\footnote{See Appendix \ref{app:proof-classification} for formal definitions.
Here, we restrict our attention to cases in which the agglomeration forces are neither too weak nor too strong, so that the symmetric equilibrium is stable for some values of~$\phi$ and unstable for others.} 

With $\{c_k\}$ being model-dependent constants, we can summarize as follows:
\begin{itemize}
    \item 
    \textbf{Type~L} emphasizes local dispersion forces as the dominant dispersion mechanism. 
    The symmetric equilibrium is stable if transport costs are low, and unstable if transport costs are high. 
    $\VtV = C(\VtD)\cdot( c_0 \VtI + c_1 \VtD )$ with $c_0 < 0$ indicating the local dispersion force. 
    \item 
    \textbf{Type~G} emphasizes global dispersion forces as the dominant dispersion mechanism. 
    The symmetric equilibrium is stable if transport costs are high, and unstable if transport costs are low. 
    $\VtV = C(\VtD) \cdot( c_1 \VtD + c_2 \VtD^2)$, with $c_2 < 0$ representing the global dispersion force. 
    \item 
    \textbf{Type~LG} has both local and global dispersion forces as the dominant dispersion mechanisms. $\VtV = C(\VtD) \cdot (c_0 \VtI + c_1 \VtD + c_2 \VtD^2)$, where $c_0, c_2 < 0$ and $c_1 > 0$.  
    The symmetric equilibrium is stable for both high and low transport cost levels. 
\end{itemize}
\Cref{fig:G.three-classes} illustrates the typical shapes of the quadratic term $c_0 + c_1 \Theta + c_2 \Theta^2$ in each class, and \cref{tab:model-class-examples} lists representative examples. 

The model-dependent coefficients ${c_0,c_1,c_2}$ are functions of the model's structural parameters, with transport costs entering only through $\VtD$.
The constant $c_0$ captures the forces operating \emph{within} regions and therefore represents the local component.
The coefficients $c_1$ and $c_2$ represent the forces that operate \emph{across} regions.
The first-order term $c_1$ reflects direct interregional effects, such as agglomeration spillovers in the Beckmann model.
The quadratic term $c_2$ captures higher-order spillovers.\footnote{In many-region settings, $c_2$ reflects indirect interactions mediated by third regions. For example, in models with interregional trade, agents in the region $i$ are affected by the population of the region $j$ because agents in the region $i$ compete with those in the region $j$ for income generated in other regions $k = 1,2,\ldots$. Such effects typically arise second order in transport frictions, as they depend on the transport costs between regions $i$ and $k$, as well as between regions $j$ and $k$.}

Each coefficient $c_k$ in the ${c_0,c_1,c_2}$ representation captures the \emph{composite} general equilibrium effect associated with the corresponding order of $\VtD$.
Its sign therefore reflects the \emph{net} contribution to $\omega$ in the order of $\Theta$. 
For example, if there are both local agglomeration economies and diseconomies, the sign of $c_0$ reveals which force dominates: $c_0 > 0$ indicates net agglomeration, while $c_0 < 0$ indicates net dispersion.
In the Allen--Arkolakis framework, for example, one obtains $c_0 = \alpha - \beta - \frac{1 + \alpha}{\sigma}$, and $\alpha < \beta$ is a sufficient condition for the \emph{net} local effect $c_0$ to be negative.

\section{Many regions} 
\label{sec:N-region}

This section examines how the proposed taxonomy of spatial models maps to the \emph{endogenous spatial patterns} and their comparative statics in an $N$-region economy.  
All variables and functions (e.g., $\Vtx$, $\Vtv$, $[\phi_{ij}]$, $\VtD$) are straightforwardly extended. 
The set of regions is now denoted by $\ClI \Is \{1,2,\hdots,N\}$. 

\begin{figure}[tp]
    \centering
    \includegraphics[height=2.8cm]{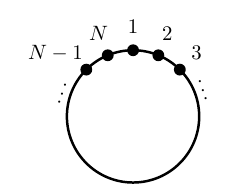}
    \caption{$N$-region symmetric circle.}
    \label{fig:RE}
\end{figure}

We focus on a stylized geography in which homogeneous regions are symmetrically placed over a circle and transport is possible only along the circumference (\cref{fig:RE}).    
\renewcommand{\theassumption}{C}
\begin{assumption}
\label{assum:racetrack-economy}
The proximity matrix is given by $\phi_{ij} = \phi^{\ell_{ij}}$, where $\phi\in(0,1)$ is the ease of transport between two consecutive regions, and $\ell_{ij} \Is \min\{|i - j|,N - |i - j|\}$ is the distance between regions $i$ and $j$ over the circumference. 
All regions are symmetric regarding their local fundamentals (e.g., innate amenity or productivity). 
In addition, $N$ is a multiple of four.\footnote{This restriction on $N$ is only for expositional simplicity. See \cref{remark:N} in \cref{app:proofs}.} 
\end{assumption}

This abstracts away the advantages from each region's unique geographic position: every region has the same level of geographic accessibility in a circle. 
Combined with the perfect symmetry in other regional fundamentals, the symmetric distribution $\BrVtx \Is (\frac{1}{N},\frac{1}{N},\hdots,\frac{1}{N})$ is always a spatial equilibrium (\cref{lem:xbar} in Appendix \ref{app:proof-classification}).

\subsection{The stability of the symmetric equilibrium}
\label{sec:agglomeration-in-RE} 

Endogenous agglomeration occurs when the symmetric equilibrium $\BrVtx$ is unstable, that is, when it does not withstand small migration shocks. 
In an economy with $N$ regions, such shocks can be represented as 
\begin{equation}
\BrVtx + \epsilon \Vtz
    = (\Brx + \epsilon z_1, \Brx + \epsilon z_2, \ldots, \Brx + \epsilon z_N)
\end{equation}
where $\epsilon$ is a sufficiently small scalar. 
The vector $\Vtz = (z_i)_{i\inI}$ is a \emph{deviation pattern}: $z_i > 0$ indicates an inflow into region~$i$ and $z_i < 0$ an outflow.  
Since the total population is fixed, we require $\sum_{i\inI} z_i = 0$. 
The $N$-region setting admits a substantially richer set of deviation patterns than the two-region case, because $(\Brx+\epsilon,\Brx-\epsilon)$ is the only possible perturbation in the two-region setting. Below, we normalize $\|\Vtz\| = 1$.

Analogous to the two-region case, the expected utility gains for marginal movers under migration shocks is closely related to the benefit matrix~$\VtV \Is \big[\PDF{v_i}{ x_j}(\BrVtx)\big]_{i\inI,j\inI}$. 
\begin{lemma}
\label{lem:gain-formula}
For any deviation pattern $\Vtz$ with $\sum_{i\inI} z_i = 0$ and $\|\Vtz\| = 1$, the expected utility 
gain of marginal movers is
$\omega(\Vtz)
    \Is 
    \Vtz^\top \VtV \Vtz$, where $\top$ denotes transpose.
\end{lemma}

Then, the value $\omega(\Vtz)$ characterizes the stability of $\BrVtx$.  
If $\omega(\Vtz) < 0$ for \emph{all} possible deviation patterns, then any migration shock reduces the utility of the movers and $\BrVtx$ is locally stable.  
If $\omega(\Vtz) > 0$ for \emph{some} $\Vtz$, the movers benefit from relocating according to that pattern, implying instability.  
Thus, stability is determined by whether the maximal attainable gain $\omega^* \Is \max_{\Vtz} \omega(\Vtz)$ is positive or negative.

To characterize $\omega^*$, let $\{\omega_k\}$ denote the eigenvalues of $\VtV$, and 
$\{\Vtz_k\}$ be their associated eigenvectors.  
Because $\omega(\Vtz_k) = \Vtz_k^\top \VtV \Vtz_k = \omega_k \Vtz_k^\top \Vtz_k = \omega_k \|\Vtz_k\|^2 = \omega_k$, we observe 
\begin{align}
    \omega^* = \max_k \{\omega_k\}, 
\end{align}
implying that the largest eigenvalue of $\VtV$ determines the stability of $\BrVtx$.\footnote{The discussion here is closely related to \cite{Allen-etal-AER2024}.
In their framework, the spectral radius of a matrix that collects key model elasticities governs the uniqueness of the equilibrium.
In our setting, their uniqueness condition [Theorem 1(a)] broadly corresponds to a sufficient condition for the stability of $\BrVtx$ for all $\phi$, which rules out endogenous agglomeration.
Our focus instead lies on environments with multiple equilibria [cf.\ their Theorem 1(c)], and we take a step toward understanding how the underlying network structure shapes the positive properties of these equilibria.} 

How can we obtain $\{\omega_k\}$? Under \cref{assum:racetrack-economy}, for all models in \cref{tab:model-class-examples}, the benefit matrix satisfies $\VtV = \Omega(\VtD)$, where the row-normalized proximity matrix $\VtD$ is replaced by its $N\times N$ counterpart. 
For each model, the base gain function $\Omega(\Theta)$ is the same as in the case of two-regions.\footnote{This is simply because, for all models we saw in \cref{sec:ge-models}, the gain functions are derived for the general number of locations under \cref{assum:racetrack-economy}, and  then specialized to the $N = 2$ case.}  
Then, $\VtV$ and $\VtD$ share the same set of eigenvectors $\{\Vtz_k\}$, and $\VtV = \Omega(\VtD)$ implies 
\begin{align}
    \omega_k = \Omega(\Theta_k), 
    \label{eq:omega_k}
\end{align}
where $\Theta_k$ is the eigenvalue of $\VtD$ corresponding to $\Vtz_k$. 
Each $\Theta_k$ is a function of $\phi$ because $\phi$ is the only parameter of $\VtD$ according to our assumptions. 

These observations yield a transparent approach to stability analysis, as the following example illustrates. 
\begin{example}
In the Beckmann model, we again have $\VtV = -\beta \VtI + \alpha \VtD$, where $\VtI$ and $\VtD$ are replaced by the $N$-region identity and row-normalized proximity matrices.\footnote{Observe that \cref{eq:v-Bm} is defined for general $N$.}  
The $k$th eigenvalue of $\VtV$ is then $\omega_k = -\beta + \alpha \Theta_k$ because $\VtD \Vtz_k = \Theta_k \Vtz_k$ implies $\VtV \Vtz_k = (-\beta + \alpha \Theta_k) \Vtz_k$.  
The stability of~$\BrVtx$ is determined by the sign of $\omega^* = \max_k \{\omega_k\}$. 
Consider a value of $\phi$ such that $\BrVtx$ is stable, i.e., $\omega_k < 0$ for all $k$ and hence $\omega^* = \max_k\{\omega_k\} < 0$. 
Suppose then that $\phi$ increases or decreases monotonically. 
Suppose \emph{some} $\omega_{k*}$ changes its sign from negative to positive at some $\phi^*$. 
Then, after that point, deviation towards the $\Vtz_{k^*}$-direction improves the utility of the movers, and $\BrVtx$ becomes unstable. 
In this sense, the corresponding eigenvector $\Vtz_{k^*}$ represents the critical deviation pattern.
\end{example}

\subsection{Proximity gains and deviation patterns}

The eigenvalues $\{\Theta_k\}$ of $\VtD$ have a clear interpretation analogous to the proximity gain $\Theta$ in the two-region case. 
Again, for the special case of the Beckmann model with $\beta = 0$ and $\alpha = 1$, the indirect utility function boils down to a simple proximity measure. 
We have $\VtV = \VtD$ and $\omega_k = \Theta_k$, which can be interpreted as follows. 
\begin{observation}
Each eigenvalue $\Theta_k$ of $\VtD$ measures the proximity gain experienced by marginal movers when migration shocks occur in the corresponding direction, $\Vtz_k$. 
\end{observation}

For example, if $N = 4$, the normalized proximity matrix under \cref{assum:racetrack-economy} is 
\begin{align}
\renewcommand{\arraystretch}{0.8}
    \VtD
    = 
    \frac{1}{1 + 2\phi + \phi^2}
     \begin{bmatrix}
     1 &\phi&\phi^2&\phi
     \\
     \phi &1&\phi&\phi^2
     \\
     \phi^2 &\phi &1&\phi
     \\
     \phi & \phi^2 & \phi & 1
     \end{bmatrix}. 
\end{align} 
There are two relevant eigenvalues: 
\begin{align}
    \Theta_1 = \frac{1 - \phi}{1 + \phi}
    \quad\text{and}\quad
    \Theta_2 = \left(\frac{1 - \phi}{1 + \phi}\right)^2. 
\end{align}

We can check that there are two eigenvectors associated with $\Theta_1$, namely $\Vtz_1^+ = (1, 0, -1, 0)$ and $\Vtz_1^- = (0, 1, 0, -1)$. 
Both represent \emph{monocentric} spatial patterns. 
For example, in $\BrVtx + \epsilon \Vtz_1^- = (\Brx, \Brx + \epsilon, \Brx, \Brx - \epsilon)$, one region grows at the expense of another that is two steps away, creating a single center of attraction. 
For $\Theta_2$, the associated eigenvector is $\Vtz_2 = (1, -1, 1, -1)$, and represents a \emph{polycentric} agglomeration pattern $\BrVtx + \epsilon \Vtz_2 = (\Brx + \epsilon, \Brx - \epsilon, \Brx + \epsilon, \Brx - \epsilon)$. 
Two regions located two steps apart grow symmetrically, while the two intermediate regions shrink. 

There are three intuitive properties about $\Theta_1$ and $\Theta_2$. 
First, both are positive. Any deviation from the full dispersion induces some form of agglomeration, and for movers this increases the proximity to others. 
Second, each $\Theta_k$ decreases as $\phi$ increases. 
The proximity gain decreases if transport costs are less important. 
Third, $\Theta_1 > \Theta_2$ at any value of $\phi$. 
Naturally, from the perspective of movers, deviation toward a monocentric pattern induces a greater proximity gain than toward a polycentric pattern. 

\begin{figure}[tb]
	\centering
	\begin{subfigure}[b]{.24\hsize}
		\centering
		\begin{overpic}[width=.85\hsize]{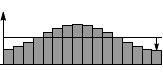}
            \put( -7,21){\footnotesize $\Brx$}
            \put( 33,-1){\scriptsize \text{Regions}}
            \put( -1,45){\footnotesize $x_i = \Brx + \epsilon z_{k,i}$}
            \put(101,18){\footnotesize $\epsilon z_{k,i}$}
        \end{overpic}
		\caption{$k = 1$\label{fig:16-z1} }
	\end{subfigure}
    \hfill
	\begin{subfigure}[b]{.24\hsize}
		\centering
		\includegraphics[width=.85\hsize]{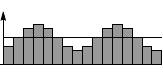}
		\caption{$k = 2$ \label{fig:16-z2} }
	\end{subfigure}
    \hfill
	\begin{subfigure}[b]{.24\hsize}
		\centering
		\includegraphics[width=.85\hsize]{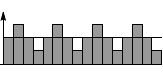}
		\caption{$k = 4$ \label{fig:16-z4} }
	\end{subfigure}
    \hfill
	\begin{subfigure}[b]{.24\hsize}
		\centering
		\includegraphics[width=.85\hsize]{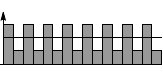}
		\caption{$k = 8$ \label{fig:16-z8} }
	\end{subfigure}
    
	\caption{Examples of spatial patterns generated by $\Vtz_k$ ($N = 16$).}
	\label{fig:16-region-patterns}
\end{figure}
These properties generalize to the $N$-region case.\footnote{\cite{Akamatsu-Takayama-Ikeda-JEDC2012}, Lemma 4.2, provides the analytical formulae for $\{\Theta_k\}$ and $\{\Vtz_k\}$, while the interpretation of $\{\Theta_k\}$ as proximity gains is newly given in the present study. \cref{lem:D_eigenpairs} in Appendix \ref{app:proofs} reproduces the relevant part of the aforementioned lemma, adapted to our context.} 
All $\{\Theta_k\}$ are positive and decrease in $\phi$. 
Each $\Vtz_k$ corresponds to a deviation pattern with $k$ symmetric peaks as illustrated in 
\cref{fig:16-region-patterns}.  
Most importantly, the maximum and minimum proximity gains are unambiguously determined: for each $\phi$, \begin{align}
    & \max_{k} \Theta_k = \Theta_1 
    \quad \text{and} \quad 
    \min_k \Theta_k = \Theta_{\frac{N}{2}}. 
    \label{eq:min-max_Theta}
    & 
\end{align}
Intuitively, at any value of $\phi$, the monocentric agglomeration pattern ($k = 1$, \cref{fig:16-z1}) yields the largest proximity gain for the movers. 
The proximity gain is the smallest if the movers agglomerates in every other region ($k = \frac{N}{2}$, \cref{fig:16-z8}). 

\subsection{Contrasting implications for spatial patterns}
\label{sec:local-disp-force} 

\begin{figure}[tb]
	\centering
    \begin{subfigure}[b]{.45\hsize}
        \centering
        \begin{overpic}[height=4cm]{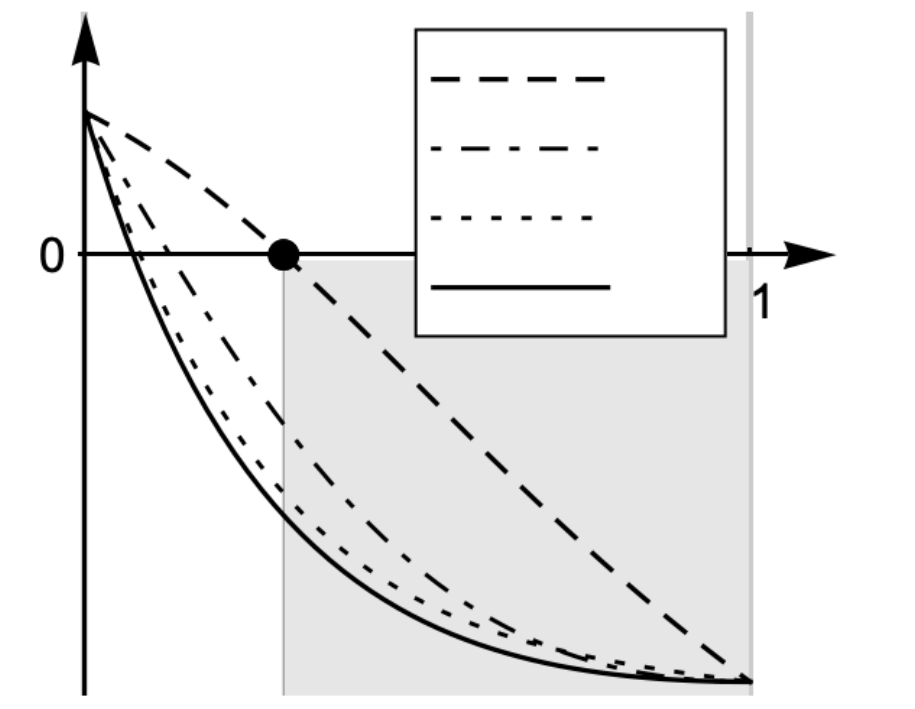}
            \put(93,50){$\phi$}
            \put(70,68){\scriptsize $\omega_1$}
            \put(70,60){\scriptsize $\omega_2$}
            \put(70,52){\scriptsize $\omega_3$}
            \put(70,44){\scriptsize $\omega_4$}
            \put(31,56){$\phi^*$}
        \end{overpic}
        \caption{The Beckmann model (Type~L)}
        \label{fig:Bm.omega.16}
    \end{subfigure}
    \begin{subfigure}[b]{.45\hsize}
        \centering
        \begin{overpic}[height=4cm]{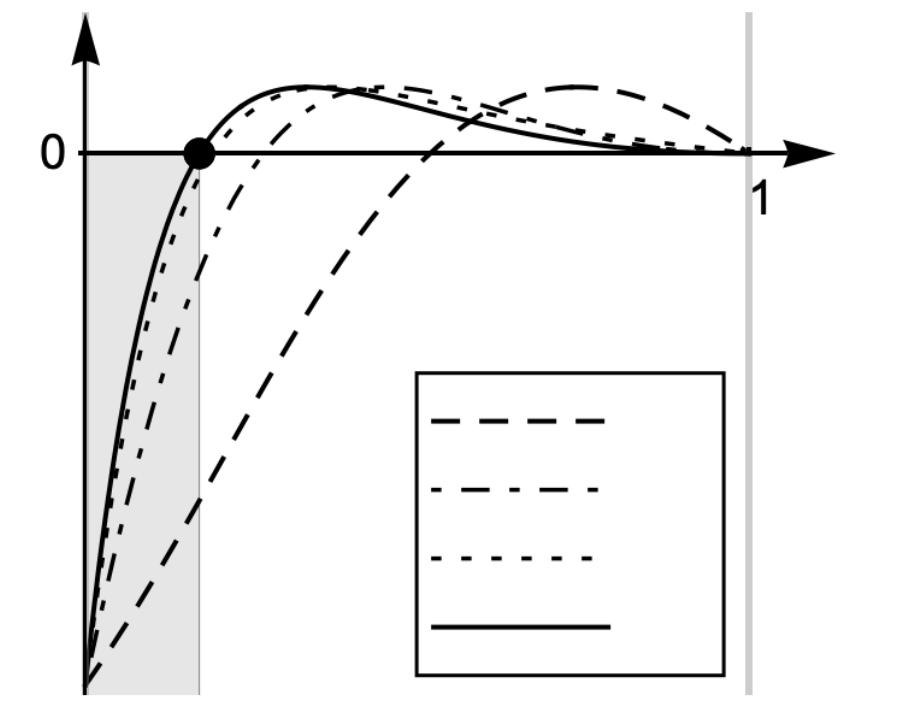}
            \put(93,60){$\phi$}
            \put(70,30){\scriptsize $\omega_1$}
            \put(70,22.1){\scriptsize $\omega_2$}
            \put(70,14.2){\scriptsize $\omega_3$}
            \put(70,6.3){\scriptsize $\omega_4$}
            \put(14,65){$\phi^*$}
        \end{overpic}
        \caption{The Braid model (Type~G)}
        \label{fig:Br.omega.8}
    \end{subfigure}
	\caption{Curves of $\omega_k = \Omega(\Theta_k)$ for the two minimal models ($N = 8$, $k = 1,2,3,4$).}
    \label{fig:Bm-Br.omega.16}
    \FigureNote{The symmetric equilibrium is stable if all curves $\{\omega_k\}_{k = 1}^4$ stay below the horizontal axis (gray region). It becomes unstable at $\phi^*$ where the largest eigenvalue cuts the axis. In the Beckmann model, $\omega_1$ is the first to cross the axis, whereas in the Braid model, it is $\omega_4$. }
\end{figure}

The relationship \eqref{eq:min-max_Theta} has important implications for endogenous spatial patterns. 
This is because the \emph{maximum} eigenvalue of $\VtD$ is critical if the model incorporates only local dispersion forces. 
For instance, since $\omega_k = - \beta + \Theta_k$ in the Beckmann model, we see 
\begin{align}
    \omega^* 
    < 
    0 
    \quad\Leftrightarrow \quad 
    \max_k 
    \left\{
    - \beta + \alpha \Theta_k 
    \right\} 
    = 
    - \beta 
    + 
    \alpha 
    \max_k 
    \left\{
        \Theta_k 
    \right\} 
    < 
    0 
    \quad\Leftrightarrow \quad 
    \Theta_1 
    < 
    \frac{\alpha}{\beta}.  
\end{align}
\Cref{fig:Bm.omega.16} illustrates this for the $N = 8$ case, where we draw curves of $\omega_k$ as a function of $\phi$. 
The symmetric equilibrium is stable if $\Theta_1$ is sufficiently small, i.e., if $\phi$ is sufficiently large. 
If $\phi$ monotonically decreases from a high level and crosses $\phi^*$, then $\BrVtx$ becomes unstable at $\phi^*$ because a deviation of the form $\BrVtx + \epsilon \Vtz_1$ induces a positive utility gain for the movers. Monocentric agglomeration should form at such a point. 

By contrast, if the model has only global dispersion forces, the \emph{minimum} eigenvalue of $\VtD$ is critical. 
In the Braid model, $\omega_k = \alpha \Theta_k  - \Theta_k^2 = \Theta_k (\alpha - \Theta_k)$, which implies that 
\begin{align}
    \omega^* 
    <  
    0 
    \quad\Leftrightarrow \quad 
    \max_k 
    \left\{
    \alpha - \Theta_k 
    \right\} 
    = 
    \alpha 
    -
    \min_k 
    \left\{
     \Theta_k 
    \right\} 
    < 
    0 
    \quad\Leftrightarrow \quad 
    \alpha
    < 
    \Theta_{\frac{N}{2}}.  
    \label{eq:Br-omegamax}
\end{align}
\Cref{fig:Br.omega.8} illustrates this for the $N = 8$ case. 
The symmetric equilibrium is stable if $\Theta_{\frac{N}{2}}$ is sufficiently large, that is, if $\phi$ is sufficiently small. 
If $\phi$ gradually increases from a very small value, at  $\phi^*$, the $\frac{N}{2}$-centric deviation pattern $\Vtz_\frac{N}{2}$ becomes attractive for movers.

The two minimal examples show that the \emph{spatial pattern} emerging at the onset of instability is critically dependent on the spatial scale of the dispersion force in the model.
In particular, polycentric patterns arise only when the dispersion force operates at a global scale. 
This insight from the reduced-form models can be extended to cover general equilibrium models discussed in \cref{sec:ge-models}: 
\begin{proposition}
    \label{prop:classification}
    Suppose \cref{assum:racetrack-economy}.
    \begin{enumerate}
        \item Consider a model of Type~L or LG with local dispersion forces. 
        Then, the symmetric equilibrium $\BrVtx$ is stable for large $\phi$. 
        Suppose that the model parameters are set so that $\BrVtx$ is stable for all $\phi > \phi^{*}$ with some threshold value $\phi^{*}\in(0,1)$, and becomes unstable at $\phi^{*}$, i.e., $\BrVtx$ is unstable for $\phi$ slightly smaller than $\phi^*$. 
        Then, a single-peaked monocentric spatial equilibrium path branches from $\BrVtx$ at $\phi^{*}$. 
        \item Consider a model of Type~G or LG with global dispersion forces. 
        Then, $\BrVtx$ can be stable for small $\phi$. 
        Suppose that the model parameters are set so that $\BrVtx$ is stable for all $\phi < \phi^*$ with some threshold value $\phi^*\in(0,1)$, and becomes unstable at $\phi^*$, i.e., $\BrVtx$ is unstable for $\phi$ slightly larger than $\phi^*$. 
        Then, a polycentric spatial equilibrium path with $\frac{N}{2}$ symmetric peaks branches from $\BrVtx$ at $\phi^*$. 
    \end{enumerate}
\end{proposition}

\cref{prop:classification} considers settings in which multiple equilibria may arise, while $\BrVtx$ remains stable for some values of $\phi$.
This condition need not always hold.
For example, in a Type~L model with sufficiently strong local dispersion forces, $\BrVtx$ is stable for all $\phi \in (0,1)$, and the threshold $\phi^*$ in \cref{prop:classification}~(b) does not exist.
In such cases, the equilibrium is typically unique, a convenient feature that makes unambiguous counterfactual analysis feasible in quantitative spatial models. 
Equilibrium uniqueness implies that, absent exogenous geographical asymmetries, $\BrVtx$ is the only possible outcome.
At the opposite extreme, in Type~G models with sufficiently strong agglomeration forces, $\BrVtx$ can be unstable for all $\phi$, leading all agents to concentrate in a single location.
\cref{prop:classification} deliberately excludes both of these extremal cases. 

\subsection{Evolution of spatial patterns\label{sec:evolution}}

Because \cref{prop:classification} relies on a local stability analysis around $\BrVtx$, it does not establish whether an instability toward polycentric deviations actually leads to stable polycentric equilibria. 
To trace how stable spatial equilibria evolve as transport costs change beyond \cref{prop:classification}, one must specify an indirect utility function, and the resulting characterizations are therefore model dependent.\footnote{For example, \cite{Kucheryavyy-etal-JIE2024} focused on a specific but flexible two-region model that encompasses \cite{Allen-Arkolakis-QJE2014} and \cite{Krugman-JPE1991} as special cases. They essentially showed that the agricultural sector \`a la \cite{Krugman-JPE1991} produces a global dispersion force that stabilizes the symmetric equilibrium at high transport costs, which is consistent with our results.}

\begin{figure}[tb]
	\centering
    \begin{subfigure}[b]{.48\hsize}
        \centering
        \includegraphics[height=6.5cm,clip]{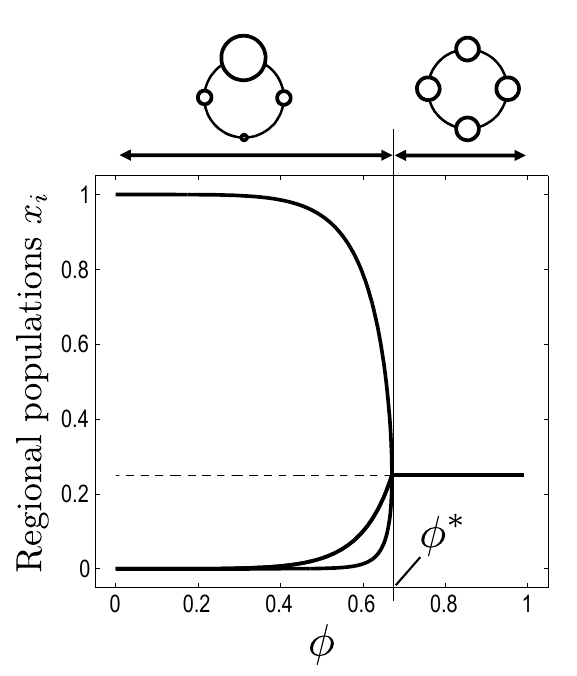}
        \caption{The Helpman model (Type~L) \label{fig:Hm} }
    \end{subfigure}
    \begin{subfigure}[b]{.48\hsize}
        \centering
        \includegraphics[height=6.8cm,clip]{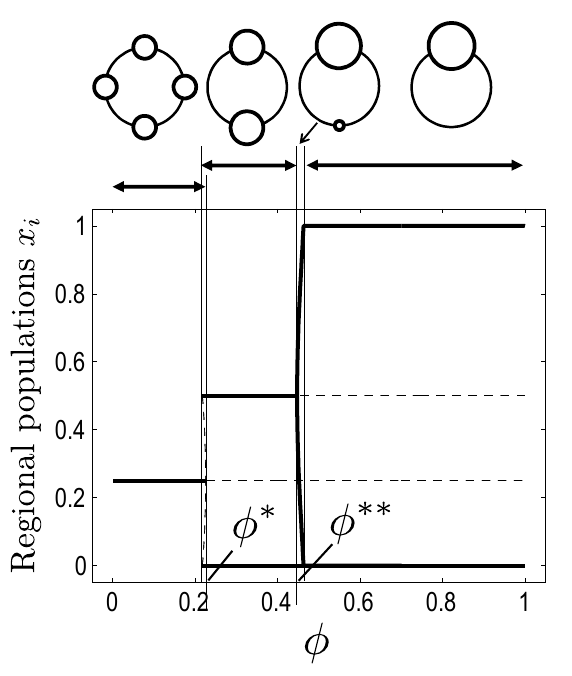}
        \caption{The Forslid--Ottaviano model (Type~G) \label{fig:FO} }
    \end{subfigure}
    \caption{Numerical illustration of \cref{prop:class-I,prop:class-II}.}
    \label{fig:FO-Hm}
\end{figure}

Nonetheless, in specific models, we can show that polycentric spatial patterns become stable only when global dispersion forces are sufficiently strong, consistent with \cref{prop:classification}.
This is illustrated by the following formal results and \cref{fig:FO-Hm}.  
\begin{proposition}
    \label{prop:class-II}
Consider the Type~L model of \cite{Helpman-Book1998} on a symmetric four-region circle. 
Suppose that the full dispersion $\BrVtx = (\tfrac{1}{4},\tfrac{1}{4},\tfrac{1}{4},\tfrac{1}{4})$ is stable for large $\phi$, but that multiple equilibria may exist. 
Then there exists a threshold $\phi^* \in (0,1)$ such that $\BrVtx$ is unstable for all $\phi \in (0,\phi^*)$. 
For this range, any duocentric equilibrium of the form $(m_1,m_2,m_1,m_2)$ with $m_1 > m_2$, if it exists, is also unstable, implying that all stable equilibria must be single-peaked. 
For $\phi \in (\phi^*,1)$, the fully dispersed allocation $\BrVtx$ is stable, and a single-peaked equilibrium path connects to $\BrVtx$ at $\phi^*$ [cf.\@ \cref{prop:classification}~(a)].
\end{proposition}
\begin{proposition}
    \label{prop:class-I}
Consider the Type~G model by \cite{Forslid-Ottaviano-JoEG2003} on a four-region symmetric circle.  
Assume that the full dispersion $\BrVtx = (\frac{1}{4},\frac{1}{4},\frac{1}{4},\frac{1}{4})$ is stable for small $\phi$. 
Suppose that the initial state is $\BrVtx$, and $\phi$ increases monotonically from $0$. 
At some threshold $\phi^*$, $\BrVtx$ becomes unstable, and duocentric spatial equilibrium of the form $(m_1,m_2 ,m_1,m_2)$ with $m_1 > m_2$ branches from $\BrVtx$ [cf.\@ \cref{prop:classification}~(b)]. 
In particular, the new stable equilibrium is the duocentric agglomeration such as $(\frac{1}{2},0,\frac{1}{2},0)$. 
At some $\phi^{**} > \phi^*$, the duocentric equilibrium becomes unstable. 
Finally, the full agglomeration in a single region such as $(1,0,0,0)$ becomes the stable spatial equilibrium for large $\phi$. 
\end{proposition}
\begin{proof}[Proof of \cref{prop:class-I,prop:class-II}]
See \cite{AMT-2016}. 
\end{proof}

In \cref{fig:FO-Hm}, the schematic diagrams above show the corresponding spatial patterns. 
In particular, in the Helpman model, a monocentric distribution of the form $\Vtx = (m_2, m_1, m_2, m_3)$ with $m_1 > m_2 > m_3$ is the stable equilibrium for $\phi \in (0,\phi^*)$. 
It converges to the full dispersion at $\phi^*$, which is consistent with \cref{prop:classification} (a). 

\begin{figure}[tb]
    \centering
    \begin{subfigure}[b]{.32\hsize}
        \centering
        \begin{overpic}[width=5cm, height=6.5cm]{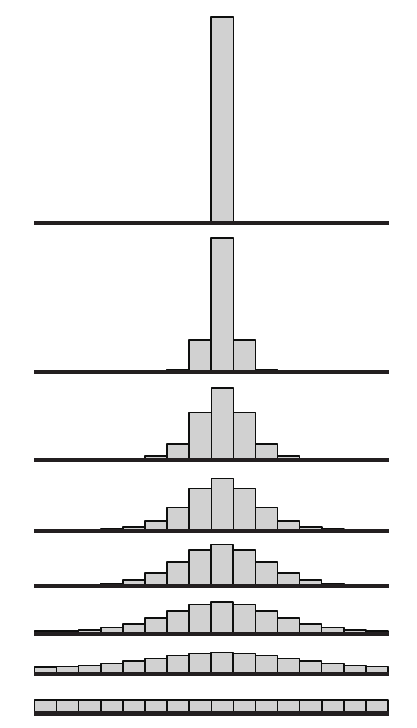}
            \put(0.5,100){\vector(0,-1){100}}
            \put( -4,  3){\rotatebox{90}{\scriptsize  $\phi\approx 1$}}
            \put( -4, 90){\rotatebox{90}{\scriptsize  $\phi\approx 0$}}
        \end{overpic}
        \caption{Type~L \label{fig:Hm-16}}
    \end{subfigure}
    \hskip 2mm
    \begin{subfigure}[b]{.32\hsize}
        \hfill 
        \includegraphics[width=5cm, height=6.5cm]{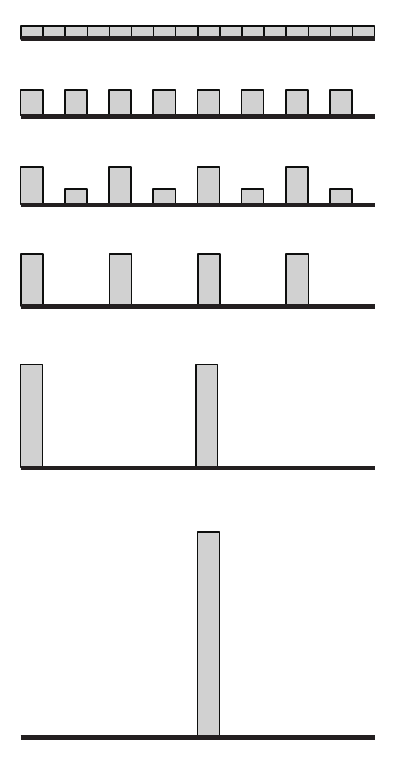}
        \caption{Type~G\label{fig:Km-16} }
    \end{subfigure}
    \begin{subfigure}[b]{.32\hsize}
        \centering
        \includegraphics[width=5cm, height=6.5cm]{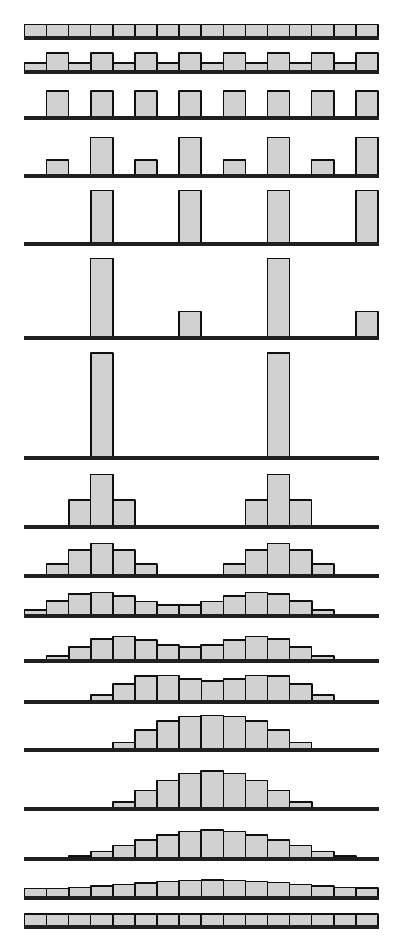}
        \caption{Type~LG \label{fig:PS-16} }
    \end{subfigure}
    \caption{Stable spatial patterns at different transport cost levels.}
    \label{fig:16-evolution-3-classes}
    \FigureNote{The spatial distributions in the circular economy are visualized as if it is on a line segment. In each figure, the leftmost region is neighboring to the rightmost one under \cref{assum:racetrack-economy}.}
\end{figure}

As a further numerical example, \Cref{fig:16-evolution-3-classes} shows the typical evolution of spatial patterns, assuming $N = 16$.
We consider a monotonic increase in $\phi$ (i.e., a monotonic decrease in transport costs), and follow a path of stable spatial equilibria. 

\Cref{fig:Hm-16} considers the Type L model by \cite{Helpman-Book1998}.
The symmetric equilibrium $\BrVtx$ is unstable if $\phi$ is small and the agents concentrate around a single peak.
As $\phi$ increases, the monotonic spread of the single-peaked distribution occurs. 
At a critical level of $\phi$, the spatial distribution converges to $\BrVtx$ [\cref{prop:classification}~(a)].
In particular, stable equilibria are single-peaked throughout the process.  

\Cref{fig:Km-16} considers the Type~G model of \cite{Krugman-JPE1991}.
When $\phi$ is low, the symmetric equilibrium is stable.
As $\phi$ increases, an endogenous transition occurs, and a polycentric equilibrium with eight agglomerations becomes stable [cf.\@ \cref{prop:classification}~(b)].
Further increase in $\phi$ triggers successive instabilities: the number of agglomerations falls, the spacing between them widens, and each remaining center grows larger (cf.\@ \cref{prop:class-I}).
As a result, the number of centers in the stable equilibrium evolves as $16 \to 8 \to 4 \to 2 \to 1$.
In Type G models, spatial adjustment generates both winners and losers. Centers that initially grow may later decline as larger agglomerations expand at their expense.
For instance, in \cref{fig:Km-16}, the fifth region from the left initially gains population as $\phi$ increases but eventually loses population. 

\Cref{fig:PS-16} considers a many-region version of the Type~LG model by \cite{Pfluger-Suedekum-JUE2008}.
The symmetric equilibrium is stable if $\phi$ is small.
As $\phi$ increases, eight-centric agglomerations emerge at some point, as in Type~G.
Multiple bell-shaped agglomerations are generated at moderate $\phi$. 
Increasing $\phi$ further causes a decrease in the number of agglomerations and the spread of each agglomeration.
When $\phi$ is close to one, the economy becomes monocentric, as in Type~L. 
Notably, in the large $\phi$ regime, the model exhibits a transformation from a two-peaked to a single-peaked pattern accompanied by local spreading. 
This broadly resembles the dual evolution of cities discussed in \cref{sec:introduction}.\footnote{While the real-world dynamics appear to unfold simultaneously, the model generates these changes somewhat sequentially: first through a reduction in the number of peaks, and then through the flattening of the remaining agglomeration. 
This discrepancy may reflect limitations of the model, the most fundamental issues being the absence of inter-location commuting and dynamic decisions.}

\section{Asymmetries}

Real-world geography differs markedly from the stylized benchmarks discussed so far.
Geographic accessibility and other region-fixed attributes can vary between locations, generating distortions absent in idealized settings.
Yet, as a brief exploration in this section demonstrates, these exogenous asymmetries do not alter the core insights: the spatial scale of dominant dispersion mechanisms can fundamentally shape the spatial distribution and their response to transport costs.

\begin{figure}[t]
    \centering
    \begin{subfigure}[c]{.45\hsize}
    \centering
    \includegraphics[width=4.5cm, bb=0 0 715 365]{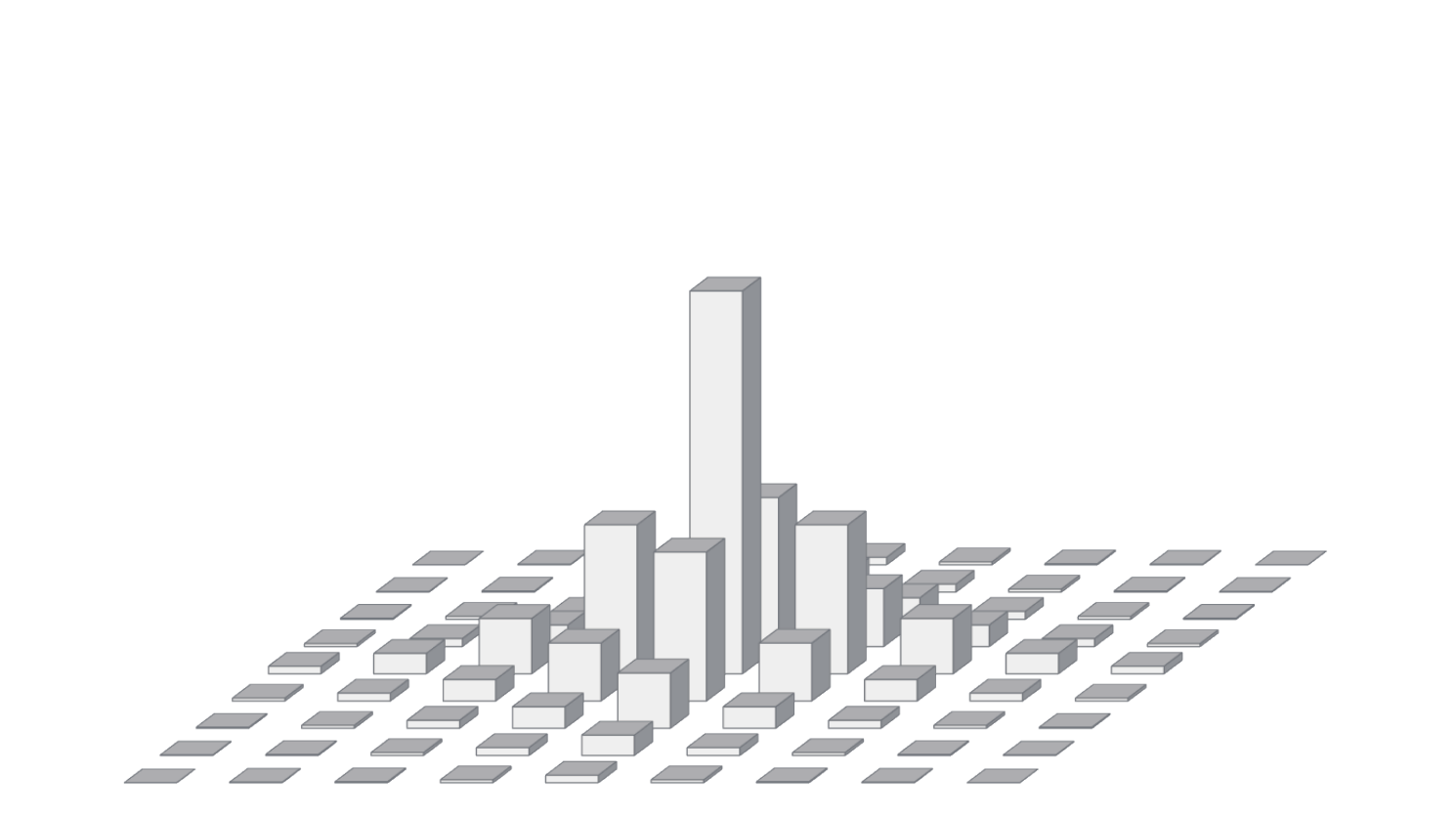}
    
    \includegraphics[width=4.5cm, bb=0 0 715 365]{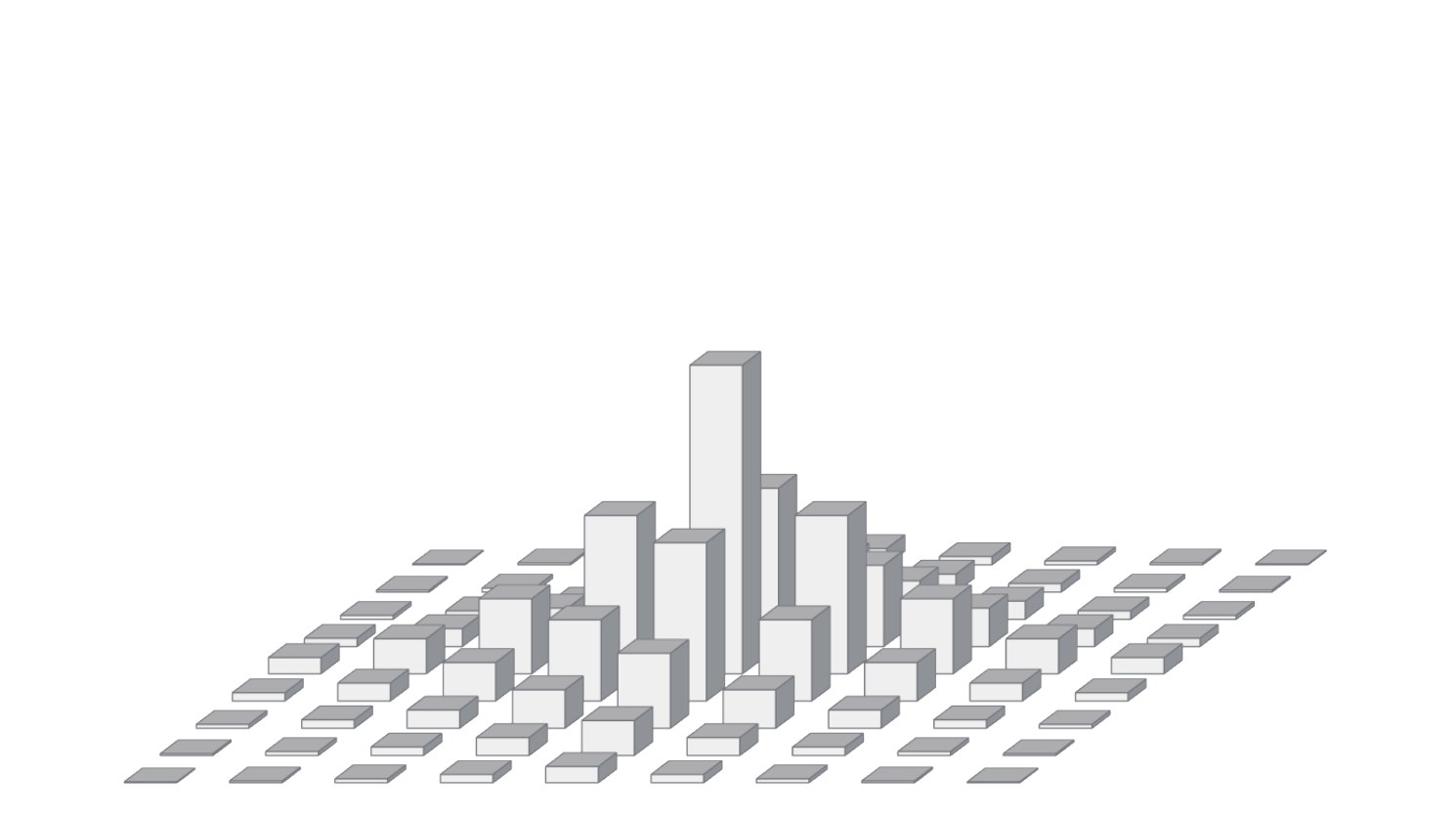}
    
    \includegraphics[width=4.5cm, bb=0 0 715 365]{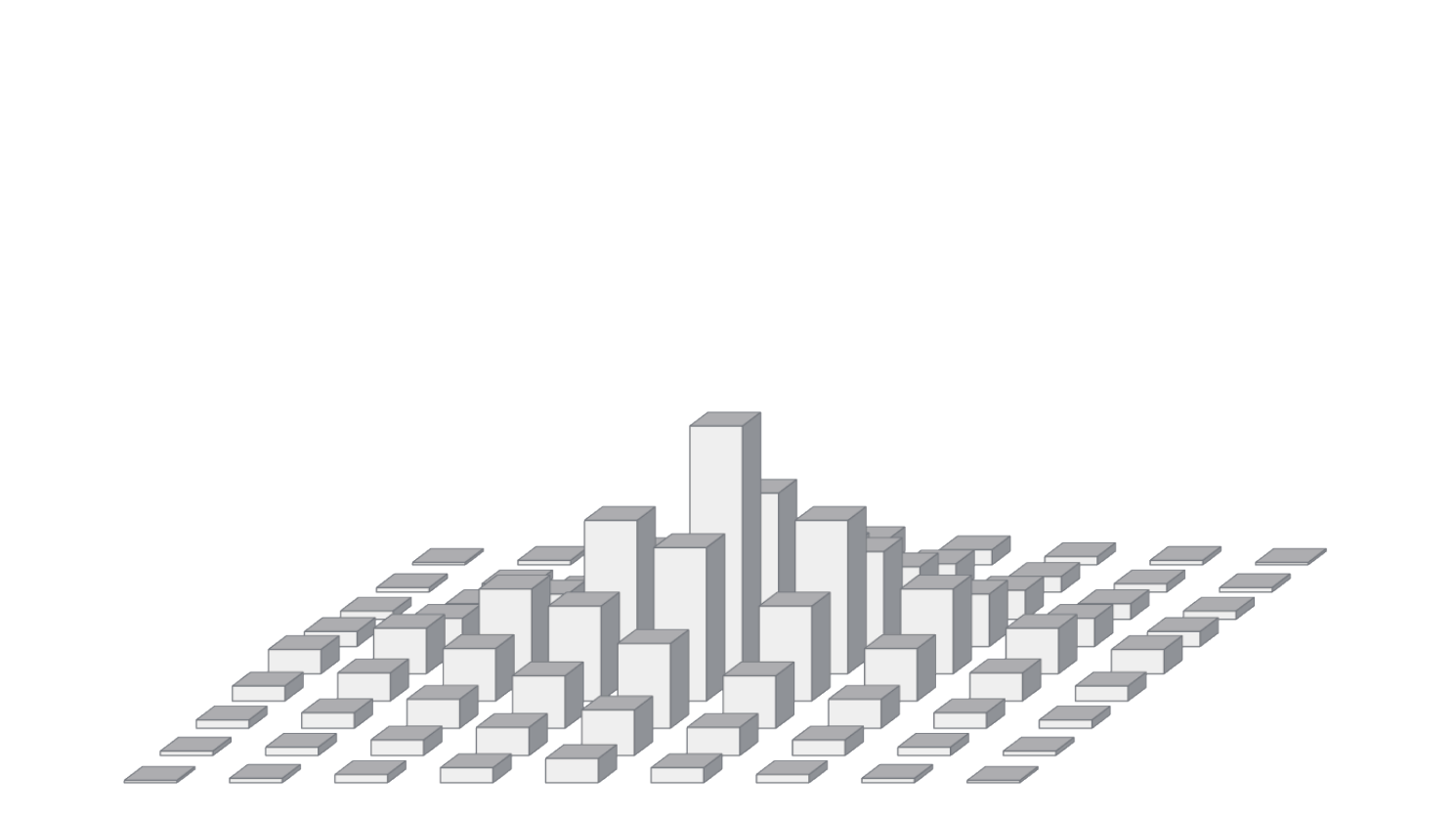}
    
    \includegraphics[width=4.5cm, bb=0 0 715 365]{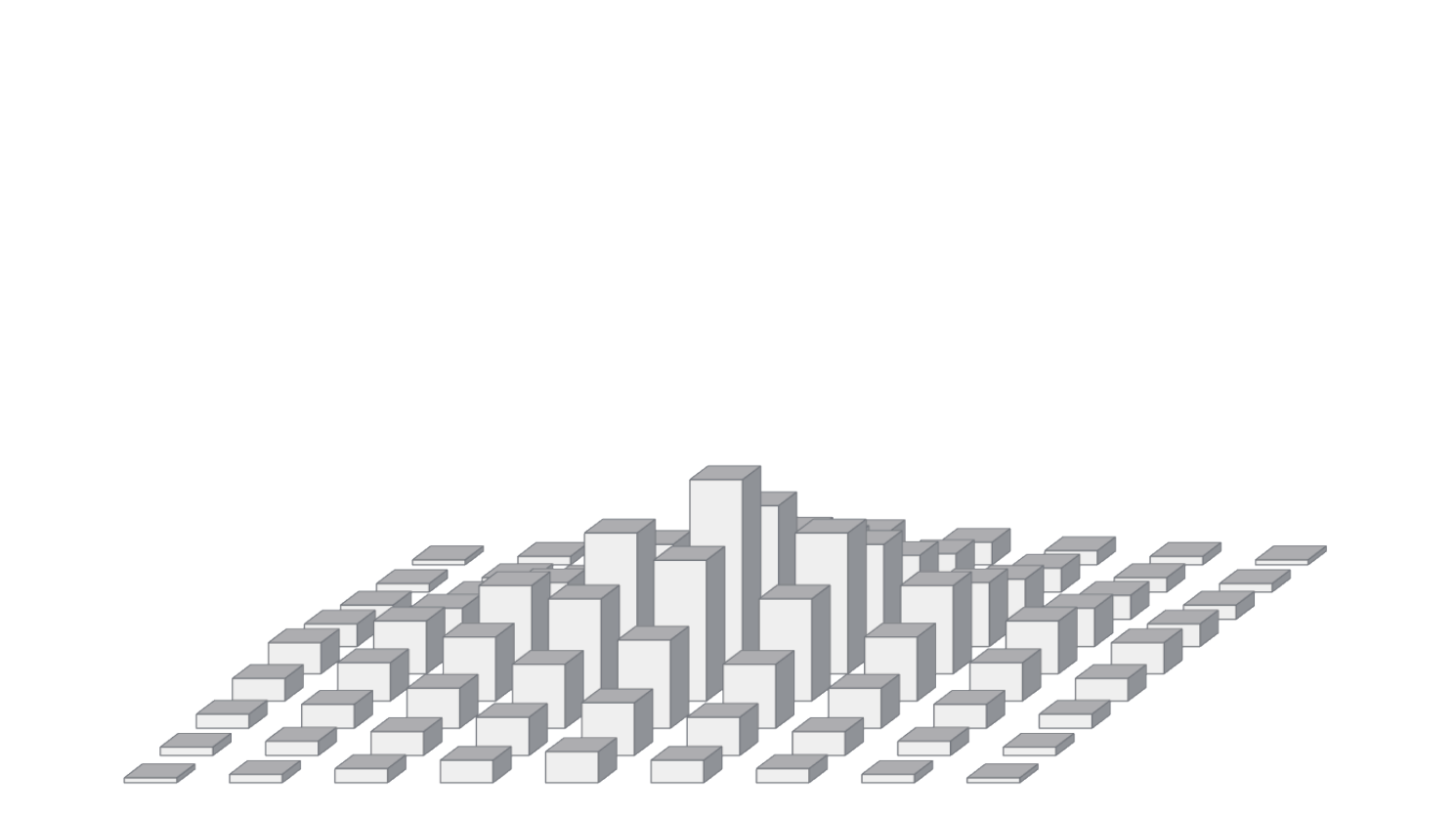}
    
    \begin{overpic}[width=4.5cm, bb=0 0 715 365]{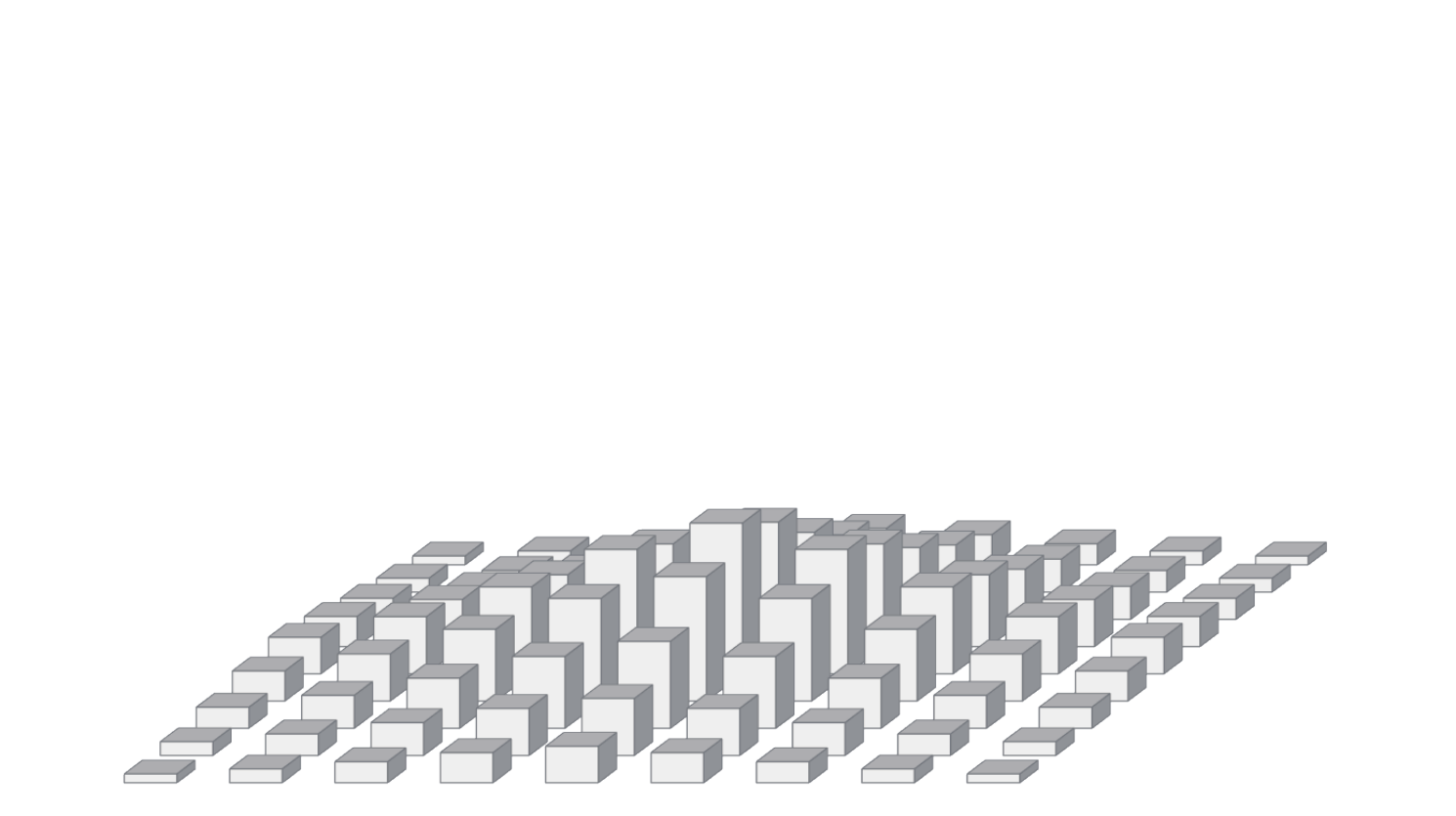}
        \put(0.5,240){\vector(0,-1){240}}
        \put(-10, 180){\rotatebox{90}{\scriptsize High transport costs}}
        \put(-10,   5){\rotatebox{90}{\scriptsize Low transport costs}}
    \end{overpic}
    \caption{The Allen--Arkolakis model (Type L)}
    \end{subfigure}
    \hskip 2em 
    \begin{subfigure}[c]{.45\hsize}
    \centering
    \includegraphics[width=4.5cm, bb=0 0 715 365]{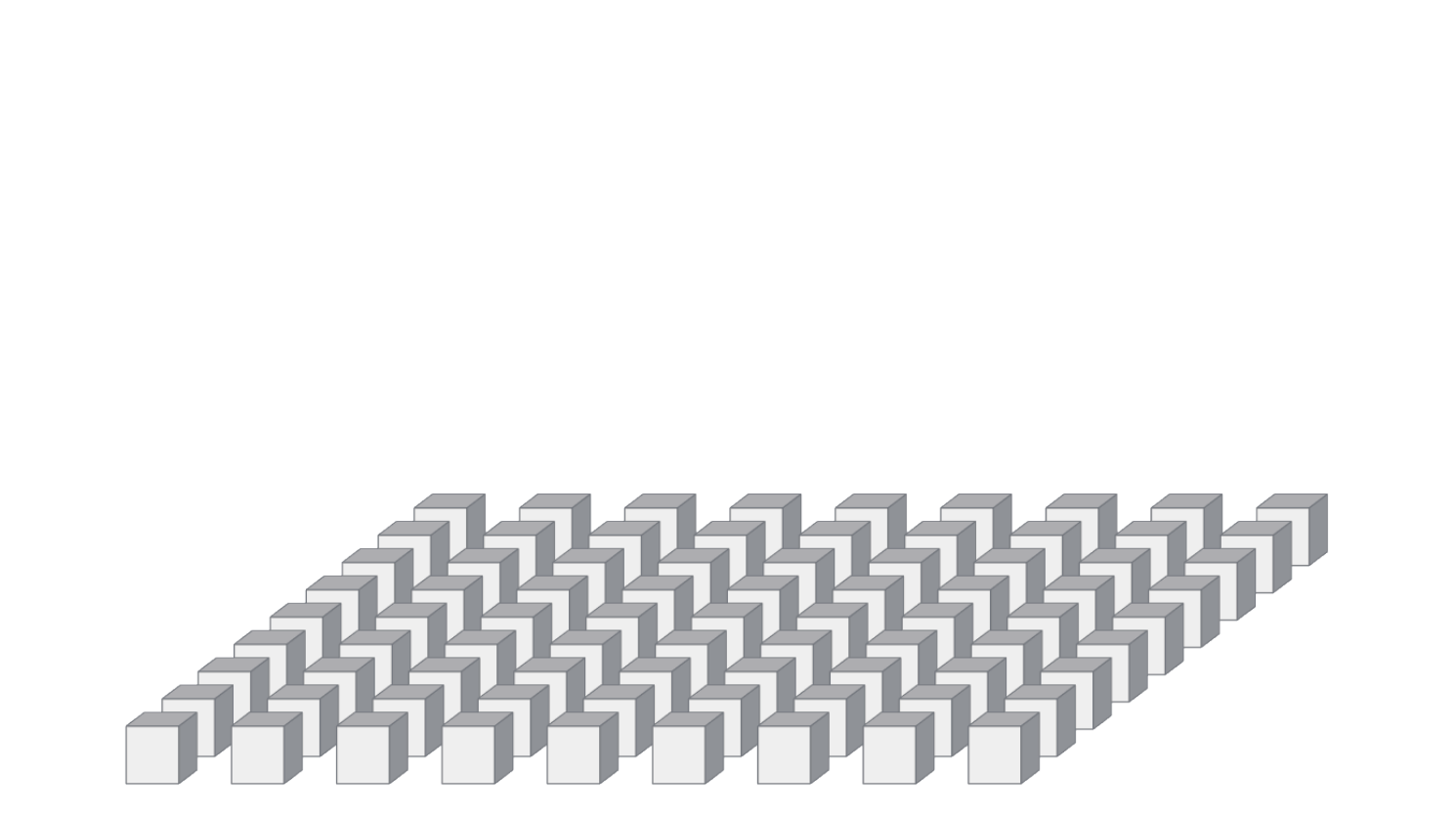}
    
    \includegraphics[width=4.5cm, bb=0 0 715 365]{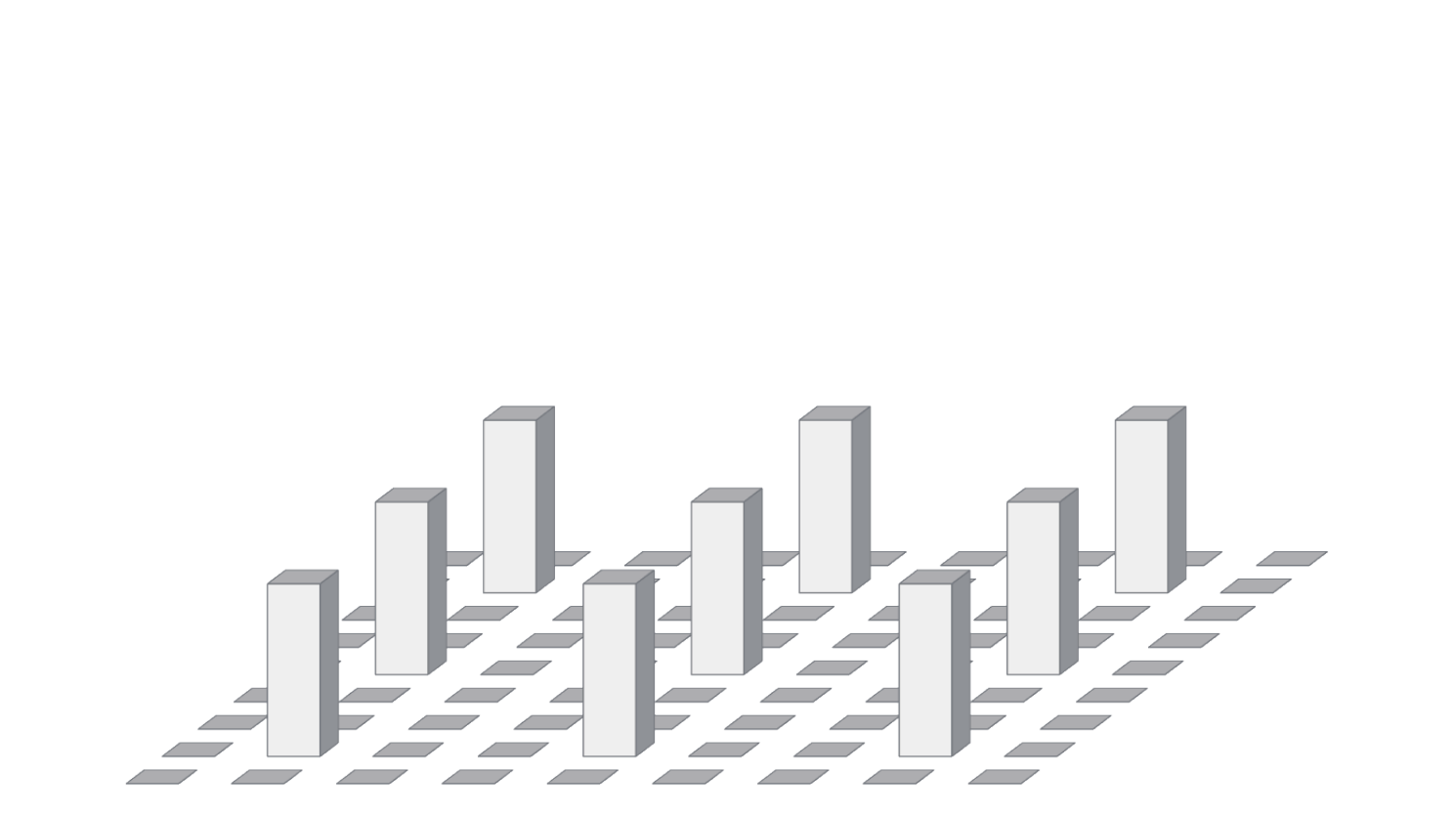}
    
    \includegraphics[width=4.5cm, bb=0 0 715 365]{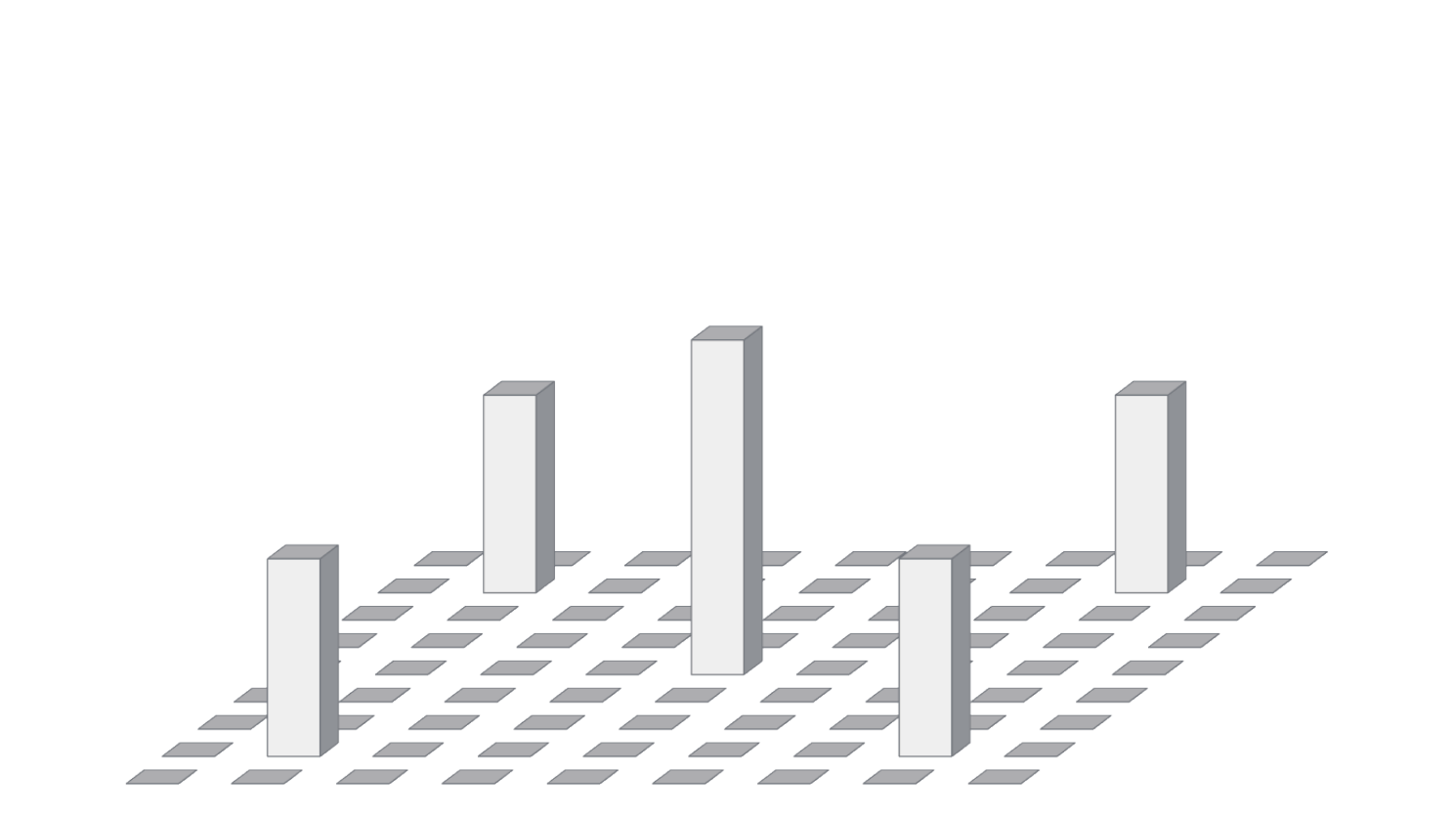}
    
    \includegraphics[width=4.5cm, bb=0 0 715 365]{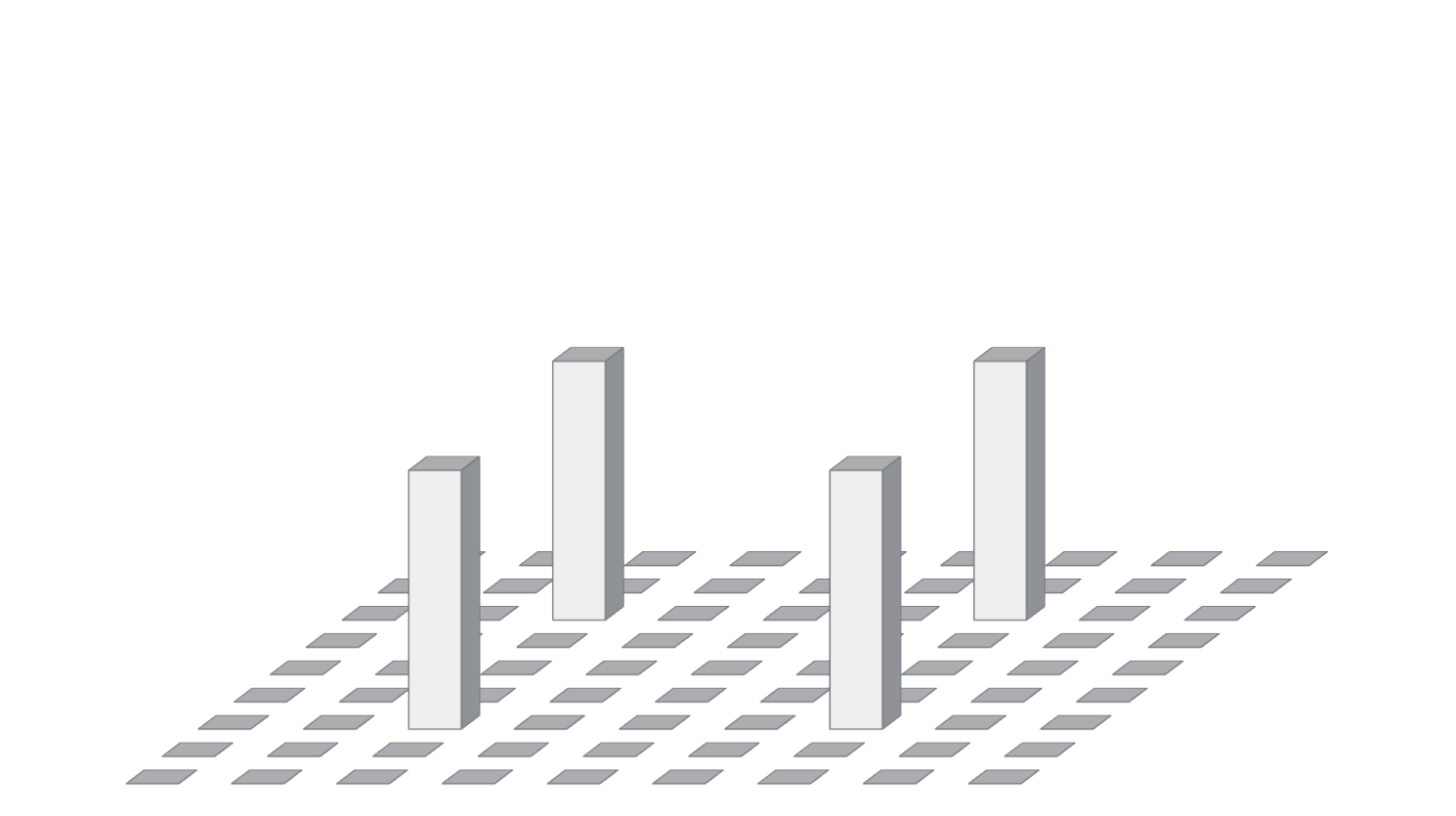}
    
    \includegraphics[width=4.5cm, bb=0 0 715 365]{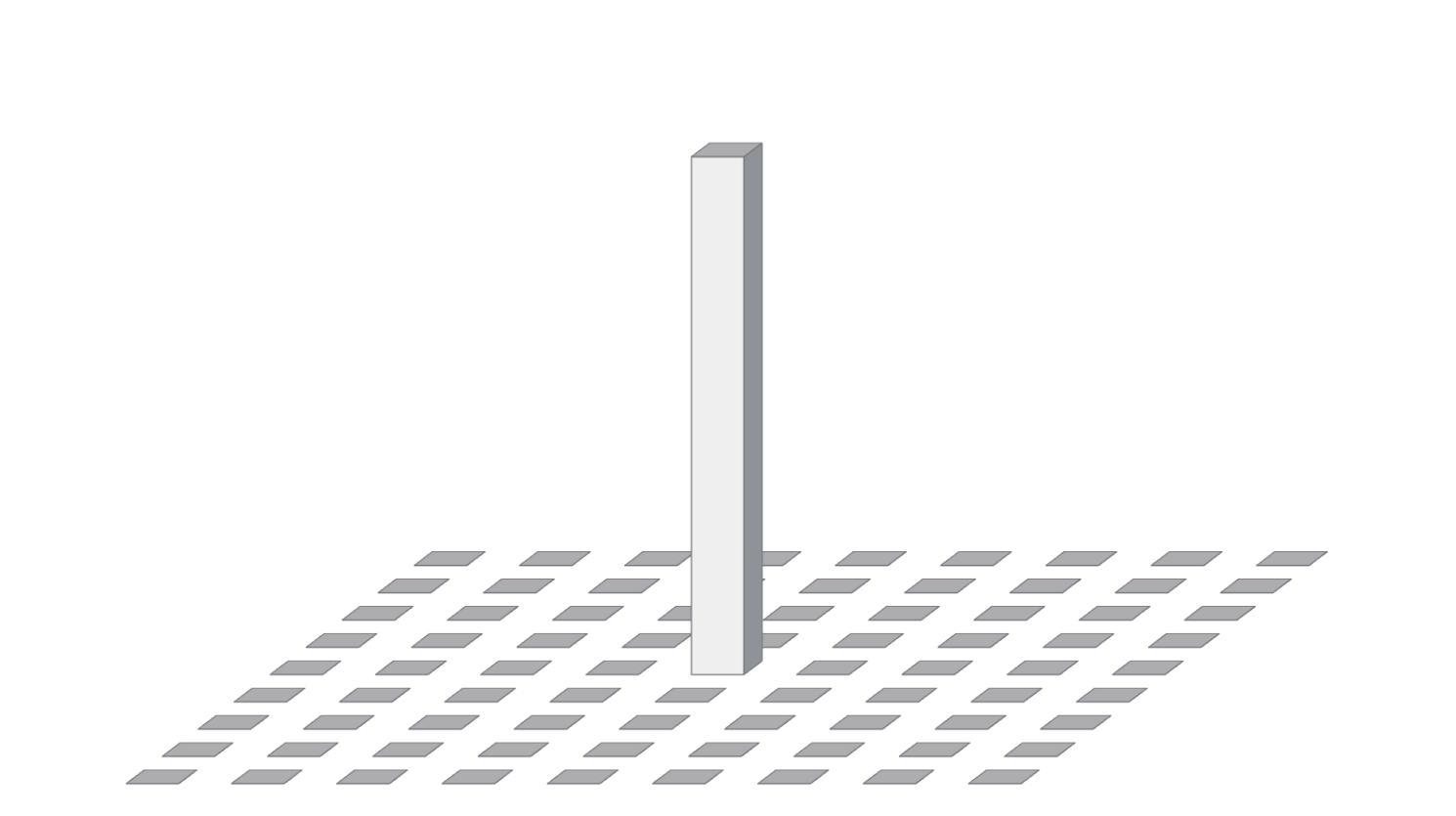}
    \caption{The Krugman model (Type G)}
    \end{subfigure}
    \caption{Stable spatial patterns in a square economy ($9^2 = 81$ locations).}
    \label{fig:sq-main}
\end{figure}

\subsection{Geographic accessibility}

Asymmetries can arise solely from the underlying distance structure between regions.
To illustrate this point, \Cref{fig:sq-main} considers a square geography with homogeneous local fundamentals and compares stable equilibria under the Type~L and Type~G models.
Unlike in a circular economy, some regions are ``central'' and therefore enjoy an inherent accessibility advantage.
Nonetheless, our theoretical result continues to hold: the spatial distribution is monocentric under Type~L, whereas it is polycentric under Type~G.
Moreover, comparative statics with respect to reductions in transport costs are qualitatively similar to those in \cref{fig:Hm-16,fig:Km-16}: under Type~L, population spreads monotonically, while under Type~G, agglomeration proceeds through successive concentration into fewer locations.
See Appendix \ref{app:geo-advantage} for further examples.
To obtain formal results for such asymmetric settings, one approach is to select tractable models representative of each type and study their implications across alternative network structures, following the strategy of \cite{Matsuyama-RIE2017} in a trade context.

\subsection{Local fundamentals} 
\label{sec:local-fundamentals}

Other than innate accessibility advantages, region-specific parameters such as exogenous productivity and amenity levels are equally fundamental, especially in quantitative spatial models (QSMs). 
For example, in the model of \citet{Allen-Arkolakis-QJE2014}, indirect utility can be written as
\begin{align}
v_i(\Vtx) = \underbrace{u_i x_i^{-\beta}}_{\mathllap{\text{\shortstack[c]{Congestion in amenities\\(local dispersion force)}}}} 
\cdot 
\Big(\sum_{k\inI} w_k^{1 - \sigma} \left(b_k x_k^{\alpha}\right)^{\sigma - 1} \phi_{ki} \Big)^{1/(\sigma - 1)}
w_i,
\label{eq:AA-utility}
\end{align}
where $\{w_i\}$ are the nominal wages determined in the market equilibrium given $\Vtx$: 
\begin{align}
& w_i x_i
= \sum_{j\inI}
\frac{
w_i^{1-\sigma} \big(\overbrace{b_i x_i^{\alpha}}^{\mathclap{\text{Within-region productivity spillover (local agglomeration force)}}}\big)^{\sigma - 1} \phi_{ij}
}{
\sum_{k\inI} w_k^{1-\sigma} \left(b_k x_k^{\alpha}\right)^{\sigma - 1} \phi_{kj}
}
w_j x_j
& \forall i \inI,
\label{eq:AA-weq}
\end{align}
The key elasticities are $\alpha > 0$, $\beta > 0$, and $\sigma > 1$.
The equilibrium is unique if $\alpha - \beta < 0$. 
In this case, given the observed population vector $\HtVtx$, we can uniquely solve for the region-specific parameters ${u_i}$ and ${b_i}$ that rationalize $\HtVtx$ as the model's equilibrium.
A natural question then is how such regional differences affect our results.

To address this question, the circular economy remains useful.
Consider the proximity structure as in \cref{assum:racetrack-economy}, but allow for variations in region-specific amenities, which we denote by $\Vta = (a_i)_{i\inI}$ with $a_i > 0$. 
When $a_i = \Bra$ for all $i$, we recover the symmetric racetrack economy of \cref{sec:N-region}, for which $\BrVtx$ is an equilibrium. 
If we slightly perturb $\Vta$ from $\BrVta \Is (\Bra,\Bra,\hdots,\Bra)$, then $\BrVtx$ is also slightly perturbed to form a new equilibrium $\Vtx(\Vta)$, which can be seen as a function of $\Vta$. 

To summarize the overall effect of such heterogeneities in $\Vta$ on the spatial distribution, we can use the covariance between each region's relative advantage and its deviation in population share from $\BrVtx$:
\begin{align}
\rho
\Is
\sum_{i\inI}
\ 
\overbrace{(a_i - \Bra)}^{\mathclap{\text{Exogenous regional (dis)advantage}}}
\ 
\underbrace{(x_i(\Vta) - \Brx)}_{\mathclap{\text{\shortstack[c]{Endogenous deviation from $\BrVtx$}}}}.
\label{eq:rho-definition.main}
\end{align}
If $\rho = 0$, heterogeneity in $\Vta$ does not affect the spatial distribution.
We assume $\rho > 0$ without loss of generality, as more advantaged regions should attract more population.

Since the mapping $\Vtx(\Vta)$ is model dependent, the magnitude of $\rho$ at a given transport cost level captures how the model's endogenous forces translate variation in $\Vta$ into variation in regional population distribution. 
Appendix \ref{app:local_advantage} formally shows that the sensitivity of $\rho$ to transport costs differs markedly between models, depending on the spatial scale of the dispersion forces (\cref{prop:local-a,prop:rho-sensitivity}).

\begin{figure}[t!]
    \centering
    \begin{subfigure}[b]{.48\hsize}
        \centering 
        \includegraphics[width=7.5cm]{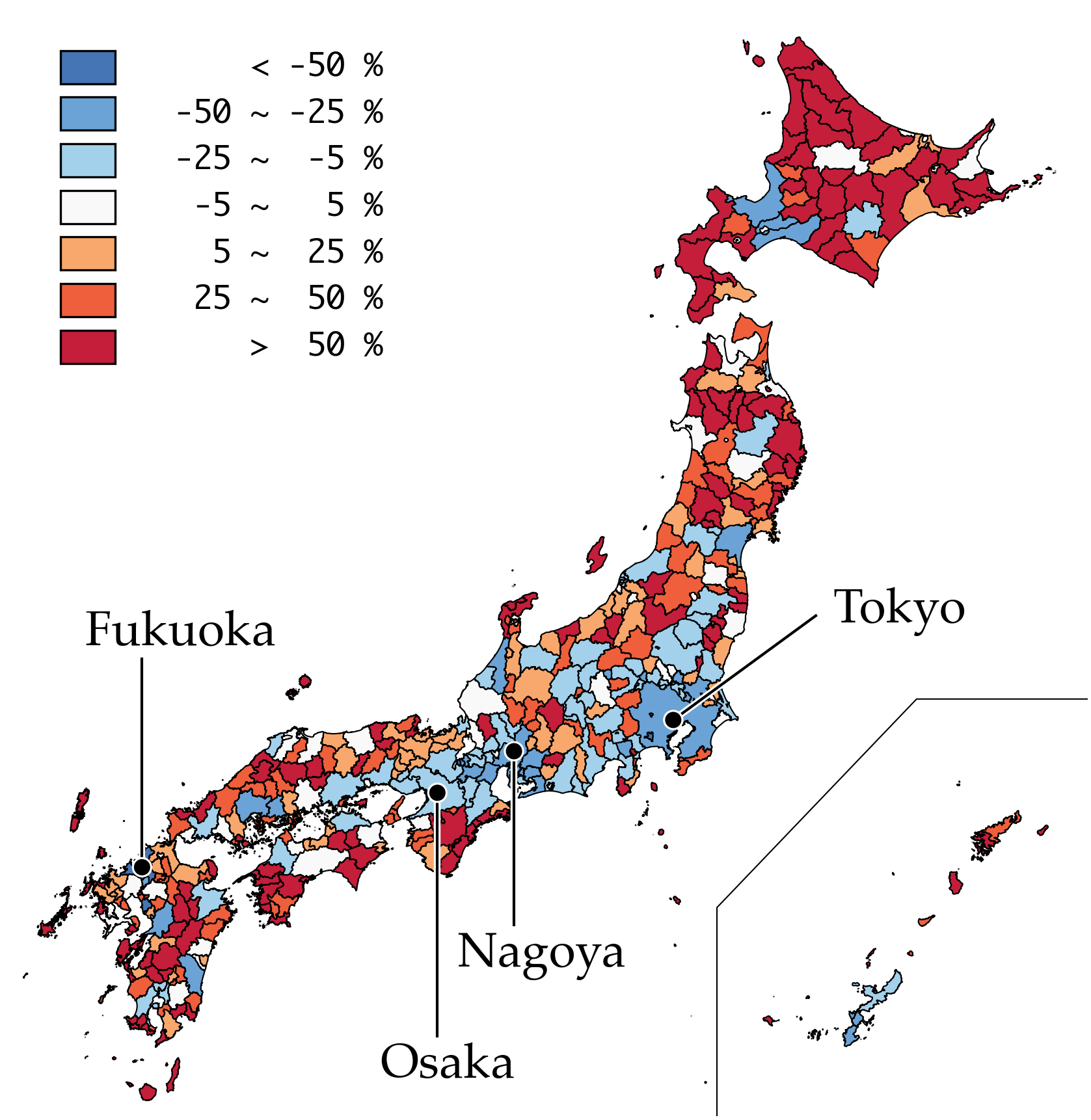}
        \caption{Observed (from 2020 to 1970)\label{fig:JP-ac-spatial}}
    \end{subfigure}

    \begin{subfigure}[b]{.48\hsize}
        \centering 
        \includegraphics[width=7.5cm]{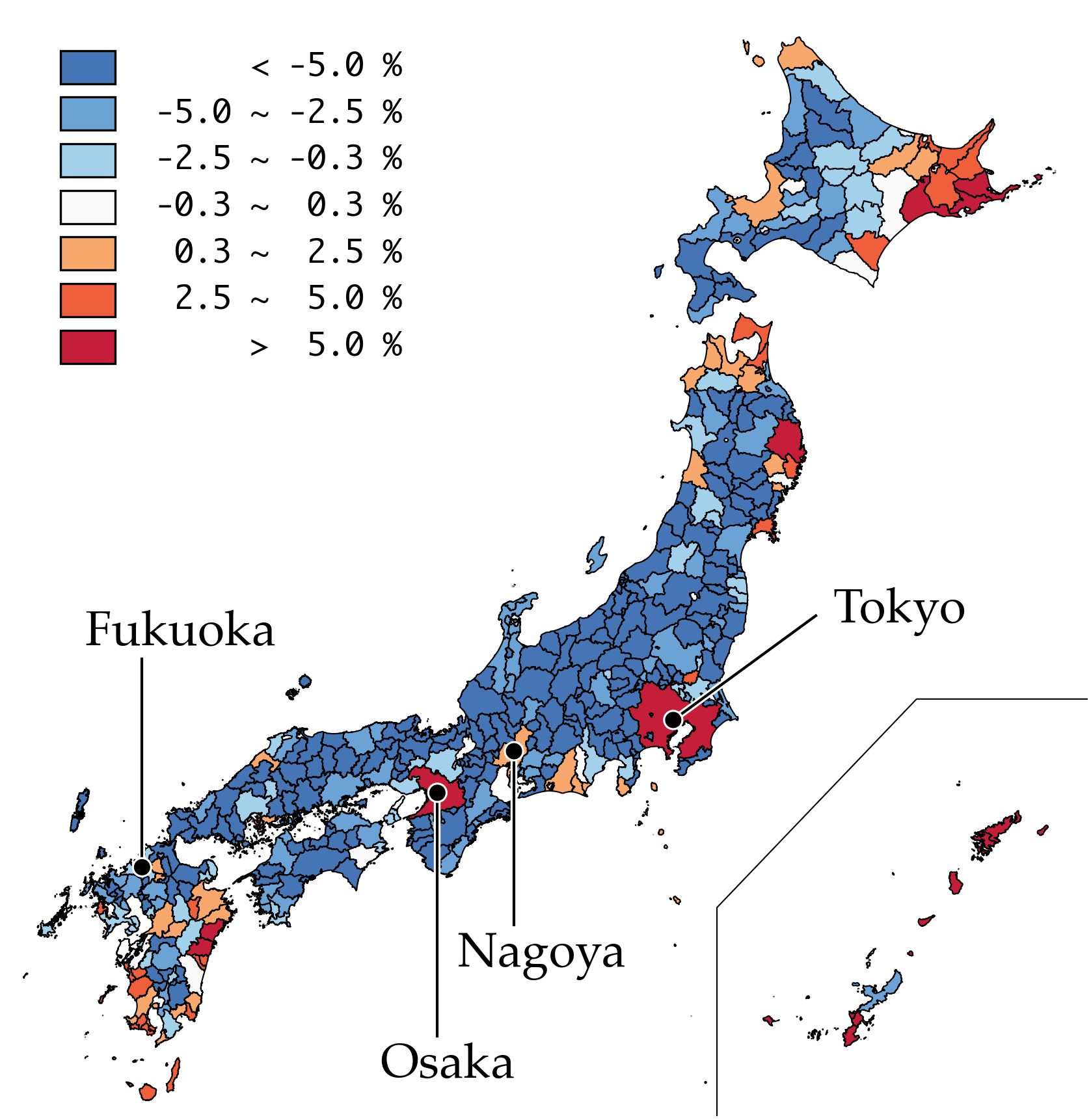}
        \caption{Removal of highways (Type~L)\label{fig:JP-TypeL}}
    \end{subfigure}
    \begin{subfigure}[b]{.48\hsize}
        \centering 
        \includegraphics[width=7.5cm]{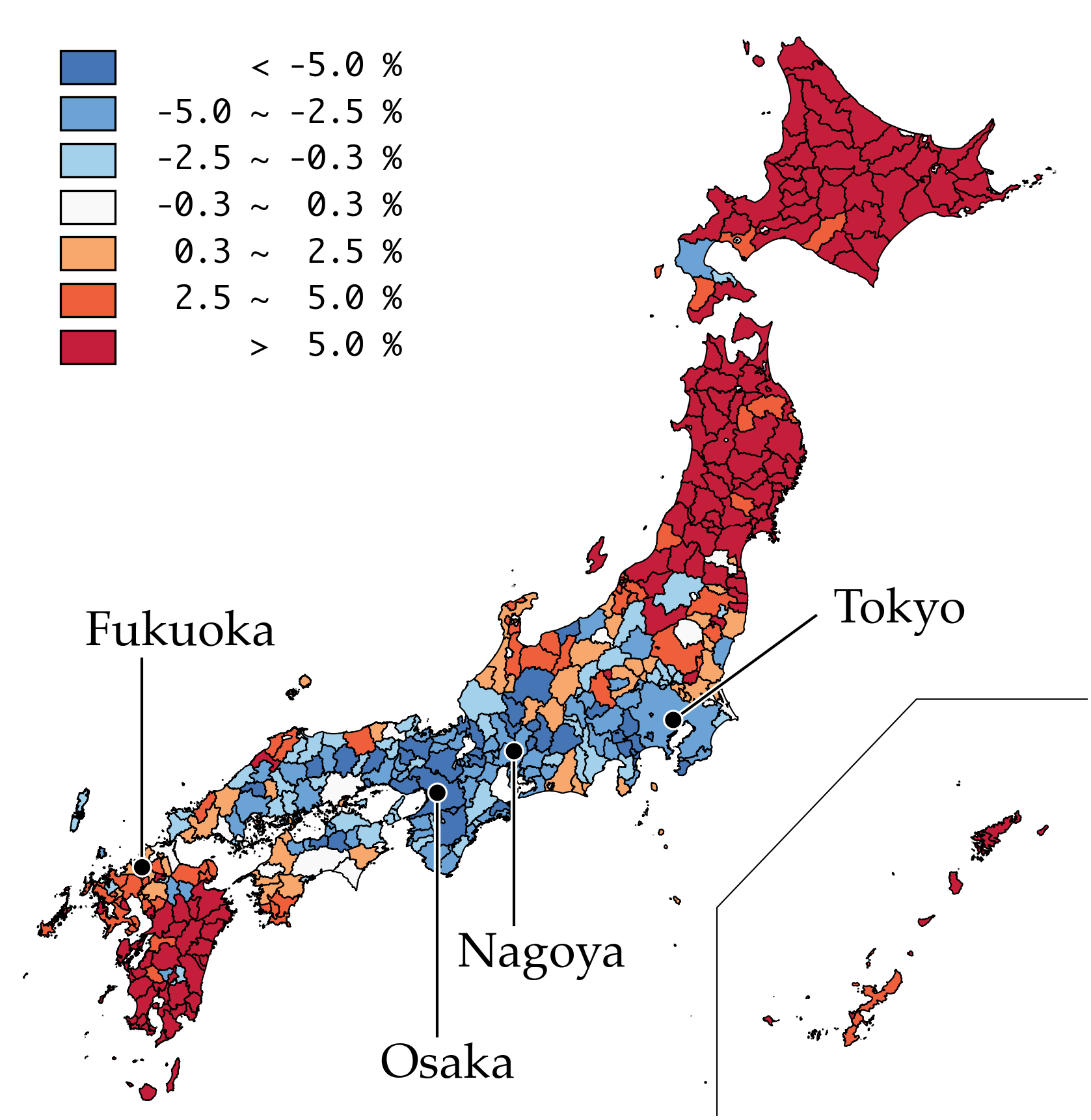}
        \caption{Removal of highways (Type~LG)\label{fig:JP-TypeLG}}
    \end{subfigure}
    
    \caption{Observed and counterfactual population change rates of Japanese regions.} 
    \label{fig:JP-QSM}
    \FigureNote{Figures are based on the simulation data of \cite{Sugimoto-etal-DP2025}. Panel (B) reports a many-region version of the Type~L model of \cite{Helpman-Book1998}. Panel (C) reports the Type~LG model of \cite{Sugimoto-etal-DP2025}, which extends the Helpman framework by incorporating land and intermediate goods as additional inputs. Although Sugimoto et al. does not assume immobile workers, land as an immobile factor generates a global dispersion force, as in \cite{Pfluger-Tabuchi-RSUE2010}.}
\end{figure}
Specifically, in models with pronounced global dispersion forces, improved interregional access tends to \emph{magnify} initial local advantages: $\rho$ increases as the freeness of transport $\phi$ increases and the population becomes more concentrated in the regions favored due to exogenous advantages. 
By contrast, in models with strong local dispersion forces, the same transport improvement tends to \emph{dampen} the role of innate heterogeneity, resulting in a flatter distribution, and $\rho$ decreases as $\phi$ increases. 
These patterns are consistent with the two-region examples in \cref{fig:Toy-1-asym,fig:Toy-2-asym}.

\subsection{The combination: A quantitative example} 
\label{sec:japan}

In reality, both geographic accessibility and local fundamentals vary between regions. 
We briefly discuss \cite{Sugimoto-etal-DP2025}'s results that compare Type~L and Type~LG over the Japanese geography. 
\Cref{fig:JP-QSM} reports the observed and counterfactual population growth rates. 
Panel~(A) shows actual changes from 2020 to 1970 (i.e., backward in time), highlighting the shift to major centers such as Tokyo, Osaka, and Nagoya. 
Panels~(B) and~(C) present counterfactual simulations based on calibrated Type~L and Type~LG models, respectively. 
Each model is calibrated to the population distribution observed in 2020 under the assumption of a unique equilibrium and then used to evaluate the counterfactual effects of removing the highway network. 
\begin{figure}[t!]
    \centering
    \begin{subfigure}[b]{.48\hsize}
        \centering 
        \includegraphics[width=.9\hsize]{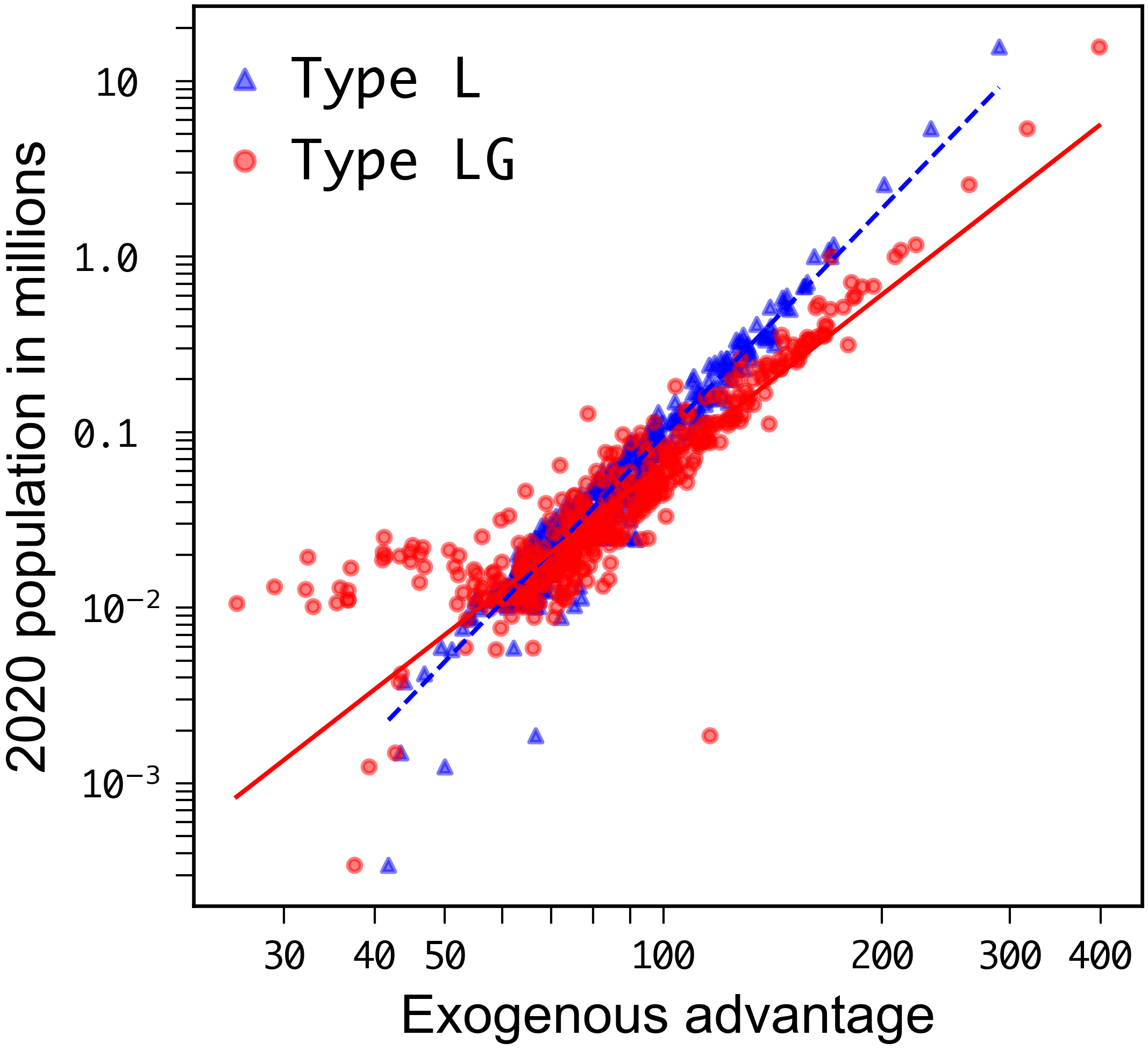}
        \caption{Calibration to the population in 2020\label{fig:JP-Am}}
    \end{subfigure}
    \hfill 
    \begin{subfigure}[b]{.48\hsize}
        \centering 
        \includegraphics[width=.9\hsize]{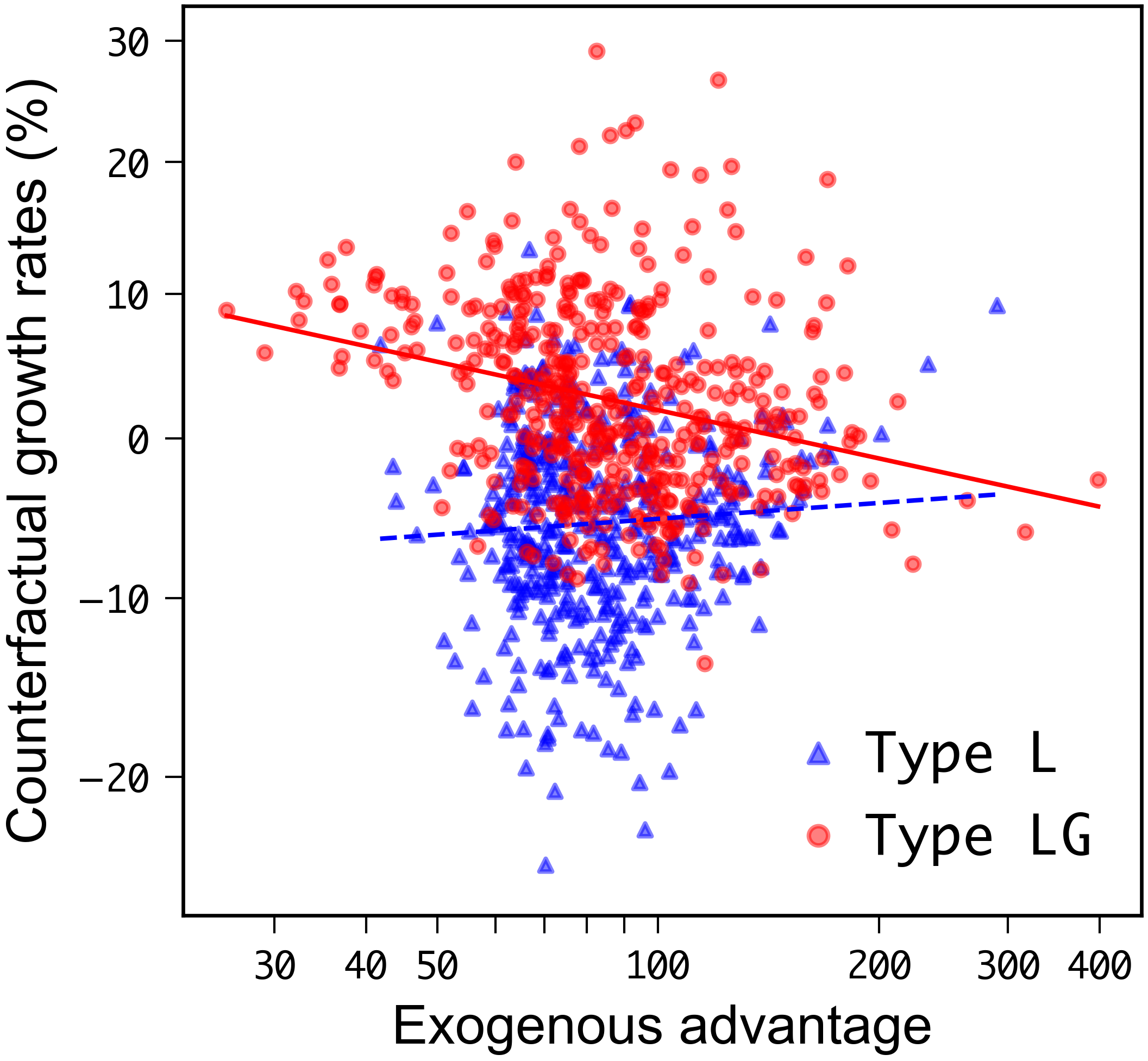}
        \caption{Counterfactual growths\label{fig:JP-Am-vs-Growth}}
    \end{subfigure}
    
    \caption{Comparison of model behaviors in \cref{fig:JP-QSM}} 
    \label{fig:JP-QSM-Error-Analysis}
    \FigureNote{Panel (A): In both models, beyond $90$\% of log population variation is explained by exogenous region-fixed fundamentals (a composite of observed land area and unobserved amenities). 
    Panel (B): In the Helpman model (Type L), growth rates and exogenous fundamentals are positively associated, i.e., highway removal induces growths at high-amenity regions, which are, as Panel (A) shows, essentially more populated regions in 2020. The converse holds true for Sugimoto et al.'s Type LG model as we observe negative association between exogenous fundamentals and growth rates. 
    }
\end{figure}

There is a notable contrast between the two models.
In the Type~L model, the removal of highways leads to further concentration in the core regions (\cref{fig:JP-TypeL}).
This aligns with the theoretical property of Type~L models that higher transport costs induce greater centralization.
\Cref{fig:JP-TypeLG} shows the opposite pattern: in the Type~LG model, the same shock produces a substantial population shift toward peripheral regions. 

\Red{\Cref{fig:JP-Am-vs-Growth} uses the same simulation data as \cref{fig:JP-TypeL,fig:JP-TypeLG} to show that, as transport costs rise, more advantaged regions grow faster in Type~L, whereas the opposite occurs in Type~LG.
This pattern is consistent with the discussion in \cref{sec:local-fundamentals}.
These qualitative reversals illustrate that the embedded forces can fundamentally shape the counterfactual implications of spatial models in asymmetric geographies.}

\subsection{On quantitative spatial models}

These results call for research on the role of endogenous economic forces in QSMs. 
A central premise of the QSM literature is that it ``does not aim to provide a fundamental explanation for the agglomeration of economic activity, but rather to provide an empirically relevant quantitative model to perform general equilibrium counterfactual policy exercises'' \citep[p.~23]{Redding-Rossi-Hansberg-ARE2017}. 
Building on this premise, QSMs typically impose equilibrium uniqueness and interpret unexplained interregional variation (i.e., structural residuals) as innate regional fundamentals. 
As a result, much of the observed variation is attributed to structural residuals (cf. \cref{fig:JP-Am}), and endogenous forces play a more limited role than in stylized theories of agglomeration.\footnote{For example, in the seminal works of \cite{Redding-Sturm-AER2008} and \cite{Allen-Arkolakis-QJE2014}, structural residuals account for 90\% and 78\% of the logarithmic variation of the city size.} 

However, as the Japan example in \cref{sec:japan} illustrates, seemingly innocuous choices regarding endogenous forces in QSMs can lead to markedly different counterfactual predictions, especially for distributional outcomes across regions.  
Most regional QSMs are Type~L and rely on local dispersion forces for tractability \citep[][Fn.~8]{Redding-Dict2025}, which leads them to predict decentralization as transport costs fall.\footnote{A subtle point is that the local--global distinction reflects detailed modeling choices rather than broad economic mechanisms. For example, a ``congestion'' force can be global if it arises from crowding of facilities that are accessible across regions with positive transport costs.} 
This narrows the range of distributional outcomes such models can generate and, in turn, constrains the policy conclusions that can be drawn. 

Further, some empirical contexts may be at odds with the unique-equilibrium assumption.  
For large transport projects such as highway systems, winners and losers may be uncertain ex-ante, making multiple equilibria potentially relevant.
Evidence from the Chinese urban system indicates that urban hierarchies can be based on strong agglomeration forces or immobile factors \citep{Baum-Snow-etal-JUE2020}, both of which can generate multiple equilibria. 
The persistence of spatial patterns further indicates that strong agglomeration forces and path dependence shape long-run regional outcomes \citep{Lin-Rauch-RSUE2022}. 
A central challenge is therefore to assess whether incorporating such forces improves the empirical fit and counterfactual performance of QSMs. 
Relatedly, \cite{Graham-Horcher-RP2024} argue that while QSMs hold promise for applied transport policy analysis, they are not yet practice-ready, citing model validation and uncertainty quantification as key obstacles.
These challenges are closely tied to how endogenous forces are specified in QSMs, since the strength and nature of agglomeration and dispersion mechanisms critically determine equilibrium spatial structure and, ultimately, the robustness of counterfactual predictions. 

\section{Concluding remarks}

We briefly discuss several examples to illustrate that our theoretical framework also offers a unifying interpretation of seemingly heterogeneous empirical findings. 
See \cite{Duranton-Turner-2025} for a more comprehensive survey of empirical evidence on how transport infrastructure affects urban and regional growth. 

Evidence on regional growth from 
\cite{Faber-REStud2014} is consistent with the predictions of Type~G or~GL models. He examines peripheral cities in China and finds negative effects on economic output. 
This may reflect a tendency for economic activity to concentrate in relatively larger or more central regions as transport access improves.\footnote{\cite{Duranton-Turner-REStud2012} document that transport infrastructure in neighboring MSAs negatively affects the employment growth rate of an MSA (Table E2), a pattern consistent with global agglomeration.} 
\cite{Baum-Snow-etal-JUE2020} provide complementary evidence for China, documenting slower growth in the hinterland prefectures compared to regional primates following the expansion of the highway system. 

In the intra-urban context, \cite{Baum-Snow-QJE2007} and \cite{Baum-Snow-et-al-REStat2017} provide evidence for the US metropolitan areas from 1950 to 1990 and for Chinese prefectures from 1990 to 2010.
Both studies examine how the share of population or production in the central area within a larger region changes as the transportation infrastructure expands, and both report negative effects in the central area. 
This is consistent with the behavior of Type~L or~LG models after transport investments. 
Such local spread can also be viewed as suburbanization driven by improved intra-urban transport infrastructure or by the diffusion of motorized transportation in the Alonso--Muth--Mills framework.\footnote{The structural transformation away from agriculture frees land around cities and also contributes to the decline of urban density \citep{Coeurdacier-etal-DP2022}. This can also be interpreted through the AMM framework as a reduction in the opportunity cost of land.} 

As seen above, a unified theoretical framework can help synthesize and interpret empirical findings.
Since this study focuses on static models with a single agent type, further scrutinies are essential for providing a bird's-eye view of both the empirical evidence and the now vast quantitative spatial economics literature.
We conclude by outlining two directions that merit further theoretical investigation.

First, it is important to consider multiple types of mobile agents that differ in their proximity matrices and/or the degree of increasing returns they experience. 
Such heterogeneity is ubiquitous in multi-sector models \citep{Fujita-et-al-EER1999, Hsu-EJ2012, Gaubert-AER2018, Davis-Dingel-JIE2020} and in intracity models with multiple types of agents \citep[e.g.,][]{Fujita-Ogawa-RSUE1982, Lucas-Rossi-Hansberg-ECTA2002, Ahlfeldt-et-al-ECTA2015, Heblich-etal-QJE2020}. 
For example, \cite{Duranton-Morrow-Turner-REStud2014} studied the impact of new highway connections on intercity trade in the US and showed that heavier industries are more sensitive to improved access. 
\Red{\cite{Allen-etal-AER2024} considered a quite general spatial model with multiple spatial interactions, but the characterization of endogenous equilibrium spatial structure has yet to be done, in particular for the cases with multiple equilibria.}  
 Circular geography provides a tractable starting point for analysis of many-locations under such structures \citep{Tabuchi-Thisse-JUE2011, Osawa-Akamatsu-DP2019}. 
As \cite{Hsu-EJ2012} suggests, the incorporation of sectoral heterogeneity can be particularly important for understanding the mechanisms behind the remarkable regularities in the size and spatial variation of cities \citep{Mori-etal-PNAS2020}. 

Second, models with a continuum of agents as considered in this study are complementary to ``granular'' spatial models \citep[e.g.,][]{Ahlfeldt-et-al-CEPR2022}, in which endogenous agglomeration arises from increasing returns and the indivisibility of agents.
Continuum-agent models can replicate systematic spatial regularities, such as periodic agglomeration patterns and city-size distributions that include their fractal structure \citep[e.g.,][]{Hsu-EJ2012, Tabuchi-Thisse-JUE2011, Mori-et-al-DP2022}. 
Granular spatial models are better suited to capture idiosyncratic location choices by superstar firms and large plants \citep[e.g.,][]{Greenstone-et-al-JPE2010}.
Combining these two approaches may yield a deeper understanding of the spatial patterns of economic activities as the result of endogenous forces.

\clearpage

\appendix
\crefalias{section}{appendix}
\numberwithin{equation}{section}
\numberwithin{figure}{section}
\numberwithin{table}{section}
\numberwithin{equation}{section}

\section{Proofs}
\label{app:proofs}

\subsection{Proof of \cref{lem:omega-stab}} 

First, we can represent $\Delta$ in terms of marginal migration from region $2$ to $1$: $\Delta(\epsilon) \coloneq \Delta(\BrVtx + \frac{\epsilon}{2} (1, -1) ) = v_1(\Brx + \frac{\epsilon}{2}, \Brx - \frac{\epsilon}{2} ) - v_2(\Brx + \frac{\epsilon}{2}, \Brx - \frac{\epsilon}{2} )$. 
Then, $\omega = \frac{\Brx}{\Brv} \Delta'(\epsilon)|_{\epsilon = 0}$, meaning that $\omega$ is proportional to the directional derivative of $\Delta$ with respect to the unit migration from region $2$ to $1$, as $\frac{1}{2} (1, -1)$ is a normalized vector. 
Here, we used the fact that $\PDF{v_1}{x_1}(\BrVtx) = \PDF{v_2}{x_2}(\BrVtx)$ and $\PDF{v_1}{x_2}(\BrVtx) = \PDF{v_2}{x_1}(\BrVtx)$ due to the symmetry of the regions at $\BrVtx$. 
The stability condition based on the sign of $\omega$ is valid for general $\Vtv$ provided that $\Brv> 0$.

\subsection{Proof of \cref{lem:gain-formula}}
Let $\delta v_i(\Vtz) \Is v_i(\BrVtx + \Vtz) - v_i(\BrVtx)$ be the utility difference in each region under deviation $\Vtz$. 
The utility gain for a mover from region $j$ to $i$ is 
\begin{align}
    \frac{\Brx}{\Brv}\left(\delta v_i(\Vtz)-\delta v_j(\Vtz)\right). 
\end{align}
By adding up the utility gains for all the migrants in the economy, the aggregate utility gain for migrants is measured by 
$\omega(\Vtz) = \frac{\Brx}{\Brv} \sum_{i\inI} \delta v_i(\Vtz) z_i = \frac{\Brx}{\Brv} \delta\Vtv(\Vtz)^\top \Vtz$, 
as there are $z_i$ agents migrated to region $i$ (if $z_i > 0$) or from region $i$ (if $z_i < 0$), each experiencing utility difference $\delta v_i(\Vtz)$ at both their origin and destination. 
The first-order approximation shows $\delta \Vtv(\Vtz) \approx \Vtv(\BrVtx) + \nabla \Vtv(\BrVtx) \Vtz - \Vtv(\BrVtx) = \nabla\Vtv(\BrVtx)\Vtz$ and hence gives $\omega(\Vtz) = \Vtz^\top \VtV \Vtz$. 

It is noted that $\omega(\Vtz)$ must be considered subject to $\Vtz \in T$, where $T\Is \{\Vtz\in\BbR^N \mid \sum_{i\inI} z_i = 0\}$ is the set of all feasible deviations that preserves the total population. 
To avoid technicality, the main text do not mention the constraint $\Vtz_k\in T$. This constraint excludes deviations of the form $\Vtz = (\epsilon,\epsilon,\hdots,\epsilon)$, which represents the symmetric increase or decrease of population across all regions.

\subsection{Proof of \cref{prop:classification}}
\label{app:proof-classification}

We consider spatial models described by a \emph{payoff function} (i.e., indirect utility) $\Vtv(\Vtx) \Is (v_i(\Vtx))_{i\inI}$, parametrized by a proximity matrix $[\phi_{ij}]$, along with \cref{assum:racetrack-economy}.  
Let $\VtD$ be the row-normalized proximity matrix, whose $(i,j)$th element is $ \frac{\phi_{ij}}{\sum_{k\inI} \phi_{ik}}$. 
Throughout, we assume that $\Vtv$ is differentiable if $x_i > 0$ for all $i\inI$. 
The precise version of the symmetry of exogenous local fundamentals in \cref{assum:racetrack-economy} is the following: 
\renewcommand{\theassumption}{S}
\begin{assumption}
\label{assum:equivariance-full}
For all $\Vtx$, payoff function $\Vtv$ satisfies 
$\Vtv(\VtP\Vtx) = \VtP\Vtv(\Vtx)$ for all permutation matrices $\VtP$ that satisfy $\VtP\VtD = \VtD\VtP$. 
\end{assumption}
\begin{example}
    \label{example:v-equivariance}
Suppose $N = 4$. 
If we consider regions 1 and 3, swapping their indices corresponds to applying the following permutation matrix to the spatial distribution and the payoff function:
\begin{align}
    \renewcommand{\arraystretch}{0.8}
    \VtP =
    \begin{bmatrix}
        0 & 0 & 1 & 0 \\
        0 & 1 & 0 & 0 \\
        1 & 0 & 0 & 0 \\
        0 & 0 & 0 & 1
    \end{bmatrix}.
\end{align}
This matrix simply switches the values of $x_1$ and $x_3$ in any vector $(x_1, x_2, x_3, x_4)$. 
That is, $\VtP\Vtx = (x_3, x_2, x_1, x_4)$. 
This corresponds to relabeling the regions while keeping their physical positions fixed. 
The condition $\VtP\VtD = \VtD\VtP$ ensures that the relabeling preserves the spatial relationships encoded in $\VtD$. 
If $\Vtv$ does not include any region-specific heterogeneities, then the transformed utility vector must satisfy $\Vtv(\VtP\Vtx) = \VtP\Vtv(\Vtx)$. 
This property is called \emph{equivariance}; it ensures that utility differences are determined entirely by the spatial distribution, not by arbitrary index labels. 
Equivariance allows us to employ the machineries from group-theoretic bifurcation theory \citep[see, e.g.,][]{Golubitsky-Stewart-Book2003,Golubitsky-et-al-Book2012,Ikeda-Murota-Book2014}.  
\end{example}

Under \cref{assum:racetrack-economy}, we can use the full dispersion as the initial state. 
\begin{lemma}
    \label{lem:xbar}
Under \cref{assum:racetrack-economy} (including \cref{assum:equivariance-full}), the uniform distribution of agents, $\BrVtx = (\Brx, \Brx, \hdots, \Brx)$ with $\Brx \Is 1/N$, is a spatial equilibrium. 
\end{lemma}
\begin{proof}
For any permutation matrix $\VtP$, $\BrVtx = \VtP\BrVtx$. 
Then, $\Vtv(\VtP\BrVtx) = \VtP\Vtv(\BrVtx)$ reduces to $\Vtv(\BrVtx) = \VtP\Vtv(\BrVtx)$ for all permutation matrix $\VtP$ that satisfies $\VtP\VtD = \VtD\VtP$. 
This implies that $v_i(\BrVtx) = v_j(\BrVtx)$ for any $i,j\inI$. That is, $\BrVtx$ is a spatial equilibrium. 
\end{proof}

We focus on a class of models that include all models discussed in the main text. 
As in the main text, let $\VtV = \frac{\Brx}{\Brv}\nabla\Vtv(\BrVtx)$ be the \emph{benefit matrix} for a given payoff function. 
\begin{definition}
\label{def:canonical-model}
A \emph{canonical model} is a model associated with a rational function $\Omega$ that is continuous over $[0,1]$ such that $\VtV = \Omega\left(\VtD\right)$.
We call $\Omega$ the \emph{gain function} of the model.
\end{definition}
\noindent In \cref{def:canonical-model}, a rational function $\Omega$ is a function of form $\Omega(\cdot) = \Gs(\cdot)/\Gf(\cdot)$ with polynomials $\Gs(\cdot)$ and $\Gf(\cdot) \neq 0$, where our convention is that $\Gf(\cdot) > 0$. 
Given such $\Omega$, we let $\Omega(\VtD) = \Gf(\VtD)^{-1} \Gs(\VtD)$, where, for a polynomial $P(\Theta) = c_0 + c_1 \Theta + c_2 \Theta^2 + \cdots$,
we define
$P(\VtD) = c_0\VtI + c_1 \VtD + c_2 \VtD^2 + \cdots$,
with $\VtI$ being the identity matrix.

Below, we study the stability of $\BrVtx$ in canonical models.  
Formally, we must introduce some dynamics to define the stability of $\BrVtx$ and study agglomeration from there. 
For a wide class of dynamics, however, we can focus on the analysis of the benefit matrix $\VtV$. 
\begin{lemma}
\label{lem:stability}
Assume a canonical model and assume \cref{assum:racetrack-economy}. 
For a wide class of myopic adjustment dynamics, $\BrVtx$ is stable (unstable) if the largest eigenvalue of $\VtV$, excluding the one corresponding to $\Vt1 = (1,1,\hdots,1)$, is smaller (greater) than zero. 
Furthermore, if only the sign of the largest eigenvalue turns from negative to positive at some $\phi^*\in(0,1)$, then a new spatial equilibrium branches from $\BrVtx$ at $\phi^*$, toward the direction of associated eigenvector. 
\end{lemma}
\begin{proof}
See \cref{app:proof-lem-stability}. 
\end{proof}

\noindent We can assume, for example, the replicator dynamics \citep{Taylor-Jorker-MB1978} to define local stability of spatial equilibria (see the proof of \cref{lem:stability} for more examples). 

Thus, we only need the eigenvalues of $\VtV$. 
A useful fact is that, if $\{(\Theta_k,\Vtz_k)\}$ are the eigenpairs (eigenvalue--eigenvector pairs) of the normalized proximity matrix $\VtD$, then the eigenpairs of $\VtV = \Omega(\VtD)$ are given by $\{(\Omega(\Theta_k),\Vtz_k)\}$ \citep[e.g.,][Section 1.1]{Horn-Johnson-Book2012}. 
Thus, we need the eigenpairs of $\VtD$. 
For the sake of simplicity, we assume that $N$ is a multiple of four. Then, we have the following lemma. 
\begin{lemma}[Corollary of \cite{Akamatsu-Takayama-Ikeda-JEDC2012}, Lemma 4.2]
    \label{lem:D_eigenpairs}
Assume \cref{assum:racetrack-economy}. 
The largest eigenvalue of $\VtD$ is $\Theta_0 \Is 1$, with $\Vtz_0 = (1,1,\hdots,1)$ being the associated eigenvector. 
Including $\Theta_0$, there are $\frac{N}{2} + 1$ distinct eigenvalues. 
Every eigenvalue $\Theta_k$ ($k \ne 0$) is a strictly decreasing function of $\phi$, with $\lim_{\phi\downarrow 0} \Theta_k = 1$ and $\lim_{\phi\uparrow 1} \Theta_k = 0$. 
Let $\Theta_\text{max}$ denote the largest eigenvalue and $\Theta_\text{min}$ denote the smallest eigenvalue of $\VtD$ excluding $\Theta_0$, respectively. 
Further, assume that $N$ is a multiple of four. 
Then, 
\begin{align}
    & \Theta_\text{max} = \Theta_1 \Is \frac{1 - \phi}{1 + \phi} \cdot \frac{1 - \phi^2}{1 - 2 \cos(\kappa) \phi +\phi^2}\cdot \frac{1 + \phi^{N/2}}{1 - \phi^{N/2}}\\
    & \Theta_\text{min} = \Theta_{N/2} \Is \left(\frac{1 - \phi}{1 + \phi}\right)^2 
\end{align}
at any $\phi$, with $\kappa = \frac{2\pi}{N}$, and $\Theta_\text{max} = \Theta_1$ has multiplicity two. 
For a vector $\Vtz$, let$\langle z_i \rangle_{i = 0}^{N-1} \Is \frac{1}{\|\Vtz\|}(z_i)_{i = 0}^{N-1}$ denote its normalized version. 
Then, the eigenvector associated with $\Theta_\text{max}$ is 
    $\Vtz_1^+ \Is \langle \cos(\kappa i) \rangle_{i = 0}^{N-1} 
    $ and $
    \Vtz_1^- \Is \langle \sin(\kappa i) \rangle_{i = 0}^{N-1}
$, 
and that associated with $\Theta_\text{min}$ is 
$\Vtz_{N/2} 
\Is \langle (-1)^i  \rangle_{i = 0}^{N-1} 
= \langle 1, -1, 1, -1,\hdots, 1, -1\rangle$. 
\end{lemma}

Since $\Theta_k\in(0,1)$ for all relevant $k$ and $\Omega$ is well-defined for all $[0,1]$, the eigenpairs of $\VtV = \Omega(\VtD)$ are in fact given by $\{(\Omega(\Theta_k),\Vtz_k)\}$. 
Thus, with $\omega_k \Is \Omega(\Theta_k)$, the symmetric equilibrium $\BrVtx$ is stable if $\omega_k < 0$ for all $k$.

As discussed in \cref{sec:model}, if agglomeration (dispersion) force of the model is too strong, $\Omega(\Theta) > 0$ ($\Omega(\Theta) < 0$) can happen for all $\Theta$ whereby $\BrVtx$ is unstable (stable) for all $\phi$. 
As we are interested in spatial agglomeration in the course of changing $\phi$, we assume that $\BrVtx$ can switch its stability depending on $\phi$: 
\renewcommand{\theassumption}{E}
\begin{assumption}[{Endogenous agglomeration occurs}]
\label{assum:E}
The values of the model parameters are such that $\Omega$ switches its sign at least once in $(0,1)$.
\end{assumption}

Under \cref{assum:E}, we can define three prototypical classes of canonical models (see \cref{fig:G.three-classes} in the main text for illustration). 
\begin{definition}
\label{def:model-class}
Under \cref{assum:E}, a canonical model with gain function $\Omega$ is
\begin{enumerate}
    
	\item \textbf{Type~L},
	if there can be one and only one $\Theta^{**}\in(0,1)$
	such that
	$\Omega(\Theta) < 0$
	for $\Theta \in (0, \Theta^{**})$,
	$\Omega(\Theta^{**}) = 0$,
	and
	$\Omega(\Theta) > 0$
	for $\Theta \in (\Theta^{**}, 1)$.
    
	\item \textbf{Type~G},
	if there can be one and only one root $\Theta^*\in(0,1)$ for $\Omega$
	such that
	$\Omega(\Theta) > 0$
	for $\Theta \in (0, \Theta^*)$,
	$\Omega(\Theta^*) = 0$,
	and
	$\Omega(\Theta) < 0$
	for $\Theta \in (\Theta^*, 1)$.

	\item \textbf{Type~LG},
	if there can be two $\Theta^*,\Theta^{**}\in(0,1)$
	such that
	$\Omega(\Theta^*) = \Omega(\Theta^{**}) = 0$ and $\Theta^{**} < \Theta^{*}$,
	with
	$\Omega(\Theta) < 0$
	for $\Theta\in(0,\Theta^{**})\cup(\Theta^{*},1)$
	and
	$\Omega(\Theta) > 0$
	for $\Theta\in(\Theta^{**},\Theta^{*})$.
\end{enumerate}
\end{definition}

As discussed in the main text, the classification corresponds to the composition of the consequential dispersion forces in the model. 
We focus on the three model classes defined above and consider the destabilization of $\BrVtx$. 
There can be a fourth class of models such that $\Vtx$ is stable for medium levels of $\Theta$ but not for small or large $\Theta$. However, we are not aware of any model that falls into this category. 

\begin{figure}[tb]
    \centering
    \begin{overpic}[width=14cm,clip]{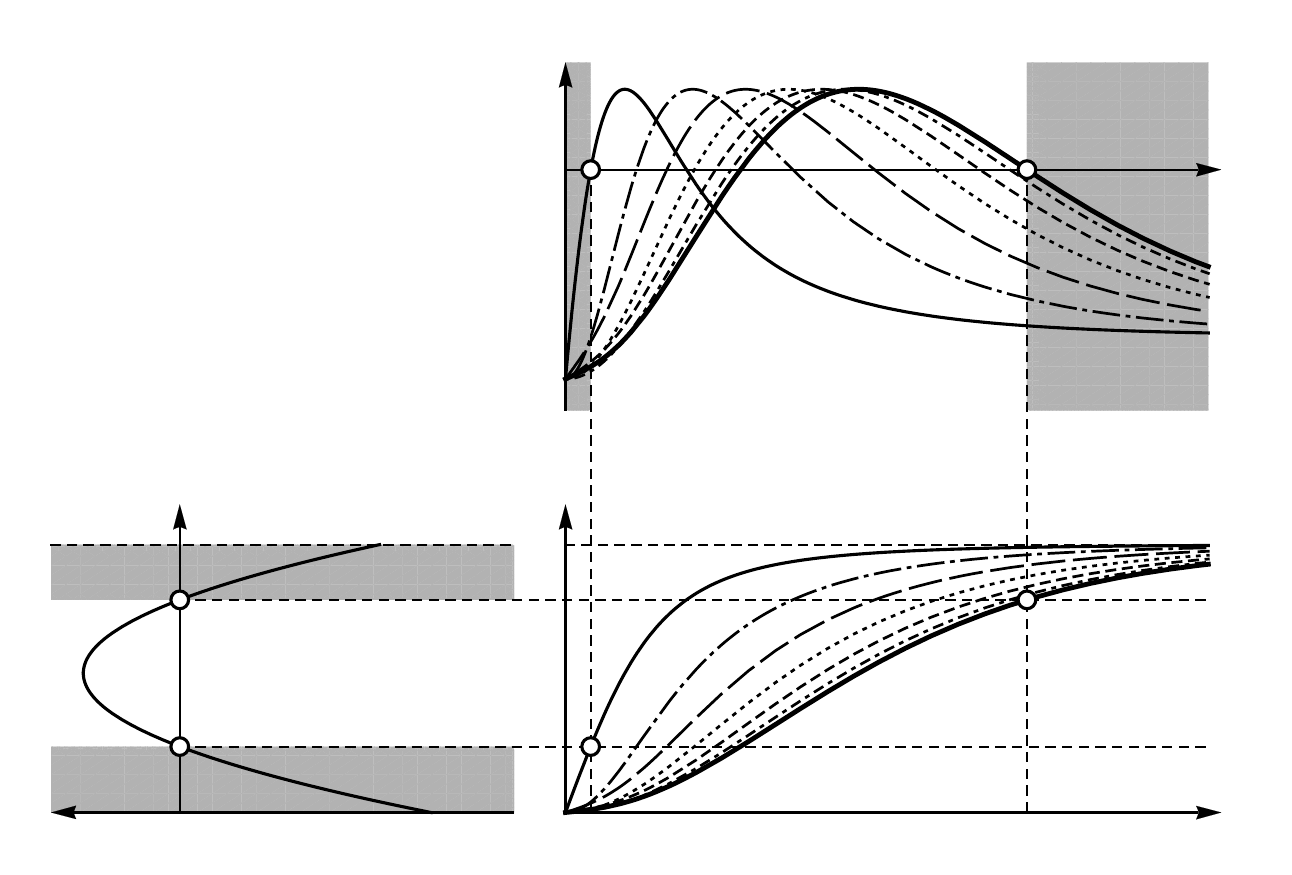}
        \scriptsize 
        \put(32,2.3){$c_0$}
        \put(45,2){$\phi^{**} = \phi_1^*$}
        \put(78,2){$\phi^{*} = \phi_{N/2}^*$}
        \put(96,4){$-\log(\phi)$}
        \put(42,30){$\Theta_k$}
        \put(42,65){$\omega_k^\sharp = \Gs(\Theta_k)$}
        \put(40.5,38){$c_0$}
        \put(41.5, 4){$0$}
        \put(41.5,25){$1$}
        \put(41.5,54){$0$}
        \put(96,54){$-\log(\phi)$}
        \put(-5,4){$\Gs(\Theta)$}
        \put(13.3,2){$0$}
        \put(13.3,30){$\Theta$}
        \put(15,11.5){$\Theta^{**}$}
        \put(15,18.5){$\Theta^{*}$}
        \put(47,19){$\Theta_{1}$}
        \put(50,16){$\Theta_{2}$}
        \put(70,15){$\Theta_{N/2}$}
        \put(57,40){$\omega_1^\sharp$}
            \put(58,43){\vector(1,2){2}}
        \put(75,64){$\omega_{N/2}^\sharp$}
            \put(77.5,61.5){\vector(-1,-2){2}}
    \end{overpic}
    \caption{The relationships between $\Gs$, $\{\Theta_k\}$, and $\{\omega_k^\sharp\}$. }
    \label{fig:omega-decomposition}
    \FigureNote{Top: Graphs of $\omega_k^\sharp = \Gs(\Theta_k)$. 
        Bottom left: Net gain function $\Gs$ for a hypothetical Type~LG model with a quadratic net gain function of the form $\Gs(\Theta) = c_0 + c_1 \Theta + c_2 \Theta^2$. 
        Bottom right: The full set of eigenvalues $\{\Theta_k\}$ of $\VtD$. 
        In the shaded regions of $\phi$ or $\Theta$, $\BrVtx$ is stable. 
        For the $\phi$ axis, the negative $\log$ scale is used for better readability, with the transport cost level being high toward the right. 
        We have $\max\{\Theta_k\} = \Theta_{\mathrm{max}}$ and $\min\{\Theta_k\} = \Theta_{\mathrm{min}}$ at any given level of $\phi$.}
\end{figure}

As we consider canonical models, there is a rational function $\Omega(\cdot) = \Gs(\cdot)/\Gf(\cdot)$ with some polynomials $\Gs$ and $\Gf(\cdot) > 0$. 
That is, $\Gs(\cdot)$ determines the sign of $\omega_k$ and thus governs the stability of $\BrVtx$. 
We will focus on $\Gs$ below, and let $\omega^\sharp \Is \Gs(\Theta_k)$ so that $\sgn[\omega_k] = \sgn[\omega^\sharp_k]$.  
\cref{fig:omega-decomposition} schematically shows connections between $\{\omega_k^\sharp\}$, $\Gs(\Theta)$, and $\{\Theta_k\}$ to help understanding the following arguments.

\paragraph{Type~L.} 
By definition, there is $\Theta^{**}$ such that  $\Gs(\Theta) < 0$ for all $\Theta\in(0,\Theta^{**})$, that $\Gs(\Theta^{**}) = 0$, 
and that $\Gs(\Theta^{**}) > 0$ for all $\Theta\in(\Theta^{**},1)$. 
Thus, $\BrVtx$ is stable if and only if $\Theta_k \in (0,\Theta^{**})$, so that $\omega^\sharp_k = \Gs(\Theta_k) < 0$,  for all $k$, i.e., if $\Theta^{**} >  \max_{k} \Theta_k = \Theta_{1}$. 
Thus, $\BrVtx$ is stable for all $(\phi^{*},1)$ where $\phi^{**}$ is the unique solution for $\Theta_1(\phi) = \Theta^{**}$. 
Because $\Gs(\Theta) > 0$ for all $\Theta\in(\Theta^{**},1)$ and $\Theta_1$ is strictly decreasing, $\BrVtx$ is unstable for all $(0,\phi^{**})$. 

\paragraph{Type~G.} 
By definition, there is $\Theta^*$ such that $\Gs(\Theta) < 0$ for all $\Theta\in(\Theta^*,1)$, that $\Gs(\Theta^*) = 0$, and that $\Gs(\Theta) > 0$ for all $\Theta\in(0,\Theta^*)$. 
By \cref{lem:D_eigenpairs}, $\{\Theta_k(\phi)\}$ are strictly decreasing from $1$. 
Thus, $\BrVtx$ is stable if and only if $\Theta_k \in (\Theta^* ,1)$, so that $\omega^\sharp_k = \Gs(\Theta_k) < 0$, for all $k$, i.e., if $\Theta^* < \min_{k} \Theta_k = \Theta_{N/2}$. 
Thus, $\BrVtx$ is stable for all $(0,\phi^*)$ where $\phi^* = (1 - \sqrt{\Theta^*})/(1 + \sqrt{\Theta^*})$ is the unique solution for $\Theta_{N/2} = \Theta^*$. 
Because $\Gs(\Theta) > 0$ for all $\Theta\in(0,\Theta^*)$ and $\Theta_{N/2}$ is strictly decreasing, $\BrVtx$ is unstable for all $(\phi^*,1)$ because $\omega_{N/2}^\sharp > 0$ for the range. 

\paragraph{Type~LG.} Via similar logic, we see $\BrVtx$ is stable if $\phi\in (0,\phi_{N/2}^{*})\cup(\phi_1^{*},1)$. 

\paragraph{Spatial patterns.} Consider a state where $\BrVtx$ is stable. 
From \cref{lem:stability}, at $\phi^*$ in Types G or LG, a polycentric pattern with $N/2$ peaks branches from $\BrVtx$, while at $\phi^{**}$ in Types L or LG, a monocentric configuration branches from $\BrVtx$. 
\hfill $\square$

\begin{remark}
The bifurcation towards the monocentric direction ($k = 1$) is a \emph{double bifurcation}, where the associated eigenvalue $\omega_1$ has multiplicity two, that is, there are two linearly independent eigenvectors. Migration patterns at this bifurcation take the form $c^+\Vtz_1^+ + c^-\Vtz_1^-$ for $c^+, c^- \in \BbR$. Under \Cref{assum:racetrack-economy}, only $(c^+, c^-) = (c, 0)$ or $(c, c)$ for some $c \in \BbR$ are admissible \citep{ikeda2012spatial}. The conclusion of \cref{prop:classification} is not affected as both combinations yield monocentric configurations.
\end{remark}

\begin{remark}
Although \cref{def:model-class} introduces three prototypical model classes, Type~LG models sometimes span multiple classes depending on parametric restrictions \citep[e.g., models by][]{Pfluger-Tabuchi-RSUE2010,Kucheryavyy-etal-JIE2024}. 
In such cases, the parameter space can be partitioned to map model behavior to the typology. 
Also, the definition of models often impose parametric restrictions that fix their class. 
In principle, flexible specifications would allow empirical identification of the class supported by data through parameter estimation. 
\end{remark}

\begin{remark}
\label{remark:N}
In \cref{lem:D_eigenpairs}, we assume that $N$ is a multiple of four to ensure $\min_{k} \{\Theta_k\} = \Theta_{\frac{N}{2}}$. 
This is inconsequential for the broad implication of \cref{prop:classification} on spatial patterns. 
If $N$ is an even, $\min_{k}\{\Theta_k\} = \min\{\Theta_{{\frac{N}{2}}-1}, \Theta_{\frac{N}{2}}\}$. 
If $N$ is an odd, $\min_{k}\{\Theta_k\} = \min\{\Theta_{\lfloor \frac{N}{2} \rfloor}, \Theta_{\lfloor \frac{N}{2} \rfloor - 1}\}$. 
Thus, $\min_{k}\{\Theta_k\}$ corresponds to a polycentric direction, except for the case $N = 2$ or $3$ in which polycentric patterns cannot occur. 
\end{remark}

\begin{remark}
\label{remark:global_behavior}
Beyond the local result of the proposition, \cite{ikeda2012spatial} characterized the possible equilibrium configurations and bifurcations in symmetric circular economy by group-theoretic analysis. 
Two formal predictions are worth mentioning. 

First, no symmetry-breaking bifurcations can occur after the emergence of a single-peaked spatial pattern. 
For Type~L models, the spatial configuration remains monocentric for the whole range of $\phi$ if the full dispersion is unstable (\cref{fig:Hm,fig:C2-evol,fig:bif-ClassII})

The other prediction is that, if $M$ same-sized agglomerations are equidistantly placed on a circle, a symmetry-breaking bifurcation may reduce their number to $K < M$, with $K$ again dividing $N$ and the agglomerations remaining equidistant.

\cref{prop:classification} (a) implies that Type~G models yield $\frac{N}{2}$ agglomerations, with further bifurcations of the form $\frac{N}{2} \to \frac{N}{4} \to \frac{N}{8} \to \cdots \to 2 \to 1$ expected if $N$ is a power of two \citep{Akamatsu-Takayama-Ikeda-JEDC2012,ikeda2012spatial,Osawa-et-al-JRS2017}. 
For Type~L, \cite{Takayama-Ikeda-Thisse-RSUE2020} confirmed the emergence of single-peaked or monocentric patterns in the \citep{Murata-Thisse-JUE2005} model. 
\cite{AMT-2016} formally compares \cite{Forslid-Ottaviano-JoEG2003} (Type G) and \cite{Helpman-Book1998} (Type L). 
All available formal results in the literature corroborates with \cref{prop:classification} and the numerical examples in this study.    
\end{remark}

\subsection{Proof of \cref{lem:stability}} 
\label{app:proof-lem-stability}

A myopic adjustment dynamic is a system of ordinary differential equations that describes the rate of change in the spatial distribution $\Vtx$. 
Denote the dynamic that adjusts $\Vtx$ over the set of all possible spatial distributions $\ClX\Is \left\{\Vtx\geq\Vt0\Mid\sum_{i\inI} x_i = 1\right\}$ by $\DtVtx = \Vtf(\Vtx)$, where $\DtVtx$ represents the time derivative satisfying $\sum_{i\inI} \Dtx_i = 0$ so that the total population is invariant. 
For example, $\Vtf(\Vtx) = \TlVtf(\Vtx,\Vtv(\Vtx))$ where $\TlVtf$ maps each pair $(\Vtx,\Vtv(\Vtx))$ of a state and its associated payoff to a motion vector $\DtVtx$. 

We require the following conditions on $\Vtf$: 
(RS) $\Vtf(\Vtx) = \Vt0$ if $\Vtx$ is a spatial distribution in which all populated regions earn the same payoff level, i.e., $v_j(\Vtx^*) = v_k(\Vtx^*)$ for all $j,k\in\{ i\inI \mid x_i^* > 0\}$, (PC) $\Vtv(\Vtx) ^\top \Vtf(\Vtx) > 0$ if $\Vtf(\Vtx)\neq\Vt0$, 
and (Sym) $\VtP \Vtf(\Vtx) = \Vtf(\VtP\Vtx)$ for \emph{all} permutation matrices $\VtP$. 
We call dynamics that satisfy (RS), (PC), and (Sym) \textit{admissible dynamics}. 
The conditions (RS) and (PC) are called \textit{restricted stationality} and \textit{positive correlation}, respectively \citep{Sandholm-Book2010}. 
Also, (Sym) requires that $\Vtf$ treats all regions symmetrically. 
Finally, we assume that $\Vtf$ admits a $C^1$ extension to an open neighborhood of $\ClX$ in $\BbR^N$ to use simple derivatives. 

Admissible dynamics include the Brown--von Neumann--Nash dynamic 
\citep{Brown-vonNeumann-Rand1950,Nash-AM1951}, 
the Smith dynamic 
\citep{Smith-TS1984}, 
and 
Riemannian game dynamics 
\citep{Mertikopoulos-Sandholm-JET2018}. 
The projection dynamic 
\citep{Dupuis-Nagurney-AOR1993} 
and 
the replicator dynamic 
\citep{Taylor-Jorker-MB1978} are representative instances of Riemannian game dynamics that satisfy (Sym), and are often applied for regional models. 

For the uniform distribution $\BrVtx$, (RS) implies $\Vtf(\BrVtx) = \Vt0$, i.e., $\BrVtx$ is a stationary point of $\Vtf$. 
Denote the Jacobian matrix of $\Vtf$ at $\BrVtx$ by $\VtF = [\frac{\partial f_i}{\partial x_j}(\BrVtx)]$. 
Assume that $\VtF$ has no eigenvalues with zero real parts. 
Then, $\BrVtx$ is \emph{linearly stable} if all the eigenvalues of $\VtF$, which we denote by $\{\eta_k\}$, have negative real parts, and \emph{linearly unstable} if some eigenvalue has positive real parts \citep[see,e.g.,][]{Hirsch-et-al-Book2012}. 
Spatial equilibrium $\BrVtx$ is said to be \emph{stable} (\emph{unstable}) if it is linearly stable (unstable) under admissible dynamics. 
The marginal case in which the largest eigenvalue has zero real parts is unimportant as it often corresponds to measure-zero subsets of the parameter space.

Under admissible dynamics, the stability of $\BrVtx$ can be determined by $\VtV$, i.e., without checking $\VtF$ explicitly. 
To exclude degenerate cases, assume that there is no other equilibrium in the neighborhood of $\BrVtx$. 
Then, (PC) implies   
that there is a neighborhood $\ClO\subset\ClX$ 
of $\BrVtx$ such that $\Vtv(\Vtx)^\top \Vtf(\Vtx) > 0$ 
for all $\Vtx\in\ClO\setminus\{\BrVtx\}$. 
For small deviation $\Vtz = \Vtx - \BrVtx$ with $\Vtx\in\ClO\setminus\{\BrVtx\}$, 
$\Vtf(\Vtx) \approx \Vtf(\BrVtx) + \nabla\Vtf(\BrVtx)\Vtz = \VtF\Vtz$, 
$\Vtv(\Vtx) \approx \Vtv(\BrVtx) + \nabla\Vtv(\BrVtx)\Vtz = \Brv\Vt1 + \frac{\Brv}{\Brx} \VtV$, 
and $0 = \Vt1^\top \DtVtx = \Vt1^\top \Vtf(\Vtx) \approx \Vt1^\top \VtF \Vtz$. 
Combined together, for all $\Vtx \in\ClO\setminus\{\BrVtx\}$, 
\begin{align}
    \Vtv(\Vtx)^\top \Vtf(\Vtx) 
    \approx 
    \left(\Vtv(\BrVtx) + \nabla\Vtv(\BrVtx)\Vtz \right)^\top
    \left(\Vtf(\BrVtx) + \VtF\Vtz \right) 
    = 
    \tfrac{\Brv}{\Brx}
    \left(\VtV\Vtz\right)^\top
    \left(\VtF\Vtz \right) > 0 \label{eq:B4.PC}. 
\end{align}
Under \cref{assum:racetrack-economy}, we can choose the same set of eigenvectors for $\VtV$ and $\VtF$ because they are both symmetric \emph{circulant} matrices. 
Let $\{\Vtz_k\}$ be the set of eigenvectors and let $\omega_k$ and $\eta_k$ be the eigenvalues of $\VtV$ and $\VtF$ associated with $\Vtz_k$, respectively. 
Then, for each eigenvector $\Vtz_k$ except for $\Vtz_0 = \Vt1$, \cref{eq:B4.PC} yields
\begin{align}
    \left(\VtV\Vtz_k\right)^\top
    \left(\VtF\Vtz_k \right)  
    =   
    \omega_k \eta_k 
    > 0. 
    \label{eq:B2_eps}
\end{align}
As $\VtF$ and $\VtV$ are both symmetric, $\eta_k$ and $\omega_k$ are both real. 
Thus, \cref{eq:B2_eps} implies $\sgn[\eta_k] = \sgn[\omega_k]$. 
Therefore, $\BrVtx$ is stable under all admissible dynamics if and only if $\omega_k < 0$ for \emph{all} $k$, excluding $k = 0$ that corresponds to $\Vtz_0 = \Vt1$. 
Likewise, $\BrVtx$ is unstable if and only if $\omega_k > 0$ for \emph{some} $k$, again excluding $k = 0$. 

Suppose exactly one $\omega_k$ changes sign from negative to positive at $\phi_k^*$. 
Then, from \cref{eq:B2_eps}, the corresponding eigenvalue of the Jacobian matrix of any admissible dynamic at $\BrVtx$ must also cross zero at $\phi_k^*$. 
Bifurcation theory shows that the system departs from $\BrVtx$ along the direction of the associated eigenvector $\Vtz_k$, as it is tangent to the ``unstable manifold'' at the bifurcation point \citep[see, e.g.,][]{Hirsch-et-al-Book2012,Kuznetsov-Book2013}.

\begin{center}
    * \quad * \quad *
\end{center}
\noindent Appendices \ref{app:ua} to \ref{app:example_models} are provided as a separate online appendix.

\clearpage

{\small\singlespacing
\ifx\undefined\bysame
\newcommand{\bysame}{\leavevmode\hbox to\leftmargin{\hrulefill\,\,}}
\fi

}
\end{bibunit}

\clearpage

\newgeometry{margin=.8in}
\setcounter{page}{1}
\renewcommand{\thepage}{A\arabic{page}}
\begin{bibunit}

\onehalfspacing
\begin{center}
{\large \TITLE}

\bigskip 
    
{\large Online Appendix}

\bigskip 
 
\smallskip
\today
\end{center}
\begin{center}
\color{myblue} 
Takashi Akamatsu, Tomoya Mori, Minoru Osawa, and Yuki Takayama
\end{center}

\medskip
This appendix collects derivations and numerical examples omitted from the main text.
Appendix \ref{app:ua} examines the evolution of Japanese cities from 1970 to 2020.
Appendix \ref{app:eight-regions} presents numerical examples for an eight-region circular economy.
Appendix \ref{app:geo-advantage} considers alternative transport network geometries while preserving symmetry in local characteristics.
Appendix \ref{app:local_advantage} studies variations in local characteristics within the circular economy.
Appendix \ref{app:example_models} contains detailed derivations.

\setcounter{section}{1}
\renewcommand\contentsname{\large Contents}

\clearpage

\section{Evolution of cities}
\label{app:ua}

\subsection{Development of high-speed transport networks in Japan}

\begin{figure}[ht]
\centering
\begin{subfigure}[b]{.9\hsize}
\includegraphics[width=\hsize]{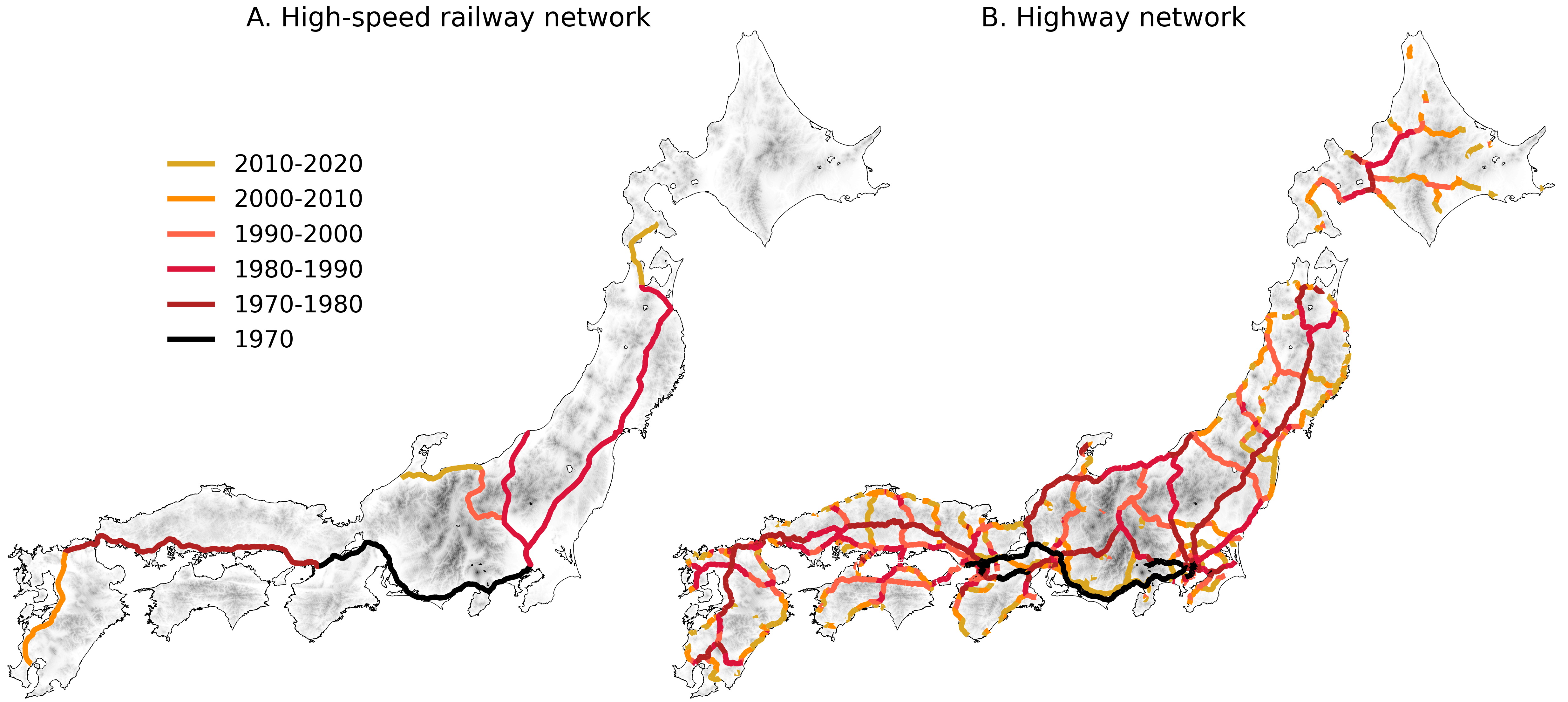}
\caption{Transport networks}
\label{fig:highways}

\medskip 

\end{subfigure}
\begin{subfigure}[b]{.5\hsize}
\includegraphics[width=\hsize]{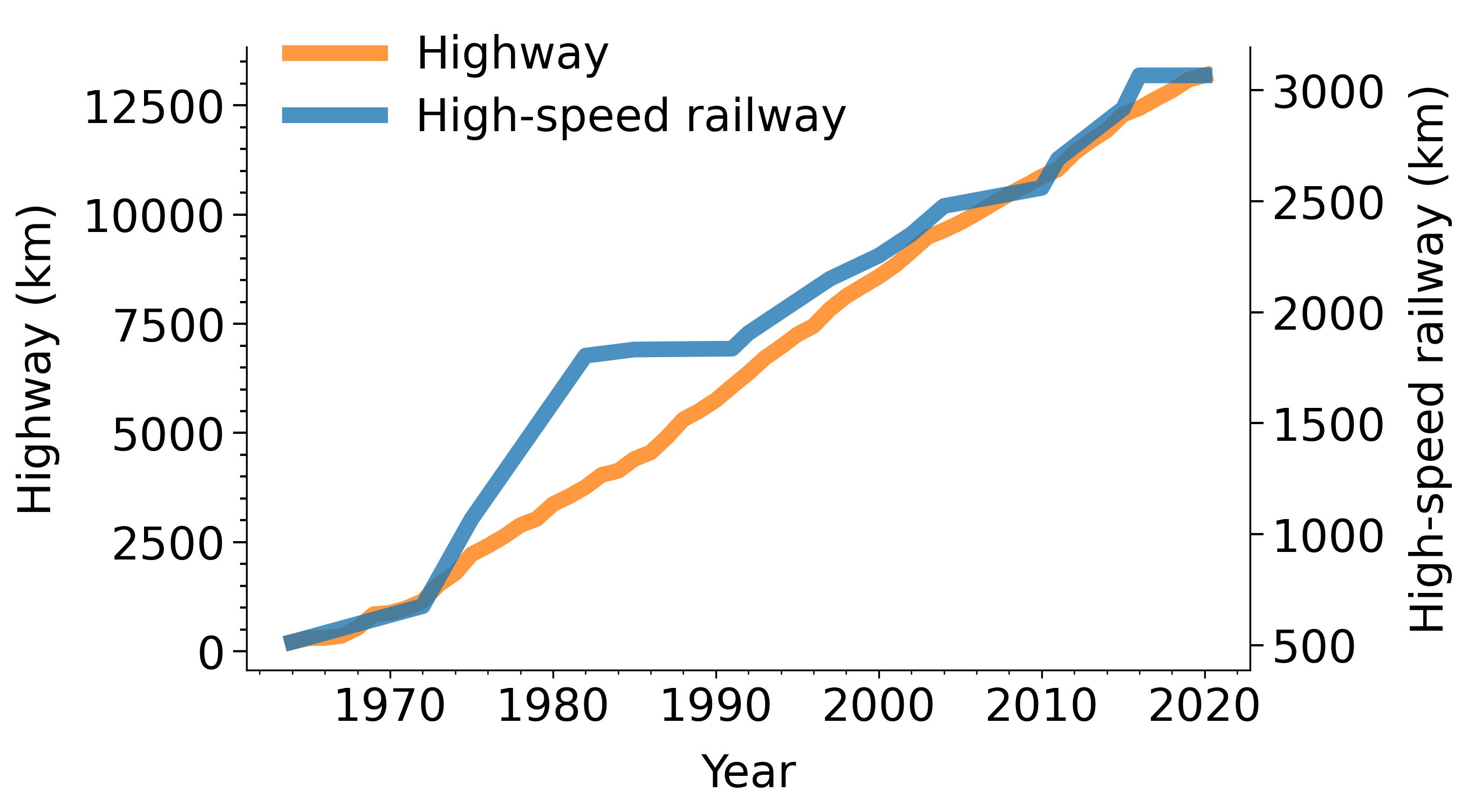}
\caption{Network lengths}
\end{subfigure}
\caption{The development of high-speed network in Japan}\caption*{\footnotesize\textit{Notes:} Data source: The Digital National Land Information Download Service (Highway: \url{https://nlftp.mlit.go.jp/ksj/gml/datalist/KsjTmplt-N06-2023.html}, High-speed railway: \url{https://nlftp.mlit.go.jp/ksj/gml/datalist/KsjTmplt-N05-2023.html}).}
\label{fig:network}
\end{figure}

\Cref{fig:network} reports the development of high-speed railway and highway networks, respectively, in Japan between 1970 and 2020. 
Both networks were initially spurred by infrastructure investments surrounding the 1964 Tokyo Olympics. 
Over this period, total highway length increased from 1,119 km to 9,050 km, while high-speed rail expanded from 515 km to 3,106 km, which are more than eightfold and sixfold increases, respectively. 
The steady, long-run expansion of these networks makes Japan a natural setting for studying comparative statics with respect to transport costs. 

\subsection{Japanese cities and their growths}
\label{app:Japanese-uas}

\begin{figure}[tb]
\centering
\includegraphics[width=.4\hsize]{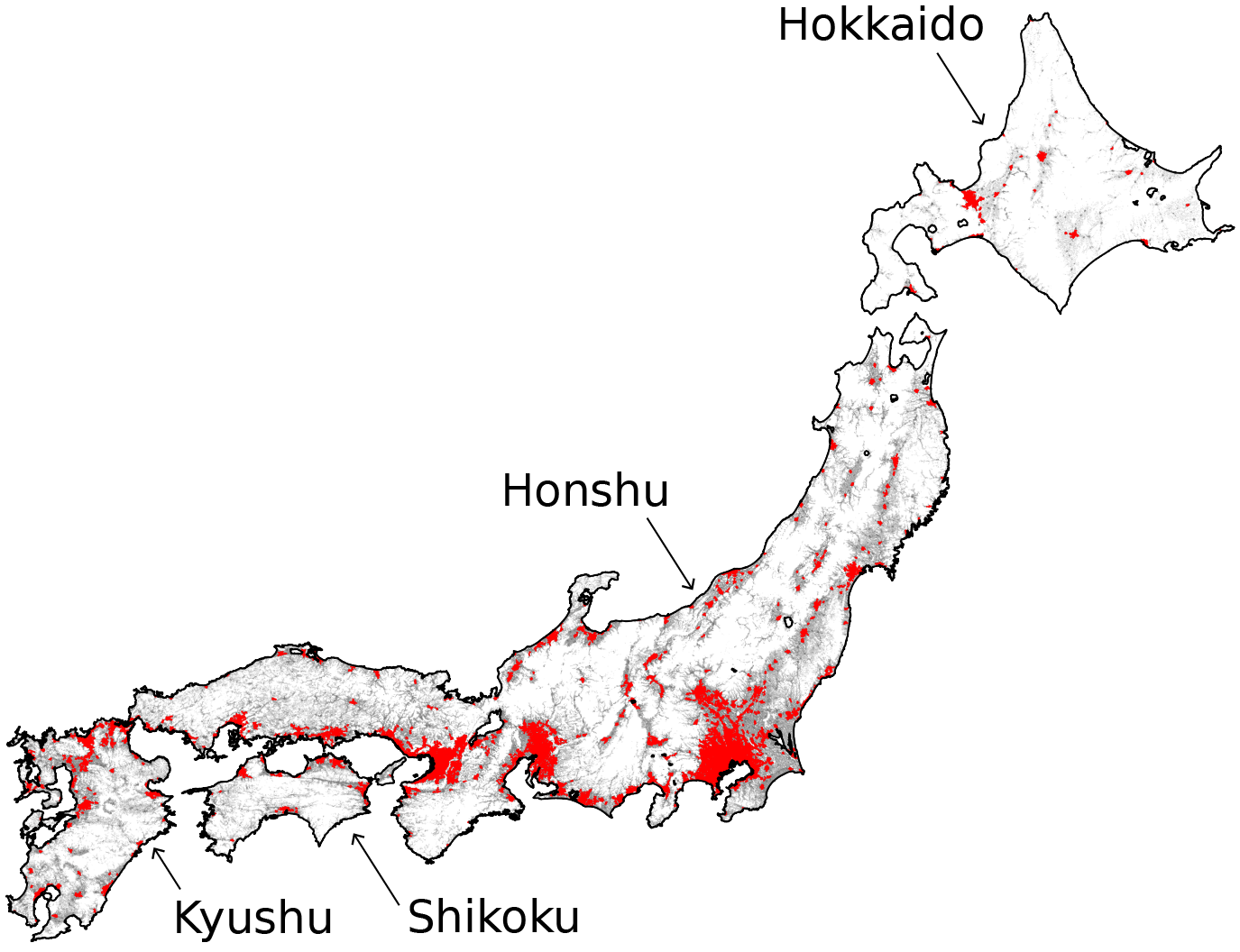}
\caption{Urban agglomerations of Japan in 2020}\label{fig:ua}
\end{figure}

We identify cities in Japan using the Grid Square Statistics from the Population Census for 1970--2020.
A \emph{city} is defined as an \emph{urban agglomeration} (\emph{UA}), consisting of contiguous 1~km~$\times$~1~km grid cells with a population density of at least 1,000~persons/km\textsuperscript{2} and a total population of at least 10,000. 
Our results are robust to alternative threshold values.
\Cref{fig:ua} displays the 431 UAs identified in 2020.
These UAs occupy 6\% of Japan's land area while containing about 80\% of the national population.
Populated cells with fewer than 1,000 residents are shown in grey, with darker shading indicating higher population counts. 
We restrict the analysis to grid cells that are reachable by road from the four major islands—Hokkaido, Honshu, Shikoku, and Kyushu.
UAs are identified separately for each census year from 1970 to 2020 (at five-year intervals), and consistent unique IDs are assigned across years to track individual agglomerations over time. 
For details on the construction of UAs, see \cite{Mori-Murakami-DP2025}.

\begin{figure}
\centering
\includegraphics[width=\hsize]{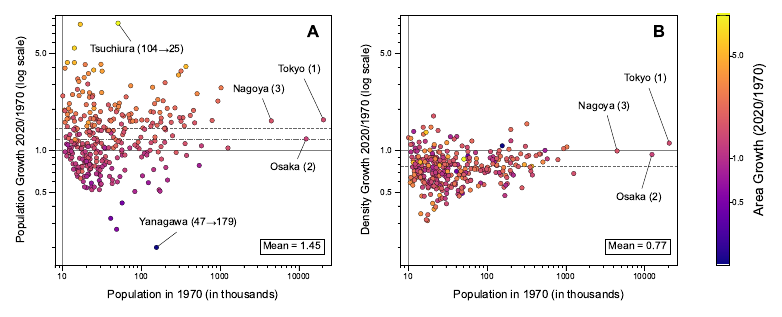}
\caption{Population and density growths from 1970 to 2020}\label{fig:ua-pop-dens-growth}
\FigureNote{In Panel A, the dashed horizontal line indicates simple mean of the growth ratios of the identified Japanese cities from 1970 to 2020 ($1.45$), and the dot-dashed line indicates the growth ratio of the total population of Japan ($1.21$). Likewise, the dashed horizontal line in Panel B shows the arithmetic mean of the density growths ratios. In both panels, marker color encodes the area growth ratio during the same period. For the city labels, number in parentheses shows the change in a city's population rank (or the invariant rank) from 1970 to 2020.}
\end{figure}

\begin{figure}
    \centering
    \begin{subfigure}[b]{\hsize}
		\centering
		\includegraphics[width=.95\hsize]{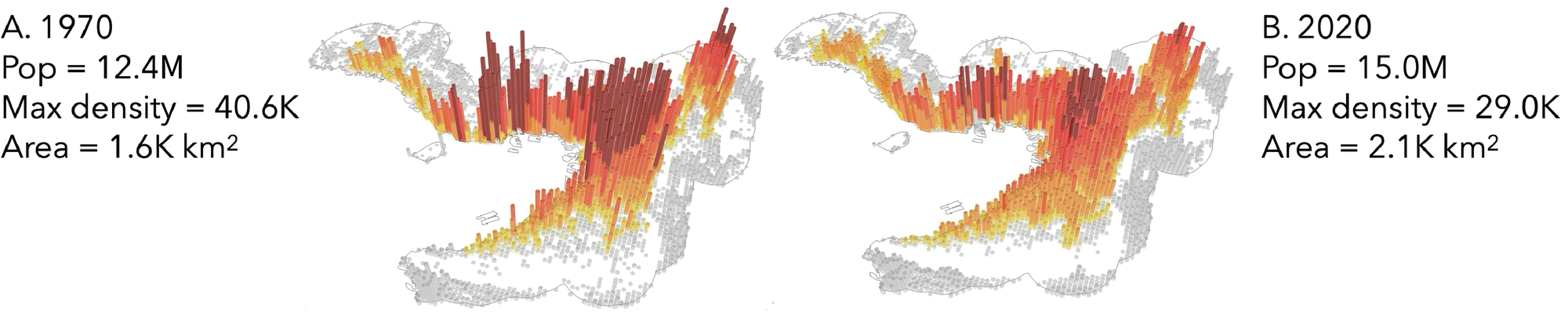}
		\caption{The population distribution within Osaka in 1970 and 2020\label{fig:JP-LD-Osaka}}
	\end{subfigure}

    \bigskip 

	\begin{subfigure}[b]{\hsize}
		\centering
		\includegraphics[width=.8\hsize]{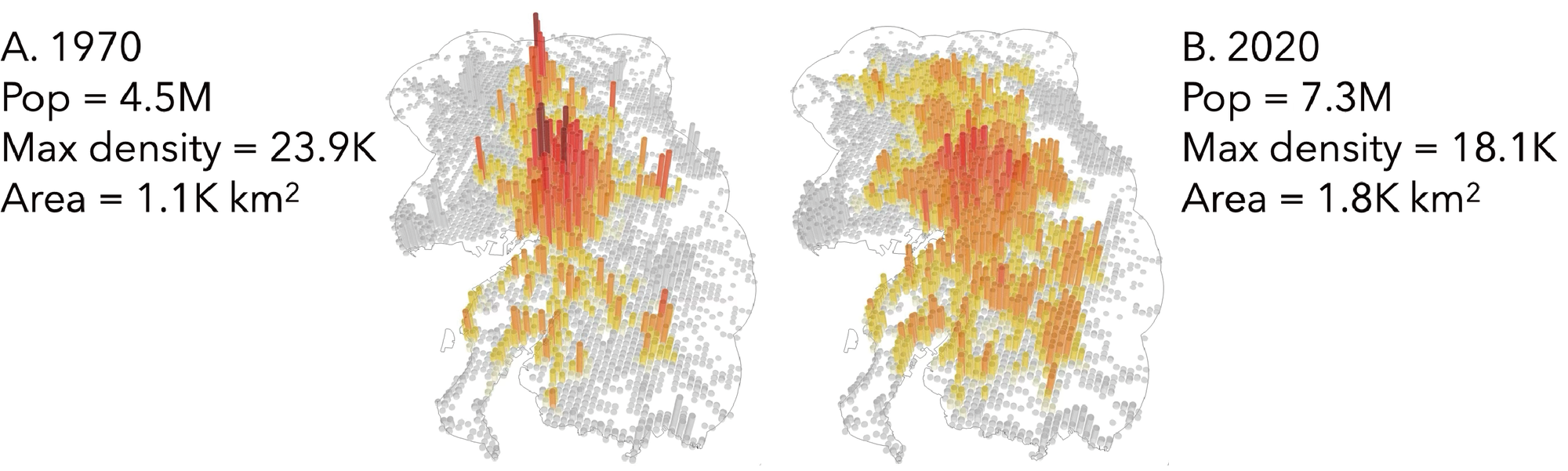}
		\caption{The population distribution within Nagoya in 1970 and 2020\label{fig:JP-LD-Nagoya}}
	\end{subfigure}

    \caption{Local dispersion of Osaka (rank $=$ 2) and Nagoya (rank $=$ 3).}
    \label{fig:JP-LD-2-3}
    \FigureNote{The warmer colors indicate larger populations. The darkest grid cells have at least 20,000 inhabitants. The other thresholds are 15,000, 10,000, 5,000, 2,000, and 1,000 inhabitants.}
\end{figure}

\begin{figure}[t!]
\includegraphics[width=.46\hsize]{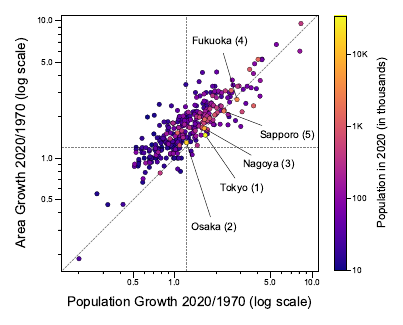}
\includegraphics[width=.46\hsize]{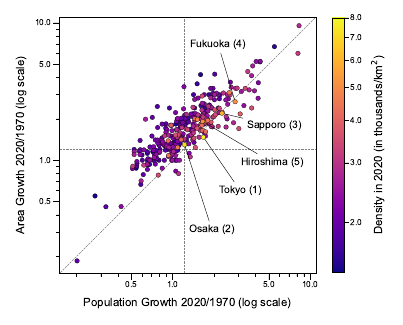}
\caption{Population and area growths from 1970 to 2020}\label{fig:ua-pop-vs-area}
\FigureNote{The diagonal line in each panel represents the locus of unchanged population density, as it indicates $\log (\text{density ratio}) = \log (\text{population ratio}) - \log (\text{area ratio}) = 0$. Marker color encodes the population size in 2020 for the left panel, and the population density in 2020 for the right panel. The number in parenthesis after the city names represents their rank in the respective senses, and top five cities are shown.}
\bigskip 

\centering
\includegraphics[width=.6\hsize]{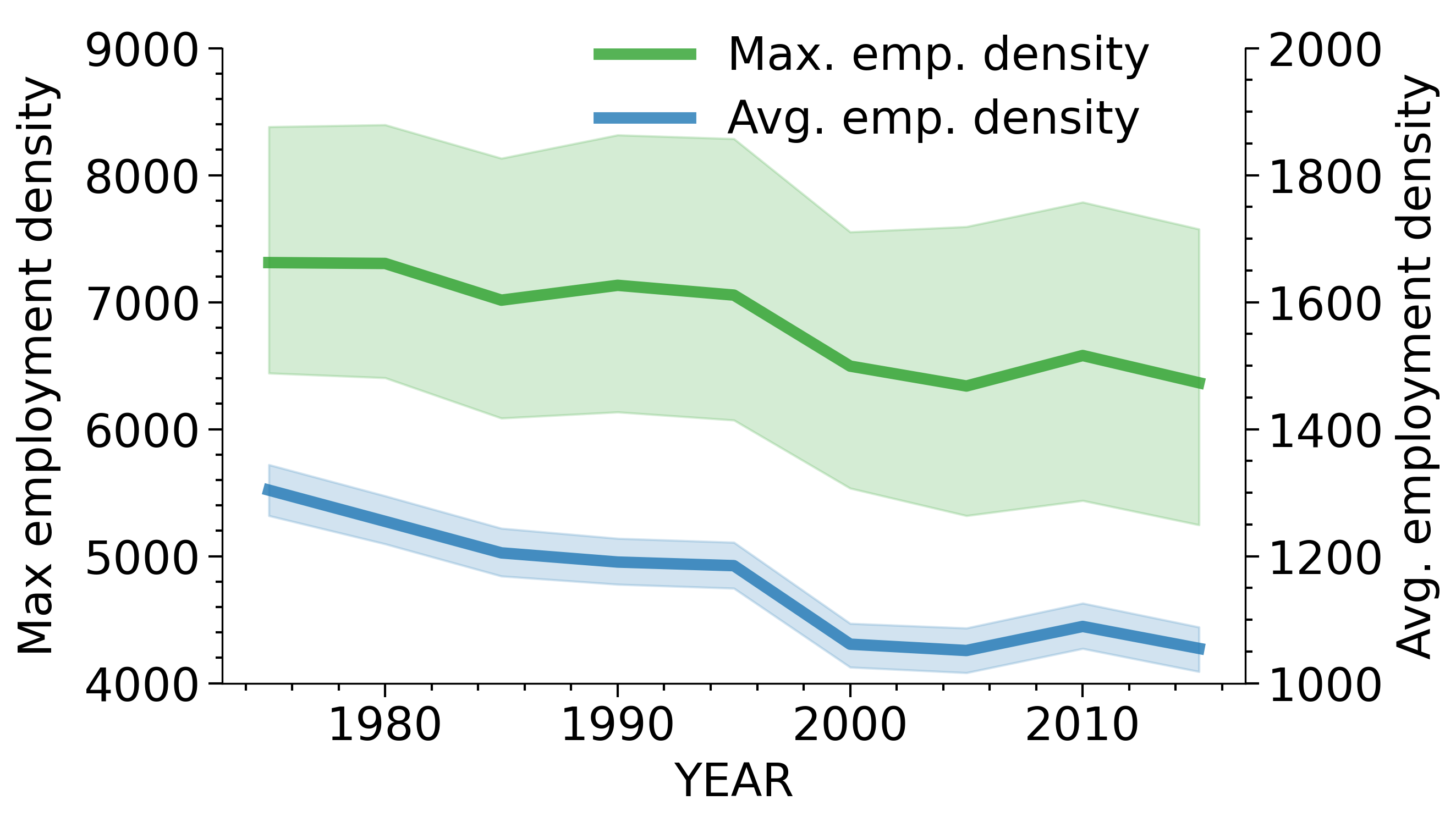}
\caption{Maximum and average employment density within a city in Japan in 1975--2014}\caption*{\footnotesize\textit{Notes:} The green and blue lines show the arithmetic means of maximum and average employment densities within a city for years 1975, 1981, 1986, 1991, 1996, 2001, 2006, 2009, and 2014.
 The shaded area indicates the range covering 90\% of the values for individual cities.
 The grid-cell data of employment are obtained from the Grid Square Statistics of the Census for Establishment (1975, 1981, 1986, 1991); Establishment and Enterprise census (1996, 2001, 2006); Economic Census for Business Frame (2009 and 2014) of Japan.}
\label{fig:flattening-emp}
\end{figure}

From 1970 to 2020, the total population in these cities grew by $55\%$, while the national population increased by only 21\%. 
The population of the largest city, Tokyo, has grown by 67\%, an increase about the same size of the second largest city, Osaka. 
\Cref{fig:ua-pop-dens-growth,fig:ua-pop-vs-area} provide closer looks. 
Population growth is typically associated with concurrent areal growth (\cref{fig:ua-pop-dens-growth}A), but population density generally decreased during the 50 years period (\cref{fig:ua-pop-dens-growth}B), in particular for small to medium-sized cities, indicating local spreading of these cities. 
It is noted that the largest cities also experienced local spreading as evident in their spatial distributions shown in \cref{fig:JP-LD-Tokyo} as well as \cref{fig:JP-LD-2-3}. 

\Cref{fig:ua-pop-vs-area} plots cities' areal growth rates against their population growth rates, both measured on a logarithmic scale.
By definition, $\log(\text{area}) = \log(\text{population}) - \log(\text{density})$, so the diagonal line in each panel represents constant population density (i.e., a density ratio of one).
Cities above this line experienced a decline in density, whereas those below experienced an increase.
The vertical and horizontal reference lines indicate Japan's aggregate population growth ratio ($1.21$), and their intersection corresponds to a hypothetical city whose population and area both grew at the national average rate.
For most cities, areal growth exceeded population growth.
Relative to these reference lines, the northeast quadrant indicates simultaneous expansion in population and area (\emph{overall growth}), while the southwest quadrant corresponds to joint decline, or \emph{urban shrinkage}.
The northwest quadrant reflects \emph{relative urban sprawl} (area growth accompanied by relative population decline) whereas the southeast quadrant indicates \emph{relative densification} (population growth accompanied by relatively slow area expansion). 
The southeast quadrant contains very few observations, suggesting that relative densification has been limited.

\Heading{Employment distribution} 
The tendency of local dispersion is observed in alternative indicator of agglomeration other than population. 
\Cref{fig:flattening-emp} shows the change in the mean values of the maximum and average employment density within a city across all cities in Japan from 1975 to 2014. 
Their long-run trend indicates that the geographical distribution of employment in a city has flattened over the past half century.

\subsection{Cities in other countries}

\begin{figure}
    \providecommand\LS{LandScan$^\text{TM}$}
    \flushleft
	\begin{subfigure}[b]{.48\hsize}
            \includegraphics[width=.49\hsize]{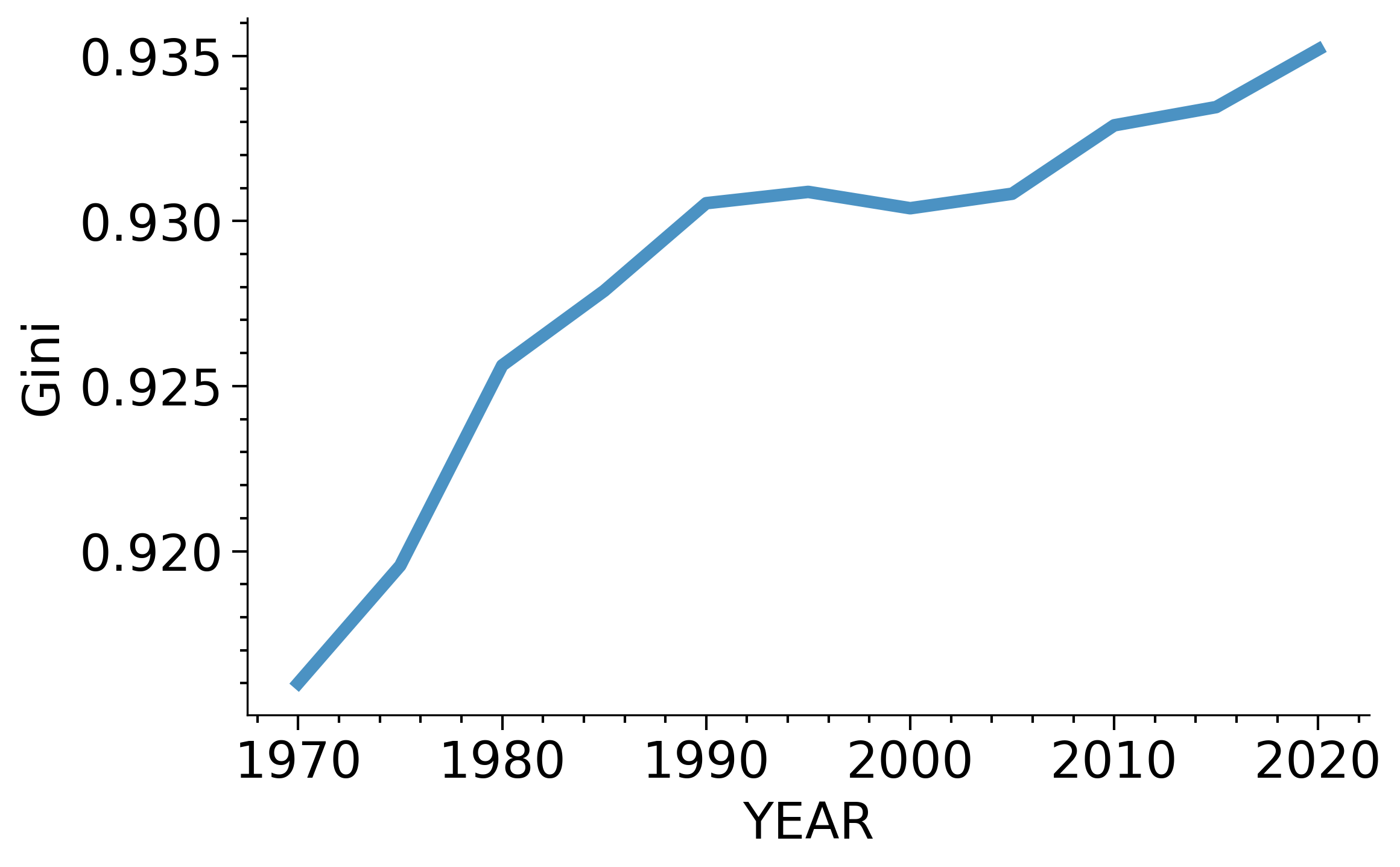}        \includegraphics[width=.49\hsize]{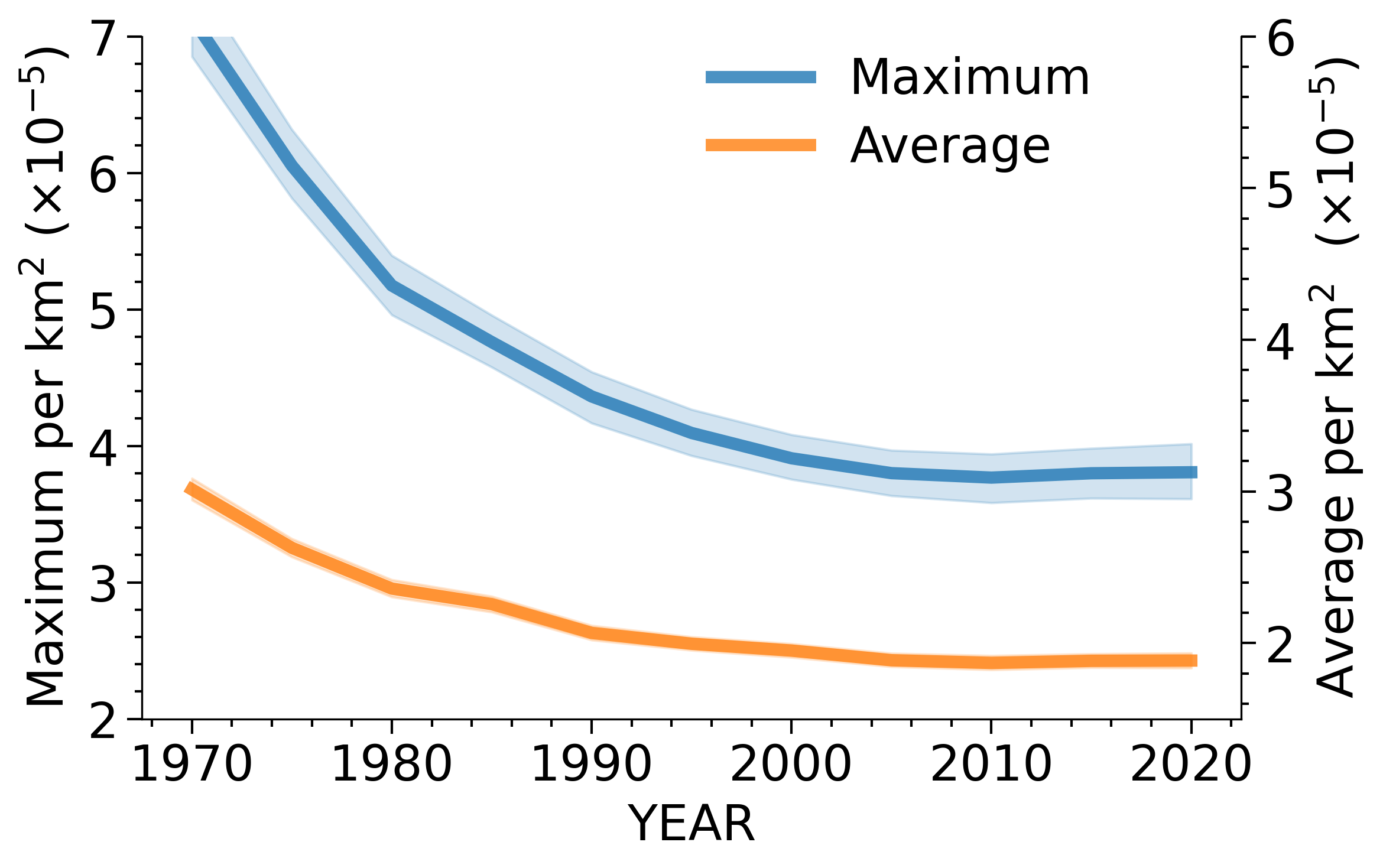}
		\caption{Japan (1970--2020), Census \label{fig:Japan(0)} }
	\end{subfigure}
	\begin{subfigure}[b]{.48\hsize}
            \includegraphics[width=.49\hsize]{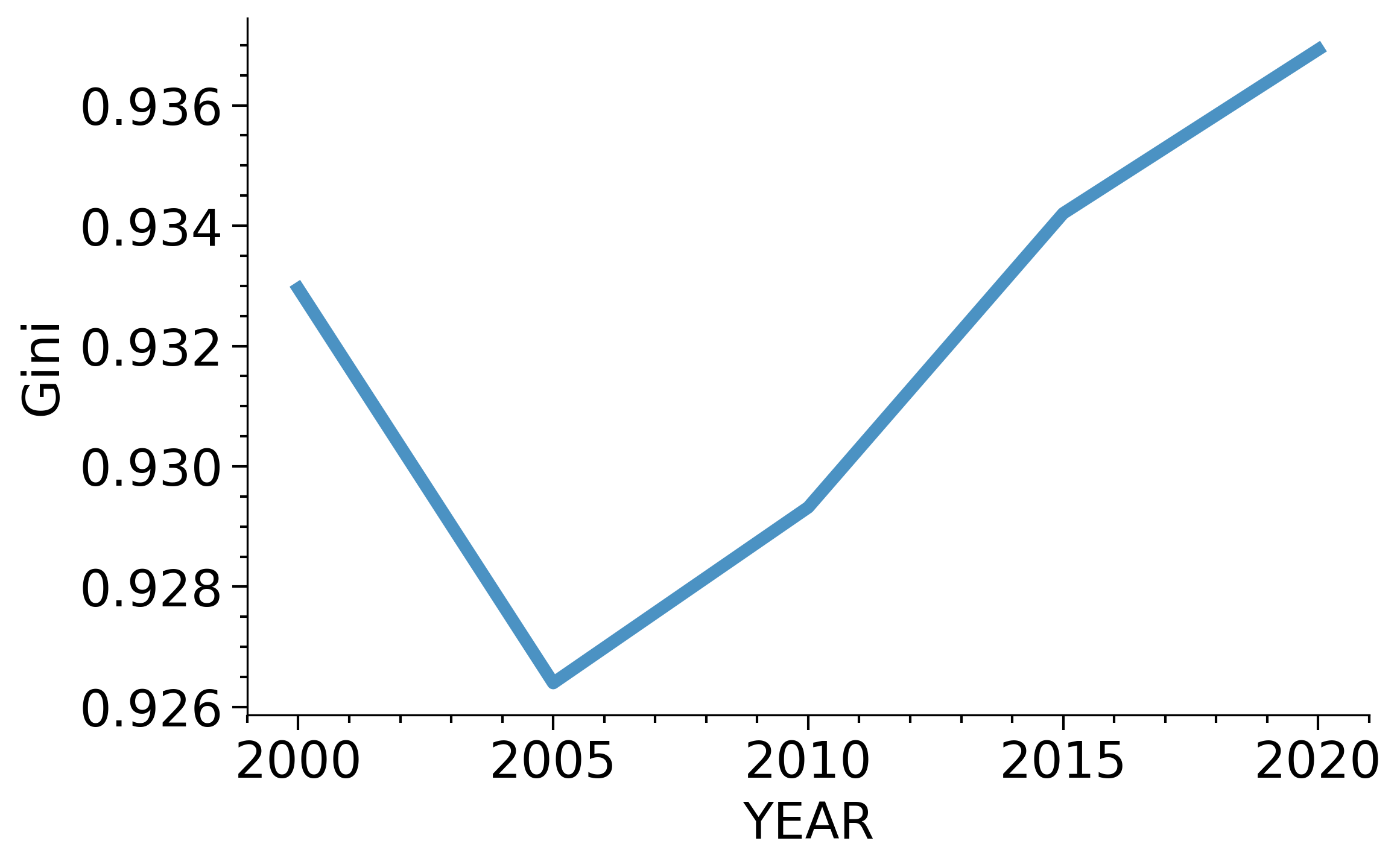}        \includegraphics[width=.49\hsize]{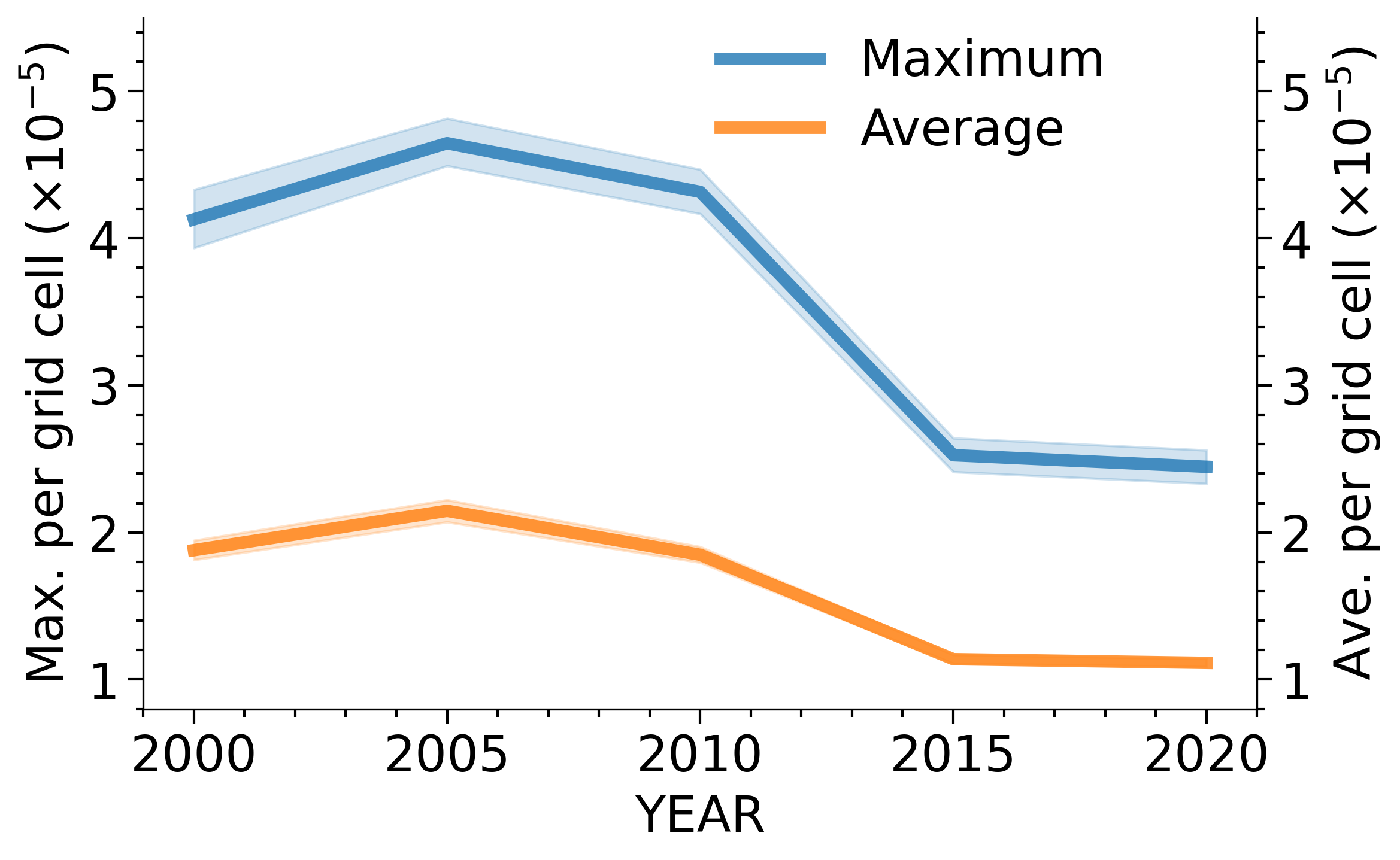}
		\caption{Japan (2000--2020), \LS \label{fig:Japan(2)} }
	\end{subfigure}
	\begin{subfigure}[b]{.48\hsize}
            \includegraphics[width=.49\hsize]{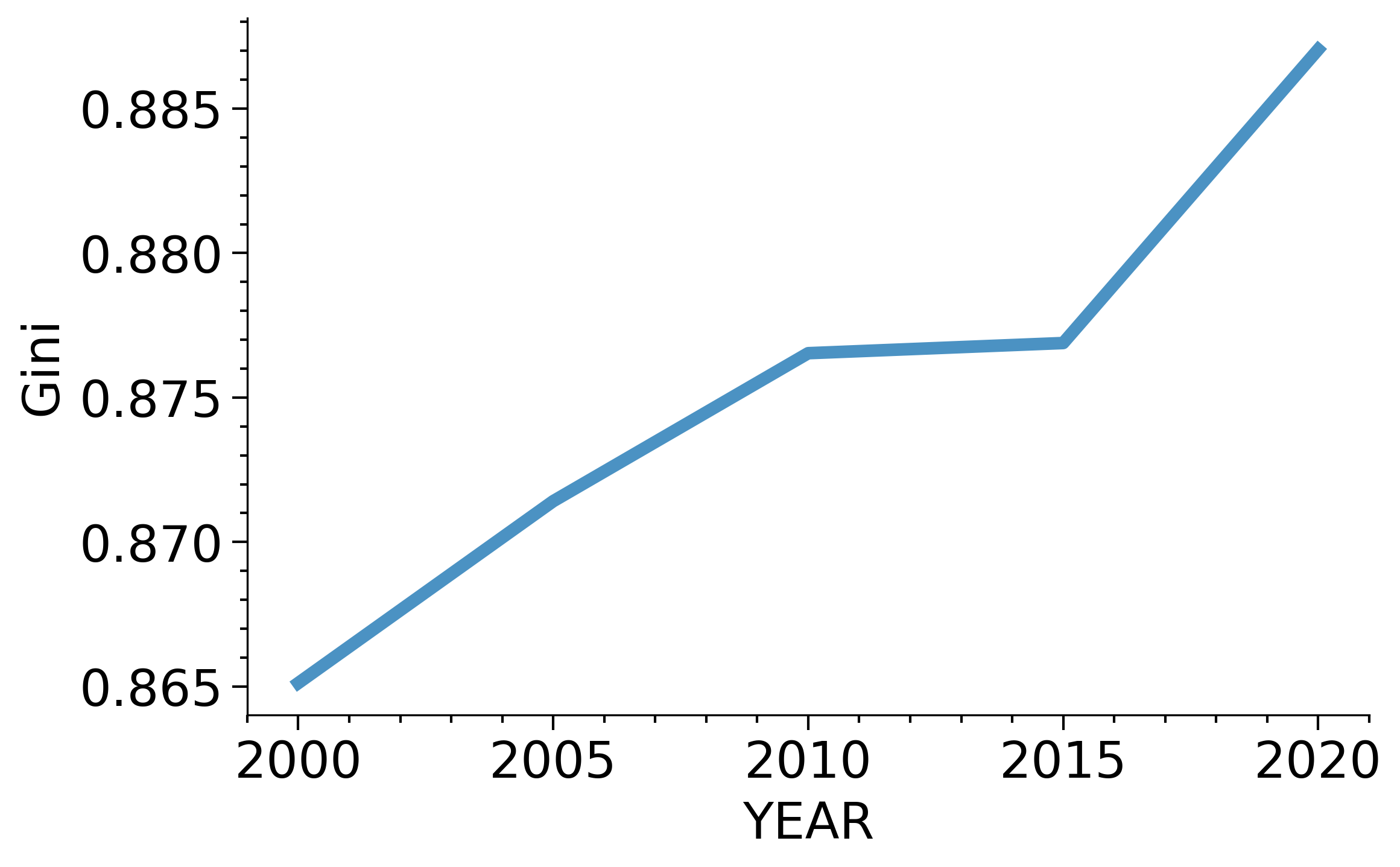}        \includegraphics[width=.49\hsize]{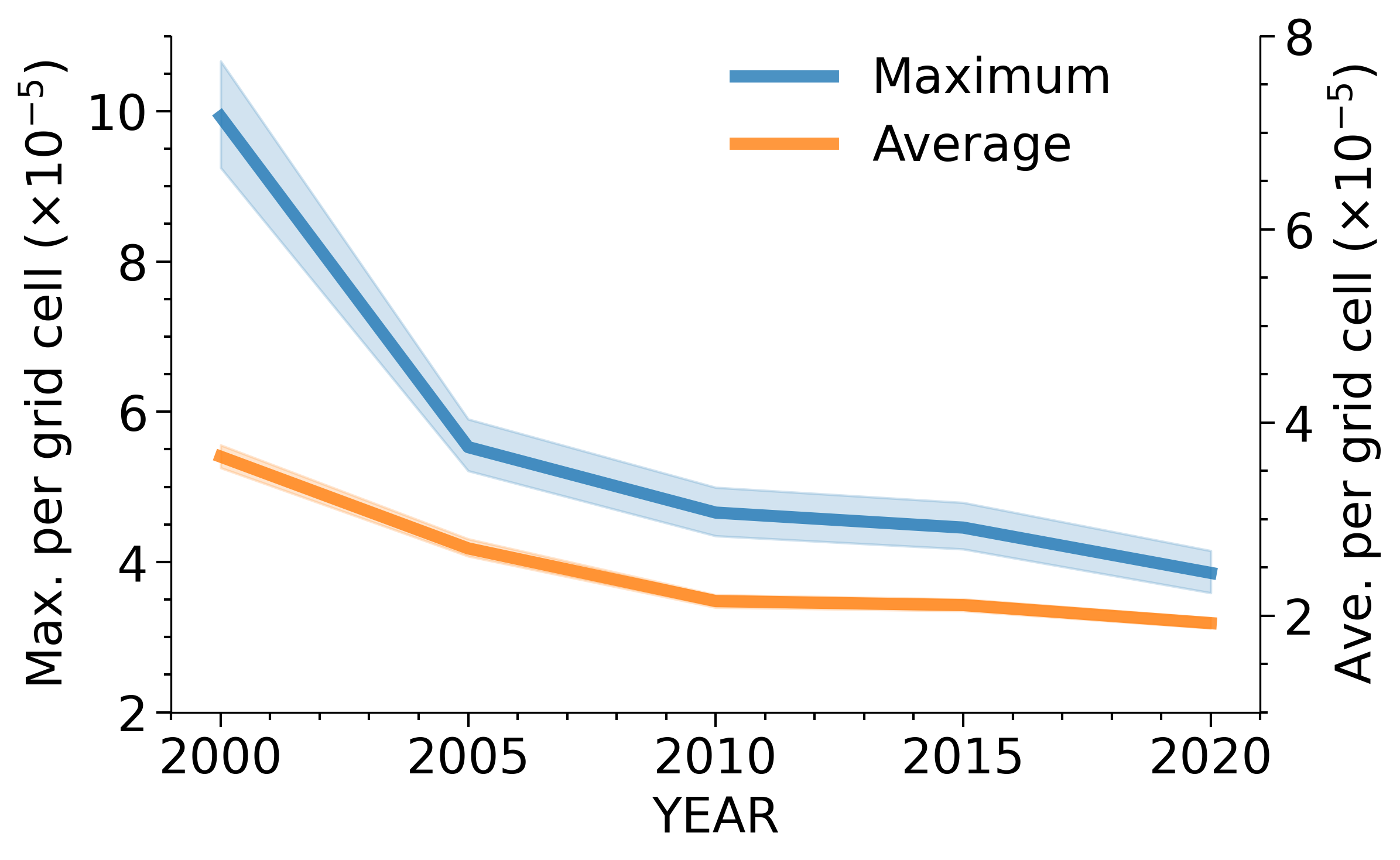}
		\caption{France (2000--2020), \LS \label{fig:France} }
	\end{subfigure}

	\begin{subfigure}[b]{.48\hsize}
            \includegraphics[width=.49\hsize]{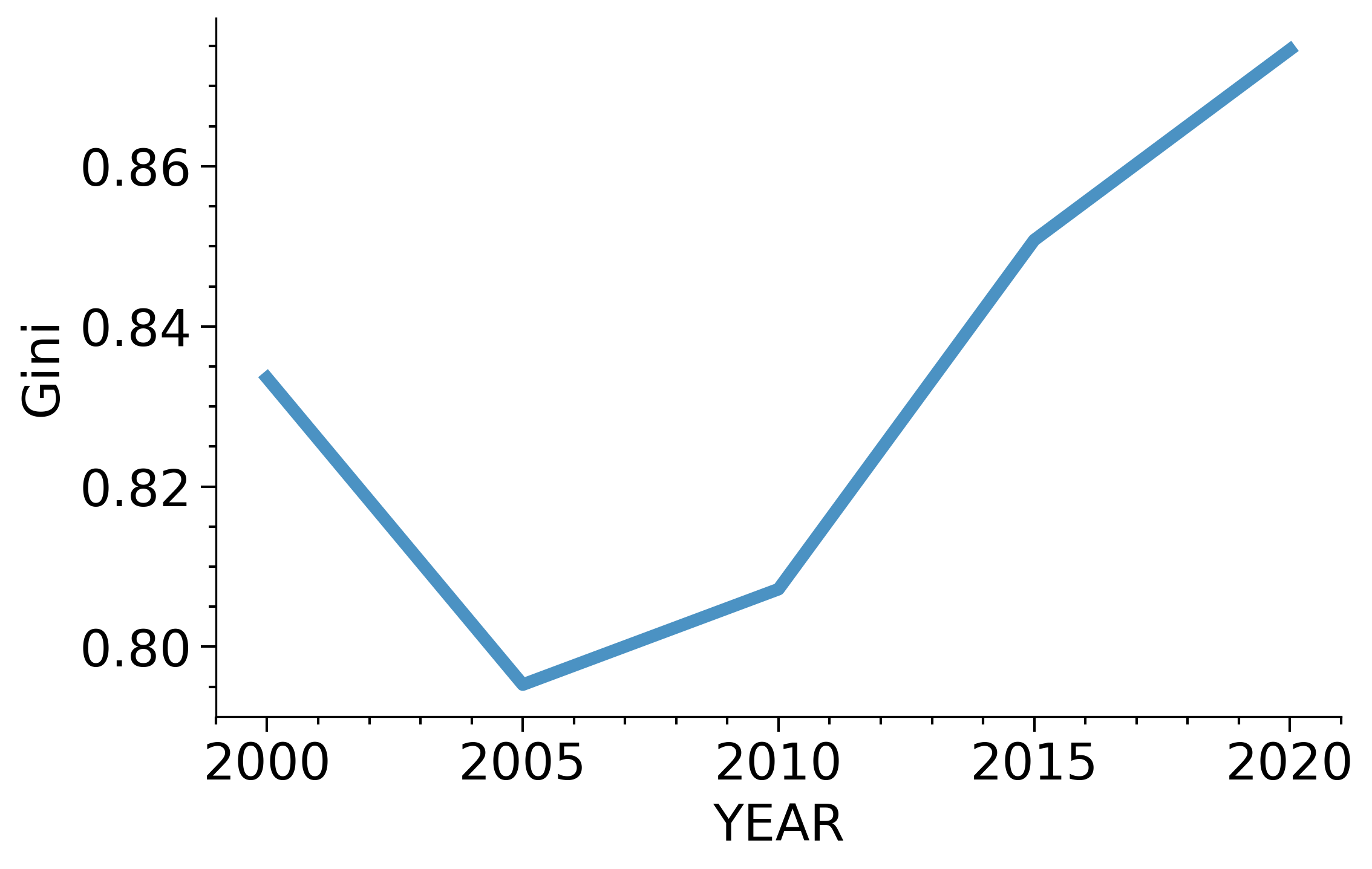}        \includegraphics[width=.49\hsize]{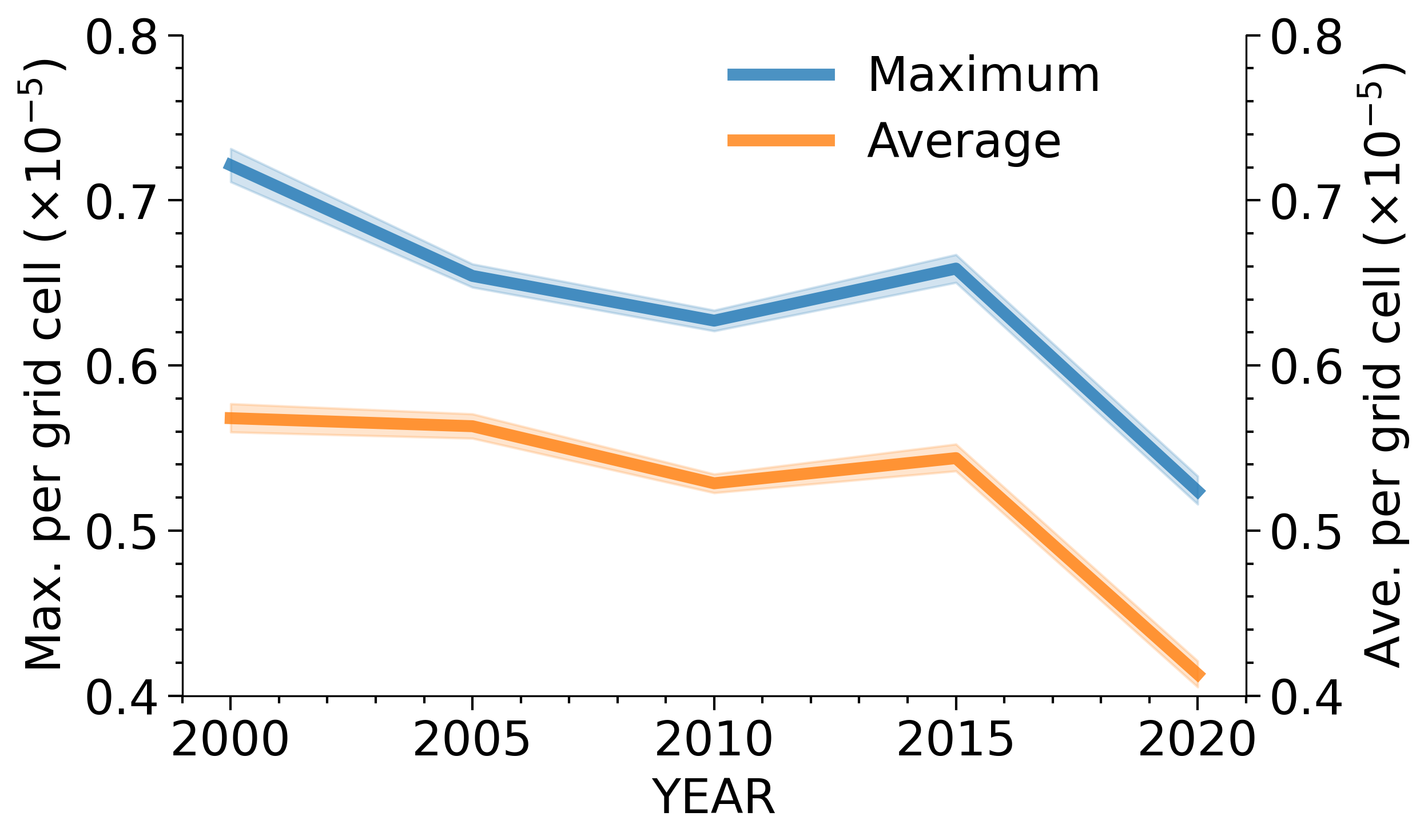}
		\caption{China (2000--2020), \LS \label{fig:China} }
	\end{subfigure}
	\begin{subfigure}[b]{.48\hsize}
            \includegraphics[width=.49\hsize]{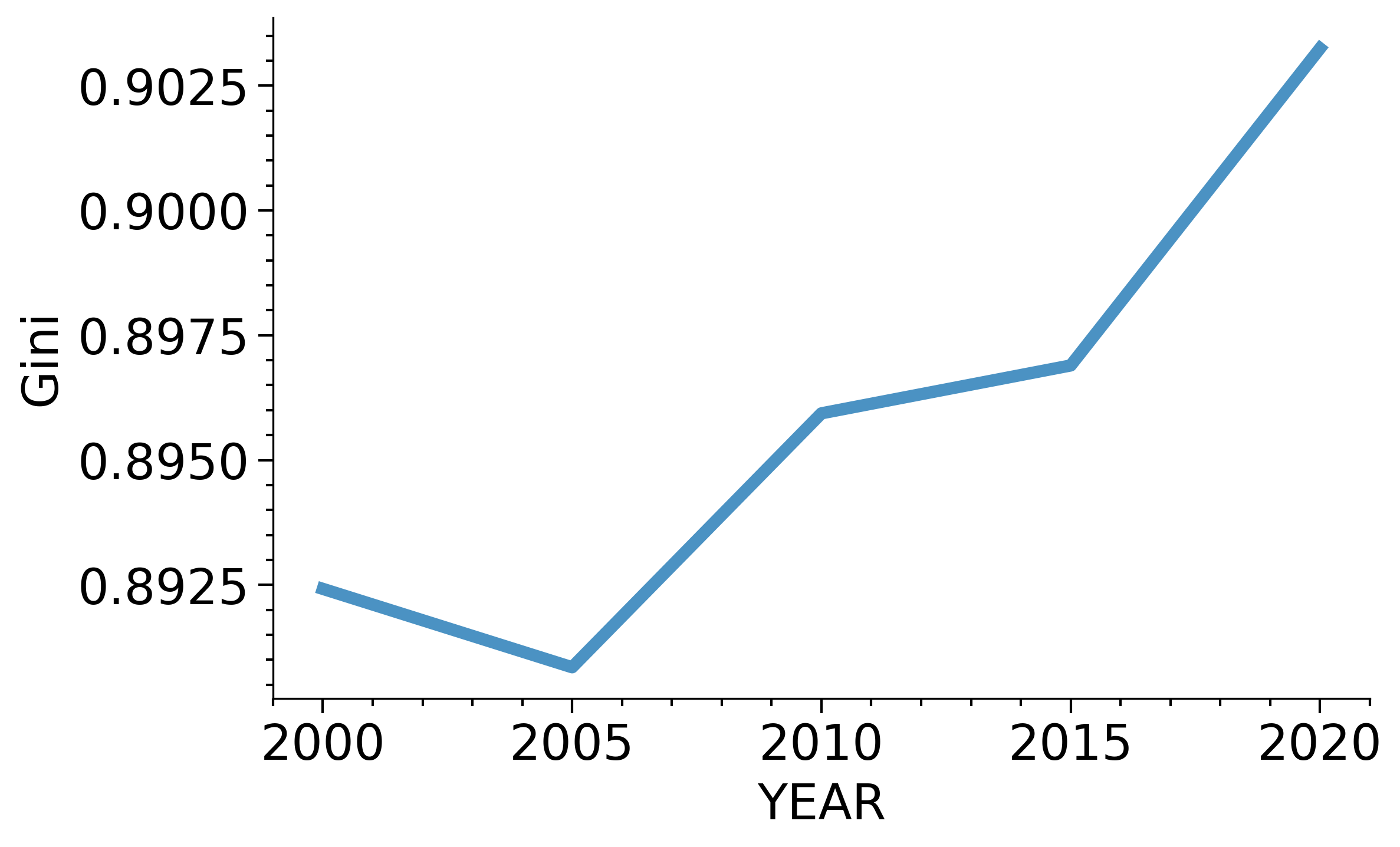}        \includegraphics[width=.49\hsize]{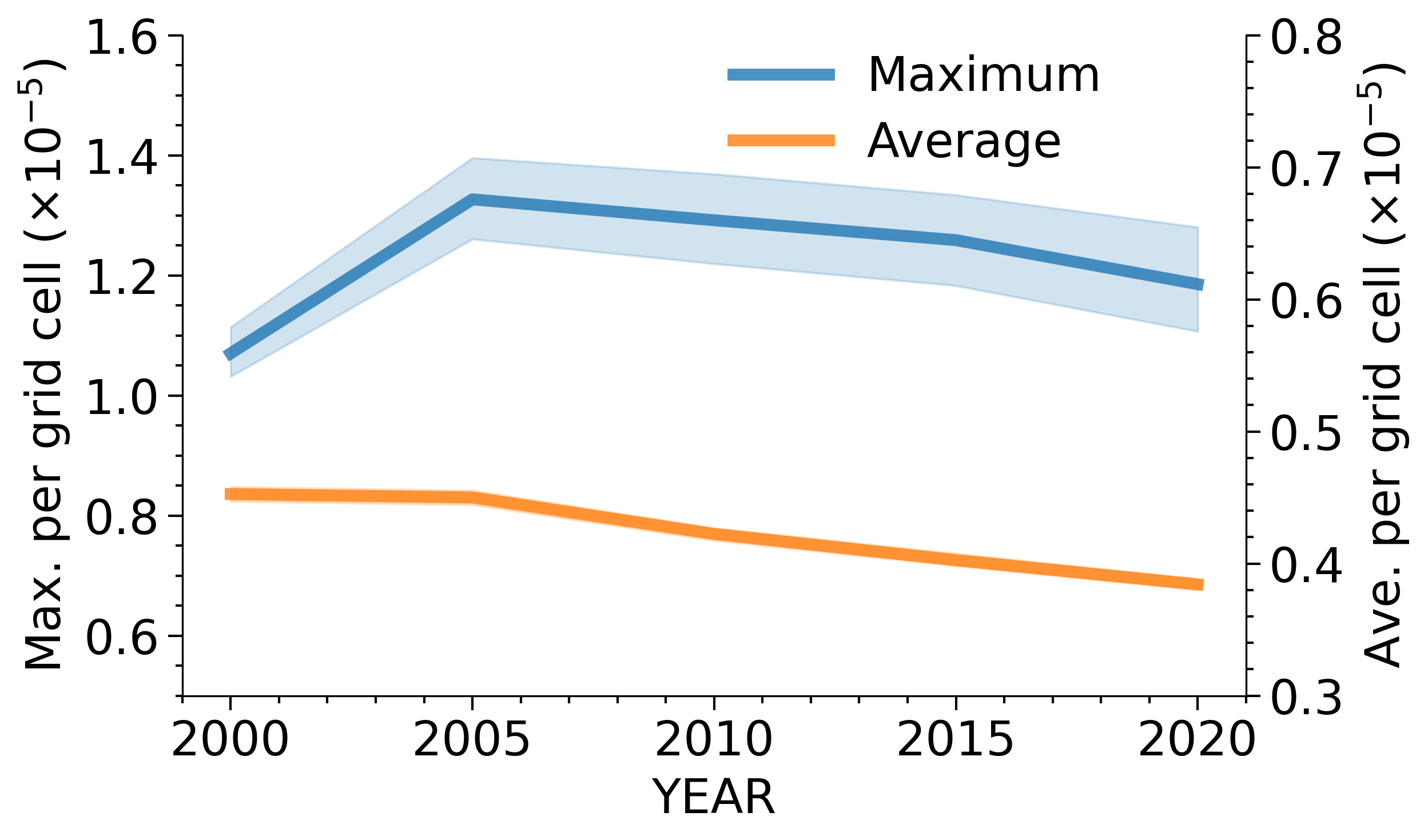}
		\caption{The US (2000--2020), \LS \label{fig:TheUS} }
	\end{subfigure}

	\begin{subfigure}[b]{.48\hsize}
            \includegraphics[width=.49\hsize]{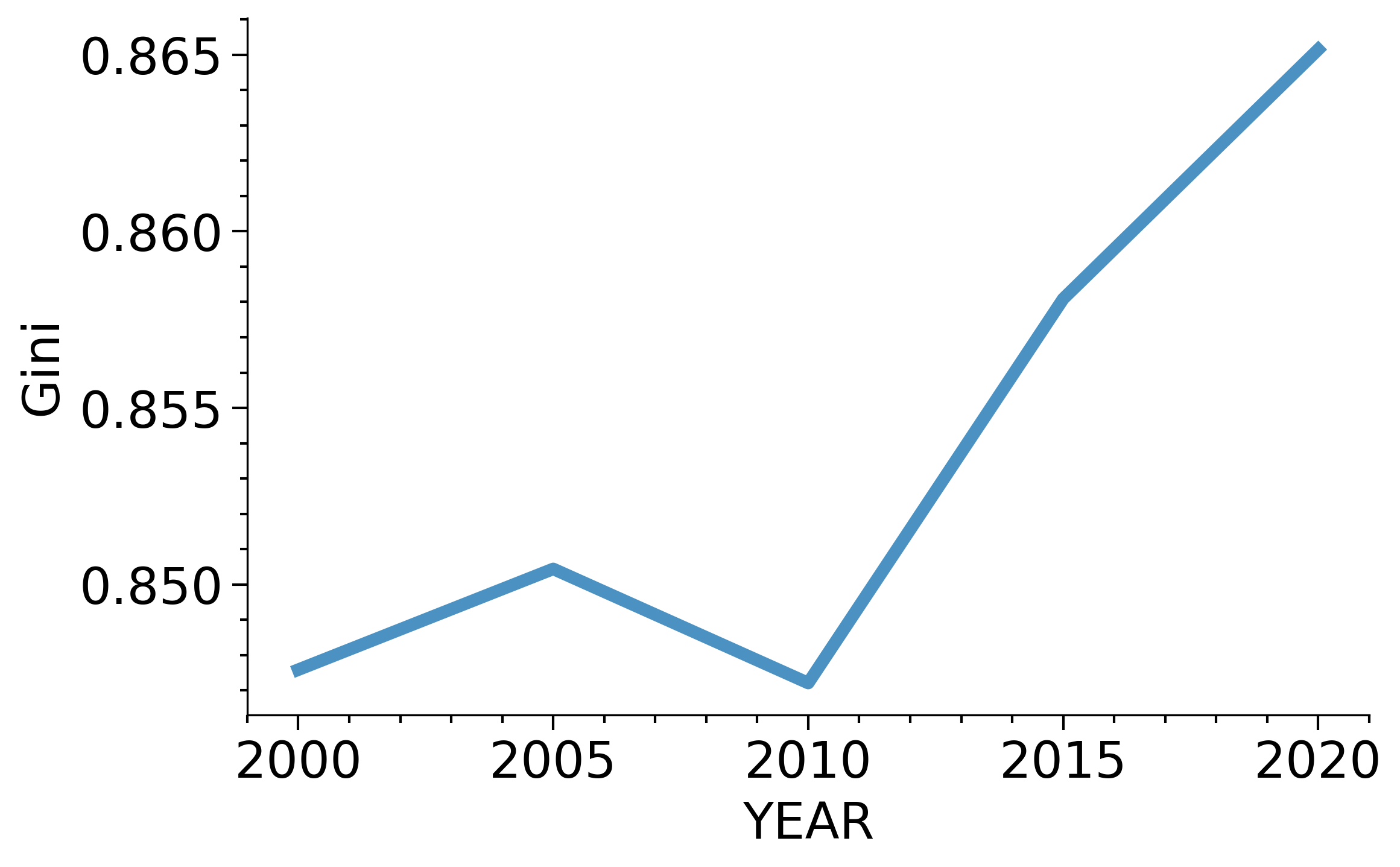}        \includegraphics[width=.49\hsize]{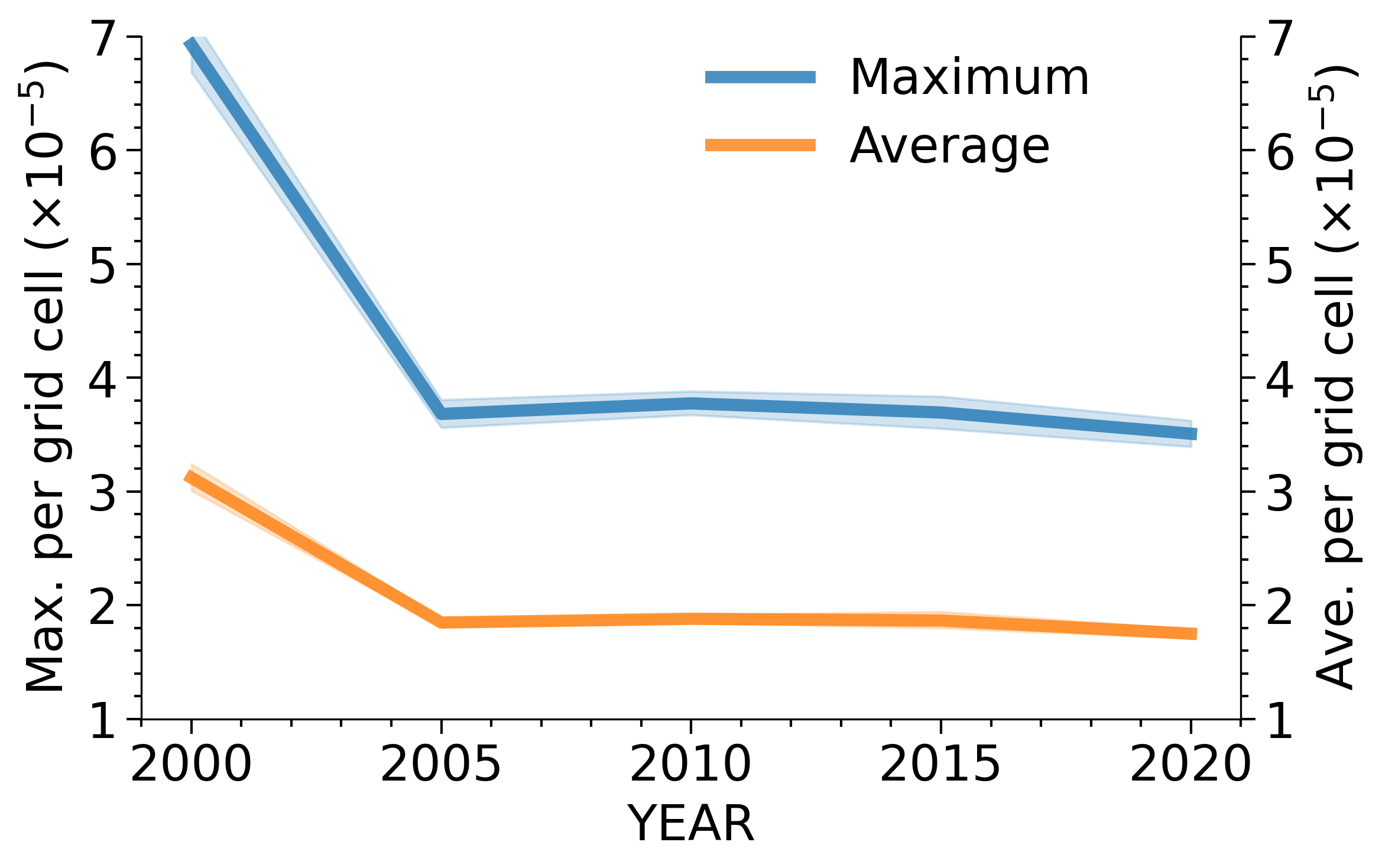}
		\caption{Germany (2000--2020), \LS \label{fig:Germany} }
	\end{subfigure}
	\begin{subfigure}[b]{.48\hsize}
            \includegraphics[width=.49\hsize]{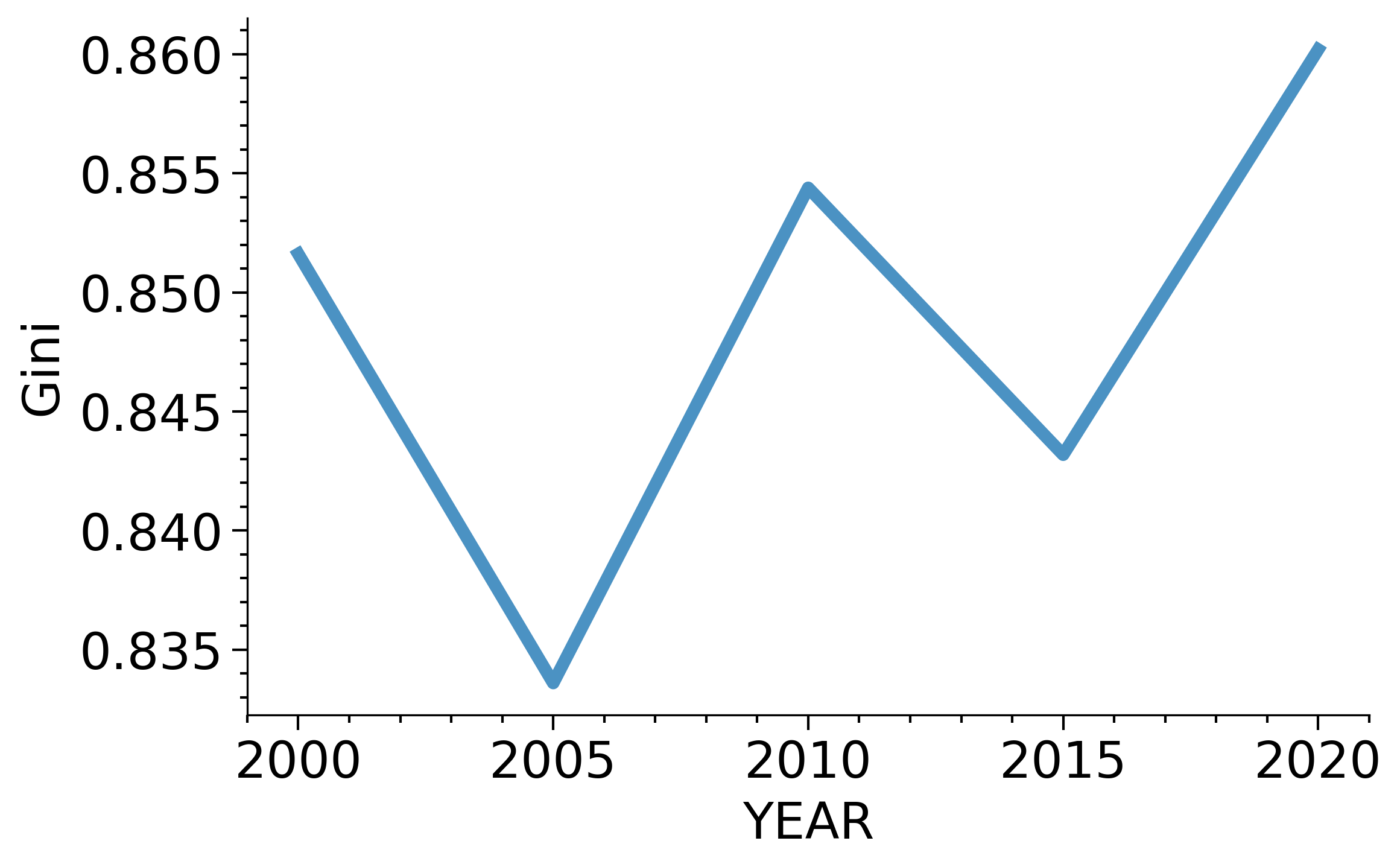}        \includegraphics[width=.49\hsize]{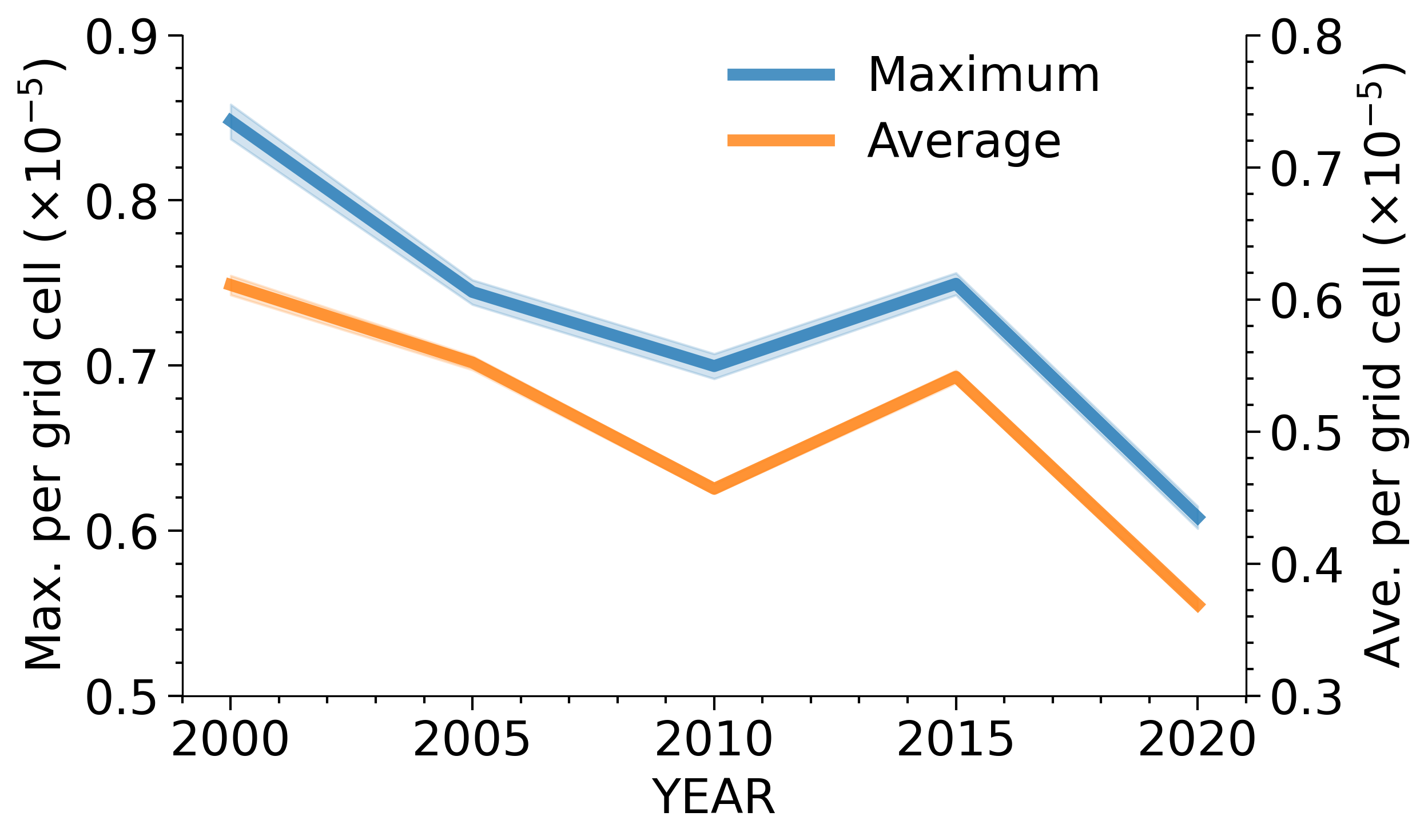}
		\caption{India (2000--2020), \LS \label{fig:India} }
	\end{subfigure}

    \caption{Global concentration and local dispersion}
    \label{fig:gc-ld-other-countries}
\end{figure}

\Cref{fig:JP} in \cref{sec:introduction} uses Japanese Census data. 
To allow a parallel comparison between different countries, we employ the the LandScan\texttrademark\  Global Population Database, developed by the Department of Energy's Oak Ridge National Laboratory (ORNL), as the basic grid population data to see the evolution of cities. 
Cities are defined in the same manner as for Japanese cities. 
To asses compatibility between the LandScan data and the census we compare Census and the LandScan data for Japan. 
\Cref{fig:Japan(0),fig:Japan(2)} are, respectively, based on the Census and the the LandScan data. 
The LandScan data is only available after 2000, and are only roughly in agreement with the precise data based on Census.
It is also noted that, while the LandScan data is available every year, we employ 5-year steps to avoid noises due to its data generation procedure that incorporates various interpolations. 
Nonetheless, \Cref{fig:gc-ld-other-countries} confirms the broad tendency of nationwide concentration and local flattening of the cities. 
To be consistent with our theory that assume a fixed total population, we normalize the total population in each country to unity. 
In the cases of France, Germany and Japan, the local flattening is apparent even without the normalization. 

\clearpage
\section{Eight regions\label{app:eight-regions}}
This section considers agglomeration processes in the $N = 8$ circular economy.
For a selected model from each model category, we follow stationary equilibria branching from $\BrVtx$ and then numerically check the local stability of those stationary equilibrium solutions under the replicator dynamic \citep{Taylor-Jorker-MB1978}. 

\begin{figure}[tb]
    \centering
    
    \centering
    \begin{subfigure}[b]{.49\hsize}
        \centering
        \includegraphics[height=5.5cm]{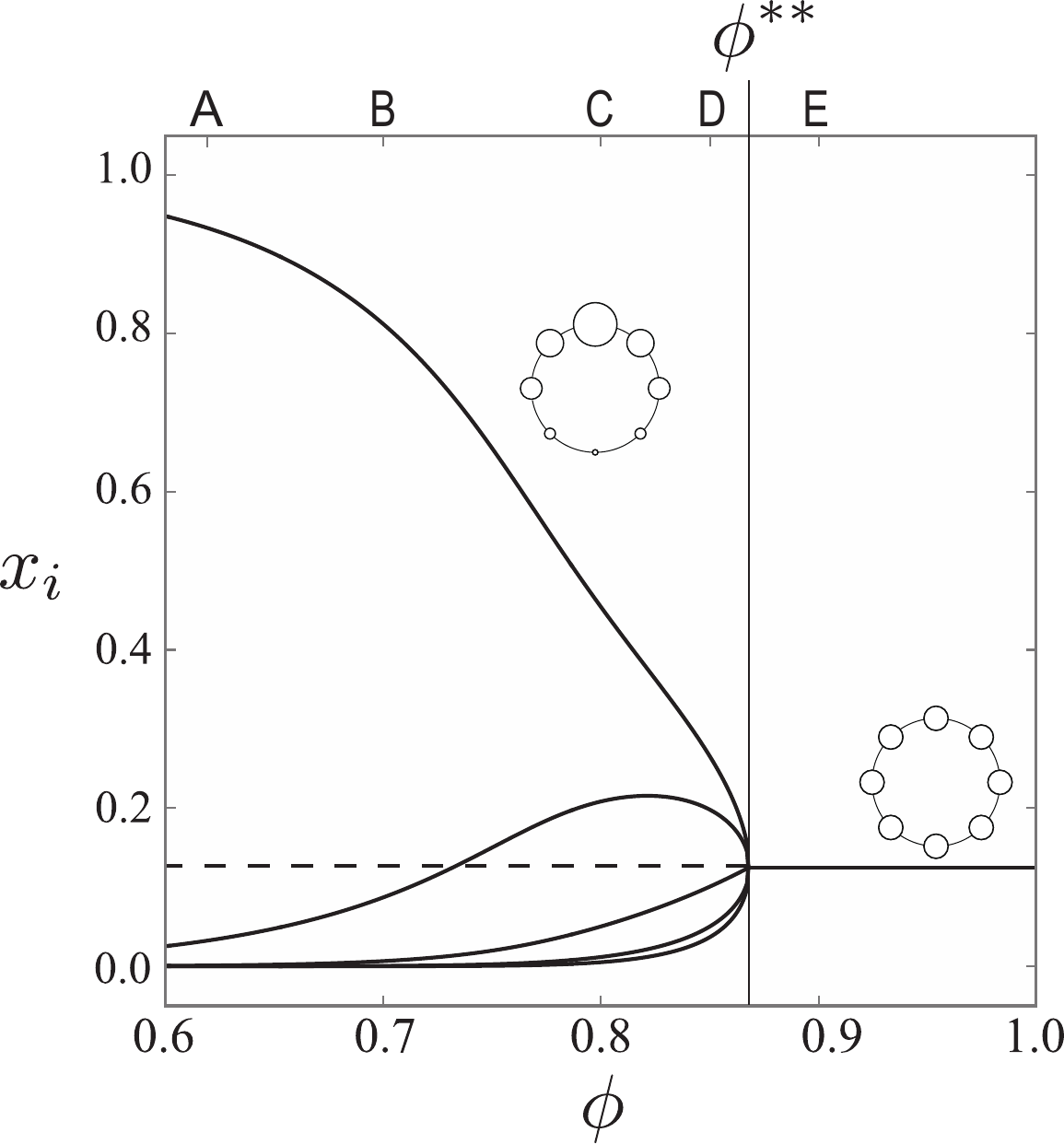}
        \caption{Bifurcation diagram \label{fig:C2-bif} }
    \end{subfigure}
    \begin{subfigure}[b]{.48\hsize}
        \centering 
        \includegraphics[height=5.5cm]{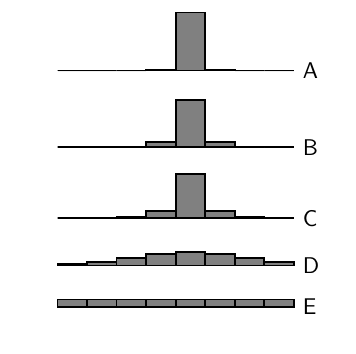}
        \caption{Spatial patterns \label{fig:C2-evol} }
    \end{subfigure}

    \caption{Type~L model \citep{Allen-Arkolakis-QJE2014}} 
    \label{fig:bif-ClassII}

    \bigskip 
    \bigskip 
    \begin{subfigure}[b]{.49\hsize}
        \centering
        \includegraphics[height=5.5cm]{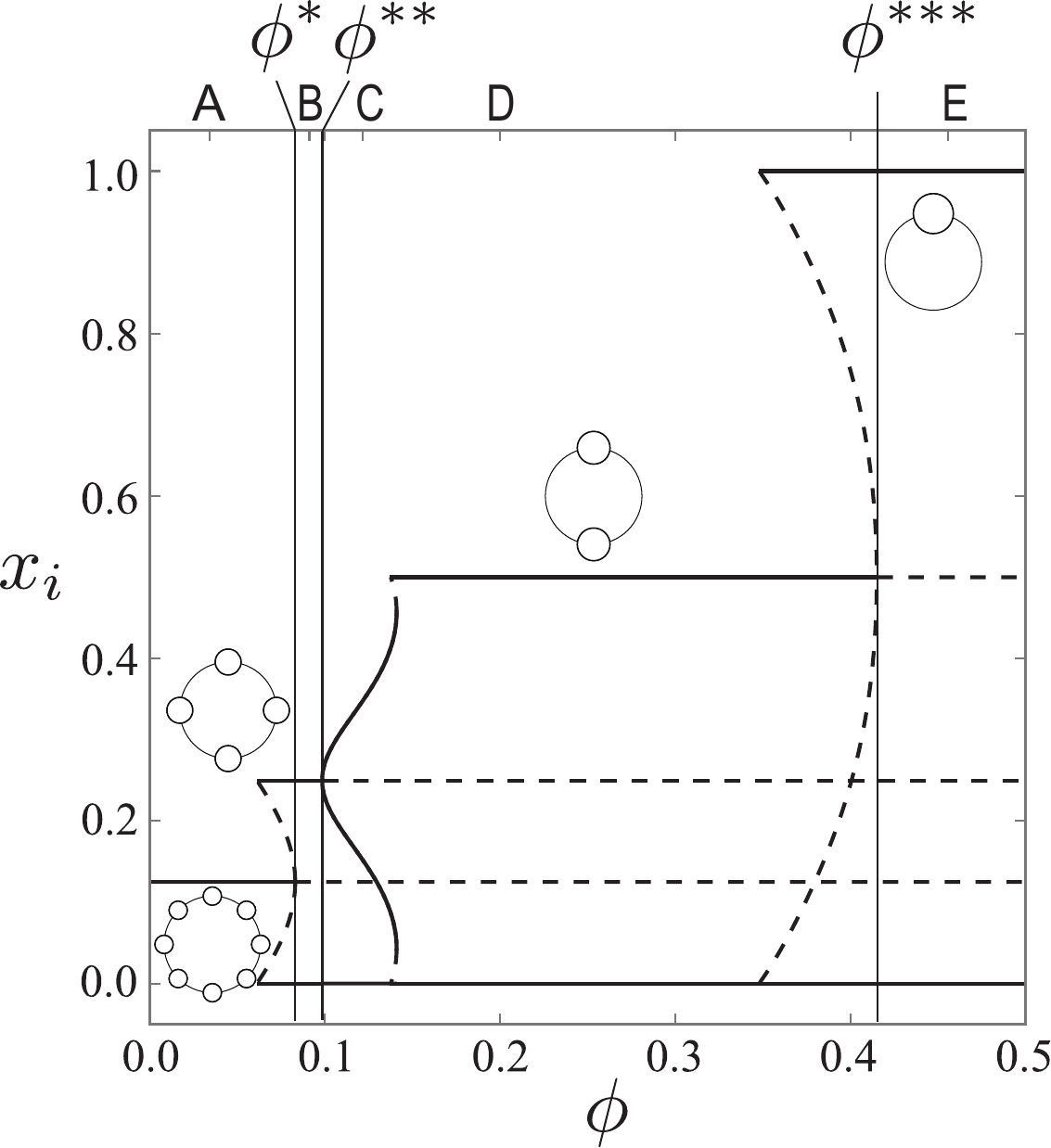}
        \caption{Bifurcation diagram \label{fig:C1-bif} }
    \end{subfigure}
    \begin{subfigure}[b]{.49\hsize}
        \centering 
        \includegraphics[height=5.5cm]{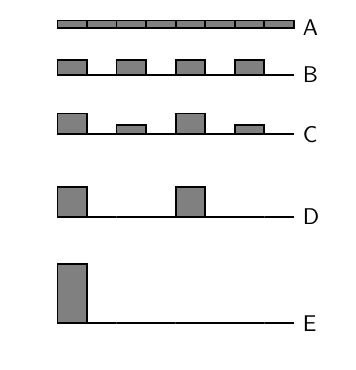}
        \caption{Spatial patterns \label{fig:C1-evol} }
    \end{subfigure}
    \caption{Type~G model \citep{Krugman-JPE1991}}
    \label{fig:bif-ClassI}

    \bigskip 
    \bigskip 

    \centering
    \begin{subfigure}[b]{.37\hsize}
    \centering
    \includegraphics[height=5.5cm]{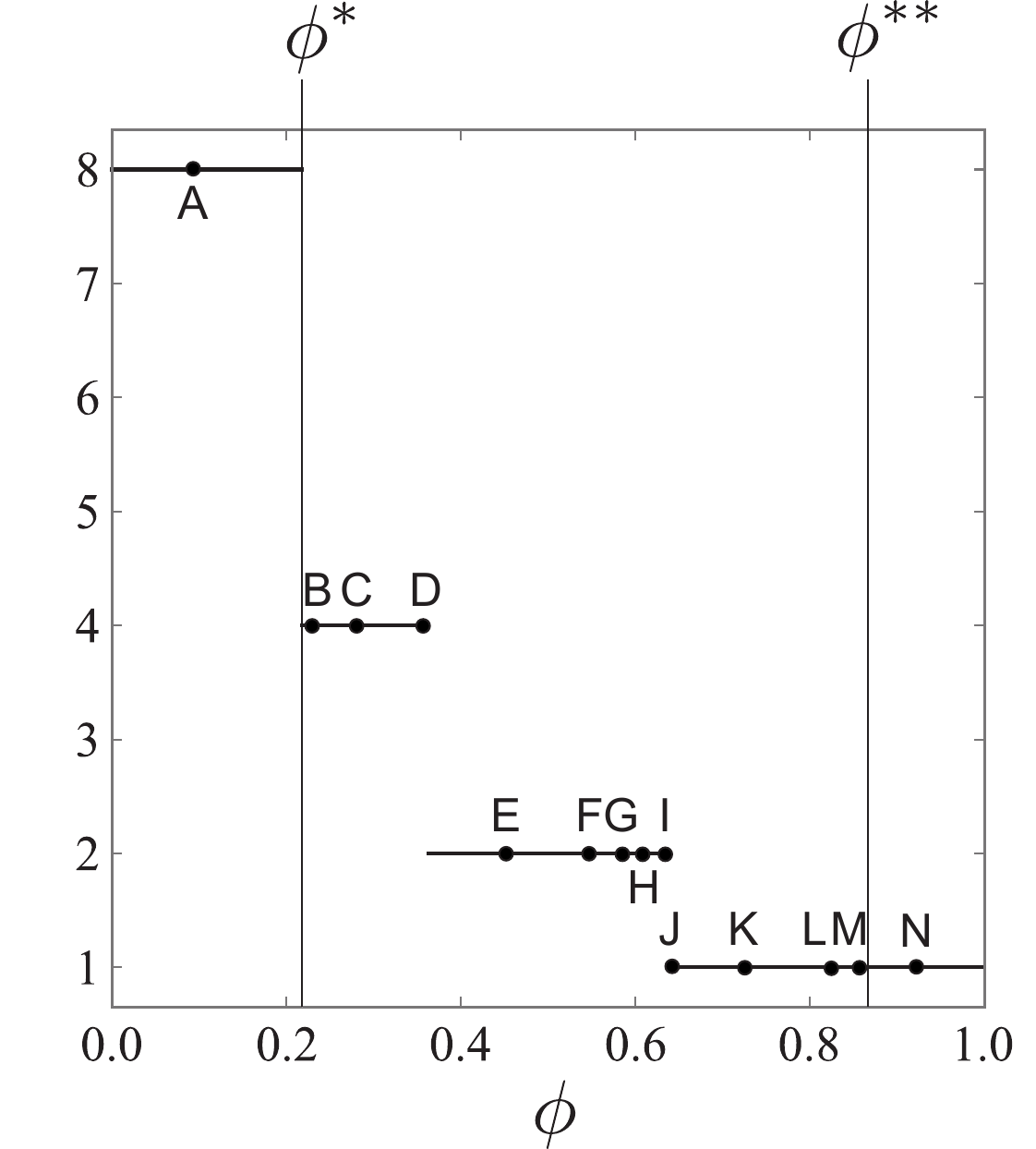}
    \caption{Number of agglomerations \label{fig:bif-both} }
    \end{subfigure}
    \hfill 
    \begin{subfigure}[b]{.6\hsize}
    \centering
    \includegraphics[height=5.5cm]{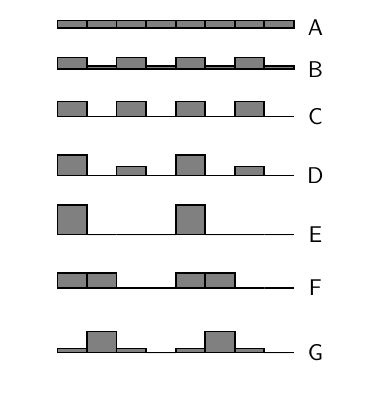}\includegraphics[height=5.5cm]{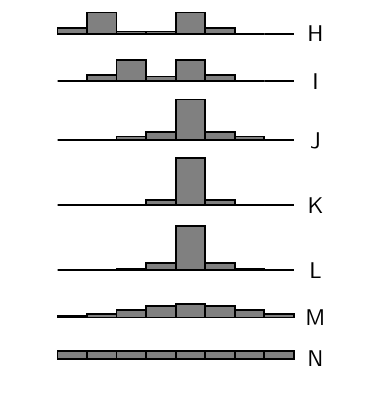}
    \caption{Spatial pattern \label{fig:bif-both-patterns} }
    \end{subfigure}
    \caption{Type~LG model \citep{Pfluger-Suedekum-JUE2008}}
    \label{fig:bif-ClassIII}
\end{figure}

\Heading{Type L model}
\Cref{fig:bif-ClassII} considers a Type~L model by \cite{Allen-Arkolakis-QJE2014} (\cref{app:allen-arkolakis}; we set $\alpha = 0.5$, $\beta = -0.3$, and $\sigma = 6.0$). 
The model incorporates a local dispersion force but no global dispersion force. 
The uniform equilibrium $\BrVtx$ is stable when transport costs are low (when $\phi$ is close to $1$). 
If we start from $\BrVtx$ and consider the process of a monotonic \emph{decrease} in $\phi$ from $\phi \approx 1$, then a unimodal pattern emerges due to the bifurcation at $\phi^{**}$ [\cref{prop:classification} (a)]. 
This is \emph{the} bifurcation in the model.  
When $\phi$ decreases further, the spatial pattern smoothly converges to a full concentration in a single region in the lower extreme ($\phi \approx 0$). 
The local dispersion force is less important than the benefits of agglomeration when interregional transportation is prohibitively costly. 
Mobile agents prefer concentrating on a smaller number of regions because of the agglomeration forces. 
As $\phi$ increases, agglomeration force due to costly transportation diminishes, and the \emph{relative} rise in the local dispersion force induces a crowding-out from the populated region to the adjacent regions. 
As a result, the spatial pattern gradually flattens and connects to $\BrVtx$ at $\phi^{**}$. 
We can interpret the region at the mode of population distribution (region $i$ such that $x_i > x_{i - 1}$ and $x_i > x_{i + 1}$ where mod $N$ for indices) as the location of an agglomeration. 
Then, this model endogenously produces at most one agglomeration. 

\Heading{Type~G model} 
\Cref{fig:bif-ClassI} 
reports stable equilibrium patterns 
in the course of increasing $\phi$ 
for the \cite{Krugman-JPE1991} model (\cref{app:krugman}; we set $\mu = 0.5$, $\sigma = 10$, and $L = 8$.). 
In \cref{fig:C1-bif}, the black solid (dashed) curves depict the stable (unstable) equilibrium values of 
$x_i$ at each $\phi$. 
\Cref{fig:C1-evol} is the schematic illustration of the stable spatial pattern on the path. 
The letters in \cref{fig:C1-evol} correspond to those in \cref{fig:C1-bif}. 
The global dispersion force in the Krugman model stems from competition between firms over consumers' demand. 
If $\phi$ is low (if transport costs are high), firms have few incentives to agglomerate, and the uniform distribution is stable. 
If we increase $\phi$, competition with firms in other regions becomes fiercer, as the markets of other regions become closer.
At some point firms are better off forming small agglomerations so that each agglomeration has its dominant market area but is relatively remote from other agglomerations of firms.
At the so-called ``break point'' $\phi^*$, a \emph{bifurcation} from $\BrVtx$ occurs and the spatial pattern is pushed towards the formation of $\frac{8}{2} = 4$ distinct agglomerations [\cref{prop:classification} (b)].  
A further increase in $\phi$ causes the second and third bifurcations at $\phi^{**}$ and $\phi^{***}$, respectively. 
These bifurcations sequentially double the spacing between agglomerations,
each time halving their number, $4\to2\to1$, in a close analogy to the first bifurcation at $\phi^*$. 
We can formally analyze the successive bifurcations if we assume a specific model \citep[see, e.g.,]{ikeda2012spatial,Akamatsu-Takayama-Ikeda-JEDC2012,Osawa-et-al-JRS2017}.  
At the higher extreme of $\phi$, agents concentrate in a single region. 
This behavior can be understood as a gradual extension of the market area of each agglomeration. 

\Heading{Type~LG model}
With both local and global dispersion forces, 
Type~LG models exhibit an interplay
between the number of agglomerations, spacing between them (as in Type~G models),  
and the spatial extent of each agglomeration (as in Type~L models). 
\Cref{fig:bif-both}
shows the evolution of 
the number of agglomerations  
in the course of increasing $\phi$ 
under the \cite{Pfluger-Suedekum-JUE2008}'s model (\cref{app:pfluger-suedekum}; we set $\mu = 0.4$, $\sigma = 2.5$, $L = 4$, $\gamma = 0.5$, and $a_i = 1$). 
The number of agglomerations in 
a spatial distribution is defined 
by that of the local maxima therein. 
\Cref{fig:bif-both} exhibits the mixed characteristics of 
 \cref{fig:bif-ClassI,fig:bif-ClassII}, as expected. 
When $\phi < \phi^*$ or $\phi > \phi^{**}$, $\BrVtx$ is stable. 
We interpret the number of agglomerations in $\BrVtx$ as either $8$ (for a low $\phi$) or $1$ (for a high $\phi$) to acknowledge that $\BrVtx$ at the low and high levels of $\phi$ are distinct. 
When $\phi$ gradually increases from $\phi \approx 0$, 
the number of agglomerations 
reduces from $8\to 4 \to 2 \to 1$ as in the Type~G models (\cref{fig:bif-ClassI}), 
whereas it is always $1$ in the latter stage as per the Type~L models (\cref{fig:bif-ClassII}). 
The initial stage is governed by a decline in the global dispersion force, while the later stage is marked by a relative rise of the local dispersion force. 

\Cref{fig:bif-both-patterns} 
illustrates the spatial patterns 
associated with \cref{fig:bif-both}. 
Uniform pattern $\BrVtx$ is initially stable (Pattern A) and    
the first bifurcation at $\phi^*$ leads to a quad-modal
agglomeration (B, C), 
whereas the second bifurcation to the formation of 
a bimodal agglomeration (D, E). 
These transitions are in line with \cref{fig:bif-ClassI} 
and are governed by the gradual decline in the global dispersion force. 
A further decline in the global dispersion force 
increases the relative importance 
of the local dispersion force. 
As a result, the bimodal agglomeration flattens out gradually (F, G).
When $\phi$ increases further, it reduces to a unimodal agglomeration (J, K).
The unimodal agglomeration flattens out as $\phi$ increases (L, M) until it converges to the complete dispersion (N) at $\phi^{**}$.

\clearpage
\section{Geographic advantages}
\label{app:geo-advantage}
The implications of \cref{prop:classification,prop:class-II,prop:class-I} qualitatively generalize to different settings such as one-dimensional line segment, two-dimensional spaces. 
The spatial distribution of agents becomes polycentric in Type G models, whereas it becomes monocentric in Type L models. 

\begin{figure}
    \centering
    \null 
    \hfill 
    \begin{subfigure}[b]{.3\hsize}
        \centering
        \begin{overpic}[width=5cm]{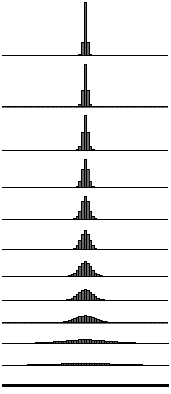}
            \put(-1,100){\vector(0,-1){100}}
            \put(-4, 78){\rotatebox{90}{\scriptsize High transport costs}}
            \put(-4,  1){\rotatebox{90}{\scriptsize Low transport costs}}
        \end{overpic}
        \caption{Type~L}
    \end{subfigure}
    \hfill
    \begin{subfigure}[b]{.3\hsize}
        \centering
        \includegraphics[width=5cm]{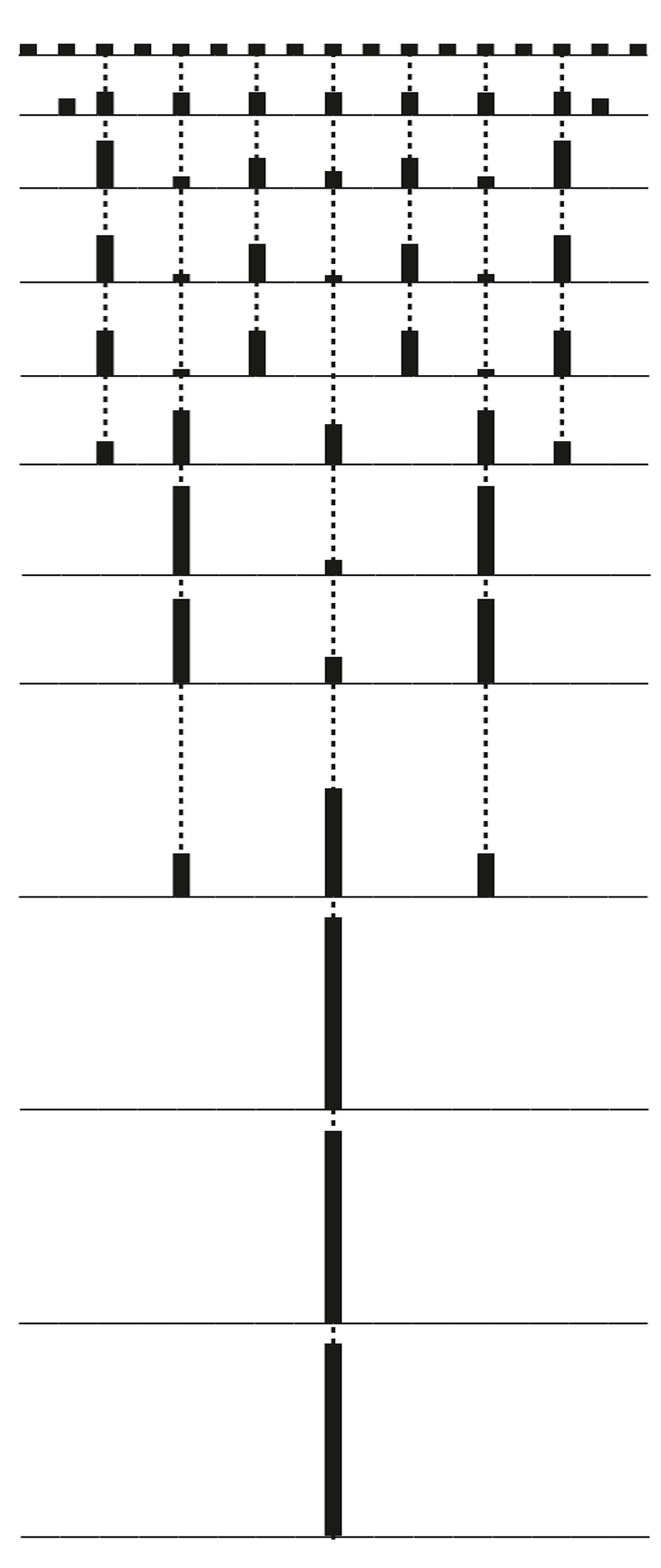}
        \caption{Type~G}
        \label{fig:Ikeda-etal}
    \end{subfigure}
    \hfill
    \begin{subfigure}[b]{.3\hsize}
        \centering
        \includegraphics[width=5cm]{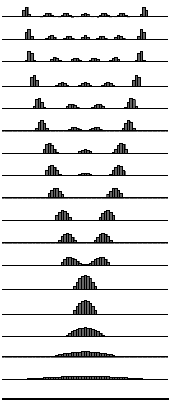}
        \caption{Type LG}
    \end{subfigure}
    \caption{Stable spatial patterns in a line segment.} 
    \label{fig:line-segment}
    \FigureNote{There are no asymmetries in regional characteristics except for geographic accessibility. The transport cost between every consecutive pair of regions is uniform. The level of transport cost monotonically decreases from top to bottom. Panels (A) and (B) consider a line segment with 65 locations. Panel (B) adapted from \cite{Ikeda-et-al-IJET2017} considers 17 locations; see the original paper for an extensive discussion.}
\end{figure}

The simplest way to introduce geographic asymmetry into our one-dimensional setting is to consider a bounded line segment, which is a standard stylized setting in urban economic theory. 
\cite{Ikeda-et-al-IJET2017} considered a Type G model \citep{Forslid-Ottaviano-JoEG2003} in a line segment. 
They showed that multiple agglomerations emerge as in the circular economy and demonstrated that the evolution of spatial structure in a line segment approximately follows the ``period doubling'' behavior \citep{Akamatsu-Takayama-Ikeda-JEDC2012,Osawa-et-al-JRS2017}. 
For Types L and LG, \cref{fig:line-segment} reports examples of endogenous agglomeration patterns in the models by \cite{Helpman-Book1998} and \cite{Pfluger-Suedekum-JUE2008}. 
For both models, qualitative properties of the spatial patterns are consistent with those discussed in \cref{app:eight-regions}. 

The two-dimensional counterpart of the symmetric circle is bounded lattices with periodic boundary conditions, for which a basic theory of spatial agglomeration is provided in \cite{Ikeda-Murota-Book2014}.  
For Type G models, they typically produce multiple disjointed agglomerations and period-doubling behavior as discussed in \cref{app:eight-regions} 
\cite[see, e.g.,][]{Ikeda-etal-IJBC2012,Ikeda-etal-JEBO2014,Ikeda-et-al-JRS2017,Ikeda-etal-JEDC2018}. 
As concrete examples, \Cref{fig:sq-main} in the main text shows endogenous equilibrium spatial patterns over a bounded square economy with $9\times 9 = 81$ regions in the course of increasing $\phi$ for the Krugman and Allen--Arkolakis models. 
The parameters are the same as \cref{fig:bif-ClassI,fig:bif-ClassII}. 
Their agglomeration processes are qualitatively consistent with \cref{prop:classification,prop:class-II,prop:class-I} and examples in \cref{app:eight-regions}, suggesting the robustness of qualitative implications of our theoretical developments. 

The implications of \cref{prop:classification} seem to extend to different assumptions on transport technology that are not formally covered by \cref{assum:racetrack-economy}. 
For example, \textit{linear} transport costs are often assumed in the literature \citep[e.g.,][]{Mossay-Picard-JET2011,Picard-Tabuchi-JET2013,Blanchet-et-al-IER2016}. 
\cite{Mossay-Picard-JET2011} considered a variant of the Beckmann model (Type L) and showed that the only possible equilibrium is a unimodal distribution in a continuous line segment. \cite{Blanchet-et-al-IER2016} considered a general Type L model over a continuous one- or two-dimensional space; they showed that the equilibrium spatial pattern for the Beckmann model is unique and given by a regular concave paraboloid, i.e., a unimodal pattern. 
\cite{Picard-Tabuchi-JET2013} also considered a Type L general equilibrium model in a two-dimensional space and showed that spatial distribution becomes unimodal. 
The numerical results of \cite{anas1996general} and \cite{anas1998urban} in line segments bear a close resemblance to, respectively, agglomeration behaviors of Type L and G models, although they assume endogenous transport costs between locations. 

\clearpage
\section{Local advantages}
\label{app:local_advantage}

This appendix introduces small region-specific asymmetries within a circular geography.
We show that improved interregional access raises population in regions with greater exogenous advantage under global dispersion forces, but lowers it under local dispersion forces. 

\subsection{Evaluating the impacts of local characteristics}
Consider a spatial model with the indirect utility function $\Vtv$.
Let $a_i > 0$ denote the innate characteristics of region $i$, and write $\Vta = (a_i)_{i\inI}$. 
For example, $a_i$ may represent the level of exogenous amenities or productivity in region $i$\footnote{All endogenous mechanisms related to agents' spatial distribution $\Vtx$, including endogenous amenities \citep[e.g.,][]{Diamond-AER2016}, are embedded in the indirect utility function $\Vtv$ in our framework.}.
All regions are perfectly symmetric if $a_i = \Bra > 0$ for all $i$. 
In this case, the uniform population distribution $\BrVtx$ is an equilibrium. 

When equilibrium is unique, counterfactual analysis proceeds by examining how the population distribution $\Vtx$ responds to changes in interregional transport costs, holding the calibrated regional characteristics fixed.
This assumption that $\Vta$ remains unchanged under the counterfactual shocks is central to quantitative predictions. 
Hence, it is important to understand how $\Vta$ influences model outcomes.

Suppose now that $\Vta$ deviates slightly from $\BrVta$, so that $\BrVtx$ is no longer an equilibrium.
If the deviation is small, the resulting equilibrium $\Vtx(\Vta)$ will remain close to $\BrVtx$ except for knife-edge cases.
Thus, we can view $\Vtx(\Vta)$ as a continuous function of $\Vta$ satisfying $\Vtx(\BrVta) = \BrVtx$.

To quantify the overall impact of variations in $\Vta$, we define the covariance between each region's relative advantage and its deviation in population share from $\BrVtx$:
\begin{align}
\rho
\Is
\sum_{i\inI}
\ 
\overbrace{(a_i - \Bra)}^{\mathclap{\text{Exogenous regional (dis)advantage}}}
\ 
\underbrace{(x_i(\Vta) - \Brx)}_{\mathclap{\text{\shortstack[c]{Population deviation from $\BrVtx$}}}}.
\label{eq:rho-definition}
\end{align}
For example, if $\rho = 0$, variations in $\Vta$ have no impact on the spatial distribution.
We assume $\rho > 0$, reflecting the natural intuition that more advantaged regions attract more population.

Importantly, the response of $\rho$ to changes in transport costs captures how the internal structure of a model shapes its counterfactual behavior.
In particular, it reveals the model's \emph{intrinsic directional bias}: whether it systematically favors the concentration of population in relatively advantaged regions ($a_i > \Bra$) under transport cost shocks.
An increase in $\rho$ implies that such regions gain population in the new equilibrium; a decrease implies the opposite. 

\subsection{Formal characterizations}
\label{sec:rho-two-regions}

We can analytically characterize the response of $\rho$ under symmetric transport cost structures. 

\Heading{Two regions}
Assume that two regions have the same local characteristics and assume that $\BrVtx = (\Brx,\Brx)$ is stable. 
Consider a marginal regional asymmetry of the form $\Vta = (\Bra + \epsilon, \Bra - \epsilon)$ with small $\epsilon$, so that $\BrVtx$ is perturbed to a new equilibrium $\Vtx = (\Brx + \xi, \Brx - \xi)$ with small $\xi$.  
We can assume that $\xi > 0$ and $\epsilon > 0$. 
Then, by definition, we have 
\begin{align}
    \rho = (a_1 - \Bra)(x_1 - \Brx) + (a_2 - \Bra)(x_2 - \Brx) = \epsilon \xi + (-\epsilon)(-\xi) = 2\epsilon \xi > 0.
\end{align}

Notably, we can obtain the analytical expression for $\rho$ for a given spatial model. 
First, we recall that the utility gain due to migration from $\BrVtx$ is negative ($\omega < 0$) as we assume $\BrVtx$ is stable. 
Next, the utility gain induced by small regional asymmetry $(\Bra + \epsilon, \Bra - \epsilon)$ can be evaluated by the following elasticity, analogous to $\omega$:
\begin{align}
    \omega^\natural
    \Is
    \frac{\Bra}{\Brv}
    \left(
	\PDF{v_1(\BrVtx,\BrVta)}{a_1}
	-
	\PDF{v_2(\BrVtx,\BrVta)}{a_1}
    \right), 
\end{align}
where $\Brv \Is v_i(\BrVtx,\BrVta)$. 
The dependence of $\Vtv$ on $\Vta$ is made explicit. 
Then, we have: 
\begin{lemma}
    \label{lem:rho-2-reg}
Assume $N = 2$. 
For the perturbed equilibrium under $\Vta = (\Bra + \epsilon, \Bra - \epsilon)$ with small $\epsilon$, we have 
$\rho  = - c \cdot \frac{\omega^\natural}{\omega}$ where $c\Is 2\epsilon^2 \frac{\Brx}{\Bra}$. 
\end{lemma}
\begin{proof}
Let $f(x,a) \Is v_1(x,a) - v_2(x,a)$ with $x \coloneq x_1$ and $a \coloneq a_1$. 
Then, $f(x,a) = 0$ since the new spatial distribution $\Vtx$ is an equilibrium. 
At $(x,a) = (\Brx,\Bra)$ and on the equilibrium curve $f(x,a) = 0$, we have $0 = f_{x}(x,a) \xi + f_{a}(x,a) \epsilon = \frac{\Brx}{\Brv} f_{x}(x,a) \frac{\Brv}{\Brx}\xi + \frac{\Bra}{\Brv}  f_{a}(x,a) \frac{\Brv}{\Brx}\epsilon = \omega \frac{\Brv}{\Brx} \xi + \omega^\natural \frac{\Brv}{\Bra} \epsilon$.  
That is, $\omega < 0$ and $\omega^\natural > 0$ should counterweight each other. 
From $
\rho = 2\epsilon\xi 
$, we obtain the desired formula. 
\end{proof}

\cref{lem:rho-2-reg} implies the following result on the response of $\rho$ to the increase in $\phi$. 
\begin{proposition}
    \label{prop:local-a}
Assume $N = 2$. 
If local regional characteristic is multiplicatively separable as $v_i(\Vtx,\Vta) = a_i v_i(\Vtx)$ with $v_i(\Vtx)$ satisfying \cref{assum:equivariance-full} (i.e., no other exogenous asymmetries),  
\begin{align}
    \sgn \rho'(\phi)
    =
    \sgn \omega'(\phi). 
    \label{eq:drho/dphi-a}
\end{align}
\end{proposition}
\begin{proof}
Under the hypothesis of the claim, $\omega^\natural = \frac{\Bra}{\Brv} (v_1(\BrVtx) - 0) = \Bra > 0$. 
Then, $\rho  = - c \cdot \frac{\omega^\natural}{\omega} =  - \frac{c \Bra}{\omega}$ and $\rho'(\phi) = \frac{c\Bra}{\omega^2} \omega'(\phi)$ where $'$ denotes the differentiation by $\phi$, implying \cref{eq:drho/dphi-a}. 
\end{proof}

\begin{figure}[tb]
    \centering
    \begin{subfigure}[b]{.45\hsize}
        \centering
        \includegraphics[height=3cm]{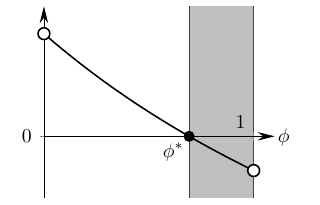}
        \caption{The Redding--Sturm model (Type~L) \label{fig:Gs-Hm-comp} }
    \end{subfigure}
    \begin{subfigure}[b]{.45\hsize}
        \centering 
        \includegraphics[height=3cm]{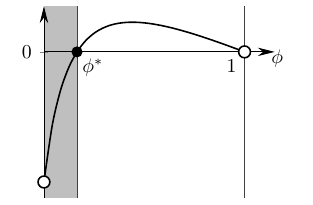}
        \caption{The Krugman model (Type~G) \label{fig:Gs-Km-comp} }
    \end{subfigure}
    \caption{The curve of the utility gain $\omega$ for the two-region case. \label{fig:Gs-Km-Hm-comp} }
\end{figure}

There is a broad connection between the spatial scale of dispersion forces and the sign of $\rho'(\phi)$. 
In Type~G models, agglomeration occurs when $\phi$ increases. 
Reflecting this, for Type~G models in the literature, we have $\omega'(\phi) > 0$ if $\BrVtx$ is stable. 
Converse is true for Type~L models, for which dispersion occurs when transport access improves and $\omega'(\phi) < 0$ if $\BrVtx$ is stable. 
\cref{fig:Gs-Km-Hm-comp} shows the curves of $\omega$ for the Krugman model (Type~G) and the Redding--Sturm model (Type~L), which demonstrates that Types~L~and~G can come to the opposite conclusions in the two-region economy. 
If a model has only a global dispersion force, its mechanisms strengthen the effects of exogenous local advantages in innate amenities, and the converse is true for a model with only a global dispersion force. 

For other forms of exogenous fundamentals, such clean characterization is not available. 
For example, heterogeneities in innate regional productivity can affect the utility level of other regions through interregional trade, and thus are not multiplicatively separable. 

\Heading{The circular geography}
Nonetheless, we have a characterization of the response of $\rho$ for general local characteristics. 
Now suppose the symmetric circle (\cref{assum:racetrack-economy}). 
Several notations and assumptions are in order. 
As we have seen in \cref{sec:N-region}, the utility elasticity matrix $\VtV = \frac{\Brx}{\Brv}[\frac{\partial v_i}{x_j}(\BrVtx)]$ at $\BrVtx$ is simply represented by the row-normalized proximity matrix $\VtD$. 
Suppose
$\VtV = \Omega(\VtD)$ 
where $\Omega$ is a scalar-valued rational function that is continuous over $[0,1]$ and the interpretation of $\Omega(\VtD)$ is the same as in \cref{app:proof-classification}. 
In an analogous manner, let $\VtA = \frac{\Bra}{\Brv}[\frac{\partial v_i}{\partial a_j}(\BrVtx)]$ be the utility elasticity matrix with respect to the local characteristic vector $\Vta$ of interest, and suppose that $\VtA$ can be represented by $\VtD$. 
Specifically, let $\VtA = \Gn(\VtD)$ where $\Gn(\cdot)$ is another rational function that is continuous over $[0,1]$. 
In fact, regional heterogeneities in $\Vta$ can be seen as deviations from $\BrVta$, so the effects of such deviations on the utility vector can be evaluated exactly the same manner as population deviations from $\BrVtx$ considered in the two-region case. 
Let 
\begin{align}
    \delta(\Theta) = - \frac{\Gn(\Theta)}{\Omega(\Theta)}, 
\end{align} 
which corresponds to $- \frac{\omega^\natural}{\omega}$ for the two-region case. 
\begin{proposition}
\label{prop:rho-sensitivity}
Suppose \cref{assum:racetrack-economy} and assume that $\BrVtx$ is stable. 
Then, 
\begin{enumerate}
    \item $\rho'(\phi) > 0$, if $\delta'(\Theta) < 0$ for all $\Theta\in(0,1)$ such that $\Omega(\Theta) < 0$.
    \item $\rho'(\phi) < 0$, if $\delta'(\Theta) > 0$ for all $\Theta\in(0,1)$ such that $\Omega(\Theta) < 0$.
\end{enumerate}
\end{proposition}
\begin{proof}
See \cref{sec:proof-rho-sensitivity}. 
\end{proof}

The multiplicatively separable case is the special case in which $\Gn(\cdot ) = \Bra$. 
For the general case with non-constant $\Gn$, our three model classes are not precisely mapped to \cref{prop:rho-sensitivity} (a) or (b). 
Still, there is a broad tendency that initial advantages are amplified in Type~G models, whereas they are diminished in Type~L models. 
For example, the regional-scale model considered in \cite{Redding-Rossi-Hansberg-ARE2017} is Type~L. 
If we consider local productivity parameters as regional characteristic vector $\Vta$, then the model satisfies $\delta'(\Theta) > 0$ for all $\Theta\in(0,1)$ when equilibrium is unique (see \cref{remark:rho-RR} in \cref{app:example_models}). 
Likewise, the Krugman model is Type~G and we show $\delta'(\Theta) < 0$ for all $\Theta\in(0,1)$ if we consider immobile demand $l_i$ as regional characteristics (see \cref{remark:rho-Km} in \cref{app:example_models}).

\subsection{Numerical examples}

This section provides numerical examples for \cref{prop:rho-sensitivity}. 
To introduce exogenous asymmetry, we multiply the utility in region $1$ by $a_1 \ge 1$, whereas we let $a_i = 1$ for all $i \neq 1$. 
We consider the Krugman model and Allen--Arkolakis model, and basic model parameters are set to be the same as \cref{fig:bif-ClassI} and \cref{fig:bif-ClassII} except that region $1$ has an exogenous advantage. 

\begin{figure}[t!]
    \centering
    \begin{subfigure}[b]{.32\hsize}\centering \includegraphics[height=6.5cm]{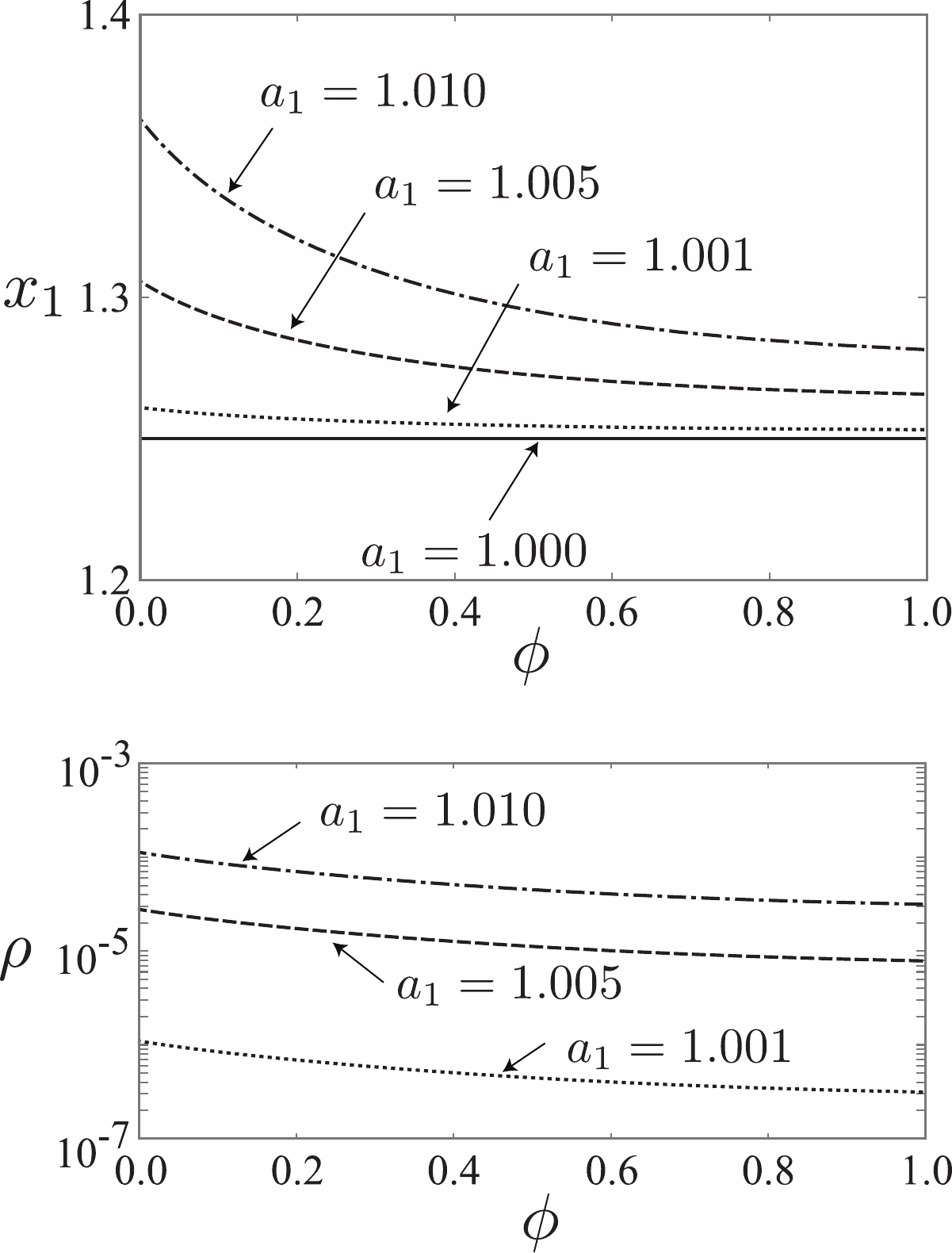}\caption{Type~L (Uniqueness)
    \label{fig:asym-AA-unique}}\end{subfigure}
    \begin{subfigure}[b]{.32\hsize}
    \centering
    \includegraphics[height=7cm]{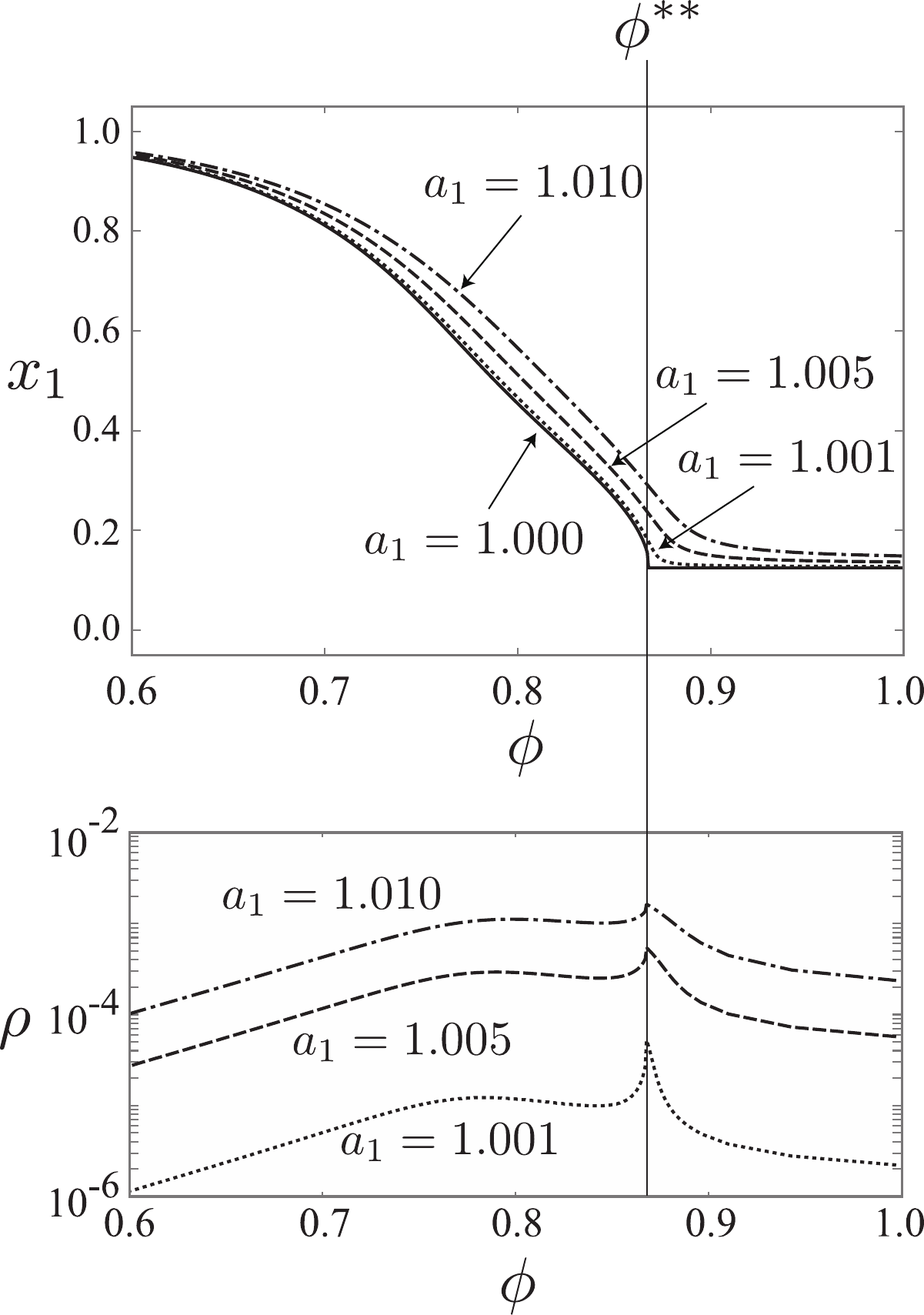}\caption{Type~L (Multiplicity)
    \label{fig:asym-AA}}\end{subfigure}
    \begin{subfigure}[b]{.32\hsize}\centering \includegraphics[height=7cm]{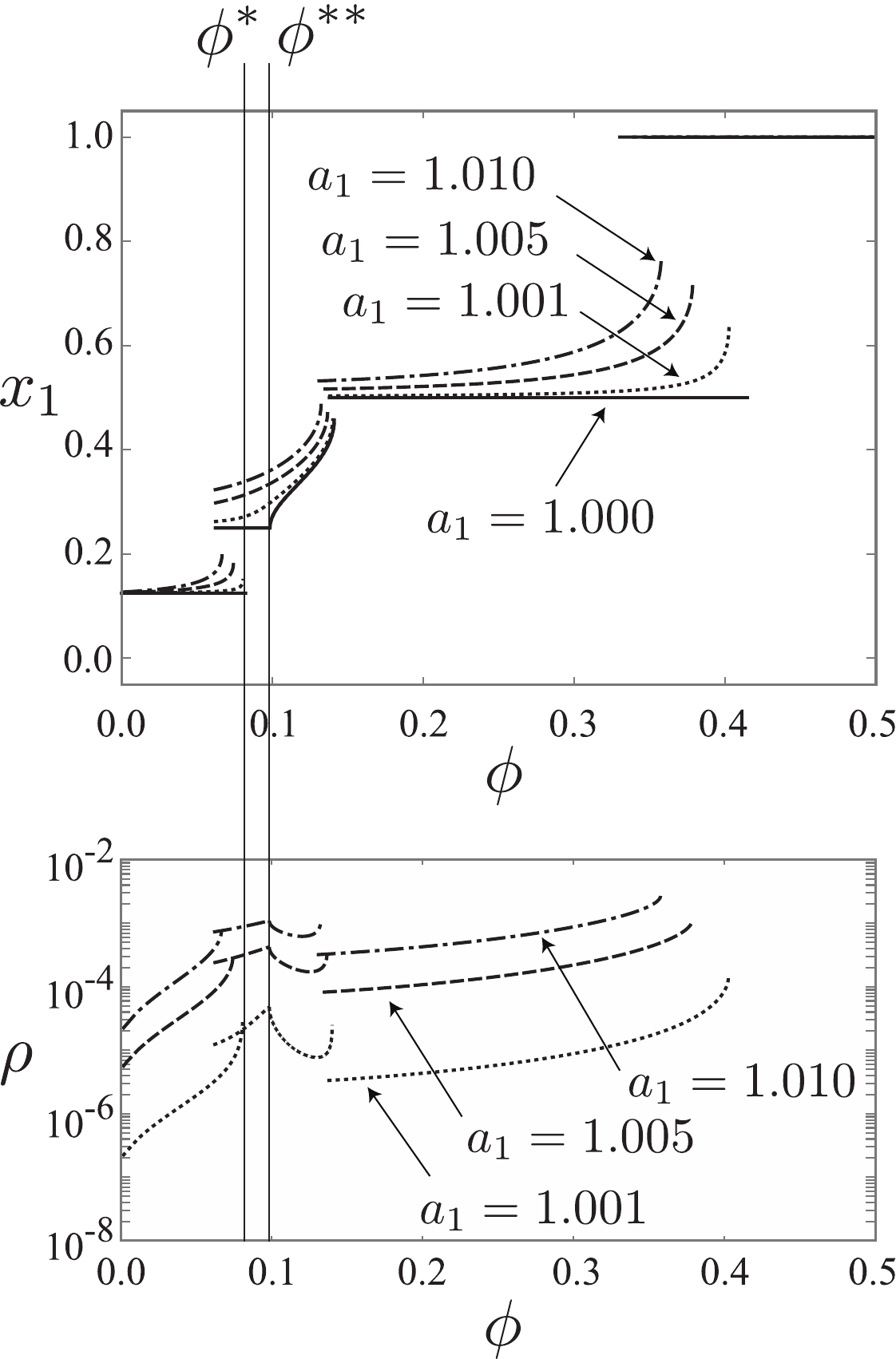}\caption{Type G \label{fig:asym-Km}}\end{subfigure}
    
    \caption{Population share of the advantageous region $1$ and covariance $\rho$}
    \label{fig:asym}
\end{figure}

\Cref{fig:asym-AA-unique} reports
equilibrium paths of $x_1$ for 
the Allen--Arkolakis model (Type~L) under the uniqueness of the equilibrium. 
The curves depict region $1$'s population share, $x_1$, at stable equilibria against $\phi$. 
Four incremental settings  
$a_1 \in\{1.000, 1.001, 1.005, 1.010\}$ are considered, 
including the baseline case with no location-fixed advantage ($a_1 = 1.000$). 
We have $\delta'(\Theta) > 0$ for all $\Theta \in (0,1)$ and see that $x_1 - \Brx > 0$ when $a_1 > 1$ and $x_1 - \Brx$ increases as $a_1$ increases, which are intuitive. 
Additionally, $x_1 - \Brx$ decreases as $\phi$ increases. 
We confirm that $\rho(\phi) > 0$ and $\rho'(\phi) < 0$ for all $\phi$. 

\Cref{fig:asym-Km,fig:asym-AA} consider the Krugman and Allen--Arkolakis models under a multiplicity of equilibria, respectively. 
Unlike the Allen--Arkolakis model, the Krugman model admits multiple equilibria for some $\phi$ for \textit{any} pair of the structural parameters $(\mu,\sigma)$. 
\cref{prop:rho-sensitivity} correctly predicts the sign of $\rho'(\phi)$ for the range of $\phi$ such that $\BrVtx$ is stable when $a_1 = 1$; 
we have $\rho'(\phi) > 0$ when $\phi \in (0,\phi^*)$ for the Krugman model, whereas $\rho'(\phi) < 0$ when $\phi\in(\phi^{**},1)$ for the Allen--Arkolakis model.

In \cref{fig:asym-Km}, the definition of $\rho$ is modified for spatial patterns with unpopulated regions. 
For the range $\phi \in (\phi^*, \phi^{**})$, $\rho$ is evaluated with respect to the four-centric pattern $(2\Brx,0,2\Brx,0,2\Brx,0,2\Brx,0)$: 
$\rho \Is \sum_{i\inI(\Vtx)} (x_i - 2\Brx)(a_i - \Bra(\Vtx))$, 
where $\ClI(\Vtx) = \{i\inI\mid x_i > 0\}$ is the set of populated regions and $\Bra(\Vtx) \Is \frac{1}{|\ClI(\Vtx)|} \sum_{i\inI(\Vtx)} a_i$. 
We define $\rho$ for two-centric pattern $(4\Brx,0,0,0,4\Brx,0,0,0)$ similarly. 
For the transitional phase after $\phi^{**}$ we let 
\begin{align}
    \rho \Is \sum_{i\inI(\Vtx)} (x_i - x_i^*)(a_i - \Bra(\Vtx)), 
    \label{eq:rho-general}
\end{align}
where $x_i^*$ corresponds to the stable solution for the symmetric case ($a_1 = 1$). 

For \cref{fig:asym-AA}, we employ \cref{eq:rho-general} as the definition of $\rho$ for the case $\phi \in (0,\phi^{**})$, i.e., we consider the deviation from the baseline equilibrium ($a_1 = 1$). 
We observe that $\rho'(\phi) < 0$ does not necessarily hold true for $\phi \in(0,\phi^{**})$. 
For instance, $\rho'(\phi) > 0$ when $\phi$ is small. 
Nonetheless, for the range of $\phi$ under which $\BrVtx$ is stable, $\rho$ decreases in $\phi$, consistent with \cref{prop:rho-sensitivity}.

\subsection{Proof of \texorpdfstring{\cref{prop:rho-sensitivity}}{Proposition 2}}
\label{sec:proof-rho-sensitivity}

We derive the analytical expression of $\rho$. 
If all regions are populated in equilibrium, we have 
\begin{align}
    \label{eq:int-equilibrium-condition}
    \Vtv(\Vtx,\Vta) - \Brv(\Vtx,\Vta) \Vt1 = \Vt0, 
\end{align}
where we make the dependence of $\Vtv$ on $\Vta$ explicit, $\Brv(\Vtx,\Vta) \Is \sum_{i\inI} v_i(\Vtx,\Vta) x_i$ is the average utility, and $\Vt1$ is $N$-dimensional all-one vector. 
The pair $(\BrVtx,\BrVta)$ is a solution to \cref{eq:int-equilibrium-condition}. 
Suppose that there is a spatial equilibrium nearby $\BrVtx$ when $\Vta$ is marginally different from $\BrVta$. 
Let $\Vtx(\Vta)$ denote the perturbed version of $\BrVtx$, which is a function in $\Vta$. 
We assume $\BrVtx$ is stable so that studying a perturbed version of it makes sense. 

The covariance $\rho$ is represented as follows:
\begin{align}
    \rho 
    \Is 
        (\Vta - \BrVta)^\top\left(\Vtx(\Vta) - \BrVtx\right)
    =
        (\VtC\Vta)^\top\VtC\Vtx(\Vta)
    =
        \Vta^\top\VtC\Vtx(\Vta)
\end{align}
where $\VtC \Is \VtI - \frac{1}{N}\Vt1\Vt1^\top$ is the centering matrix. 
Let 
$\VtX \Is [\frac{\partial x_i}{\partial a_j}(\BrVta)]$ be the Jacobian matrix of $\Vtx$ with respect to $\Vta$ at $(\BrVtx,\BrVta)$. 
Then, 
$\Vtx(\Vta) \approx \BrVtx + \VtX(\Vta - \BrVta) 
=
\BrVtx + \VtX\VtC\Vta$ and thus 
$\rho = \Vta^\top \VtC\VtX\VtC \Vta$
as $\VtC\BrVtx = \Vt0$. 
The implicit function theorem regarding \cref{eq:int-equilibrium-condition} at  $(\BrVtx,\BrVta)$ gives:
\begin{align}
    \VtX 
    &
    =
    -
    \left(
        \VtV_x 
        -
        \Vt1\BrVtx^\top
        \VtV_x
        -
        \Vt1\Vtv(\BrVtx)^\top
    \right)^{-1}
    \left(
        \VtV_a
        -
        \Vt1\BrVtx^\top
        \VtV_a
    \right)
    \\ 
    &
    =
    \left(
        \tfrac{\Brv}{\Brx}
        \tfrac{1}{N} 
        \Vt1\Vt1^\top
        -
        \left(\VtI - \tfrac{1}{N}\Vt1\Vt1^\top
        \right)
        \VtV_x 
    \right)^{-1}
    \left(\VtI - \tfrac{1}{N}\Vt1\Vt1^\top  \right)
        \VtV_a
    \\ 
    &
    =
    \tfrac{\Brx}{\Brv}
    \left(
        (\VtI - \VtC)
        -
        \VtC
        \tfrac{\Brx}{\Brv}
        \VtV_x 
    \right)^{-1}
    \VtC
    \tfrac{\Brv}{\Bra}
    \tfrac{\Bra}{\Brv}
    \VtV_a
    \\ 
    &
    =
    \tfrac{\Brx}{\Bra}
    \left( 
        (\VtI - \VtC)
        - 
        \VtC\VtV 
    \right)^{-1}
    \VtC\VtA 
    \label{eq:X-formula}
\end{align}
where $\Brv$ is the utility level, 
$\VtV_x \Is [\frac{\partial v_i}{\partial x_j}(\BrVtx,\BrVta)]$, 
$\VtV_a \Is [\frac{\partial v_i}{\partial a_j}(\BrVtx,\BrVta)]$, 
$\VtV \Is \frac{\Brx}{\Brv} \VtV_x$, and
$\VtA \Is \frac{\Bra}{\Brv} \VtV_a$. 

For tractability, we focus on a specific form of $\VtA$ which covers many relevant cases. 
\renewcommand{\theassumption}{A}
\begin{assumption}
\label{assum:advantageous-characteristics}
Suppose \cref{assum:racetrack-economy}.
Let $\VtA \Is \frac{\Bra}{\Brv} [\frac{\partial v_i}{\partial a_j}]$
be the elasticity matrix of the utility with respect to the local characteristic $\Vta$, evaluated at $(\BrVtx,\BrVta)$.
There is a rational function $\Gn$ that is continuous over $[0,1]$, positive whenever $\BrVtx$ is stable, and satisfies $\VtA = \Gn(\VtD)$.
\end{assumption}
\begin{example}
\label{example:local-a}
Suppose $v_i(\Vtx, \Vta) = a_i v_i(\Vtx)$, where $a_i> 0$ is the exogenous level of local amenities and $\Vtv (\Vtx) = (v_i(\Vtx))_{i\inI}$ is the symmetric component of the utility function (i.e., $\Vtv(\Vtx)$ satisfies \cref{assum:equivariance-full}).  
Then, $\VtA = \Bra \VtI$ and $\Gn(\cdot) = \Bra > 0$.
\end{example}

Under Assumptions \ref{assum:racetrack-economy} and \ref{assum:advantageous-characteristics}, $\VtX$ is real, symmetric, and circulant. 
Thus, the set of eigenvectors of $\VtC\VtX\VtC$ can be chosen as in \cref{lem:D_eigenpairs}~(a) because it is a circulant matrix of the same size as $\VtD$. 
Let $\{\lambda_k\}_{k = 0}^{M}$ be the distinct eigenvalues of $\VtC\VtX\VtC$. 
As $\VtC\VtX\VtC$ is symmetric, it admits the eigenvalue decomposition 
\begin{align}
    \VtC\VtX\VtC
    = 
    \lambda_0 \Vt1\Vt1^\top 
        + 
        \sum_{k = 1}^{M - 1} 
        \lambda_k 
        \left(
            \Vtz_k^+{\Vtz_k^{+}}^\top 
            +
            \Vtz_k^-{\Vtz_k^{-}}^\top 
        \right) 
        +
        \lambda_{M}
        \Vtz_{M}
        \Vtz_{M}^\top.  
\end{align}
This fact yields the following representation of $\rho$: 
\begin{align}
    \label{eq:X-eigendecomposition}
    \rho 
    = \Vta^\top\VtC\VtX\VtC\Vta 
    = \sum_{k \neq 0} \Tla_k^2 \lambda_k, 
\end{align} 
where $\TlVta \Is (\Tla_k)$ is the representation of $\Vta$ in the new coordinate system $\{\Vtz_k\}$. 
We can drop $k = 0$ because $\lambda_0 = 0$, reflecting that $\Vtz_0 = \Vt1$ represents a uniform increase in $\Vta$ and thus does not affect spatial equilibria. 
All the matrices in \cref{eq:X-formula} are circulant and hence shares the same set of eigenvectors. 
Thus, $\lambda_k$ is obtained from \cref{eq:X-formula} as follows: 
\begin{align}
    & 
    \lambda_k
    = 
    \frac{\Brx}{\Bra} 
    \frac{\kappa_k \omega^\natural_k
    }{
        ((1 - \kappa_k)
        - 
        \kappa_k \omega_k)
    }
    = 
    - 
    \frac{\Brx}{\Bra} 
    \frac{\omega^\natural_k
    }{\omega_k
    }
    &
    \forall k \inK
    ,     
\end{align}
where $\kappa_k$, $\omega_k$, and $\omega^\natural_k$ are the $k$th eigenvalues of $\VtC$, $\VtV$, and $\VtA$, respectively, with $\kappa_0 = 0$ and $\kappa_k = 1$ for all $k \ne 0$. 
As $\omega_k  = \Omega(\Theta_k)$ and $\omega^\natural_k = \Gn(\Theta_k)$ with $\{\Theta_k\}_{k\inK}$ are the eigenvalues of $\VtD$ because we assume $\VtG = \Omega(\VtD)$ and $\VtA = \Gn(\VtD)$, 
\begin{align}
    &\lambda_k = - \frac{\Brx}{\Bra}\dfrac{\Gn(\Theta_k)}{\Omega(\Theta_k)}
    = \frac{\Brx}{\Bra}\delta(\Theta_k)
        &\forall k\inK 
    \label{eq:lambda}
\end{align}
and $\lambda_0 = 0$ where $\delta(\Theta) \Is - \frac{\Gn(\Theta)}{\Omega(\Theta)}$. 

From \cref{eq:X-eigendecomposition}, $\rho > 0$ for all $\Vta$ if all $\{\lambda_k\}$ are positive except for $\lambda_0 = 0$. 
The denominator of \cref{eq:lambda}, $\Omega(\Theta_k)$, must be negative for all $k$ because $\BrVtx$ is stable by assumption. 
Thus, we see that $\rho > 0$ if $\Gn(\Theta) > 0$ for all $\Theta$ since $\Theta_k \in (0,1)$ for all $k\inK$.

\cref{prop:rho-sensitivity} follows by noting 
\begin{align*}
    \rho'(\phi) 
    = 
        \sum_{k \neq 0} \Tla_k^2 \ODF{\lambda_k}{\phi}
    = 
        \frac{\Brx}{\Bra}
        \sum_{k \neq 0} \Tla_k^2 
        \delta'(\Theta_k)
        \ODF{\Theta_k}{\phi}
    = 
        -
        \frac{\Brx}{\Bra}
        \sum_{k \neq 0} \Tla_k^2  
        \delta'(\Theta_k)
        \left| \ODF{\Theta_k}{\phi} \right|. 
\end{align*}
From \cref{lem:D_eigenpairs}, $\{\Theta_k\}_{k\inK}$ are strictly decreasing in $\phi$. 
Thus, for $\rho'(\phi) > 0$ ($\rho'(\phi) < 0$), it is sufficient that $\delta'(\Theta) < 0$ ($\delta'(\Theta) > 0$) for all $\Theta$ such that $\Omega(\Theta) < 0$.  
\clearpage
\section{Derivations\label{app:example_models}}
This section collects omitted derivations. 
The expression $\VtF_x$ denotes the Jacobian matrix of a vector-valued function $\Vtf(\Vtx)$ with respect to $\Vtx$, that is, $\VtF_x  = [\frac{\partial f_i}{\partial x_j}]$. 
For example, 
$\VtV_x \Is [\frac{\partial v_i}{\partial x_j}]$, 
$\TlVtV_w \Is [\frac{\partial \Tlv_i}{\partial w_j}]$, 
and 
$\VtW_x \Is [\frac{\partial w_i}{\partial x_j}]$. 
Throughout, $\Brv$, $\Brw$, $\Bre$ and so on represent $v_i$, $w_i$, $e_i$ evaluated at $\BrVtx$. 
$\VtD$ denotes the row-normalized proximity matrix.

\subsection{General derivations}

\subsubsection{The benefit matrix}  
The indirect utility function $\Vtv$ of a spatial model often reduce to the following implicit form:
\begin{align}
	& \Vtv(\Vtx) = \TlVtv(\Vtx,\Vtw), 
		& \label{eq:gen-payoff} \\
	& \Vts(\Vtx,\Vtw) = \Vt0. 
		& \label{eq:gen-shortrun-equilibrium} 
\end{align}
The condition \cref{eq:gen-shortrun-equilibrium} represents, e.g., the general equilibrium conditions for a given $\Vtx$ that defines endogenous variable $\Vtw$ (e.g., wages) other than $\Vtx$ as an implicit function of $\Vtx$. 
We assume that \cref{eq:gen-shortrun-equilibrium} admits a unique solution of $\Vtw$ at each $\Vtx$ for $\Vtv(\Vtx)$ to be well-defined. 
Suppose $\Vts$ and $\Vtv$ are continuously differentiable. Then, we have
\begin{align}
	& 
	\VtV_x(\Vtx) 
	= \TlVtV_x(\Vtx) + \TlVtV_w(\Vtx)\VtW_x(\Vtx), 
	\\
	& 
	 \VtW_x(\Vtx) = - \VtS_w(\Vtx)^{-1}\VtS_x(\Vtx), 
	\label{eq:general-Wx}
\end{align} 
where $\VtW_x(\Vtx)$ is obtained by applying the implicit function theorem to \cref{eq:gen-shortrun-equilibrium}. 

Under \cref{assum:racetrack-economy}, all relevant matrices commute at $\Vtx = \BrVtx$ because they are real, symmetric, and circulant at $\BrVtx$. 
Thus, $
\VtV_x 
	= 
	\VtS_w^{-1}
	( 
		\VtS_w\TlVtV_x - \TlVtV_w\VtS_x
	) 
$ at $\BrVtx$. 

\begin{example}
\label{example:gen-s}
\cref{eq:gen-shortrun-equilibrium} is often given by 
\begin{align}
	s_i(\Vtx,\Vtw) 
	= 
	w_i x_i 
	- 
	\sum_{j\inI}
		m_{ij}
		e_j
	= 0, 
	\label{eq:gen-weq-1}
\end{align}
where regional expenditure is $e_i = e(w_i,x_i)$ with some nonnegative function $e$ and 
$\VtM = [m_{ij}]$ is the expenditure share matrix. 
For example, in the Krugman and Helpman models, 
\begin{align}
m_{ij} 
	= 
	\dfrac
	{x_i w_i ^{1 - \sigma} \phi_{ij}}
	{\sum_{k\inI} x_k w_k ^{1 - \sigma} \phi_{kj}}.
	\label{eq:mij-Km}
\end{align} 
In matrix form, we can write $\Vty - \VtM \Vte = \Vt0$ where $\Vty = (w_i x_i) _{i\inI}$.   
Then, in general, 
\begin{subequations}
\label{eq:gen-Wx-1}
\begin{align}
	&
		\VtS_x(\Vtx) = 
		\diag{\Vtw} 
		- 
		\left(
			\diag{\VtM\Vte}
			-
			\VtM
			\diag{\Vte}
			\VtM^\top
		\right) 
		\diag{\Vtx}^{-1}
		-
		\VtM
		\VtE_x, 
	\\
	&
		\VtS_w(\Vtx) = 
		\diag{\Vtx} 
		+ 
		(\sigma - 1)
		\left(
			\diag{\VtM\Vte}
			-
			\VtM
			\diag{\Vte}
			\VtM^\top
		\right) 
		\diag{\Vtw}^{-1}
		-
		\VtM
		\VtE_w. 
\end{align}
\end{subequations}

Suppose \cref{assum:racetrack-economy}. 
Suppose $\Vtx = \BrVtx$ and 
let $\Brw$ be the uniform level of $\{w_i\}$ at $\BrVtx$. 
Then, we have $\VtM = \VtD$ at $\Vtx = \BrVtx$. 
Suppose the scalars $\epsilon_x$ and $\epsilon_w$ are chosen to satisfy $\VtE_x = \epsilon_x\Brw\VtI$ and $\VtE_w = \epsilon_w \Brx \VtI$ at $\BrVtx$. 
Let $\Bre = e(\Brw,\Brx)$ and 
$\zeta \Is \frac{\Bre}{\Brw\Brx}$. 
We see that 
\begin{subequations}
\label{eq:gen-Wx-2}
\begin{align}
	&
		\VtS_x 
		= 
		- 
		\Brw
		\left(
			(\zeta - 1)
			\VtI
			+
			\epsilon_x
			\VtD
			- 
			\zeta
			\VtD^2
        \right), 
        \label{eq:gen-Sx}
	\\
	&
		\VtS_w 
		= 
		\Brx
		\left( 
			\left( 
				1 + \zeta(\sigma - 1)
			\right) 
			\VtI
			-
			\epsilon_w\VtD
			-
			\zeta 
			(\sigma - 1)
			\VtD^2
		\right). 
        \label{eq:gen-Sw}
\end{align}
\end{subequations}
If $e(w_i,x_i) = w_i x_i$, 
then $\epsilon_x = \epsilon_w = 1$ and $\zeta = 1$, thereby 
$\VtW_x = \frac{\Brw}{\Brx} (\sigma \VtI + (\sigma - 1)\VtD )^{-1}\VtD$.
\end{example}

\subsubsection{The payoff elasticity matrix with respect to local characteristics} 

In \cref{eq:X-formula}, $\VtX = [\frac{\partial x_i(\BrVta)}{\partial a_i}] = \VtX_a$ acts as 
$\HtVtX \Is - \VtV_x^{-1} \VtV_a$ for $\Vtz$ such that $\Vtz^\top \Vt1 = \Vt0$. Thus, $\VtV_a$ is of interest. 

For purely local characteristics (\cref{example:local-a}),  
since $v_i(\Vtx,\Vta) = a_i v_i(\Vtx)$, it follows that $\VtV_a = \diag{\Vtv(\Vtx)}$. 
At $\BrVtx$, we have $\VtV_a = \Brv\VtI$. 
Thus, $\HtVtX = - \Brv \VtV_x^{-1}$. 

For regional characteristics that affect trade flows, the payoff function and the market equilibrium condition are, respectively, modified to $\Vtv(\Vtx,\Vta) = \TlVtv(\Vtx,\Vtw,\Vta)$ and $\Vts(\Vtx,\Vtw,\Vta) = \Vt0$. By applying the implicit function theorem, we see
$\VtV_a = \TlVtV_a + \TlVtV_w \VtW_a = \TlVtV_a - \TlVtV_w \VtS_w^{-1} \VtS_a$. 
As all matrices commute at $\BrVtx$ under \cref{assum:racetrack-economy}, it is equivalent to consider 
\begin{align}
	\HtVtX 
	& 
	= 
	- 
	\left(\TlVtV_x - \TlVtV_w \VtS_w^{-1} \VtS_x\right)^{-1} 
	\left(\TlVtV_a - \TlVtV_w \VtS_w^{-1} \VtS_a \right) 
	\\
	& 
	= 
	\left(\VtS_w \TlVtV_x - \TlVtV_w \VtS_x\right)^{-1} 
	\left(\TlVtV_w \VtS_a - \VtS_w \TlVtV_a\right). 
\end{align}
\begin{example}
\label{example:rho-RR_app}
For the regional model by \cite{Redding-Rossi-Hansberg-ARE2017}, we have
\begin{align}
	s_i(\Vtx,\Vtw,\Vta) 
	= 
	w_i x_i - 
	\sum_{j\inI}
		\dfrac
			{x_i a_i w_i ^{1 - \sigma} \phi_{ij}}
			{\sum_{k\inI} x_k a_k w_k ^{1 - \sigma} \phi_{kj}}
		e_j 
	= 0. 
\end{align}
Thus, 
$\VtS_a = - \left(\diag{\VtM\Vte} - \VtM \diag{\Vte} \VtM^\top \right) \diag{\Vta}^{-1} = - \frac{\Bre}{\Bra} \left(\VtI - \VtD^2 \right)$. See \cref{app:helpman}. 
\end{example}
\begin{example}
\label{example:rho-Km_app}
For the Krugman model, we have 
\begin{align}
	s_i(\Vtx,\Vtw,\Vta) 
	= 
	w_i x_i - 
	\sum_{j\inI}
		\dfrac
			{x_i w_i ^{1 - \sigma} \phi_{ij}}
			{\sum_{k\inI} x_k w_k ^{1 - \sigma} \phi_{kj}}
		e(w_j, x_j, a_j)
	= 0
\end{align}
where $e$ maps the tuple $(w_j,x_j,a_j)$ to the regional expenditure. 
Then, we have $\VtS_a = - \VtM\VtE_a$, or $\VtS_a = - \epsilon_a\VtD$ at $\BrVtx$ where $\epsilon_a = \frac{\partial e(\Brx,\Brw,\Bra)}{\partial a_i}$. 
See \cref{app:krugman}. 
\end{example}

\subsection{Model-specific derivations}

We provide omitted derivations of the \emph{gain functions} $\Omega$, as defined in \cref{app:proof-classification}, for the examples in the main text. 
For derivations for other models mentioned in \cref{sec:ge-models,sec:model-class}, see \cite{Akamatsu-et-al-DP2017}, an earlier draft of the current paper. 

\subsubsection{\texorpdfstring{\cite{Krugman-JPE1991}}{Krugman (1991)} model\label{app:krugman}}

There are two types of workers, mobile and immobile, and their total masses are $1$ and $L$, respectively.  
$\Vtx \Is (x_i)_{i\inI}$ is the distribution of mobile workers.
Each worker supplies one unit of labor inelastically. 

There are two industrial sectors: 
agriculture (abbreviated as A) and manufacturing (abbreviated as M).
The A-sector is perfectly competitive and 
a unit input of immobile labor is required to produce one unit of goods. 
The M-sector follows Dixit--Stiglitz monopolistic competition. 
M-sector goods are horizontally differentiated 
and produced under increasing returns to scale using mobile labor as the input. 
The goods of both sectors are transported.  
Transportation of A-sector goods is frictionless, 
while that of M-sector goods is of an iceberg form. 
For each unit of M-sector goods transported from region $i$ to $j$, 
only the proportion $1/\tau_{ij}$ arrives, 
where $\tau_{ij} > 1$ for $i\neq j$ and $\tau_{ii} = 1$.

All workers have an identical preference for both M- and A-sector goods.
The utility of a worker in region $i$ is given by a two-tier form. 
The upper tier is Cobb--Douglas 
over the consumption of A-sector goods $C_i^\RmA$ and 
that of M-sector constant-elasticity-of-substitution (CES) 
aggregate $C_i^\RmM$ with $\sigma > 1$ 
\begin{align}
    C_i^\RmM \Is 
    \Big(
        \sum_{j\inI} \int_0^{n_j} q_{ji}(\xi)^{\frac{\sigma - 1}{\sigma}}\Rmd\xi
    \Big)^{\frac{\sigma}{\sigma - 1}}, 
\end{align}
that is, $u_i = (C_i^\RmM)^{\mu}(C_i^\RmA)^{1 - \mu}$ 
where $\mu \in (0,1)$ is the constant expenditure of the latter. 
With free trade in the A-sector, the wage of the immobile worker is equalized, 
and we normalize it to unity by taking A-sector goods as the num\'eraire. 
Consequently, region $i$'s expenditure on the M-sector goods is given by $e_i = \mu(w_i x_i + l_i)$ where $l_i$ denotes the mass of immobile workers in region $i$.

In the M-sector, to produce $q$ units, 
a firm requires $\alpha + \beta q$ units of mobile labor. 
Profit maximization of firms yields the price of differentiated goods produced in region $i$ and exported to $j$ as $p_{ij} = \frac{\sigma \beta}{\sigma-1} w_i \tau_{ij}$, which in turn determines gravity trade flow from $j$ to $i$. 
That is, 
when $X_{ij}$ denotes the price of M-sector goods produced in region $i$ and sold in region $j$, 
$X_{ij} = m_{ij} e_j$ 
where the share $m_{ij}\in(0,1)$ is defined by \cref{eq:mij-Km} with 
$\phi_{ij} \Is \tau_{ij}^{1 - \sigma}$. 
The proximity matrix is thus $[\phi_{ij}] = [\tau_{ij}^{1 - \sigma}]$.

Given $\Vtx$, we determine the market wage $\Vtw \Is (w_i)_{i\inI}$ 
by 
the M-sector product market-clearing, 
zero-profit, and
mobile labor market-clearing conditions. 
These conditions are summarized by 
the trade balance $w_i x_i = \sum_{j\inI} X_{ij}$, or 
\cref{eq:gen-weq-1} with $e(x_i, w_i) = \mu(w_i x_i + l_i)$. 
By adding up \cref{eq:gen-weq-1} for the Krugman model, we see $\sum_{i\inI} w_i x_i = \frac{\mu}{1 - \mu} L$, 
which constrains the total income of mobile workers at any configuration $\Vtx$. 
The existence and uniqueness of the
solution for \cref{eq:gen-weq-1} follow from standard 
arguments \citep[e.g.,][]{Facchinei-Pang-Book2007}. 
Given the solution $\Vtw(\Vtx)$ of \cref{eq:gen-weq-1}, 
we have the indirect utility of mobile workers, which is given by 
	$v_i = 
			\Delta_i^{\frac{\mu}{\sigma - 1}}
			w_i$, 
where $\Delta_i \Is \sum_{k\inI} x_k w_k^{1-\sigma} d_{ki}$. 

Let $l_i = l \Is \frac{L}{N}$ for all $i\inI$.  
We have 
\begin{align}
	\nabla \log\Vtv(\BrVtx)
	& = 
	\dfrac{\mu}{\sigma - 1}
	\VtM^\top\diag{\Vtx}^{-1}
	- 
	\mu \VtM^\top \diag{\Vtw}^{-1}\VtW_x
	+ 
	\diag{\Vtw}^{-1}\VtW_x
	\label{eq:Km.Vx-0}
	\\
	& 
    = 
	\frac{1}{\Brx}
	\dfrac{\mu}{\sigma - 1}
	\VtD
	+ 
	\frac{1}{\Brw}\left(\VtI - \mu\VtD\right) \VtW_x
	, 
	\label{eq:Km.Vx}
\end{align}
where \cref{eq:general-Wx,eq:gen-Wx-2} give $\VtW_x$. 
By plugging $\delta = \frac{\mu(\Brw\Brx + l)}{\Brw\Brx} = 1$ and $\epsilon_x = \epsilon_w = \mu$ to \cref{eq:gen-Wx-2},  
\begin{align}
	\VtW_x 
	= 
	\dfrac{\Brw}{\Brx}
	\left( 
		\sigma\VtI
		- \mu\VtD 
		- (\sigma - 1)\VtD^2
	\right)^{-1}
		\left(
			\mu \VtD - \VtD^2
	    \right).\label{eq:Km.Wx}
\end{align}
Then, 
\cref{eq:Km.Vx,eq:Km.Wx} 
imply 
\begin{align}
	\VtV 
	=
	\Brx 
	\nabla \log \Vtv(\BrVtx)
	= 
		\frac{\mu}{\sigma - 1} 
		\VtD 
		+
		\left(\VtI - \mu\VtD\right)
		\left( 
			\sigma\VtI
			- \mu\VtD 
			- (\sigma - 1)\VtD^2
		\right)^{-1}
		\left(
			\mu \VtD - \VtD^2
	    \right), 
\end{align}
or equivalently, $\VtV = \Omega(\VtD)$ where 
\begin{align}
	\Omega(\Theta) 
	& 
	= 
		\underbrace{
		\frac{\mu}{\sigma - 1}
		\Theta}_{\text{(a)}}
		+
		\underbrace{(1 - \mu \Theta)}_{\text{(b)}} 
		\underbrace{\left(\frac{1}{\sigma}\right) \frac{\mu\Theta - \Theta^2}{1 - \frac{\mu}{\sigma} \Theta - \frac{\sigma - 1}{\sigma}\Theta^2}}_{\text{(c)}}. 
	\label{eq:Km_G_explanation}
\end{align}
From \cref{eq:Km_G_explanation} we have 
$
\VtV 
	=
	\Gf(\VtD)^{-1} 
	\Gs(\VtD)
$,
where we define 
\begin{align}
&
	\Gs(\Theta) \Is
		\mu\left(\dfrac{1}{\sigma - 1} + \dfrac{1}{\sigma}\right)\Theta
		- 
		\left(\dfrac{\mu^2}{\sigma - 1} + \dfrac{1}{\sigma}\right)
		\Theta^2, 
		\label{eq:Km.Gs.app}
\\ 
&
	\Gf(\Theta) \Is 
		1
		- \frac{\mu}{\sigma}\Theta 
		- \frac{\sigma - 1}{\sigma} \Theta^2. 
\end{align} 

\begin{remark}
\label{remark:Km-explanation}
Using the Krugman model as an example, we discuss how economic forces in a model are embedded in $\Omega$. 
We recall that 
positive (negative) terms in $\Omega$ represent agglomeration (dispersion) forces. 
In \cref{eq:Km_G_explanation}, (a) corresponds to the elasticity of price index with respect to agents' spatial distribution $\Vtx$, (b) to the elasticity of payoff with respect to nominal wage $\Vtw$, (c) to the elasticity of wage with respect to agents' spatial distribution. 
Here, (a) and the second term in (b) corresponds to the so-called cost-of-living effect through price index; 
(a) is positive, i.e., it is an agglomeration force, as the price index in a region becomes lower when more agents (firms) locate geographically close regions; the second term in (b) (i.e., $-\mu\Theta$) is negative because higher wage in a region implies higher goods prices in its nearby regions. 
Also, (b) as a whole is positive, meaning that the payoff of a region is increasing in wages even with the negative effect through price index. 
The last component (c) includes both positive and negative terms; in its numerator, the first term ($\mu \Theta$) is demand linkage where firms' profits rise when they are close to regions with high total income, and the second term ($-\Theta^2$) is the market-crowding effect due to competition between firms.
The sign of (c) is $\Theta$-dependent;  
for example, it is negative when $\Theta$ is high ($\phi$ is low) and positive otherwise. 
The denominator of (c) represents the general equilibrium effects through the so-called short-run equilibrium condition under given $\Vtx$, i.e., \cref{eq:gen-weq-1}. 
As $\Gs$ is obtained by combining these components and collecting terms according to the order of $\Theta$, these economic forces affect both the first- and second-order coefficients of $\Gs$. 
Concretely, in \cref{eq:Km.Gs.app}, 
$\frac{\mu}{\sigma - 1} \Theta$ comes from (a), 
$\frac{\mu}{\sigma} \Theta$ comes from (b) $\times$ (c), 
$-\frac{1}{\sigma}\Theta^2$ comes from (b) $\times$ (c),  
and $- \frac{\mu^2}{\sigma - 1} \Theta^2 = - \left(\frac{\mu^2}{\sigma} + \frac{\mu^2}{\sigma(\sigma - 1)}\right)\Theta^2$ comes from all three components while its leading term $-\frac{\mu^2}{\sigma}\Theta^2$ comes from (b) $\times$ (c). 
Thus, by considering $\Gs$ for a model, one can examine the net effect of \emph{all} economic forces in the model at once, and the net effect is decomposed according to its spatial scale (i.e., the order of $\Theta$). 
\end{remark}
\begin{remark}
\label{remark:rho-Km}
To obtain $\Gn$ for $\Vtl = (l_i)_{i\inI}$, 
we evaluate $\VtV_l = -\TlVtV_w\VtS_w^{-1}\VtS_l$ as $\VtA = \frac{l}{\Brv} \VtV_l$.  
From \cref{example:rho-Km_app}, 
$\VtS_l = - \mu \VtD$. 
Also, 
$
	\TlVtV_w 
	= 
	\Brv
	\frac{\partial}{\partial\Vtw} \log\Vtv(\BrVtx) 
	=
	\frac{\Brv}{\Brw}
	(\VtI - \mu \VtD)$ and $\TlVtV_l = \Vt0$. 
Thus, 
\begin{align}
\Gn(\Theta) = 
	c
	\frac{\Theta(1 - \mu\Theta)}{\Gf(\Theta)} > 0 
\end{align}
where $c = \frac{l}{\Brw}
	\frac{\mu}{\sigma} = \frac{1 - \mu}{\sigma} \Brx > 0$. 
It then follows that
\begin{align}
	\delta(\Theta) 
	= -\frac{\Brx}{\Bra}\frac{\Gn(\Theta)}{\Omega(\Theta)} 
	= 
	-
	\frac{c\Brx}{\Bra} 
	\frac{\Theta(1 - \mu\Theta)}{\Gs(\Theta)}. 
\end{align}
Straightforward algebra verifies that $\delta'(\Theta) < 0$ if $\Gs(\Theta) > 0$. 
\end{remark}

\subsubsection{\texorpdfstring{\cite{Helpman-Book1998}}{Helpman (1998)} and \texorpdfstring{\cite{Redding-Sturm-AER2008}}{Redding and Sturm (2008)} model\label{app:helpman}}

\cite{Helpman-Book1998} removed the A-sector in the Krugman model and assumed that all workers are mobile, and introduced the housing sector (abbreviated as H). 
Each region $i$ is endowed with a fixed stock $a_i$ of housing. 
Workers' preference is Cobb--Douglas of M-sector CES aggregate $C_i^\RmM$ and H-sector goods $C_i^\RmH$, 
$u_i = (C_i^\RmM)^\mu(C_i^\RmH)^\gamma$, 
where $\mu\in(0,1)$ is the expenditure share of the former and $\gamma = 1 - \mu \in(0,1)$ is that for the latter. 

There are two variants for assumptions on how housing stocks are owned: 
\emph{public landownership} (PL) 
and \emph{local landownership} (LL).
\cite{Helpman-Book1998} supposes PL in which housing stocks are equally owned by all workers; the income of a worker in region $i$ is 
the sum of the wage and an equal dividend $r > 0$ of the total rental revenue in the economy. 
However, \cite{ottaviano2002agglomeration}, 
\cite{Murata-Thisse-JUE2005}, 
and \cite{Redding-Sturm-AER2008} 
assumed that housing stocks are locally owned (i.e., LL). 
The income of a worker in region $i$ is 
the sum of the wage and an equal dividend of rental revenue \emph{in each region}. 
In fact, the model by \cite{Redding-Sturm-AER2008} is the LL version of the Helpman model.

Regarding the market equilibrium conditions, 
the only difference from the Krugman model 
is regional expenditure $e_i$ on M-sector goods in each region: 
\begin{align} 
	\text{[PL]}\quad 
	& e_i = \mu\left(w_i + r\right) x_i, \\
	\text{[LL]}\quad 
	& e_i = w_i x_i.  
\end{align}
Also, the market wage is given as the solution for \cref{eq:gen-weq-1}. 
For the PL case, we set $r = 1$ for normalization. 
For the LL case, $\Vtw(\Vtx)$ is uniquely given up to normalization. 
The indirect utility function is, with $\Delta_i \Is \sum_{j\inI} x_j w_j^{1 - \sigma} \phi_{ji}$ and $r > 0$,  
\begin{align}
	\text{[PL]}\quad
	& 
	v_i
        = 
            \left(\frac{x_i}{a_i}\right)^{-\gamma}
			(w_i + r)^\mu
			\Delta_i^{\frac{\mu}{\sigma - 1}},
			\\
	\text{[LL]}\quad
	& 
    v_i
        = 
            \left(\frac{x_i}{a_i}\right)^{-\gamma}
            w_i^\mu
            \Delta_i^{\frac{\mu}{\sigma - 1}}
		.
\end{align}

Let $a_i = 1$ for all $i\inI$. 
We compute that 
\begin{align}
	& 
	\VtV
        = 
        \Brx
        \left(
			\frac{\mu}{\sigma - 1}
			\VtM^\top\diag{\Vtx}^{-1}
			+
			\HtVtV_w 
			\VtW_x
			- 
			\gamma
            \diag{\Vtx}^{-1}
        \right),\\
\text{where}\quad
	& 
\text{[PL]}\quad
	\HtVtV_w \Is 
		\mu
		\left(
			\diag{\Vtw + r\Vt1}^{-1} 
			- 
			\VtM^\top\diag{\Vtw}^{-1}
		\right), 
	\\
	& 
\text{[LL]}\quad
	\HtVtV_w \Is \mu
		\left(
			\VtI - \VtM^\top
		\right)
		\diag{\Vtw}^{-1}, 
		\label{eq:Hm.v.LL}
\end{align}
and $\VtM$ is defined by \cref{eq:mij-Km}. 
For the PL case, we obtain $\VtV = \Omega(\VtD)$ with 
\begin{align}
	\Omega(\Theta) 
	= 
	- \gamma + \frac{\mu}{\sigma - 1} \Theta 
	+ \frac{\mu \left(\mu - \Theta \right) \Theta(1 - \Theta)}{\sigma - \mu \Theta - (\sigma - 1)\Theta^2}
	\label{eq:Hm.PL.G.0}
\end{align}
where we compute $\VtW_x$ from \cref{eq:gen-Wx-1} with $\zeta = \frac{\mu(\Brw + 1)}{\Brw}$, $\epsilon_x = \frac{\Brw + 1}{\Brw}$, $\epsilon_w = \mu$; we note that $\frac{\Brw}{\Brw + 1} = \mu$ under our normalization. 
Thus, for the PL case, we can choose $\Gs$ and $\Gf$ that satisfy 
$\VtV = \Gf(\VtD)^{-1} \Gs(\VtD)$ as follows: 
\begin{align}
	&
	\Gs(\Theta)
	\Is
			-
			\gamma 
			+ 
			\left( 
				\frac{\mu}{\sigma - 1}
				+ 
				\dfrac{\mu(\mu + \gamma)}{\sigma}
			\right) 
			\Theta 
			-
			\left(
				\frac{\mu^2}{\sigma - 1} 
				+ 
				\dfrac{\mu + \gamma}{\sigma}
				-
				\gamma 
			\right) 
			\Theta^2, \label{eq:Hm.PL.Gs.0}\\
	& 
	\Gf(\Theta) 
		\Is 
			1 
			- 
			\dfrac{\mu}{\sigma} \Theta 
			- 
			\dfrac{\sigma - 1}{\sigma}\Theta^2 
\end{align} 
where we recall $\mu + \gamma = 1$.

For the LL case, $\VtW_x$ is given in \cref{example:gen-s} and we obtain 
\begin{align}
	\Omega(\Theta) = - \gamma + \frac{\mu}{\sigma - 1} \Theta + \frac{\mu(1 - \Theta)\Theta }{\sigma + (\sigma - 1)\Theta}. 
\end{align}
We can choose $\Gs$ and $\Gf$ as follows:
\begin{align}
	&
	\Gs(\Theta)
		\Is
			-
			\gamma 
			+ 
			\left(
			\frac{\mu}{\sigma - 1} 
			+ 
			\frac{\mu + \gamma}{\sigma}
			- \gamma
			\right) 
			\Theta, 
			\\
	& 
	\Gf(\Theta) 
		\Is 
			1 
			+ 
			\dfrac{\sigma - 1}{\sigma}
			\Theta. 
	\label{eq:Hm.Gs.LL}
\end{align}

\begin{remark}
\cref{def:model-class} classifies canonical models based on the spatial scale of the ``effective'' dispersion forces. 
The Helpman model with public landownership \cref{eq:Hm.PL.Gs.0} has global dispersion forces because $c_2 < 0$ under \cref{assum:E}. 
However, unlike the Krugman model, the global dispersion forces in the Helpman model are not ``effective'' in the sense that, under any admissible values of $\mu$ and $\sigma$, this force does not stabilize the uniform distribution for any level of transport costs. 
If we drop the local dispersion force $c_0 < 0$ from $\Gs$, we have $c_1 \Theta + c_2 \Theta^2 > 0$ for all $\Theta$ and $\BrVtx$ is always unstable. 
Thus, the only dispersion force in the Helpman model that can stabilize $\BrVtx$ is its local dispersion force.
\end{remark}

\begin{remark}
\label{remark:Hm.uniqueness}
Equilibrium is unique when 
$\gamma\sigma = (1 - \mu) \sigma > 1$ \citep{Redding-Sturm-AER2008}.
For both PL and LL, this condition implies that $\Gs(\Theta) < 0$ for all $\Theta\in(0,1)$. 
\end{remark}

\begin{remark}
\label{remark:rho-RR}
The regional model in \S3 of \cite{Redding-Rossi-Hansberg-ARE2017} is a variant of the Helpman model with LL, 
in which variable input of mobile labor depends on region $i$ 
(i.e., productivity differs across regions). 
The cost function of firms in region $i$ is 
$C_i(q) = w_i(\alpha + \beta_i q)$. 
The market equilibrium condition for this case is, 
with $a_i \Is \beta_i^{1 - \sigma} > 0$, 
given by 
\begin{align}
	s_i(\Vtx,\Vtw)
	= 
	w_i x_i 
	- 
	\sum_{j\inI} 
	 		\dfrac
	 		{
	 			x_i 
	 			a_i
				w_i^{1 - \sigma}
	 			\phi_{ij} 
	 		}
	 		{
				\sum_{k\inI} x_k a_k w_k^{1-\sigma} \phi_{kj}
			}
		 	w_j x_j
	= 0. 
	\label{eq:rr.wage_eq}
\end{align}
The payoff function is given by \cref{eq:Hm.v.LL}  
with $\Delta_i = \sum_{k\inI} x_k a_k w_k^{1-\sigma} \phi_{ki}$. 

From \cref{example:rho-RR_app}, 
$\VtS_a = - \frac{\Brw\Brx}{\Bra}\left(\VtI - \VtD \right)\left(\VtI + \VtD \right)$ as $\Bre = \Brw \Brx$. 
Also, we have $\TlVtV_w = \frac{\Brv}{\Brw} \mu (\VtI - \VtD)$, 
$\TlVtV_a = \frac{\Brv}{\Bra} \frac{\mu}{\sigma - 1}\VtD$, and $\VtS_w = \sigma \Brx\Gf(\VtD)$. 
As $\VtV_a = \TlVtV_a - \TlVtV_w\VtS_w^{-1}\VtS_a$ and $\VtA = \frac{\Bra}{\Brv}\VtV_a = \Gn(\VtD)$, we compute 
\begin{align}
	\Gn(\Theta) 
	= 
	c
	\frac{(\sigma - 1) + \sigma\Theta}{\Gf(\Theta)}
	> 0
\end{align}
where $c \Is \frac{\Brv}{\Bra}\frac{\mu}{\sigma} > 0$.
This in turn implies 
\begin{align}
	\delta(\Theta) 
	= - \frac{\Brx}{\Bra}\frac{\Gn(\Theta)}{\Omega(\Theta)}
	= 
	-
	\frac{c\Brx}{\Bra} 
	\frac{(\sigma - 1) + \sigma\Theta}{\Gs(\Theta)}
\end{align}
where $\Gs(\Theta)$ is that for the LL case \cref{eq:Hm.Gs.LL}. 
If $(1 - \mu)\sigma > 1$, $\delta'(\Theta) > 0$ for all $\Theta$. 
\end{remark}

\subsubsection{\texorpdfstring{\cite{Pfluger-Suedekum-JUE2008}}{Pfluger and Sudekum (2008)} model}
\label{app:pfluger-suedekum}

The Pfl\"uger--S\"udekum model builds on \cite{Pfluger-RSUE2003} and introduces the housing sector 
(denoted by H). 
A quasi-linear upper-tier utility is assumed: $u_i = C_i^{\RmA} + \mu \log C_i^{\RmM} + \gamma \log C_i^{\RmH}$. 
The production cost for a firm in $i\inI$ is $\alpha w_i + \beta q$. 
Then, $\Vtw$ is given as follows:
\begin{align}
	w_i 
		=
			\dfrac{\mu}{\sigma}
			\sum_{j\inI} 
				\dfrac
					{\phi_{ij}}
					{\sum_{k\inI} \phi_{kj}x_k}
				(x_j + l_j). 
	\label{eq:Pf.w}
\end{align}
The indirect utility of a mobile worker in region $i$ 
is 
\begin{align}
	v_i(\Vtx) = \frac{\mu}{\sigma - 1} \ln[\Delta_i] - \gamma \ln \frac{x_i + l_i}{a_i} + w_i,
\end{align}
where $\Delta_i = \sum_{j\inI} \phi_{ji} x_j$, 
and $l_i$ and $a_i$ denote the mass of immobile workers and 
amount of housing stock in region $i$, respectively. The nominal wage in region $i$ is given by \cref{eq:Pf.w}.  
Let $l_i = l$ and $a_i = a$ for all $i$. 
Then, we see that $\VtV = \frac{1}{\Brv} \Gs(\VtD)$ with 
\begin{align}
	\Gs(\Theta) 
	 &
	 = 
		-
		\dfrac{\gamma}{1 + L}
		+
		\mu
		\left(
			\frac{1}{\sigma - 1}
			+
			\frac{1}{\sigma}
		\right) 
		\Theta
	 	- 
	 	\frac{\mu}{\sigma}(1 + L)
	 	\Theta^2
	 . 
\end{align}

\subsubsection{\texorpdfstring{\cite{Allen-Arkolakis-QJE2014}}{Allen and Arkolakis (2014)} model\label{app:allen-arkolakis}}

The Allen--Arkolakis model is a perfectly competitive \cite{Armington-IMF1969}-based framework with positive and negative local externalities. 
We abstract away all exogenous differences in regional fundamentals. 
The productivity of region $i$ is proportional to $x_i^\alpha$ with $\alpha > 0$, representing positive externalities. 
The market equilibrium condition is
\begin{align}
	s_i(\Vtx,\Vtw) = 
	w_i x_i
		- \sum_{j\inI} 
			\dfrac
				{w_i^{1-\sigma} x_i^{\alpha(\sigma - 1)} \phi_{ij}}
				{\sum_{k\inI} w_k^{1-\sigma} x_k^{\alpha(\sigma - 1)} \phi_{kj}}
			w_j x_j
		= 0. 
		\label{eq:wage_eqn_base}
\end{align}
With market wage $\Vtw$, we have  
$v_i(\Vtx) =
		x_i^{-\beta} w_i \Delta_i^{\frac{1}{\sigma - 1}}$ 
with $\Delta_i \Is \sum_{k\inI} w_k^{1-\sigma} x_k^{\alpha(\sigma - 1)} \phi_{ki}$. 
With $\beta < 0$, $x_i^{\beta}$ represents negative externalities from congestion. 
Here, we follow the original study in terms of the sign of $\beta$, while the main text uses $\beta$ as the magnitude of this externality to streamline exposition. 
We have   
$\VtV = \Gf(\VtD)^{-1}\Gs(\VtD)$ with 
\begin{align}
&
	\Gs(\Theta) = 
			\left(\alpha + \beta - \frac{1 + \alpha}{\sigma}\right)
			+ \left(\alpha + \beta + \frac{1 - \beta}{\sigma}\right) \Theta, \\
&
	\Gf(\Theta) 
	= 
	\left( \sigma + (\sigma - 1)\Theta \right) 
			\left( 1 - \Theta \right). 
\end{align}

The case with no externalities ($\alpha = 0$ and $\beta = 0$) reduces to the \cite{Armington-IMF1969} framework, and $\omega = - \frac{1}{\sigma (\sigma + (\sigma - 1)\Theta)} < 0$. 
The intrinsic working of general equilibrium effects induced by love for variety in the Armington model is in creating a dispersion force, and it is in a sense ``global'' because it depends on $\Theta$. 
However, in the context of \emph{net} agglomeration incentive at the symmetry with nonzero spillovers, this force is mainly related to the denominator $\Gf$, and $\Gs$ summarizes the net trade-off between agglomeration and dispersion forces that govern the stability of $\BrVtx$. 

\begin{figure}[tp]
    \centering
    \includegraphics[width=.8\hsize]{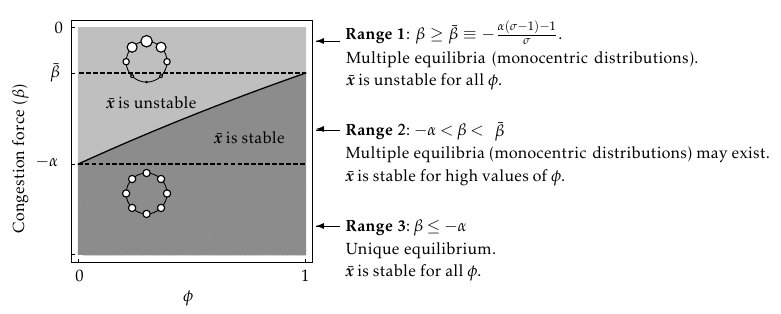}
    \caption{Uniqueness and stability of equilibria in the Allen--Arkolakis model.}
    \label{fig:AA-classification}
\end{figure}

\Cref{fig:AA-classification} classifies spatial patterns and their stability under \cref{assum:racetrack-economy}, which can be seen as a refinement of Figure I in \cite{Allen-Arkolakis-QJE2014} under \cref{assum:racetrack-economy}.    
A sufficient condition for the equilibrium uniqueness is $\alpha + \beta \leq 0$ 
(Range~3), which means there is \emph{net} local dispersion force and $\Gs(\Theta) < 0$ for all $\Theta$. 
In Range~3, $\BrVtx$ is the only equilibrium.

\subsubsection{\texorpdfstring{\cite{Krugman-Venables-QJE1995}}{Krugman and Venables (1995)} model\label{app:krugman-venables}} 

\cite{Krugman-Venables-QJE1995} considers intermediate inputs. 
When we interpret the regional share of manufactured good production as the state variable, the model is Type~G. 
With input share of manufactured good in production being $\alpha$, \cite{Fujita-Krugman-Venables-Book1999}, Appendix 14.1, derives a concise formula for $\Omega$. 
Namely, in the case that ``agricultural'' good is produced by a constant returns technology, 
\begin{align}
    \Omega(\VtD) = 
    \Gf(\VtD)^{-1} 
    \left(
        \alpha \left(
        1 + \frac{\sigma - 1}{\sigma}\right)
        \VtD 
        - 
        (1 + \alpha)^2 
        \VtD^2
    \right) 
\end{align}
with an appropriately defined $\Gf$, showing that the model is Type~G.  
Also, if the ``agricultural'' sector exhibits decreasing return, an extra local dispersion force (a term $c_0\VtI$ such that $c_0 < 0$) emerges, so that the model becomes Type~LG \citep[][Section 14.4]{Fujita-Krugman-Venables-Book1999}.

\subsubsection{\texorpdfstring{\cite{Kucheryavyy-etal-JIE2024}}{{Kucheryavyy et al. (2024)}} model}

\cite{Kucheryavyy-etal-JIE2024} considers a unified framework that nests \cite{Allen-Arkolakis-QJE2014} and \cite{Krugman-JPE1991} as special cases. 
For this model in the symmetric two-region economy, the sign of the utility gain $\omega$ coincides with the following function of $\Theta$: 
\begin{align}
    \Gs(\Theta) 
    \Is 
    - (1 - \alpha) 
    +
    (1 - (1 - \beta) \zeta +\mu)
    \Theta 
    -  
    (\alpha + (1 - (1 - \beta) \zeta )\mu)
    \Theta ^2 , 
    \label{eq:Kucheryavyy}
\end{align}
where the notations follow the original paper except for the transport cost index $\Theta$. 

For deriving the above $\Gs$, we note that, when evaluated at $x = 1$, $V'(x)$ in Appendix D of \cite{Kucheryavyy-etal-JIE2024} is essentially a negative constant multiple of the (net) utility gain, as the stability condition for the uniform distribution in their notation is $V'(1) > 0$ (see their Appendix B.4). 

As seen, the model is Type~LG in its most general form.  
Importantly, the case $(\alpha, \zeta, \mu) = (1, 1, \beta)$ corresponds to a Krugman-type model, as we observe \cref{eq:Kucheryavyy} reduces to a function of the form $c_1\Theta - c_2 \Theta^2$. 
Also, $\alpha < 1$ and $\zeta = 1$ corresponds to the Allen--Arkolakis framework, for which we confirm 
$- (1 - \alpha) < 0$ represents a local dispersion force. 
It is noted that $\Gs(1) = (1 - \mu)(1 - \beta)$. 
If $\mu < 1$, then $\Gs(1) > 0$, so that $\Gs$ must have one and one zero for $\Theta \in (0,1)$, so that $\BrVtx$ is stable only for low transport costs.  

\clearpage 
{\small\singlespacing
\ifx\undefined\bysame
\newcommand{\bysame}{\leavevmode\hbox to\leftmargin{\hrulefill\,\,}}
\fi

}
\end{bibunit}

\end{document}